\documentclass[12pt]{article}
\pdfoutput=1

\usepackage{geometry,amsmath,amsfonts}
\usepackage{slashed}
\usepackage{epsfig}
\usepackage{latexsym}
\usepackage{graphicx}
\usepackage{amssymb}
\usepackage[caption=false]{subfig}

\usepackage{multirow}
\usepackage{color}
\usepackage{rotating}
\usepackage{ifthen}
\usepackage{epsfig}
\usepackage{lineno}
\usepackage{cite}

\newcommand{\beq}{\begin{equation}}
\newcommand{\eeq}{\end{equation}}
\newcommand{\bea}{\begin{eqnarray}}
\newcommand{\eea}{\end{eqnarray}}

\newcommand{\gsim}{\lower.7ex\hbox{$\;\stackrel{\textstyle>}{\sim}\;$}}
\newcommand{\lsim}{\lower.7ex\hbox{$\;\stackrel{\textstyle<}{\sim}\;$}}

\begin{document}

\vspace*{10mm}

\begin{center}

{\large\bf  The 2HD+a model:}

\vspace*{3mm}

\mbox{\large\bf  collider, dark matter
and gravitational wave signals} 

\vspace*{12mm}

\mbox{ {\sc Giorgio Arcadi$^1$},  {\sc Nico Benincasa$^2$}, {\sc Abdelhak~Djouadi$^{2,3}$} and {\sc Kristjan Kannike$^2$} }

\vspace*{12mm}

{\small
$^1$ Dipartimento di Scienze Matematiche e Informatiche, Scienze Fisiche e Scienze della Terra, \\ Universita degli Studi di Messina, Via Ferdinando Stagno d'Alcontres 31, I-98166 Messina, Italy.\\ \vspace{0.2cm}

$^2$ NICPB, R{\"a}vala pst. 10, 10143 Tallinn, Estonia.\\ \vspace{0.2cm}

$^3$ CAFPE and Departamento de Fisica Te\'orica y del Cosmos,\\ Universidad de Granada, E--18071 Granada, Spain.\\ \vspace{0.2cm}
}

\end{center}

\vspace*{12mm}

\begin{abstract} 

We perform a comprehensive study of a model in which the Higgs sector is extended to contain two Higgs doublet fields, with the four types of possibilities to couple to standard fermions,  as  well as an additional light pseudoscalar Higgs boson which mixes with the one of the two doublets. This 2HD+a model includes also a stable isosinglet massive fermion that has the correct thermal relic abundance to account for the dark matter in the Universe. We summarize the theoretical constraints to which the model is subject and then  perform a detailed study of the phenomenological constraints. In particular, we discuss the bounds from the LHC in the search for light and heavy scalar resonances and invisible states and those from high precision measurements in the Higgs, electroweak and flavor sectors, addressing the possibility of explaining the deviation from the standard expectation of the anomalous magnetic moment of the muon and the $W$-boson mass recently observed at Fermilab. We also summarize the astrophysical constraints from direct and indirect detection dark matter experiments.  We finally  conduct a thorough analysis of the cosmic phase transitions and the gravitational wave spectrum that are implied by the model and identify the parameter space in which the electroweak vacuum is reached after single and multiple phase transitions. We then discuss the prospects for observing the signal of such gravitational waves in near future experiments such as LISA, BBO or DECIGO.
  
\end{abstract}

\newpage

\section{Introduction} 

The discovery of a new type of particle at the LHC a decade ago, the scalar Higgs boson  with a mass of 125 GeV \cite{ATLASdisc,CMSdisc}, has completed the spectrum of the Standard Model (SM) of particle physics and established it as a correct description of three of Nature's fundamental interactions at present energies \cite{ATLAS:2016neq}. It opened and even encouraged the possibility that additional Higgs particles may also exist. Such extensions of the SM Higgs sector, with its unique doublet of complex scalar fields to spontaneously break the electroweak symmetry \cite{Djouadi:2005gi}, are in fact  predicted in a plethora of new physics extensions. This is particularly the case of supersymmetric theories in which the most economical version, the minimal supersymmetric SM (MSSM) \cite{Drees:2004jm} requires the existence of two Higgs doublet fields that lead to five Higgs states in the particle spectrum: two CP-even $h$ and $H$, a CP-odd or pseudoscalar $A$ and two charged $H^\pm$ states \cite{Gunion:1989we,Djouadi:2005gj}. But, in fact, two Higgs doublets models (2HDM), independently of Supersymmetry, have been intensively discussed in the literature and their phenomenology studied in great detail; see e.g.  Ref.~\cite{Branco:2011iw} for a comprehensive review. Other extensions have also been considered in which the Higgs sector involves additional scalar multiplets, from singlet to several doublets, to triplet Higgs fields \cite{Gunion:1989we}. 

The original motivation of the extension of the SM Higgs sector to include two doublets of complex scalar fields and a singlet pseudoscalar field was to alleviate the strong constraints on the particle physics candidates for the dark matter (DM) \cite{Bertone:2004pz,Arcadi:2017kky,Arcadi:2019lka} that is expected to form about 25\% of the energy budget of the Universe \cite{Planck:2018vyg}.   Indeed, this model \cite{Ipek:2014gua,Goncalves:2016iyg,Bauer:2017ota,Tunney:2017yfp,Abe:2018bpo,Robens:2021lov} offers the possibility to induce a direct coupling of the singlet pseudoscalar state to an isosinglet fermionic DM particle, as well as a coupling between this singlet and the SM fermions, via the mixing of the new pseudoscalar $a$ with the pseudoscalar state of the two-Higgs-doublet model. This allows for an efficient annihilation  of the DM state into pairs of fermions in  order for it to have the correct cosmological relic density. At the same time, the absence of couplings between the DM state and the two CP-even Higgs bosons of the model, including the SM-like one,  forbids tree-level spin-independent interactions for the DM, allowing it to evade the stringent constraints from direct detection in astroparticle physics experiments. 

The model with two Higgs doublets and a pseudoscalar $a$  field, that we will coin here as the 2HD+a model, has a very rich phenomenology. In particular, the presence of a possibly light $a$ boson has far reaching consequences. First, as it should substantially couple to SM fermions in order to generate the correct DM relic density, it could be produced and detected in  collider experiments. In particular, searches for such a state have been performed at the Large Hadron Collider (LHC) and even earlier, and some constraints have been set on its mass and couplings; see for instance Refs.~\cite{Haisch:2018kqx,Argyropoulos:2022ezr}. On the other hand, a light $a$ state could address and resolve some anomalies that have been observed in recent experimental data, in particular,  the significant discrepancy from the standard expectation of the anomalous magnetic moment of the muon recently measured at Fermilab \cite{Abi:2021gix}. Indeed, for a range of masses and couplings of the new light pseudoscalar and the fermionic states, it has been shown that the DM problem and the Fermilab value of the $(g-2)_\mu$ could be simultaneously explained while satisfying all other constraints from astroparticle physics and collider searches, including the constraints from flavor physics \cite{Arcadi:2021zdk}. Furthermore, another recent puzzling feature could also be simply addressed in the context of this 2HD+a model \cite{Arcadi:2022dmt}, namely the large deviation from the SM expectation of the $W$ boson mass -- recently observed by the CDF experiment at the Tevatron \cite{CDF:2022hxs}. Hence, the scenario has multiple advantages, addresses several issues  and is thus a good option for physics beyond the SM. 

The model requires further scrutiny and, in particular, one should  simultaneously apply and update all the constraints to which it is subject. This is the case of the ones that apply on the heavy 2HDM and the possibly light pseudoscalar $a$ bosons from the LHC searches, especially that the ATLAS and CMS experiments have recently released their updated results with the full data set collected at an energy of 13 TeV \cite{ATLAS-new,CMS-new}. This is the first purpose of the present paper, to perform a comprehensive analysis of all the possible constraints on the 2HD+a model, from the high precision measurements of the electroweak observables including the $W$-boson mass and of the couplings of the SM-like Higgs boson at the LHC, the constraints from flavor physics, in particular the muon $(g-2)$ and $B$-meson sector observables, and those from the search for the heavy and the possibly light Higgs bosons in the various channel to which they lead at the LHC and other colliders. We will also confront these constraints with the updated ones that come from astroparticle physics searches, in particular,  the very recent results from the LZ direct detection experiment \cite{LZnew}. 

We will show that depending on the configuration or type of the couplings of the 2HDM states to the SM fermions, the so-called Type I, II, X and Y scenarios that allow for the absence of flavor changing neutral currents at tree-level, the various constraints can be either very strong as is, for instance, the case of the Type II scenario which occurs in supersymmetric models and in which one doublet field couples to isospin down-type quarks and leptons and the other to  up-type quarks,  or rather weak, as is the case in the Type I scenario, when both the Higgs doublets couple to isospin up-type and down-type quarks and charged leptons.  For each type of scenario, including the X and Y configurations, we will delineate the parameter space of the 2HD+a model that is still allowed by collider and astrophysical data and, eventually, the one in which the $(g-2)_\mu$ and the $M_W$ anomalies could be resolved. 

Another aspect, which has not been discussed before and that we address in this paper in a comprehensive manner, is the one connected to the cosmological phase transitions related to the dynamics of electroweak  symmetry breaking and its link to gravitational waves (GW) \cite{LIGOScientific:2016aoc}. Given the properties of its extended Higgs sector, the 2HD+a model can easily admit first-order phase transitions resulting in a stochastic GW background, contrary to the SM which predicts a phase transition that is a smooth cross-over \cite{Kajantie:1996mn} and does not generate observable GW signals. Following a  recent analysis for a pure 2HDM \cite{Biekotter:2022kgf}, we perform a random scan over the 2HD+a parameter space and  determine the phase transition pattern in the plane formed by the two CP-even Higgs states. We show that a certain number of points, which incidentally also address the new measurement of the $W$-boson mass at the Tevatron,    yield a GW signal that could be within the reach of future space-based GW observatories such as LISA \cite{Caprini:2015zlo}, BBO \cite{Corbin:2005ny}, Taiji \cite{Ruan:2018tsw}, TianQuin \cite{TianQin:2015yph} or DECIGO \cite{Kawamura:2020pcg}.

The rest of the paper is organized as follows. In the next section, we will introduce the 2HD+a model, including the DM aspect, and briefly summarize the theoretical constraints to which it is subject.  In section 3,  we discuss the various phenomenological and experimental constraints from collider experiments, in particular the ones from the high precision measurements in the electroweak, $B$-meson, muon and Higgs sectors and the ones from direct searches of additional Higgs bosons at the LHC and elsewhere as well as invisible states. We then present in section 4 the salient features which make that the model, when it incorporates a fermionic stable particle, leads to the correct relic density while passing  the bounds from direct and indirect detection experiments and combine all these constraints with the collider bounds. Section 5 will be devoted to the discussion of the cosmic phase transitions and the gravitational wave spectrum, as well as  the prospects for observing the signal of such waves in future experiments. A short conclusion is given in section 6.  

\section{Theoretical aspects of the  2HD+a model}

In this section, we present the 2HD+a model with a fermionic dark matter candidate and its salient theoretical features. We first introduce our constrained two-Higgs-doublet model with the four allowed types of couplings to standard fermions. We then discuss the consequences of  including a possibly light pseudoscalar Higgs field. 
The theoretical constraints on the model, mainly from the perturbativity of the scalar quartic couplings and the stability of the electroweak vacuum  are then summarized. 

\subsection{The two-Higgs-doublet model}

In a two-Higgs-doublet model (2HDM), the scalar sector consists of two doublets of complex scalar fields $\Phi_1$ and $\Phi_2$ which, when invariance under CP symmetry is assumed, can be described by the following scalar potential \cite{Branco:2011iw}
\begin{align}
 V_{\rm 2HDM} &= m_{11}^2 \Phi_1^\dagger \Phi_1+ m_{22}^2 \Phi_2^\dagger \Phi_2 - m_{12}^2  (\Phi_1^\dagger \Phi_2 + {\rm h.c.} ) +\frac12{\lambda_1} ( \Phi_1^\dagger \Phi_1 )^2 +\frac12{\lambda_2} ( \Phi_2^\dagger \Phi_2 )^2  \nonumber \, \\ &
+\lambda_3 (\Phi_1^\dagger \Phi_1) (\Phi_2^\dagger \Phi_2) +\lambda_4 (\Phi_1^\dagger \Phi_2 ) (\Phi_2^\dagger \Phi_1) +\frac12 {\lambda_5} [\,   (\Phi_1^\dagger \Phi_2 )^2 + {\rm h.c.} \, ] \, .
\label{eq:scalar_potential}
\end{align}
From the very beginning, we assume the presence of a discrete symmetry~\cite{Davidson:2005cw} which forbids the introduction of two additional couplings\footnote{CP violation would impact the dynamics of the phase transitions to  be discussed later only very weakly \cite{Fromme:2006cm} and it will be ignored here as we need to distinguish between scalar and pseudoscalar Higgs states.} $\lambda_{6}$ and  $\lambda_{7}$.  Electroweak symmetry breaking is achieved when the fields $\Phi_1$ and $\Phi_2$ acquire the vacuum expectation values (vevs) $v_1$ and $v_2$, respectively. These vevs have to satisfy the relation $\sqrt{v_1^2 + v_2^2} = v$, with  $v \simeq 246$ GeV being the standard one, and their ratio defines the very important parameter $\tan\beta =v_2/v_1$. After symmetry breaking, one obtains five physical states in the spectrum: two CP-even $h$ and $H$  bosons, a CP-odd $A^0$ and two charged Higgs bosons $H^\pm$.  

In addition to the four Higgs boson masses $M_h, M_H, M_{A^0}, M_{H^\pm}$ and the angle $\beta$ defined above, at least another input parameter is needed to entirely characterize the model: the angle $\alpha$ which describes the mixing between the two CP-even $h,H$ bosons. The $h$ state will be identified by convention to be the scalar particle with a 125 GeV mass observed at the LHC, and the $H$ boson will be considered to be heavier, $M_H > M_h$ (we ignore the unlikely reverse possibility discussed e.g. in Ref.~\cite{Biekotter:2019kde}).  The five quartic couplings $\lambda_i$ of the scalar potential above can be more conveniently expressed in terms of the physical state masses and the angles $\alpha$ and $\beta$ introduced above. Using the abbreviation $M^2 \equiv m_{12}^2/(\sin {\beta} \cos {\beta})$ (with $M$ being possibly positive or negative), the five couplings read 
\begin{align}
\label{eq:quartic_phys}
\lambda_1 v^2 &=  - M^2 \tan^2 \beta +\frac{\sin^2 \alpha}{\cos^2 \beta} M_h^2 +\frac{\cos^2 \alpha}{\cos^2\beta}M_H^2 \, , \nonumber \\
\lambda_2 v^2 &=  -\frac{M^2 } {\tan^2 \beta}+\frac{\cos^2 \alpha}{\sin^2 \beta}M_h^2+\frac{\sin^2 \alpha}{\sin^2 \beta}M_H^2 \, , \nonumber \\
\lambda_3 v^2 &= -M^2 +2 M_{H^{\pm}}^2 +\frac{\sin 2\alpha}{\sin 2\beta}( M_H^2-M_h^2) \, , \nonumber \\
\lambda_4 v^2 &=  M^2 + M_{A^0}^2 - 2 M_{H^{\pm}}^2 \, ,   \nonumber \\
\lambda_5 v^2 & =  M^2 - M_{A^0}^2 \, . 
\end{align} 
The additional parameter $m_{12}$ will enter only in the trilinear and quartic couplings among the physical Higgs states and, as we will see shortly, it can be ignored in most of the present discussion together with the mass parameters $m_{11}$ and $m_{22}$. 

The mixing between the neutral CP-even Higgs bosons of the model make that $h$ and $H$ share the coupling of the standard Higgs particle $H^0$ to the massive gauge bosons $V=W,Z$  
\begin{eqnarray}
g_{hVV}= g^{\rm 2HDM}_{hVV}/g^{\rm SM}_{H^0VV}= \sin(\beta-\alpha) \ ,   \ \ \  
g_{HVV}= g^{\rm 2HDM}_{HVV}/g^{\rm SM}_{H^0VV}= \cos(\beta-\alpha) \, . 
\end{eqnarray}
As a result of CP invariance,  there is no coupling of the CP-odd $A^0$ state to the massive $W,Z$ bosons, $g_{A^0VV}= 0$. The couplings between two Higgs bosons and a massive vector boson $V$ are complementary to the previous ones. Up to normalization factors, one has for instance
\begin{eqnarray}
g_{hA^0Z} = g_{h H^\pm W}=  \cos(\beta-\alpha) \ , \ \ \  
g_{HA^0Z} = g_{H H^\pm W} =  \sin(\beta-\alpha) . 
\label{HV-couplings}
\end{eqnarray}
There are also couplings of the charged Higgs boson to gauge bosons which simply read 
\begin{eqnarray}
g_{A^0 H^\pm W}= 1 \, , \ \  g_{H^+ H- \gamma} =  -e \, , \ \   g_{H^+ H^- Z} = -e \cos2\theta_W/ (\sin\theta_W \cos\theta_W) . 
\end{eqnarray}

Finally, there are couplings of the various Higgs bosons to the standard fermions. They  are slightly more involved and can be described by the following Yukawa-type Lagrangian
\begin{eqnarray}
-{\cal L}_{\rm Yuk}^{\rm SM}& =& \sum\limits_{f=t,b,\tau} \frac{m_f}{v} \bigg(  g_{hff} \bar f f h +g_{Hff} \bar f f H-i g_{A^0ff} \bar f \gamma_5 f A^0 \bigg) \nonumber \\
&-& \frac{\sqrt{2}}{v} \bigg( \bar{t} ( m_t g_{A^0tt} P_L + m_b g_{A^0bb} P_R ) b  H^+ +  m_\tau g_{A^0\tau\tau} \bar \nu_\tau  P_R \tau H^+  + \mathrm{h.c.} \bigg) \, ,
\end{eqnarray}
with the usual projectors $P_{L/R}= \frac12 (1 \mp \gamma_5)$.  $g_{\phi ff}$ are the reduced couplings of the $\phi$ boson to quarks and leptons and we will take into account here only those of the third generation which are the only relevant ones (except in the case of the muon $g\!-\!2$ as will be seen later). They have been normalized to the couplings of the SM $H^0$ boson, $g_{\phi ff}=g^{\rm 2HDM}_{\phi ff}/g^{\rm SM}_{H^0 ff}$. 

\begin{table}[h!]
\renewcommand{\arraystretch}{1.6}
\begin{center}
\begin{tabular}{|c|c|c|c|c|}
\hline
~~~~~~ &  Type I & Type II & Type X & Type Y \\ \hline \hline 
$g_{htt}$ & $ \frac{\cos \alpha} { \sin \beta} \rightarrow 1$ & $\frac{ \cos \alpha} {\sin \beta} \rightarrow 1$ & $\frac{ \cos \alpha} {\sin\beta} \rightarrow 1$ & $ \frac{ \cos \alpha}{ \sin\beta} \rightarrow 1$ \\ \hline
$g_{hbb}$ & $\frac{\cos \alpha} {\sin \beta} \rightarrow 1$ & $-\frac{ \sin \alpha} {\cos \beta} \rightarrow 1$ & $\frac{\cos \alpha}{ \sin \beta} \rightarrow 1$ & $-\frac{ \sin \alpha}{ \cos \beta} \rightarrow 1$ \\ \hline
$g_{h\tau\tau} $ & $\frac{\cos \alpha} {\sin \beta} \rightarrow 1$ & $-\frac{\sin \alpha} {\cos \beta} \rightarrow 1$ & $- \frac{ \sin \alpha} {\cos \beta} \rightarrow 1$ & $\frac{ \cos \alpha} {\sin \beta} \rightarrow 1$  \\ \hline\hline
$g_{Htt}$ & $\frac{\sin \alpha} {\sin \beta} \rightarrow -\frac{1}{\tan\beta}$ & $\frac{ \sin \alpha} {\sin \beta} \rightarrow -\frac{1}{\tan\beta}$ & $ \frac{\sin \alpha}{\sin \beta} \rightarrow -\frac{1}{\tan\beta}$ & $\frac{ \sin \alpha}{ \sin \beta} \rightarrow -\frac{1}{\tan\beta}$ \\ \hline
$g_{Hbb}$ & $ \frac{ \sin \alpha}{\sin \beta} \rightarrow -\frac{1}{\tan\beta}$ & $\frac{\cos \alpha}{\cos \beta} \rightarrow {\tan\beta}$ & $\frac{\sin \alpha} {\sin \beta} \rightarrow -\frac{1}{\tan\beta}$ & $\frac{ \cos \alpha} {\cos \beta} \rightarrow {\tan\beta}$ \\ \hline
$g_{H\tau\tau}$ & $\frac{ \sin \alpha} {\sin \beta} \rightarrow -\frac{1}{\tan\beta}$ & $\frac{\cos \alpha} {\cos \beta} \rightarrow {\tan\beta}$ & $\frac{ \cos \alpha} {\cos \beta} \rightarrow {\tan\beta}$ & $\frac{\sin \alpha} {\sin \beta} \rightarrow -\frac{1}{\tan\beta}$ \\ \hline\hline
$g_{A^0tt}$ & $\frac{1}{\tan\beta}$ & $\frac{1}{\tan\beta}$ & $\frac{1}{\tan\beta}$ & $\frac{1}{\tan\beta}$ \\ \hline
$g_{A^0bb}$ & $-\frac{1}{\tan\beta}$ & ${\tan\beta}$ & $-\frac{1}{\tan\beta}$ & ${\tan\beta}$ \\ \hline
$g_{A^0\tau\tau}$ & $-\frac{1}{\tan\beta}$ & ${\tan\beta}$ & ${\tan\beta}$ & $-\frac{1}{\tan\beta}$
\\ \hline
\end{tabular}
\vspace*{2mm}
\caption{Couplings of the 2HDM Higgs bosons to third generation fermions, normalized to the SM-Higgs ones,  as a function of the angles $\alpha$ and $\beta$ for the four types of 2HDM scenarios. For the CP-even $h,H$ states,  the values in the alignment limit $\alpha\! \to\! \beta\!-\!\frac{\pi}{2}$ are also shown.}
\label{table:2hdm_cplgs}
\end{center}
\vspace*{-6mm}
\end{table}

In the 2HDM with a discrete symmetry that we are considering here, the absence of flavor-changing neutral currents (FCNCs), which are experimentally constrained to be very small, is enforced by coupling in a specific manner the original $\Phi_1$ and $\Phi_2$ fields to isospin up-type quarks and isospin down-type quarks and charged leptons.  There are four configurations or Types \cite{Glashow:1976nt}. The most discussed ones \cite{Branco:2011iw} are the so-called  Type II model, in which the field $\Phi_1$ couples to isospin down-type quarks and leptons and $\Phi_2$ to up-type quarks, and the Type I model, in which the field $\Phi_2$ couples to both isospin up- and down-type fermions.  To be more general, we will also study the two   additional options in which the charged leptons will have a different coupling compared to down-type  quarks, namely the Type X or lepton-specific model in which the Higgs couplings to quarks are as in the Type I case but those to leptons are as in Type II, and the Type Y or flipped model in which the Higgs couplings are as in the previous model but with the Type I and Type II couplings reversed. 

The neutral Higgs couplings to fermions in these four flavor-conserving types of 2HDMs, as functions of the angles $\beta$ and $\alpha$, are listed in Table~\ref{table:2hdm_cplgs}. The couplings of the charged Higgs bosons follow those of the pseudoscalar $A^0$ state. In the case of the CP-even $h$ and $H$ bosons, we also give for completeness, the values of these couplings in the alignment limit in which the $h$ state is SM-like. This alignment limit is strongly favored  by LHC Higgs data as will be seen in the next section, and is achieved by simply setting $\alpha = \beta-\frac{\pi}{2}$. 

\subsection{The pseudoscalar sector of the 2HD+a model}
 
 In our study, we will consider the extension of the 2HDM previously introduced by an additional singlet pseudoscalar Higgs field $a^0$ \cite{Ipek:2014gua,Goncalves:2016iyg,Bauer:2017ota,Tunney:2017yfp,Abe:2018bpo}. The most general scalar potential for such a 2HD+a model is given by~\cite{Abe:2018bpo}
\begin{eqnarray} 
V_{\rm 2HD\!+\!a} &=& V_{\rm 2HDM}\!+\! \frac{1}{2} m_{a^0}^2 (a^0)^2\!+\! \frac{\lambda_a}{4} (a^0)^4
\!+\! \left(i \kappa a^0 \Phi^{\dagger}_1\Phi_2\!+\!\mbox{h.c.}\right) \nonumber \\
&+& \left(\lambda_{1P}(a^0)^2 \Phi_1^{\dagger}\Phi_1 \!+\! \lambda_{2P}(a^0)^2 \Phi_2^{\dagger}\Phi_2\right), 
\label{eq:V2HDa}
\end{eqnarray} 
where $V_{\rm 2HDM}$ is the 2HDM potential given in Eq.~(\ref{eq:scalar_potential}) and $\kappa,\lambda_{1P}, \lambda_{2P}$  are the new trilinear couplings between the two Higgs doublets and the pseudoscalar $a^0$ state and $\lambda_a$ the quartic $a^0$ coupling (we assume that $\kappa$ is real for simplicity).  

After electroweak symmetry breaking, the physical content of the  Higgs sector of the theory will consist of again two CP-even $h,H$ states, two charged $H^{\pm}$ bosons, but two CP-odd $a^0$ and $A^0$ states which could mix. Hence, in addition to the usual mixing angles $\alpha$ and $\beta$ of a 2HDM, there will be an extra  mixing angle $\theta$ which allows to transform the $(A^0,a^0)$ current eigenstates to the $(A,a)$  physical CP-odd eigenstates
\begin{equation}
\left(
\begin{array}{c} A^0 \\ a^0 \end{array} \right)= \left( \begin{array}{cc}
\cos\theta & \sin\theta \\ -\sin\theta & \cos\theta \end{array}
\right)  \left(
\begin{array}{c} A \\ a \end{array} \right) \, . 
\end{equation} 
This mixing angle is given,  in terms of $\kappa$ and the physical masses $M_a$ and $M_A$, by 
\beq
\tan2\theta=\frac{2 \kappa v}{M_{A}^2-M_{a}^2}\;.
\eeq
The CP-odd mixing will modify two of the quartic couplings of the 2HDM given in Eq.~(\ref{eq:quartic_phys}) when the replacement $A^0= \cos\theta A + \sin\theta a$ is made. More explicitly, one would have  
\begin{align} 
    & \lambda_4 v^2 = M^2+M_A^2 \cos^2 \theta+M_a^2 \sin^2 \theta-2 M_{H^{\pm}}^2 \, , \nonumber\\
    & \lambda_5 v^2 =M^2-M_A^2 \cos^2\theta-M_a^2 \sin^2 \theta \, .
    \label{eq:quartic-phys-A}
    \end{align}

In the case of the Higgs couplings to fermions, one can consider simply those discussed in the context of the 2HDM with the four configurations, Type I, II, X and Y, but modify the neutral Higgs sector to introduce the additional pseudoscalar Higgs state. The neutral current part of the Lagrangian $\mathcal{L}_{\rm Yuk}$ which contains the Yukawa interactions with the SM fermions will then read 
\begin{equation}
\mathcal{L}_{\rm Yuk}=\sum_f \frac{m_f}{v}\bigg[ g_{hff} h \bar f f+g_{Hff}
H\bar f f- i g_{Aff}  \bar f \gamma_5 f-i g_{aff} a \bar f \gamma_5 a \bigg] \, , 
\end{equation}
where the couplings $g_{\phi ff}$ of the 2HDM CP-even $h,H$ (as well as implicitly those of the charged Higgs bosons $H^\pm$) are given in Table \ref{table:2hdm_cplgs} in the four types of configurations, while the Yukawa couplings of the pseudoscalar Higgs bosons will be given by 
\begin{equation}
g_{Aff}=\cos\theta \, g_{A^0ff} \ , \ \ \   g_{aff}=\sin\theta \, g_{A^0ff} \, , 
\end{equation}
with the reduced couplings  $g_{A^0ff}$ again given in Table \ref{table:2hdm_cplgs} in the four 2HDM configurations. 

Finally, there are also trilinear interactions between the Higgs states which could be relevant. Here, we will be interested only in the interactions of the SM-like $h$ boson whose couplings to two pseudoscalar fields  are given by the Lagrangian $\mathcal{L}_{\rm scal}$ 
\begin{equation}
\mathcal{L}_{\rm scal}=\lambda_{haa}  h aa+\lambda_{haA}  h aA+\lambda_{hAA} h AA\, ,
\end{equation} 
where, using the abbreviations $s_X, c_{X}=\sin(X),\cos(X)$ and $t_\beta=\tan\beta$, one would have
\begin{align}
    \lambda_{haa}=& -\frac{2 M_a^2}{v}s_{\beta-\alpha} s^2_\theta-\frac{M_h^2}{v}s^2_\theta \frac{s_{\beta-\alpha} t_\beta+c_{\beta-\alpha}(1-t^2_\beta)}{t_\beta}
   +\frac{M^2}{v}s^2_\theta \frac{2 s_{\beta-\alpha}t_\beta+c_{\beta-\alpha}(1-t^2_\beta)}{t_\beta}\nonumber\\
   & -2 \lambda_{1P}v c^2_\theta \frac{s_{\beta-\alpha}-c_{\beta-\alpha}t_\beta}{1+t^2_\beta}-2 \lambda_{2P}v c^2_\theta \frac{t_\beta (s_{\beta-\alpha}t_\beta+c_{\beta-\alpha})}{1+t^2_\beta} \, , \nonumber\\
    \lambda_{haA}&=\frac{M_A^2}{2v}s^2_{\beta-\alpha}s_{ 2\theta} +\frac{M_a^2}{2v}s_{\beta-\alpha}s_{2\theta} +\frac{M_h^2}{2v}s_{2 \theta} \frac{s_{\beta-\alpha}t_\beta+c_{\beta-\alpha}(1-t^2_\beta)}{t_\beta}  \nonumber\\
   & -\frac{M^2}{2v}s_{2\theta} \frac{2 s_{\beta-\alpha}t_\beta+c_{\beta-\alpha}(1-t^2_\beta)}{t_\beta}\nonumber\\
   & - \lambda_{1P}v s_{2\theta} \frac{s_{\beta-\alpha}-c_{\beta-\alpha}t_\beta}{1+t^2_\beta}- \lambda_{2P}v s_{2 \theta} \frac{t_\beta (s_{\beta-\alpha}t_\beta+c_{\beta-\alpha})}{1+t^2_\beta} \, , \nonumber\\
    \lambda_{hAA}&=-\frac{2M_A^2}{v}s_{\beta-\alpha}c^2_ \theta-\frac{M_h^2}{v}c^2_\theta \frac{s_{\beta-\alpha}t_\beta+c_{\beta-\alpha}(1-t^2_\beta)}{t_\beta}
    +\frac{M^2}{v}c^2_\theta \frac{s_{\beta-\alpha}t_\beta-c_{\beta-\alpha}(1-t^2_\beta)}{t_\beta}\, \nonumber\\
   & - 2\lambda_{1P}v s^2_\theta \frac{s_{\beta-\alpha}-c_{\beta-\alpha}t_\beta}{1+t^2_\beta}- 2\lambda_{2P}v s^2_\theta \frac{t_\beta (s_{\beta-\alpha}t_\beta+c_{\beta-\alpha})}{1+t^2_\beta} \, . 
   \label{eq:haa}
\end{align}

\subsection{Theoretical constraints on the model}

We now summarize the theoretical constraints that one can impose on the 2HD+a model. These generally apply on the quartic couplings of the scalar potential  which can be translated into bounds on the Higgs masses $M_a,M_A,M_H, M_{H^{\pm}}$ as functions of the angles $\alpha$ and $\beta$, using for instance Eqs.~(\ref{eq:quartic_phys}) and (\ref{eq:quartic-phys-A}). The most relevant bounds can be obtained following those derived in the context of a 2HDM only \cite{Kanemura:2004mg,Becirevic:2015fmu,Barroso:2013awa} assuming $\lambda_{P1}, \lambda_{P2} >0$. 

There is first the requirement of perturbative unitarity which leads to the following bounds on the combinations of the couplings $\lambda_i$
\begin{align}
\label{eq:unitarity}
    & |x_i| < 8\pi \, , \ |\lambda_{1,2P}|< 4\pi,\,\,\,\,|\lambda_3\pm \lambda_4|< 4 \pi
    \, , \nonumber\\ & 
    \left \vert \frac{1}{2}\left(\lambda_1+\lambda_2 \pm \sqrt{(\lambda_1-\lambda_2)^2+4\lambda_k^2}\right)\right \vert < 8 \pi \, , \  k=4,5 \, , \nonumber\\
    & |\lambda_3+ 2 \lambda_4 \pm 3 \lambda_5|< 8 \pi,\,\,\,\,|\lambda_3 \pm \lambda_5| < 8 \pi \, , 
\end{align}
where the $x_i$'s are the solutions of the equation
\begin{align}
& 0=x^3-3 (\lambda_a+\lambda_1+\lambda_2)x^2  + (9 \lambda_1 \lambda_a+9 \lambda_2 \lambda_a -4 \lambda_{1P}^2-4 \lambda_{2P}^2-4 \lambda_3^2-4 \lambda_3 \lambda_4-\lambda_4^2+9 \lambda_1 \lambda_2)x \nonumber\\
& +12 \lambda_{2P}^2 \lambda_1+12 \lambda_{1P}^2\lambda_2 
 -16 \lambda_{1P}\lambda_{2P}\lambda_3-8 \lambda_{1P}\lambda_{2P}\lambda_4 +(-27 \lambda_1 \lambda_2+12 \lambda_3^2+12 \lambda_3 \lambda_4+3 \lambda_4^2)\lambda_a \, . 
\end{align}
In addition, there is the requirement that the scalar potential should be  bounded from below, which leads to  the following constraints on the scalar quartic couplings (with the assumption $\lambda_{P1}, \lambda_{P2} >0$, the last two lines are always satisfied) \cite{Kannike:2012pe}:
\begin{equation}
\label{eq:boundedbelow}
\begin{split}
  & \lambda_{1} > 0, \quad \lambda_{2} > 0, \quad \lambda_{a} > 0,   \\
  & \bar{\lambda}_{12} \equiv \sqrt{\lambda_{1} \lambda_{2}}+\lambda_{3}+\min (0, \lambda_{4}- |\lambda_{5}|) > 0,   \\
  & \bar{\lambda}_{1P} \equiv \sqrt{\frac{\lambda_{1} \lambda_{a}}{2}} + \lambda_{1P} > 0,   \ \ \  
  \bar{\lambda}_{2P} \equiv \sqrt{\frac{\lambda_{2} \lambda_{a}}{2}} + \lambda_{2P} > 0,   \\
  & \sqrt{\frac{\lambda_{1} \lambda_{2} \lambda_{a}}{2}}   + \lambda_{1P} \sqrt{\lambda_{2}} + \lambda_{2P} \sqrt{\lambda_{1}} + [\lambda_{3} + \min (0, \lambda_{4} - |\lambda_{5}|)] \sqrt{\frac{\lambda_{a}}{2}}
  + \sqrt{2} \sqrt{\bar{\lambda}_{12} \bar{\lambda}_{1P} \bar{\lambda}_{2P}} > 0.
  \end{split}
\end{equation}

It is interesting to further discuss this requirements for the specific case of the coupling $\lambda_3$.  Defining the SM Higgs self-coupling as $\lambda={M_h^2}/{(2 v^2)}$, one has $\lambda_3 > 2 \lambda$ and under the assumption $M_A \gg M_a$ which will be intensively used later on, one would have 
\begin{align}
 \lambda_3 > \frac{M_A^2-M_a^2}{v^2}\sin^2 \theta -2 \lambda \cot^2 2\beta \, .
\end{align}
When combining this equation with the perturbativity requirement $\lambda_3 < 4 \pi$, one realizes that it is not possible to have an arbitrary mass splitting between the $a$ and $A$ bosons when mixing is present, i.e.  $\sin\theta \neq 0$. This impossibility of decoupling the pseudoscalar Higgs state is also enforced by the requirement of perturbative unitarity in the $aa, aA$ and $AA$ scattering amplitudes into massive gauge bosons. These indeed give the constraint \cite{Goncalves:2016iyg}
\begin{align}
\label{eq:uni}
|\Lambda_{\pm} | = \bigg| \frac{1}{v^2} \bigg( \Delta_H^2 - \frac{\Delta^2_a}{8} (1-\cos 4\theta) \pm \sqrt{ {\Delta_H^2}{v^2}+ \frac{\Delta_a^4}{8} (1-\cos4 \theta)  } \bigg) \bigg| \leq 8 \pi, 
\end{align}
where 
\begin{align}
\Delta_a=M_A^2-M_a^2 \ , \ \ \Delta_H =M^2-M_{H^{\pm}}^2+2 M_W^2- \frac12
M_h^2  \, , 
\end{align}

In the limit  $M\gg M_a$ and with maximal mixing  $\sin2\theta=1$, there is an upper bound on $M_A$ of about $M_A \lsim 1.4$ TeV. This limit is, however, weakened if one lowers the value of $\sin2\theta$ and it disappears completely  in the absence of mixing, $\sin\theta\!=\!0$.  

\begin{figure}
    \centering
    \subfloat{\includegraphics[width=0.46\linewidth]{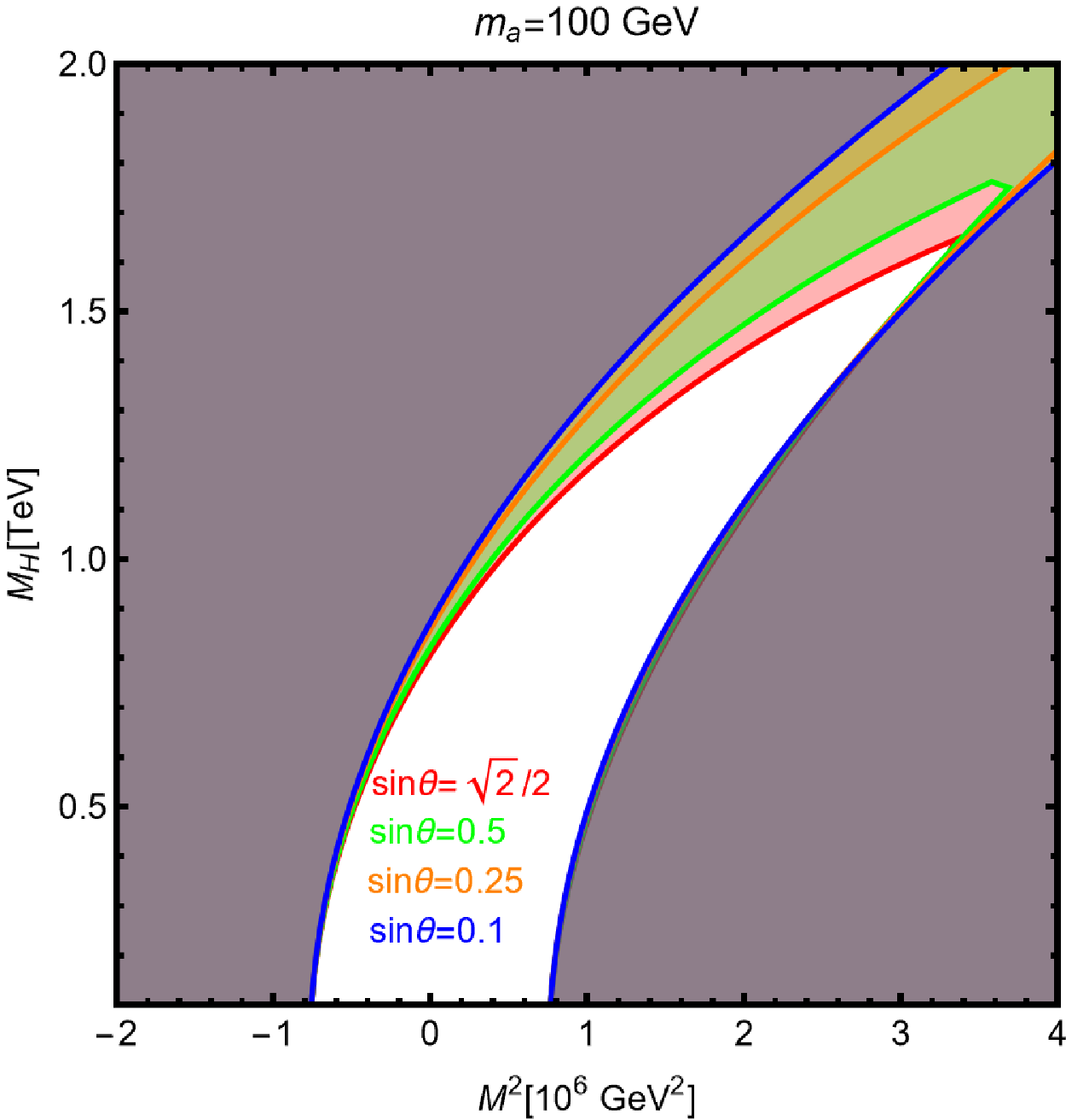}}~~~    \subfloat{\includegraphics[width=0.47\linewidth]{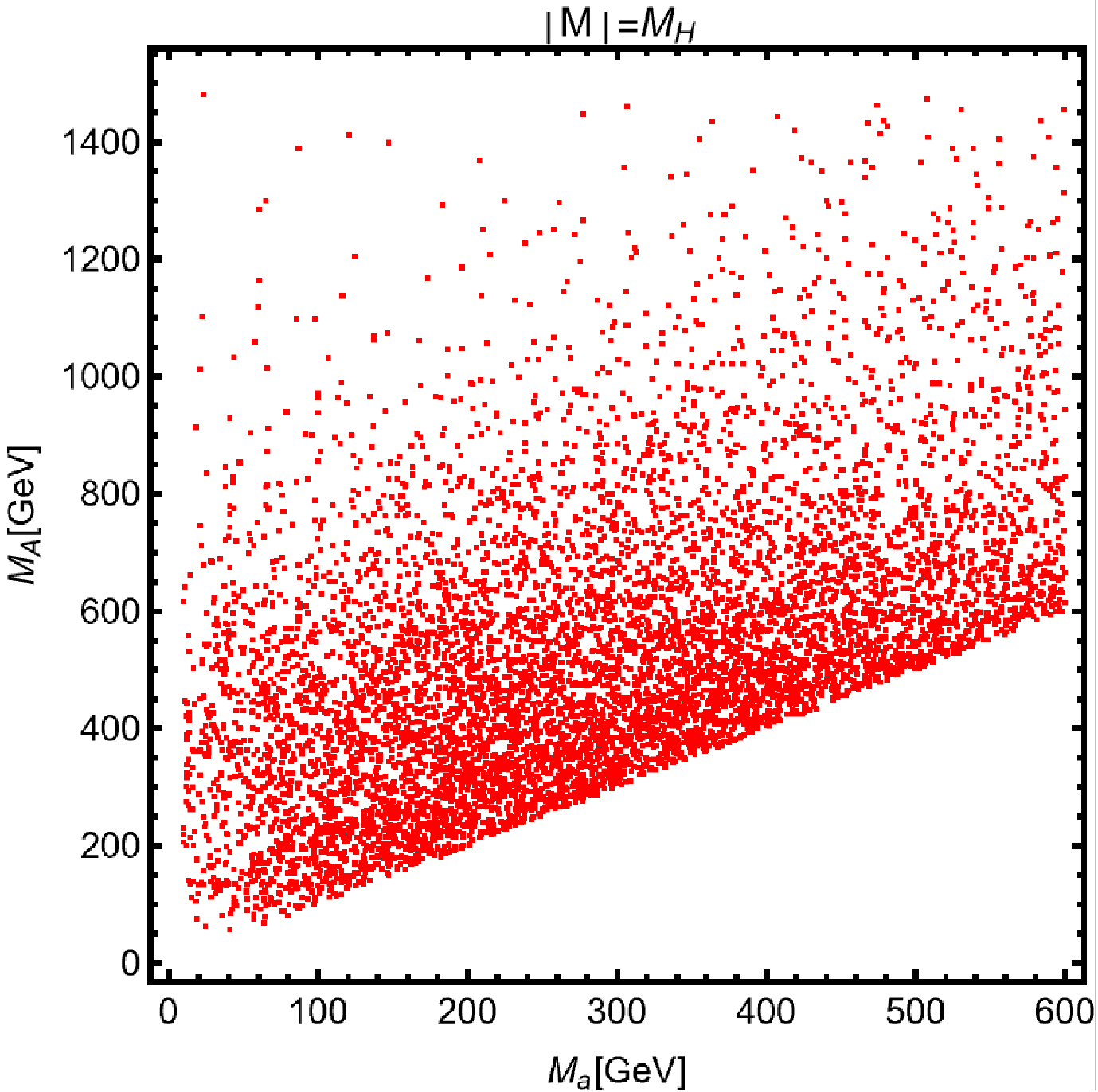}}
    \caption{Illustration of the impact of the unitarity bounds on the 2HD+a parameter space. The left panel shows for the different values of the sine of the mixing angle $\theta$ reported on the plot, the excluded regions in the $[M^2,M_{H^{\pm}}=M_A=M_H]$ plane for a light pseudoscalar Higgs boson of $M_a\!=100\,\mbox{GeV}$ mass. The right panel shows  in the $[M_a,M_A]$ plane, the outcome of a scan of the other 2HD+a parameters assuming also $M_H\!=\!|M|$ when including also the other theoretical constraints discussed in the main text.}
    \label{fig:plotUNI}
\end{figure}

The impact of these theoretical constraints from perturbative unitarity in the $aa, aA$ and $AA$ scattering amplitudes is illustrated in Fig. \ref{fig:plotUNI}. In the left panel of the figure, we illustrate the effect of the  bound $|\Lambda_{\pm}|\leq 8 \pi$ taken individually in the $[M^2,M_H]$ plane with $|M|=M_{H^{\pm}}=M_A$ for a light pseudoscalar $a$ boson with a mass $M_a=100\,\mbox{GeV}$ and several values of the sine of the mixing angle $\theta$.  For each value, we have marked with a color,  namely red $(\sin\theta=\sqrt{2}/2)$, green $(\sin\theta=0.5)$, orange $(\sin\theta=0.25)$ and blue $(\sin\theta=0.1)$, the region in which the unitarity bound is violated. 

The right panel of Fig.~\ref{fig:plotUNI} combines, instead, the unitarity bound with the other theoretical constraints discussed in this section. To achieve this result, a scan on the 2HD+a parameters has been made restricting to the mass range  $|M|=M_H=M_{H^\pm}=M_A$. The model points passing all the constraints is displayed in the $[M_a,M_A]$ plane. Notice that the bottom right region of the plot is empty because we have assumed the $M_a < M_A$ hierarchy. Despite that the plot shows that one could have a value of $M_A$ as high as 1.4 TeV, the density of model points gives a preference for mass values below 1 TeV. To improve the efficiency of the numerical analyses and the one of the scans, we will limit in the  rest of the paper, the masses of the additional heavy Higgs bosons to values less than  about 1 TeV.

\section{Collider constraints} 

\subsection{Higgs signal strengths}

We come now to the phenomenological constraints on the 2HD+a model and first discuss the ones that emanate from the high precision measurements of the properties of the  125 GeV Higgs state performed at the LHC. Indeed, precise  measurements of the $h$ boson production and decay rates strongly constrain its couplings to massive gauge bosons $g_{hVV}$ and fermions $g_{hff}$ given in the upper part of Table \ref{table:2hdm_cplgs} and,  hence, the values of the angles $\alpha$ and $\beta$. 

It has become now customary to study these $h$ couplings by looking at their deviation from the SM expectation, which is achieved when considering a specific search channel $X$, by means of the signal strength modifier $\mu_{XX}$ \cite{ATLAS:2016neq,LHCHiggsCrossSectionWorkingGroup:2016ypw}. This quantity characterizes the $h$ production cross section times  its decay branching ratio into the $X$ states, normalized to the expected SM values. One would then have, in the narrow width approximation,  the relation  
\beq
\mu_{XX}=\frac{\sigma( pp \to h \to XX)}{\sigma( pp \to H^0 \to XX)|_{\rm SM}} =   \frac{ \sigma( pp \to h)\times {\rm BR} (h \to XX)} {\sigma( pp \to H^0)|_{\rm SM} \times {\rm BR} (H^0 \to XX)|_{\rm SM} }.
\eeq
For instance, assuming that the $h$ boson is produced in the by far dominant gluon-gluon fusion process $gg\to h$ and focusing on the $h \to XX$ decay channel, one can relate the signal strength $\mu_{XX}$ to the coupling modifier $\kappa_X^2$ which measures the deviations of the $h$ coupling to the particle $X$, $g_{hXX}$, from its value as predicted in the SM 
\beq
\kappa_X^2 = \frac{\Gamma( h\to XX) }{\Gamma (H^0 \to XX) |_{\rm SM}} \  \ \simeq \frac{g^2_{hXX}} {g^2_{H^0 XX}|_{\rm SM}} \, .
\eeq
The measurement of the various $h$ couplings have been recently updated by the ATLAS and CMS collaborations for the 10th anniversary of the Higgs discovery, using the full available set of data, about 139 fb$^{-1}$,  collected at the energy of 13 TeV \cite{ATLAS-new,CMS-new}. The corresponding signal strengths measured by ATLAS and CMS are summarized in Table \ref{tab:h_signal} in the case of Higgs decays into gauge boson ($\mu_{\gamma \gamma}, \mu_{WW}, \mu_{ZZ}$) as well as bottom quark and tau lepton $(\mu_{bb}, \mu_{\tau \tau})$ final states. Together with the central values, the total (theoretical, statistical and systematical) uncertainties as estimated by the collaborations are also shown. In the last line, we also give the measured signal strength from the cross section for $h$ production in the dominant production channel, the gluon-fusion process $gg\to h$, which is dominantly mediated by top quark loops and is thus directly related to the $h$ coupling to top quarks. 

\begin{table}[h!]
\renewcommand{\arraystretch}{1.2}
\begin{center}
\begin{tabular}{|c|c|c|}
\hline
signal strength &  ~~~~ATLAS~~~~ & ~~~~~CMS~~~~~  \\ \hline\hline 
$\mu_{\gamma\gamma}$ & $1.04^{+0.10}_{-0.09} $ & $1.13 \pm 0.09$ \\ \hline 
$\mu_{ZZ}$  & $1.01 \pm 0.11$  & $0.97^{+0.12}_{-0.11}$  \\ \hline 
$\mu_{WW}$  & $1.09 \pm 0.11$   & $0.97 \pm 0.09$  \\  \hline 
$\mu_{bb}$  & $1.02^{+0.12}_{-0.11}$  & $1.05^{+0.22}_{-0.21}$ \\  \hline 
$\mu_{\tau\tau}$  & $0.93^{+0.13}_{-0.12}$  & $0.85 \pm 0.10$  \\  \hline \hline
$\mu_{ gg\to h}$ &  $1.00 \pm 0.05$ & $0.97^{+0.08}_{-0.07} $ \\ \hline
\end{tabular}
\caption{Summary of the values of the signal strengths of the 125 GeV Higgs boson assumed to be the $h$ state as measured by ATLAS \cite{ATLAS-new} and CMS \cite{CMS-new}  using the full set of available data in the various possible decay channels; the quoted uncertainties are the total ones.} 
\label{tab:h_signal}
\end{center}
\end{table}

The table shows that $h$ has been found to have SM-like properties with an accuracy of about 10\% or less. In particular, it should have an almost SM-like coupling to $V\!=\!W,Z$ bosons which, assuming the  custodial SU(2) symmetry to which the SM as well as our model obey, are equal $g_{hWW}=g _{hZZ} \equiv g_{hVV}$.   This provides the most stringent test of the departure from the SM expectation or the alignment limit, $\cos^2(\beta-\alpha)  \equiv 1- g_{hVV}^2=\!0$. One can thus turn these measurements into constraints on the angles $\alpha$ and $\beta$ of our 2HD+a scenario.   
 
\begin{figure}[!h]
\vspace*{-2mm}
\centering
\subfloat{\includegraphics[width=0.47\linewidth]{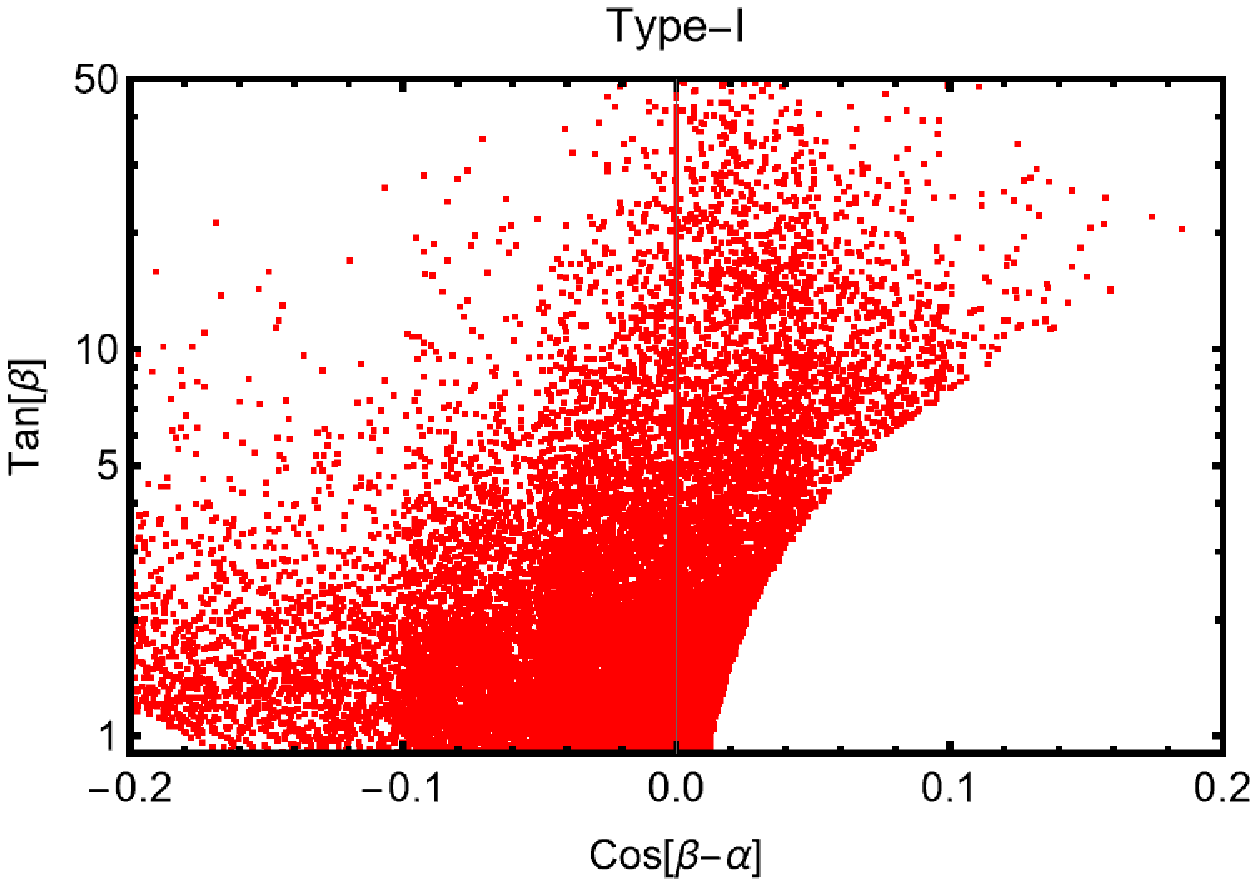}}~
\subfloat{\includegraphics[width=0.47\linewidth]{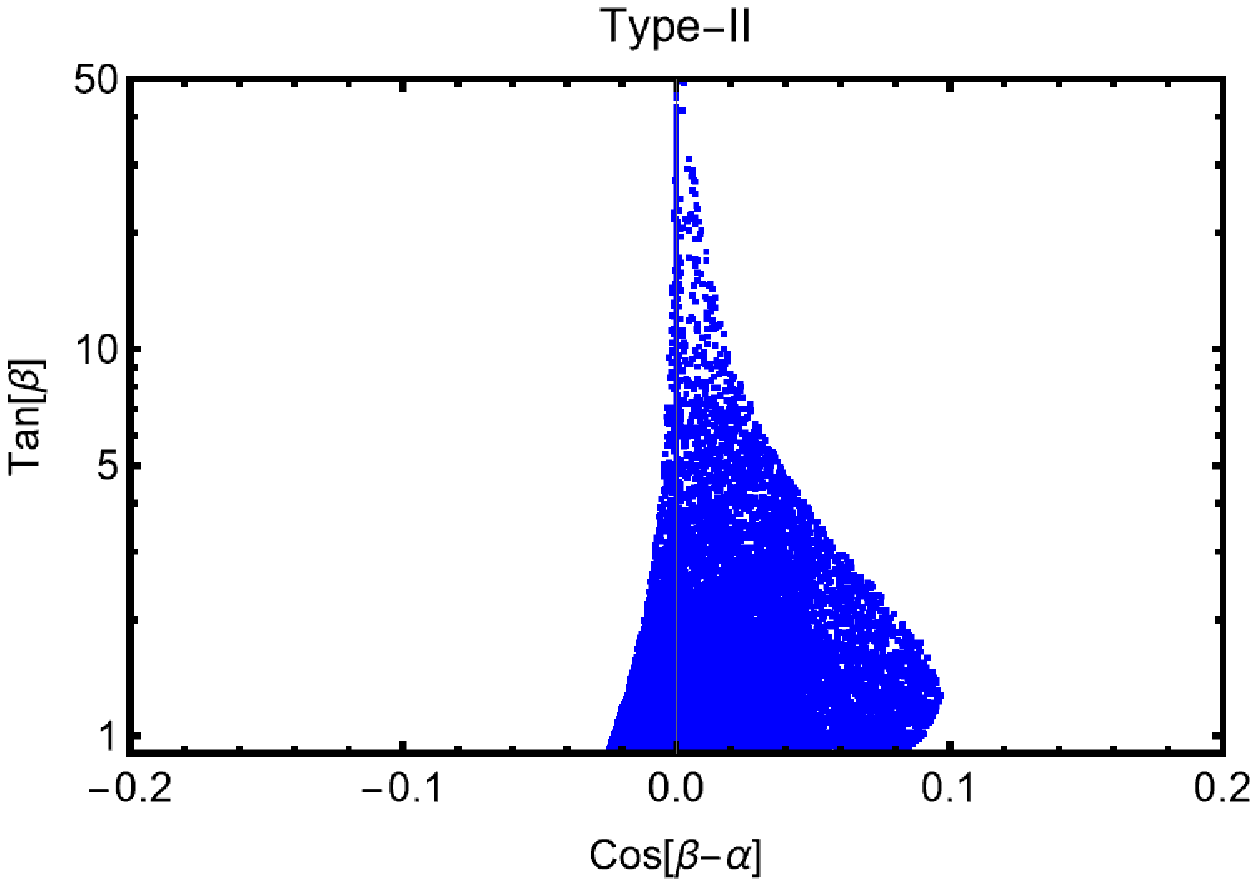}}\\
\subfloat{\includegraphics[width=0.47\linewidth]{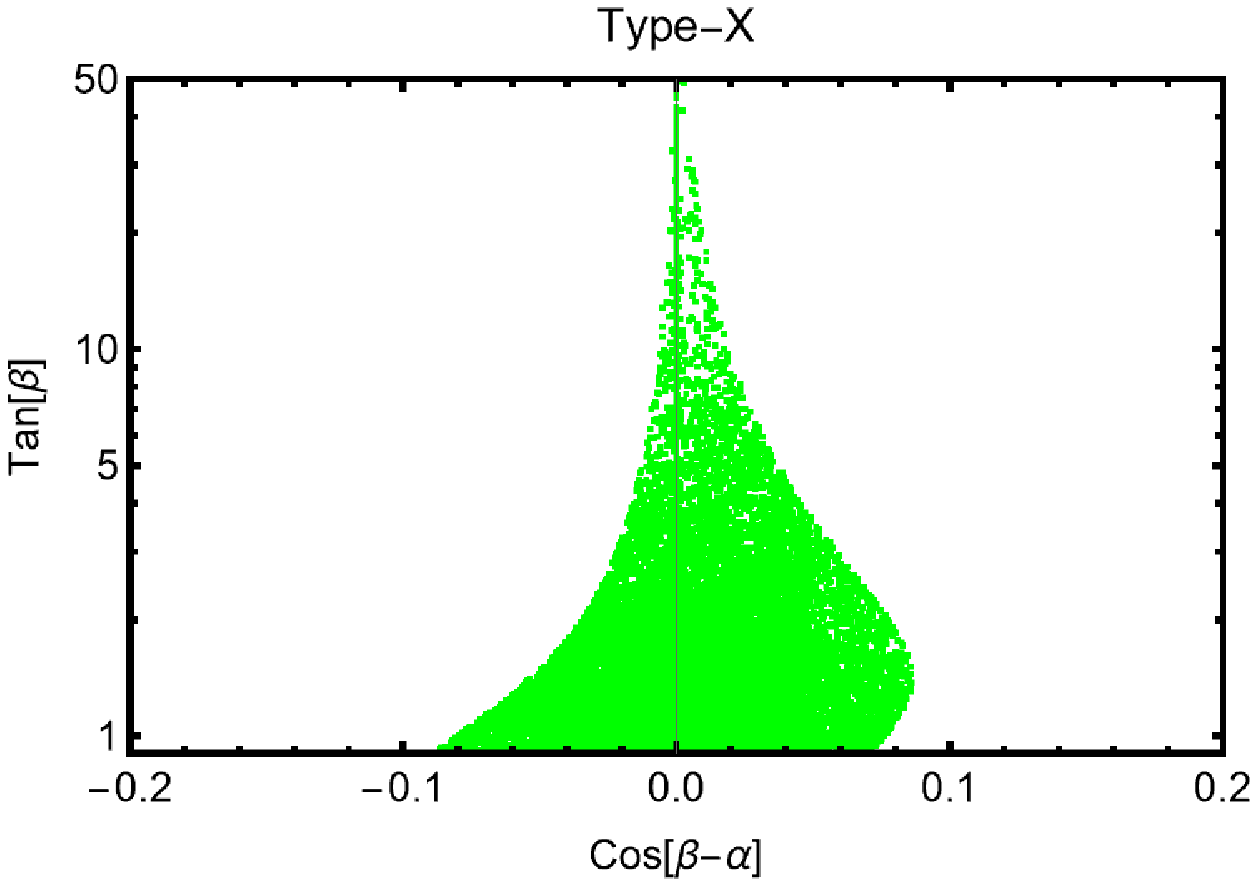}}~
\subfloat{\includegraphics[width=0.47\linewidth]{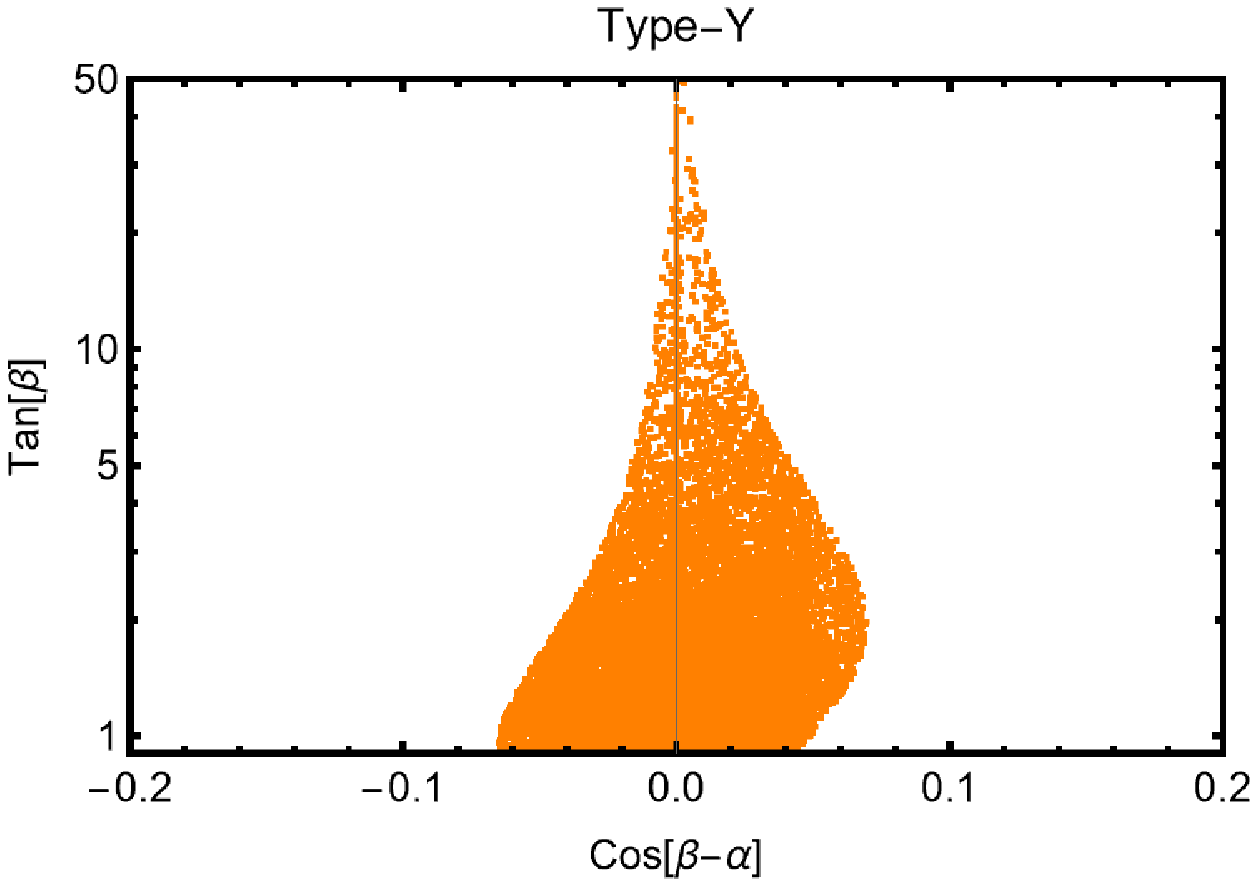}}
\vspace*{-2mm}
\caption{Allowed regions in the plane  $[\cos(\beta- \alpha), \tan\beta] $ for the $h$ signal strengths measured at the LHC in the four types of 2HD+a configurations that do not induce tree-level FCNCs.}
\label{fig:h_constr}
\end{figure}

 To avoid the risk of combining the ATLAS and CMS results given in Table \ref{tab:h_signal}, we consider only the ATLAS results (those of CMS are rather similar),  we show in Fig.~\ref{fig:h_constr} the regions in the $[\cos(\beta-\alpha),\tan\beta]$ plane which are allowed  at the 95\% confidence level (CL) by the combined constraints on the Higgs couplings to gauge bosons and fermions in the context of a 2HDM in their four specific realizations, namely Type I and II (upper row) and X and Y (lower row). As we are considering only the CP-even state $h$ whose couplings are not altered by the presence of the additional pseudoscalar Higgs boson, the results shown in the figure are also valid in our 2HD+a scenario.\footnote{Such an analysis has been performed in a complete and sophisticated way by the ATLAS collaboration in the four 2HDM types; see Ref.~\cite{ATLAS:2021vrm} and its Fig.~20. Unfortunately it cannot be used in our context since first, it involves experimental cuts (such as a rapidity cut $y_H <2.5$) and  second, it does not consider values of $\tan\beta$ higher than 10. Nevertheless, we find a qualitative agreement with the figure.}

One sees from the figure that in the Type II, Type Y and to some extent Type X scenarios, $|\cos(\beta-\alpha)|$ is constrained to be small or close to zero, $|\cos(\beta-\alpha)| \lsim 0.1$, for any value of $\tan\beta$ that we varied from a minimum of $\tan\beta=0.3$ to a maximum of $\tan\beta=50$ which are the values that allow for perturbative top  and bottom quark Yukawa couplings in the Type II and Type Y scenarios. 

The reason is that in these three models, one of the Yukawa couplings to the $b$-quark or $\tau$-lepton would be enhanced at large $\tan\beta$ values if one is not in the SM-like configuration $\cos(\beta-\alpha) \to 0$. This then forces the fermionic signal strengths $\mu_{bb}$ and/or $\mu_{\tau \tau}$ to depart from the unit values to which they are experimentally constrained to be close, as shown in Table \ref{tab:h_signal}.  In turn, in the Type I model,  no coupling to fermions is enhanced at high $\tan\beta$ and, thus,  $\cos(\beta-\alpha)$ can significantly deviate from unity for all considered values of $\tan\beta \gsim 0.3 $ without affecting too much the fermionic $h$ signal strengths. Note that at small $\tan\beta$, the deviations in $g_{hff}$ can be larger, but the measured values are below the SM expectation which forces $g_{hVV}$ to be less than unity and hence $\cos(\beta-\alpha) \neq 0$. 

Another comment to be made is that in earlier analyses of these three models, see e.g. Ref.~\cite{Arcadi:2019lka}, there were narrow  ``arms'' at  $\cos(\beta-\alpha) \gsim +0.1 $ which corresponded to the so-called ``wrong-sign'' Yukawa regime \cite{Ferreira:2014naa} in which the $h$ couplings to down-type quarks and/or leptons are equal in magnitude to those of the SM Higgs boson for $\cos(\beta-\alpha) =0$ but opposite in sign. These regions have been substantially reduced by the recent and more precise measurements and only a few such points are left in our scan.

In any case, all these constraints from the $h$ signal strengths   can be simultaneously satisfied in the so-called  alignment limit, $\alpha = \beta- \frac{\pi} {2}$ \cite{Pich:2009sp,Craig:2013hca,Carena:2013ooa,Bernon:2014nxa}.  In this case, the couplings of the CP-even $h$ and $H$ states to gauge bosons are by construction such that $g_{hVV}\!=\!1$ as for the SM Higgs and $g_{HVV}\!=\! 0$ as is the case for the pseudoscalar $A$ in our CP-conserving model. The Higgs couplings to fermions in the alignment limit, also given in Table~\ref{table:2hdm_cplgs}, are such that $g_{htt}\!=\!g_{hbb}\!=\!g_{h\tau\tau} \! \to \! 1$ again as for the SM-Higgs  and $g_{Hff} \! \to \! g_{A^0ff}$ which means that all couplings of the heavier CP-even $H$ reduce to those of the 2HDM pseudoscalar $A^0$. 

Finally, let us note that for the couplings between two Higgs and one gauge boson, those involving $h$ vanish in the alignment limit, $g_{hAZ}=g_{hH^\pm W^\mp}=0$,  while those involving $H$ become maximal, $g_{HAZ}\!=\!g_{HH^\pm W^\mp}\!=\!1$. 


\subsection{Constraints from flavor physics} 

Let us now turn to the constraints that come from flavor physics, focusing first on the heavier 2HDM Higgs bosons in the four considered configurations,  Type I, II, X and Y.  While these scenarios are free from tree-level FCNCs by construction, they are nevertheless induced at the loop level. Severe constraints come from processes in which there are $b\to s$ transitions at the basic level, which have rates that are essentially sensitive to the parameters entering the charged Higgs sector, namely $M_{H^{\pm}}$ and $\tan \beta$. The Type II and Y models are the ones that are most affected as they involve the $H^\pm$ coupling component to bottom quarks $g_{Abb}$ that is proportional to $\tan\beta$ and which can be strongly enhanced at large $\tan\beta$ values. In the configurations II and X with enhanced Higgs couplings to muons, additional constraints come from $B$-meson decays such as $B_s \to \mu^+ \mu^-$ and $B \rightarrow K \mu^+ \mu^-$~\cite{Arnan:2017lxi}.

Nevertheless, the most stringent constraints are due to the loop induced decay process $B\rightarrow X_s \gamma$. Indeed, at the fundamental level, the radiative decay $b\to s\gamma$ proceeds in a 2HDM or 2HD+a  through a triangular loop involving $W$ and $H^\pm$ bosons along with top quarks. At leading order, the contribution of the $H^\pm$ states to the amplitude is proportional to the two combinations of couplings $g_{Auu}^2$ and $g_{Auu} g_{Add}$ and hence, in the four 2HDM configurations one has contributions that are proportional to $\tan^2\beta$ as in Type II and Type Y scenarios or are proportional to $\cot^2\beta$ as in Type I and Type X scenarios. As a result, and taking into account the most up-to-date value of  the branching fraction BR($B\rightarrow X_s \gamma$) as measured by the LHCb collaboration \cite{HFLAV:2016hnz} and the most precise calculation performed at NNLO in Ref.~\cite{Misiak:2017bgg}, one obtains the following constraints, depending on the considered 2HDM type: 
\begin{eqnarray}
\text{ Type \ II \ or \ Y}\ &:& ~~~ M_{H^\pm} \gsim 800 \ \text{GeV \ for \ any \ }  \tan\beta  \, , \nonumber \\
\text{Type \ I \ or \ X}\ &:& ~~~ M_{H^\pm} \gsim 500 \  \text{GeV \ for \ } \tan\beta \lsim 1  \, . 
\end{eqnarray}
As can be seen,  these bounds are rather severe. Only in models of Type I and X and for $\tan\beta \gsim 2$ that one has the loose bound $M_{H^\pm} \gsim 80$ GeV from LEP searches (see later). 

Turning now to the case of a possibly very light pseudoscalar particle, it can affect a large variety of low energy processes, especially those involving $b$-quarks  which have enhanced couplings to $a$ at high $\tan\beta$ in Type II and Y scenarios.  This is for instance, the case of the decay rates of $B$ and $K$ mesons  which can be substantially modified by the emission of a very light $a$ state  \cite{Dolan:2014ska}. At high $\tan\beta$ values and, again in the Type II scenario, very constraining processes are the decays $\Upsilon \rightarrow a \gamma$, $B_s \rightarrow \mu^+ \mu^-$ and $B \rightarrow K \mu^+ \mu^-$. In particular, the mode $B_s \to \mu^+\mu^-$ can potentially receive large contributions from the exchange of a light $a$ state if it has large couplings to $b$-quarks and muons as is the case in the Type II model. In the case of the Type X configuration,  constraints as severe as in the Type II case could be derived from the searches of a light leptophilic scalar boson which have been performed rather recently by the BaBar collaboration \cite{BaBar:2020jma}. The corresponding limits on the mass of $a$ (even if it cannot be emitted on-shell) are included in our numerical analysis via a procedure discussed in Ref.~\cite{Arcadi:2017wqi} to which we refer for details.  Note that there are also constraints from violation of lepton universality in the decays of the $Z$ boson and the $\tau$ lepton \cite{ParticleDataGroup:2020ssz}. 

All constraints of this type  can be fulfilled at this early stage (further stronger constraints from direct LHC searches of the pseudoscalar $a$ boson will be discussed in subsection 3.5), by adopting a lower bound of 10 GeV on the mass of the $a$ state. 

\subsection{Constraints from electroweak precision measurements} 

Another set of strong constraints on the 2HD+a model emerges from electroweak high precision measurements, in particular the one of the effective electroweak mixing angle $\sin^2\theta_W$ and of the $W$ boson mass $M_W$. The by far dominant set of radiative corrections to these two quantities is the one that affects the so-called $\rho$ parameter,  which measures the strength of the neutral to charged currents ratio at zero-momentum transfer \cite{Veltman:1976rt}. It is defined by   
\beq 
\Delta \rho = {\Pi_{WW}(0)}/{M_W^2}  - {\Pi_{ZZ}(0)}/ {M_Z^2} \, , 
\eeq
where $\Pi_{VV}$ are the transverse parts of the $V\!=\!W,Z$ boson two-point functions or self-energies. This parameter strongly constrains the mass splitting between particles that belong to the same SU(2) isodoublet as they give contributions that are quadratically dependent on the masses. In a 2HDM scenario for instance, they force the masses of the additional heavy $A^0,H$ and $H^\pm$ states to be very close in mass, $M_{A^0} \approx M_H \approx M_{H^\pm}$ \cite{Baak:2011ze,Baak:2014ora}. Note that the SM-like $h$ boson also contributes to the $\rho$ parameter but the contribution is only logarithmic and it is already included in the fit of the SM data in the limit where $h$ is SM-like, i.e. in the alignment limit discussed previously.   
 
In our 2HD+$a$ model, there are not only contributions from the 2HDM extra $A^0,H,H^\pm$  bosons but also additional contributions due to the extra pseudoscalar Higgs boson $a^0$ which mixes with the 2HDM pseudoscalar state with an angle $\theta$. Outside the alignment limit that we will consider from time to time, the full contribution to the $\rho$ parameter is given by  
\begin{align}
\Delta \rho & = \frac{\alpha_{\rm QED} (M_Z^2)}{16 \pi^2 M_W^2 (1 -M_W^2/M_Z^2)}\bigg \{ \sin^2(\beta-\alpha) \big[ 
f(M^2_{H\pm},M^2_H) + \cos^2\theta f(M^2_{H\pm},M^2_A) \nonumber \\ 
&  + \sin^2\theta f(M^2_{H\pm},M^2_a) - \cos^2\theta f(M^2_A,M^2_H) - \sin^2\theta f(M^2_a,M^2_H) \big]  \nonumber\\
&  + \cos^2(\beta-\alpha) \big[ f(M^2_{H\pm},M^2_h) + \cos^2\theta f(M^2_{H\pm},M^2_A)  \nonumber \\ 
& + \sin^2\theta f(M^2_{H\pm},M^2_a) - \cos^2\theta f(M^2_A,M^2_h) - \sin^2\theta f(M^2_a,M^2_h) \big]\bigg \},
\end{align}
where $\alpha_{\rm QED}$ is the fine structure constant evaluated at $M_Z$ and the loop function $f$ reads
\beq
f(x,y) = x+y- \frac{2 x y}{x-y} \log \frac{x}{y} \, . 
\eeq
This quantity vanishes if the loop particles are mass degenerate, $f(x,x)=0$, and for a large splitting  $x \gg y$, one would have $f(x,0)=x$ and, hence, possibly large contributions. 

One can also take into account the subleading contributions to the electroweak obser\-vables beyond the $\rho$ parameter and, for instance, consider the ones of the Peskin-Takeuchi $S,T,U$ parameters \cite{Peskin:1991sw}. In this scheme, the largest contribution $T$ is in fact simply the $\Delta \rho$ contribution, $T \propto \Delta \rho -\Delta \rho|_{\rm  SM}$, while $S$ and $U$ describe new contributions to, respectively, neutral current processes at different energies and the $W$ mass from new  charged currents. 

The central values for the three variables in the case of the SM are as follows \cite{Baak:2011ze,Baak:2014ora}:
\begin{equation}
 {\cal O}^{\rm SM}= (S,T,U)^{\rm SM} =  (0.05,0.09,0.01)\, . 
\end{equation}
A global fit to all electroweak precision observables available before the new CDF measurement of the $W$ mass to be discussed later, has been made in Refs.~\cite{Baak:2011ze,Baak:2014ora} and leads to the following $\chi^2$ as a function of the departure of the three variables from their SM values
\begin{equation}
\label{eq:chi2}
\chi^2=\sum_{i,j}( {\cal O}_i-{\cal O}_i^{\rm SM}){\left(\sigma_i V_{ij} \sigma_j\right)}^{-1}( {\cal O}_j-{\cal O}_j^{\rm SM}) \, , 
\end{equation}
where the standard deviations and the covariance matrix are given by 
\begin{equation}
\sigma=(0.11,~0.13,~0.11) \ , \ \ \   V= \left(
\begin{array}{ccc} 1 & 0.9 & -0.59 \\ 0.9 & 1 & -0.83 \\ -0.59 & -0.83 & 1 \end{array} \right) \, .
\label{eq:covariance}
\end{equation}
For our numerical evaluation, we have performed a scan on the parameters of the 2HD+a model over the  following ranges,
\begin{align}
& \tan\beta \in [1,60], \ \ |\cos(\beta-\alpha)|<0.2 \, , \nonumber \\
& ~[M_H, M_A, M_{H^\pm}]  \in [ (125\,{\rm GeV}, 90\,{\rm GeV}, 80\,{\rm GeV}), {\rm 1\,TeV}] \, , \nonumber\\
& M_a \in [10,400]\, {\rm GeV}, \ \ \sin \theta \in [0.1,0.8], 
\label{eq:scans}
\end{align} 
where the 2HDM Higgs masses were taken to be such that $M_H> M_h$ (by construction) and  $M_{H^\pm} \gsim M_W, M_A \gsim M_Z$ which are the limits obtained from the negative LEP2 searches as will be seen later. In addition we have imposed the hierarchy $M_a < M_A$.

\begin{figure}[!ht]
\vspace*{-.4mm}
\begin{center}
\mbox{
\subfloat{\includegraphics[width=0.42\linewidth]{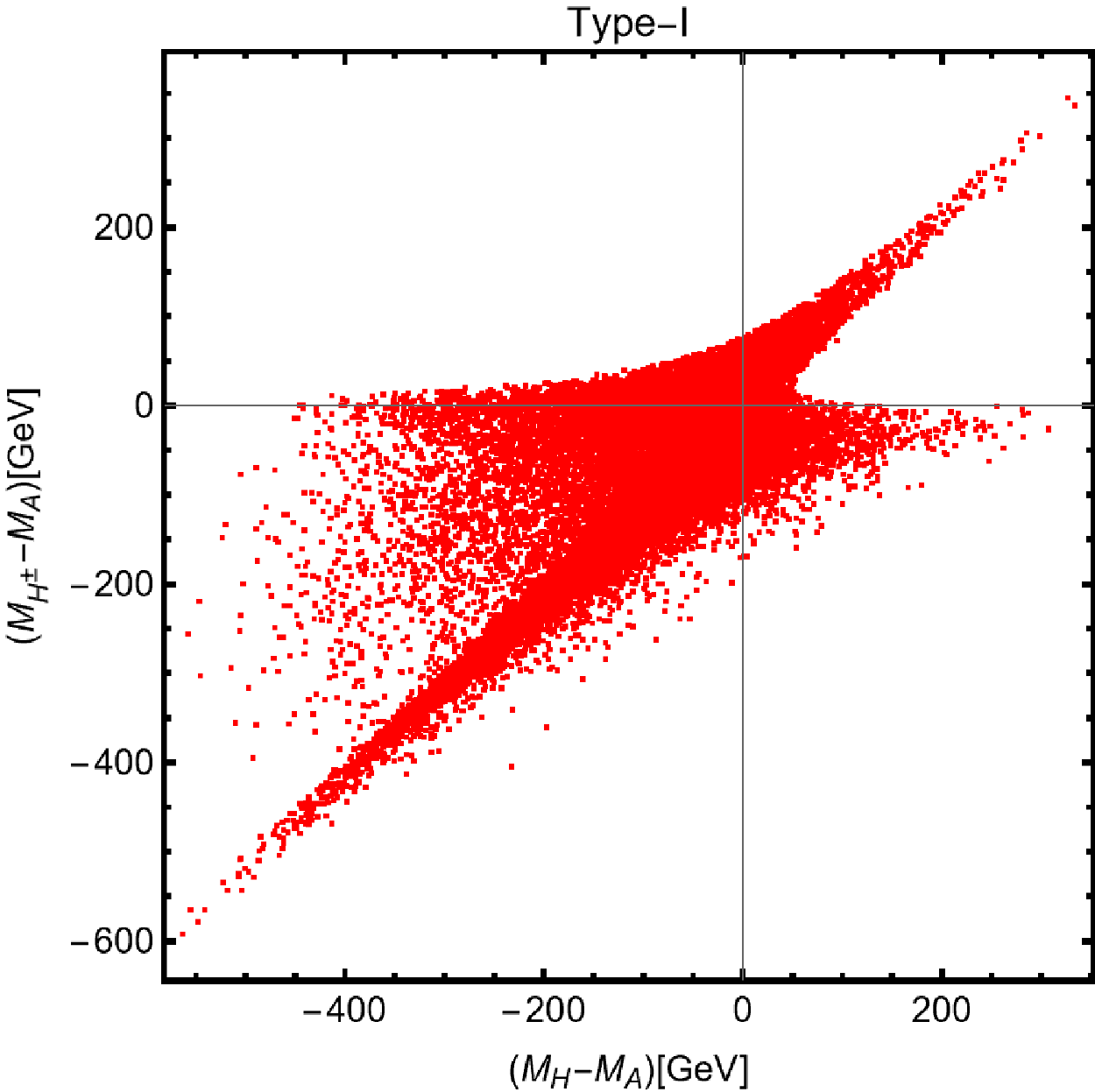}}~
\subfloat{\includegraphics[width=0.42\linewidth]{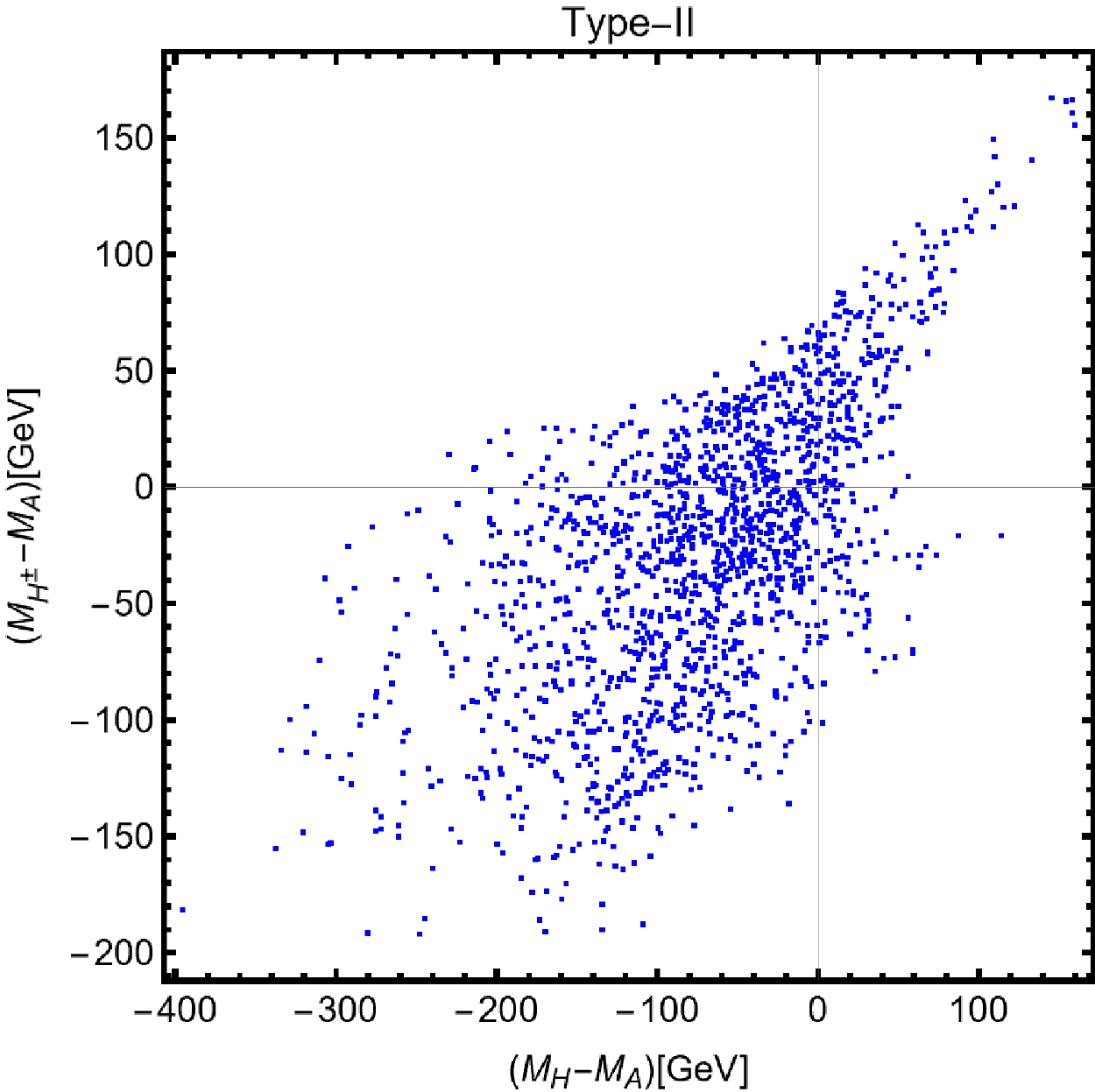}}
}\\[2mm]
\mbox{
\subfloat{\includegraphics[width=0.42\linewidth]{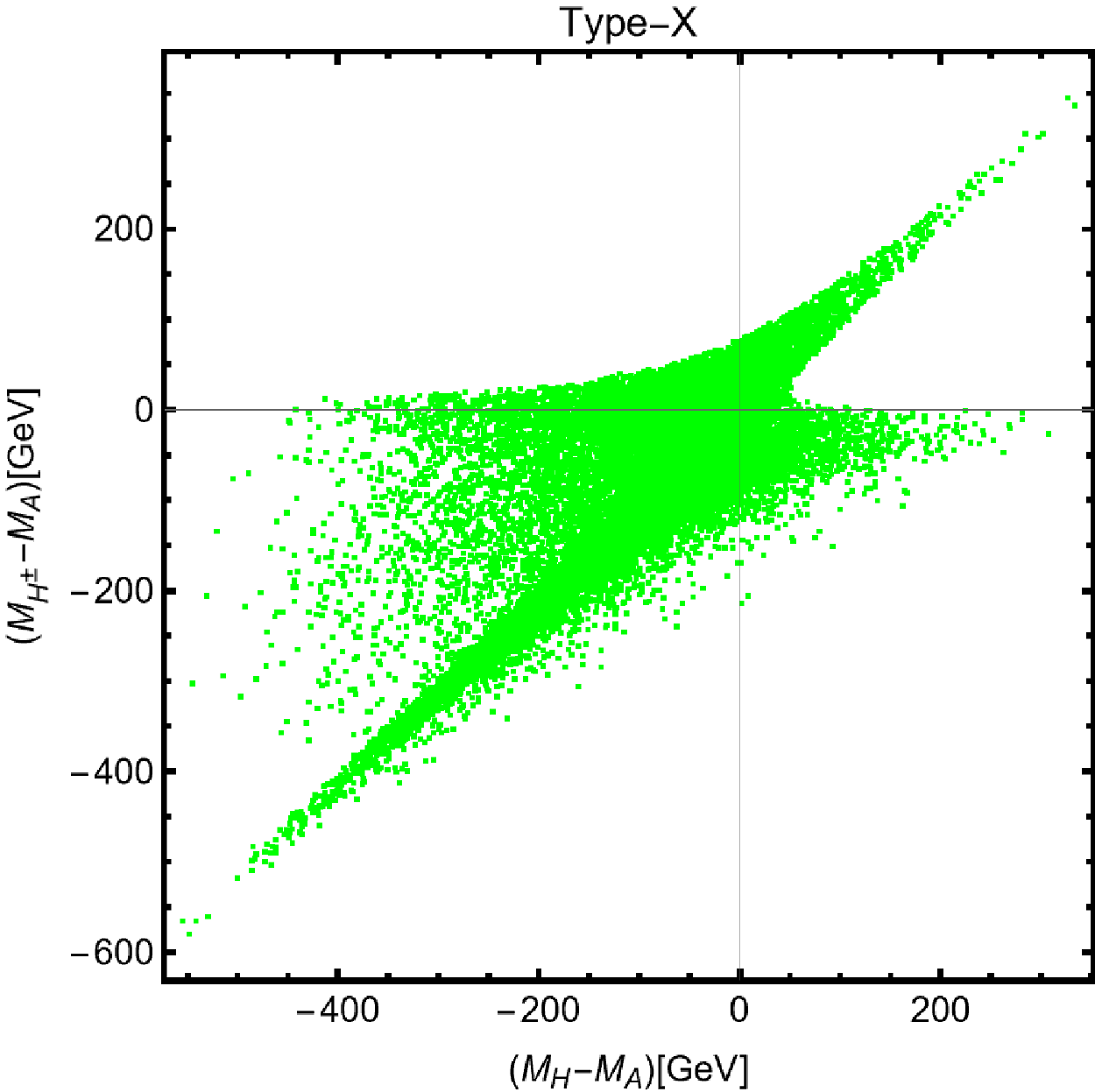}}~
\subfloat{\includegraphics[width=0.42\linewidth]{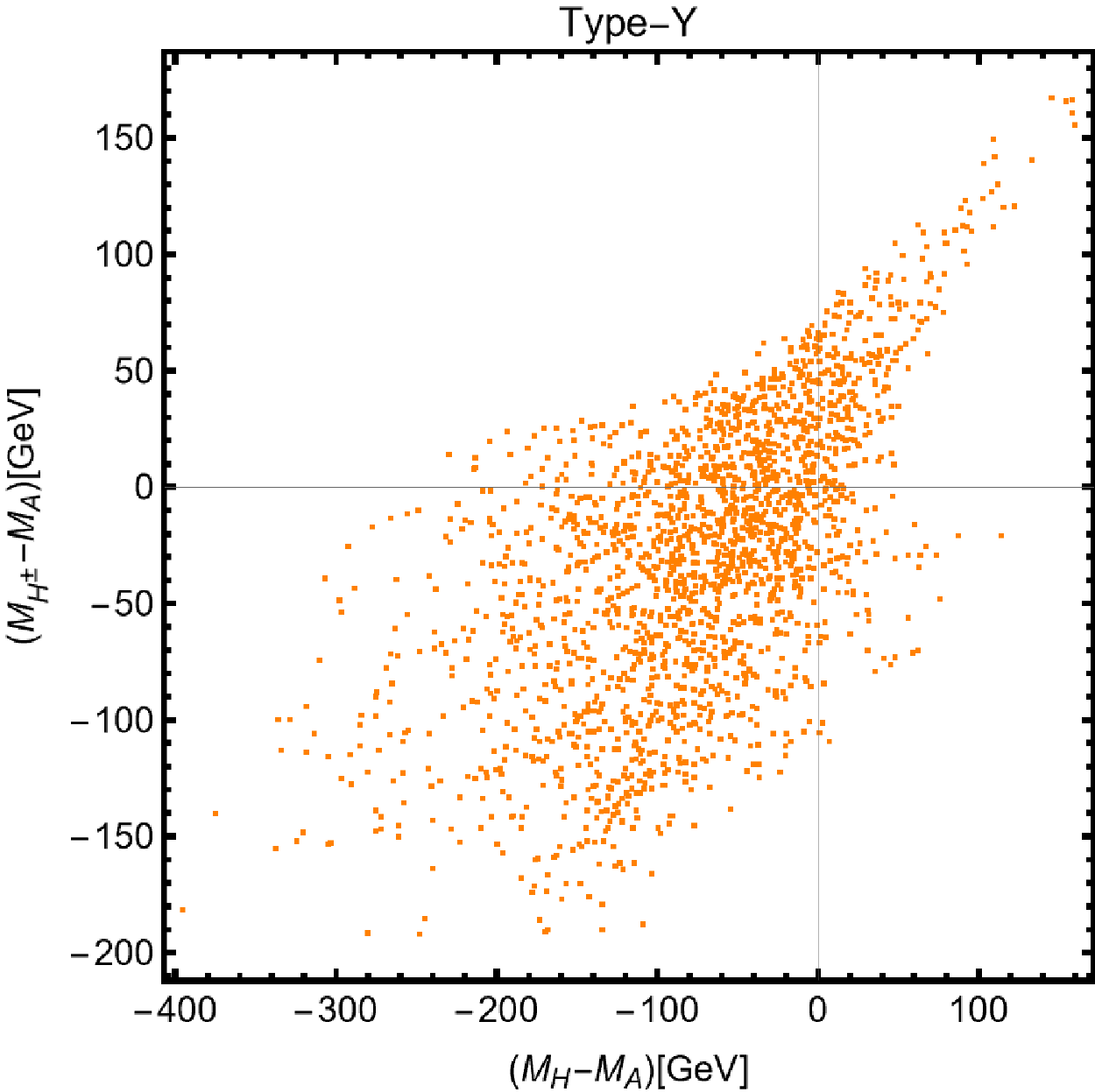}}
}
\end{center}
\vspace*{-5mm}
\caption{2HD+a model points in the $[M_H-M_A, M_H^{\pm}-M_A]$ plane in the four configurations I, II, X and Y, allowed at the 95\% CL by constraints on the high precision electroweak data (using the SM-fit for the $W$-boson mass), and including theory constraints on the quartic couplings, constraints from the $h$ signal strengths and $B$-physics constraints.} 
\label{fig:split_scan}
\vspace*{-.3mm}
\end{figure}

The amount of model points which passes the theoretical constraints discussed before, as well as the bounds on the Higgs signal strengths and flavor physics are displayed, for the Type I, II, X and Y 2HD+a model in Fig.~\ref{fig:split_scan}. The model points passing all these constraints have been shown in the $[M_{H}-M_A,M_{H^\pm}-M_A]$ planes. As can be seen, a significant Higgs mass splitting is possible, particular in Type I and X scenarios. These two configurations give similar results, as is also the case for Type II and Y scenarios.  

Let us now briefly comment on the impact of the new $M_W$ measurement performed by the CDF collaboration \cite{CDF:2022hxs} which turned out to be significantly different not only from the expectation in the SM, about $7\sigma$, but also from other measurements performed in  other experiments. In the context of the 2HDM and 2HD+a models, this deviation could be explained simply by allowing for a larger splitting between the $H,A$ and $H^\pm$ masses. In Ref.~\cite{Arcadi:2021zdk}, numerical examples have been given in the Type II and Type X scenarios to illustrate this possibility and we extend the discussion here to the other two configurations.   

\begin{figure}[!ht]
    \centering    \subfloat{\includegraphics[width=0.46\linewidth]{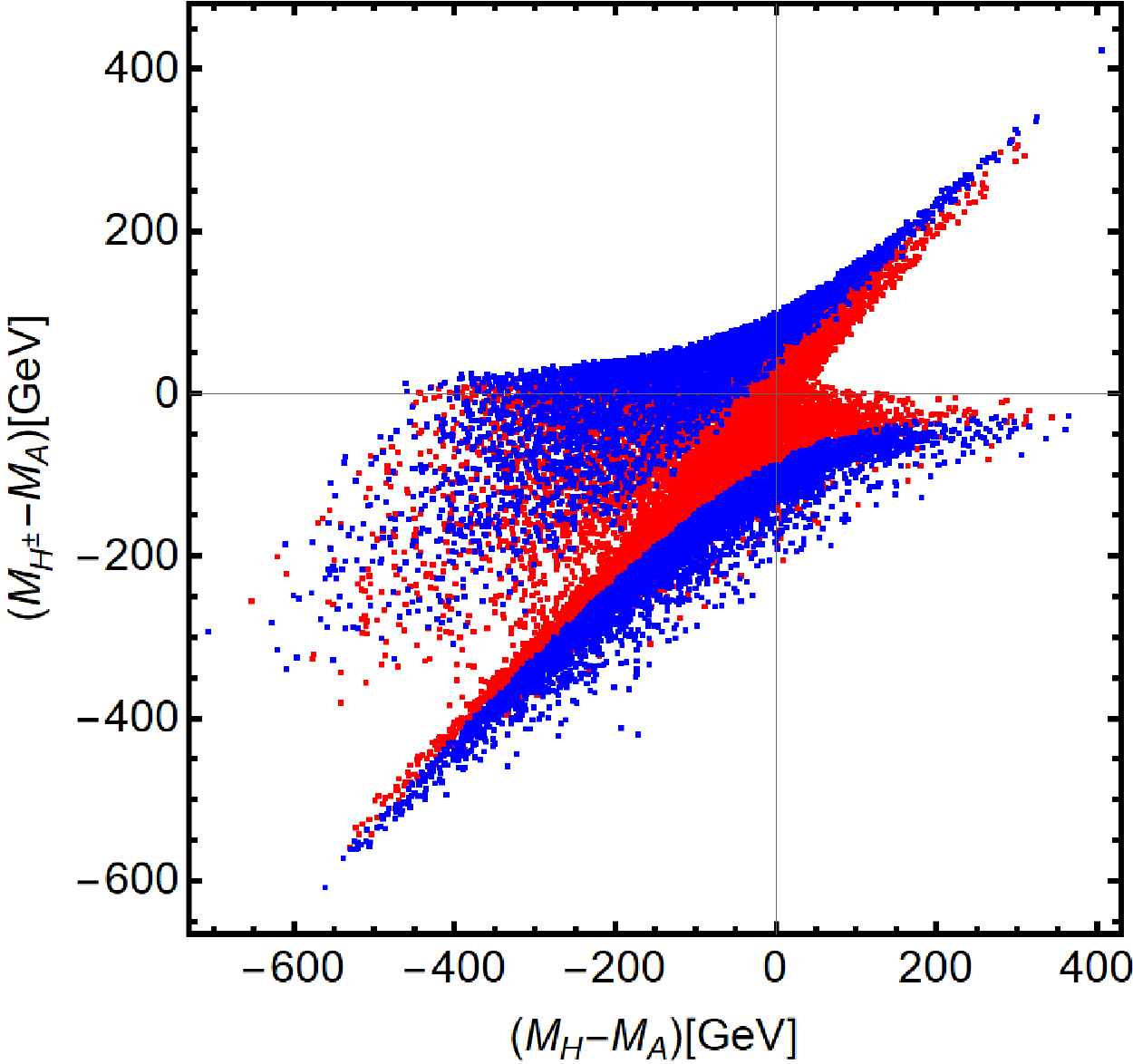}}~~    \subfloat{\includegraphics[width=0.46\linewidth]{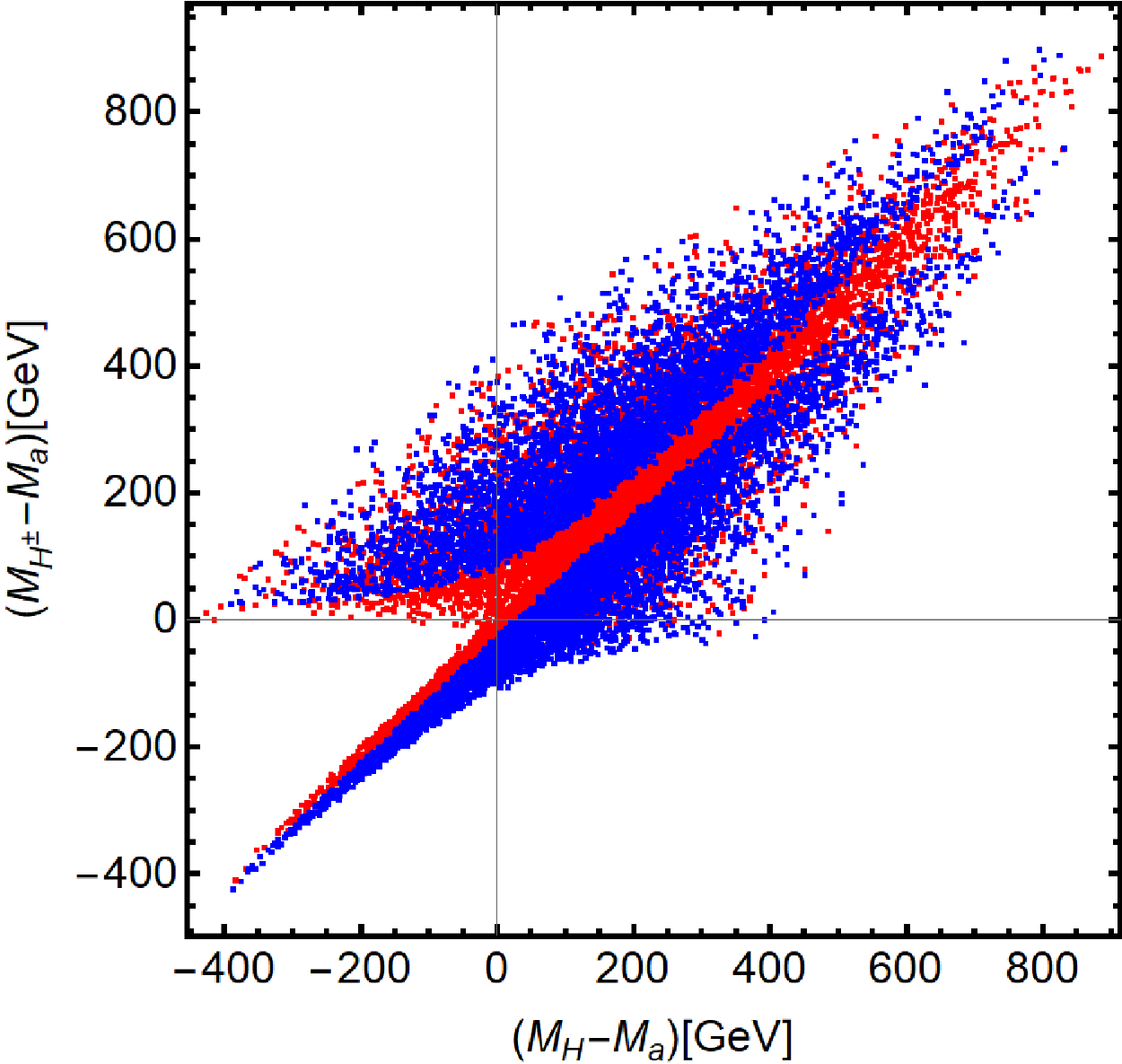}}
    \caption{The 2HD+a model points complying with theoretical and electroweak constraints assuming the SM fit (red points) or accounting for the new CDF $M_W$ value (blue points); the left/right panels show the result of the scans in the $[M_H-M_{A/a},M_{H^{\pm}}-M_{A/a}]$ planes.}
    \label{fig:plotW}
\end{figure}

We compare in Fig.~\ref{fig:plotW} the outcome of scans of the 2HD+a model, when varying the parameters  in the same range of values as before assuming either a purely SM fit for $M_W$ (as given by the red points) or those that allow for an explanation for the new CDF measurement of $M_W$ (given by the blue points). The two panels of the figure show the results in, respectively, the $[M_H-M_A,M_{H^\pm}-M_A]$ and $[M_H-M_a,M_{H^\pm}-M_{a}]$ planes. For simplicity, only the theoretical bounds on the coupling of the scalar potential have been considered in addition to the electroweak observables, to avoid to treat individually the four different Yukawa configurations. Hence, the CDF $W$-boson mass anomaly could be  easily explained in our 2HD+a scenario by simply allowing for a more significant splitting between the masses of the $H,A,H^\pm$ states.

\subsection{Impact of the muon g--2}  
 
We come now to the constraint from  the anomalous magnetic moment of the muon, $a_\mu= \frac12 (g-2)_\mu$, which has been recently 
measured by the Muon $g-2$ collaboration at Fermilab~\cite{Abi:2021gix} and which, when combined with a previous measurement at Brookhaven ~\cite{Bennett:2006fi}, gives~\cite{Abi:2021gix}
\beq \label{eq:4sigma}
    a_\mu^{\rm EXP}= (116 592 061 \pm 41)\times10^{-11}\, . 
\eeq
This results implies a $4.2 \sigma$ deviation from the consensual SM value generally adopted by theorists \cite{Aoyama:2020ynm}. It is tempting to attribute this discrepancy to new physics beyond the SM and in particular to the 2HD+a model that we are considering here, ignoring the possibility that it could partly or entirely be due to unknown uncertainties (as suggested by a debate triggered by a conflicting theoretical value obtained in a lattice calculation  \cite{Borsanyi:2020mff}).

\begin{figure}[!ht] 
\vspace*{-2.3cm}
    \centering    \subfloat{\includegraphics[width=0.99\linewidth]{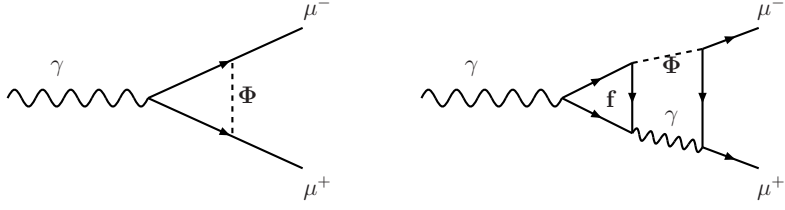}}
\vspace*{-16.5cm}
 \caption{Feynman diagrams for the one-loop (left) and two-loop (right) contributions of a Higgs boson $\Phi=h,H,H^\pm, A,a$ to the muon $(g-2)$.}
    \label{fig:g-2}
\vspace*{-.2cm}
\end{figure} 

The 2HD+a $\Phi=h,H,H^\pm, A,a$ bosons contribute to the $(g\!-\!2)_\mu$ in two ways. There is first a one-loop contribution when they are exchanged between the two muon legs in the $\gamma \mu^+\mu^-$ vertex as is shown in the left-hand side of  Fig.~\ref{fig:g-2} (in the $H^\pm$ case, the initial photon couples to the $H^+H^-$ states and a $\nu_\mu$ neutrino is exchanged between the two muon lines). Such diagrams give rise to contributions that scale like $m_\mu^4/ M^2_\Phi \times g^2_{\Phi \mu \mu}$; we assume of course the universality of the Higgs couplings and hence, $g_{\Phi \mu\mu}= g_{\Phi \tau\tau}$.  As will be discussed later, such a contribution is sizable only for enhanced $g_{\Phi \mu \mu}$ couplings, hence favoring the Type II and X scenarios, and light, namely $M_{\Phi}\lesssim \mathcal{O}(100 \,\mbox{GeV})$, exchanged neutral bosons. In our setup, only the pseudoscalar $a$ will be considered in this mass range. Within a good approximation, the one loop contribution to $(g\!-\!2)_\mu$ can be written as \cite{Dedes:2001nx,Djouadi:1989md}: \beq
\Delta a_\mu^{\rm 1\!-\!loop} = - \frac{ \alpha_{\rm QED} }{8 \pi \sin^2\theta_W}  \frac{ m^4_\mu}{M_W^2 M_a^2} \; g_{a\mu \mu}^2 \; \bigg[ {\rm log} \bigg( \frac{M_a^2}{m_\mu^2} \bigg) - \frac{11}{6} \bigg] \, .
\eeq
A comparable or even larger contribution to the one discussed above comes from Barr-Zee type diagrams \cite{Barr:1990vd,Chang:2000ii,Larios:2001ma,Ilisie:2015tra},  a representative example of which is shown in the right panel of Fig.~\ref{fig:g-2}. Their contribution is enhanced with respect to $\Delta a_\mu^{\rm 1-loop}$ by a factor $m_f^2/m_\mu^2$, compensating the higher $\alpha_{\rm QED}$ power suppression. The two-loop contribution to $\Delta a_\mu$, restricting for simplicity to the exchange of $a$, can be written as \cite{Chang:2000ii,Larios:2001ma,Ilisie:2015tra}
\beq
\Delta a_\mu^{\rm 2\!-\!loop} = \frac{\alpha^2_{\rm QED} }{8 \pi^2 \sin^2\theta_W} \; \frac{m_\mu^2}{M_W^2} g_{a\mu \mu}\; \sum_f g_{aff} N_c^f Q_f \; \frac{m_f^2}{M_a^2} \; F \bigg( \frac{m_f^2}{M_a^2} \bigg) \, , 
\eeq
where $F$ is the loop function of the mass ratio $a= m_f^2/M_a^2$ and it is defined by 
\beq
F(r) = \int_0^1 {\rm d}x \frac{ \log (r)- \log [x(1-x)] } {r -x(1-r) } \, .
\eeq
Notice that the analytical expressions provided above serve just as an illustration of the leading contributions. The numerical results illustrated below are based on a more detailed computation, adapting to the 2HD+a model the complete expressions provided in Ref.~\cite{Ilisie:2015tra}, and including all the additional Higgs bosons running in the loops.  

\begin{figure}[!ht]
    \centering    \subfloat{\includegraphics[width=0.41\linewidth]{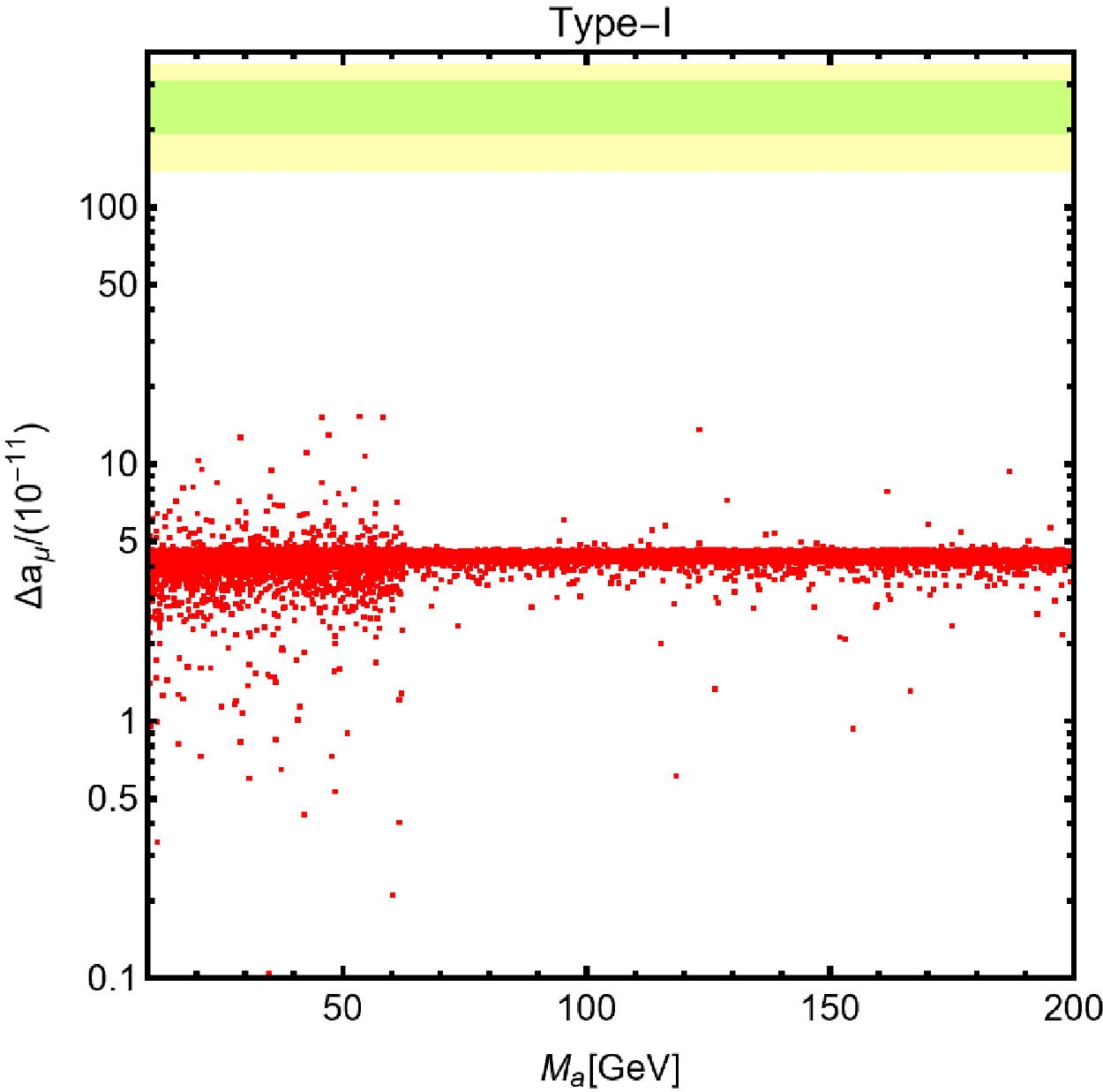}}~~    \subfloat{\includegraphics[width=0.41\linewidth]{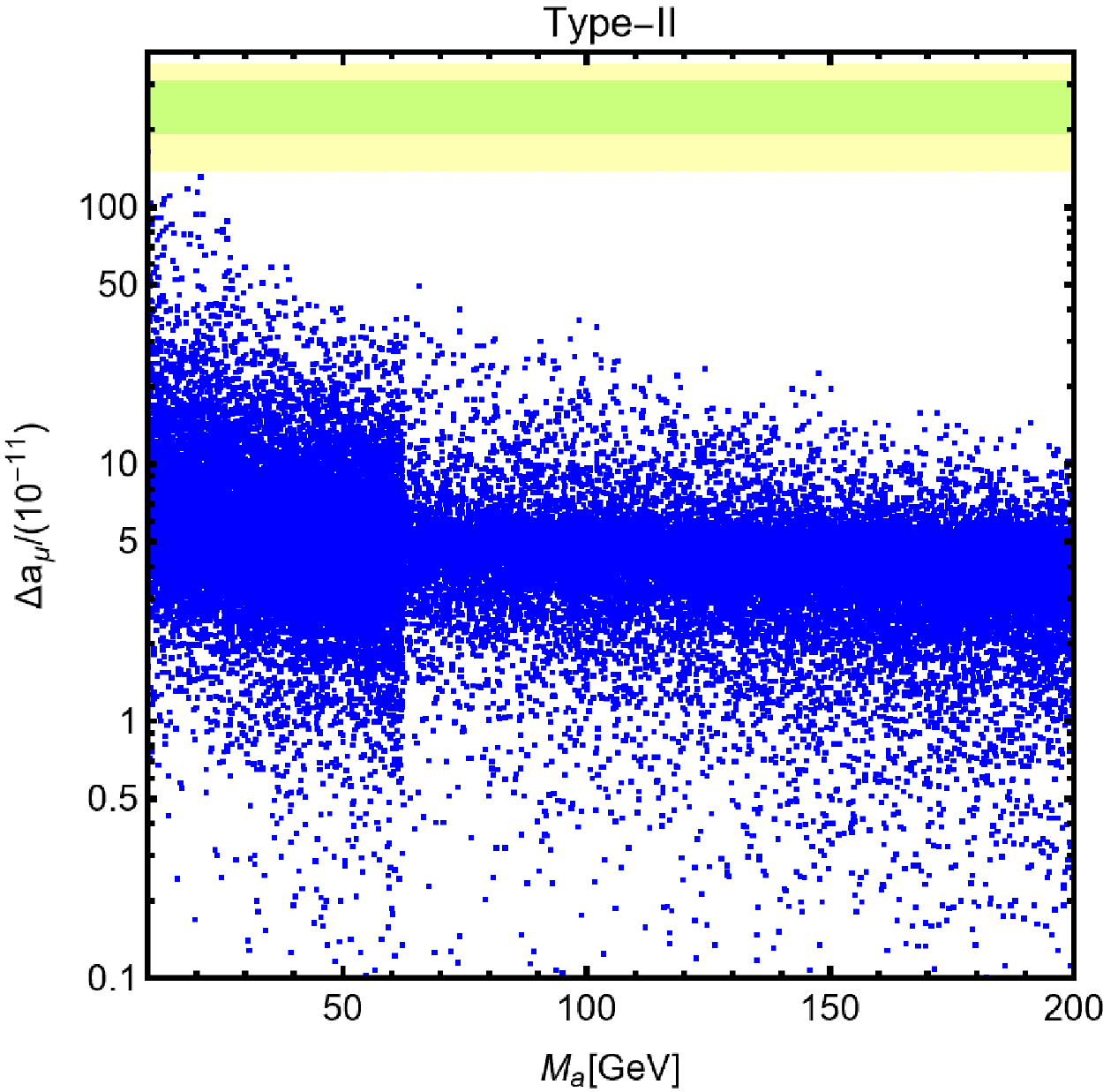}}  \\  \subfloat{\includegraphics[width=0.41\linewidth]{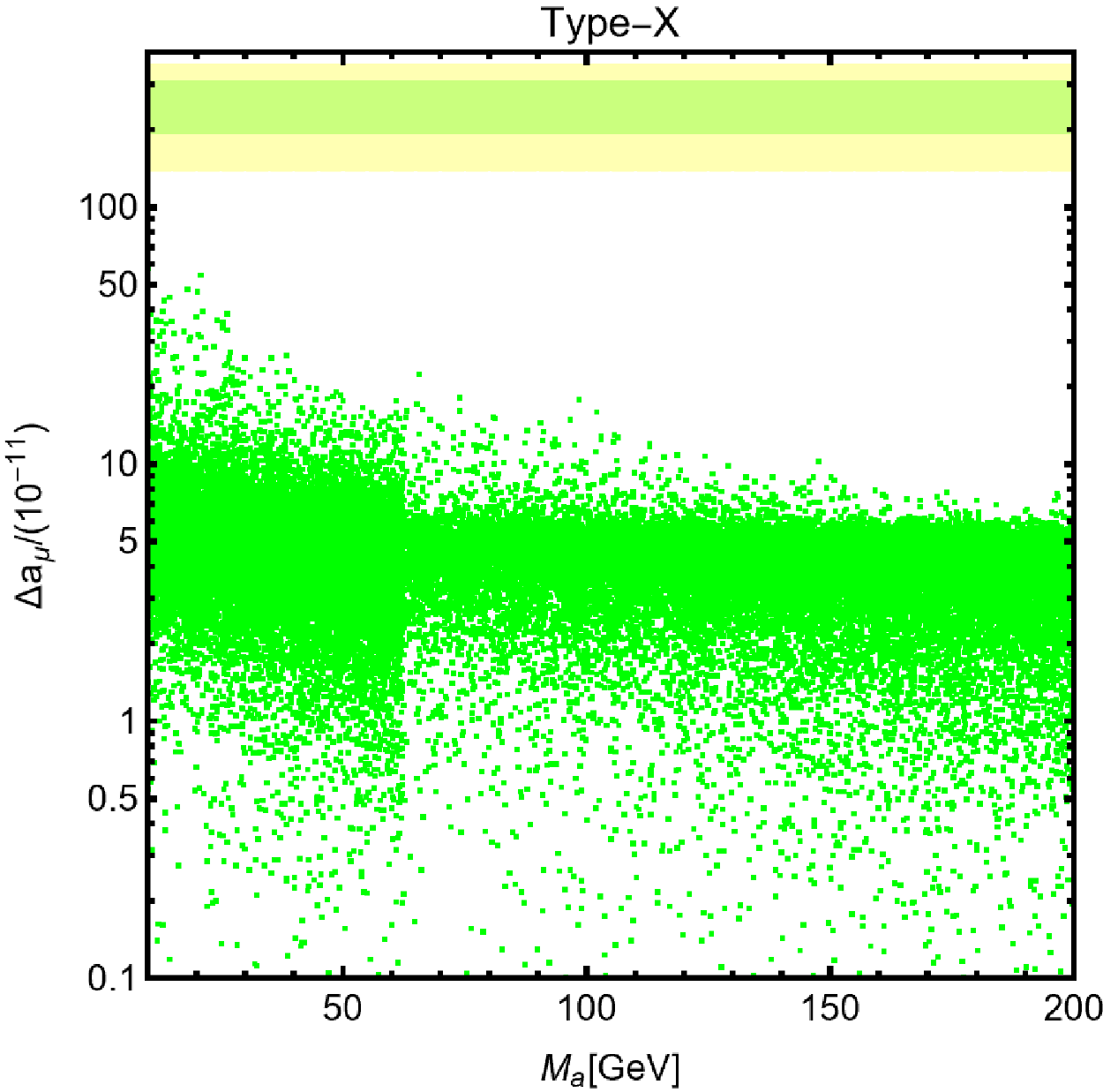}}~~    \subfloat{\includegraphics[width=0.41\linewidth]{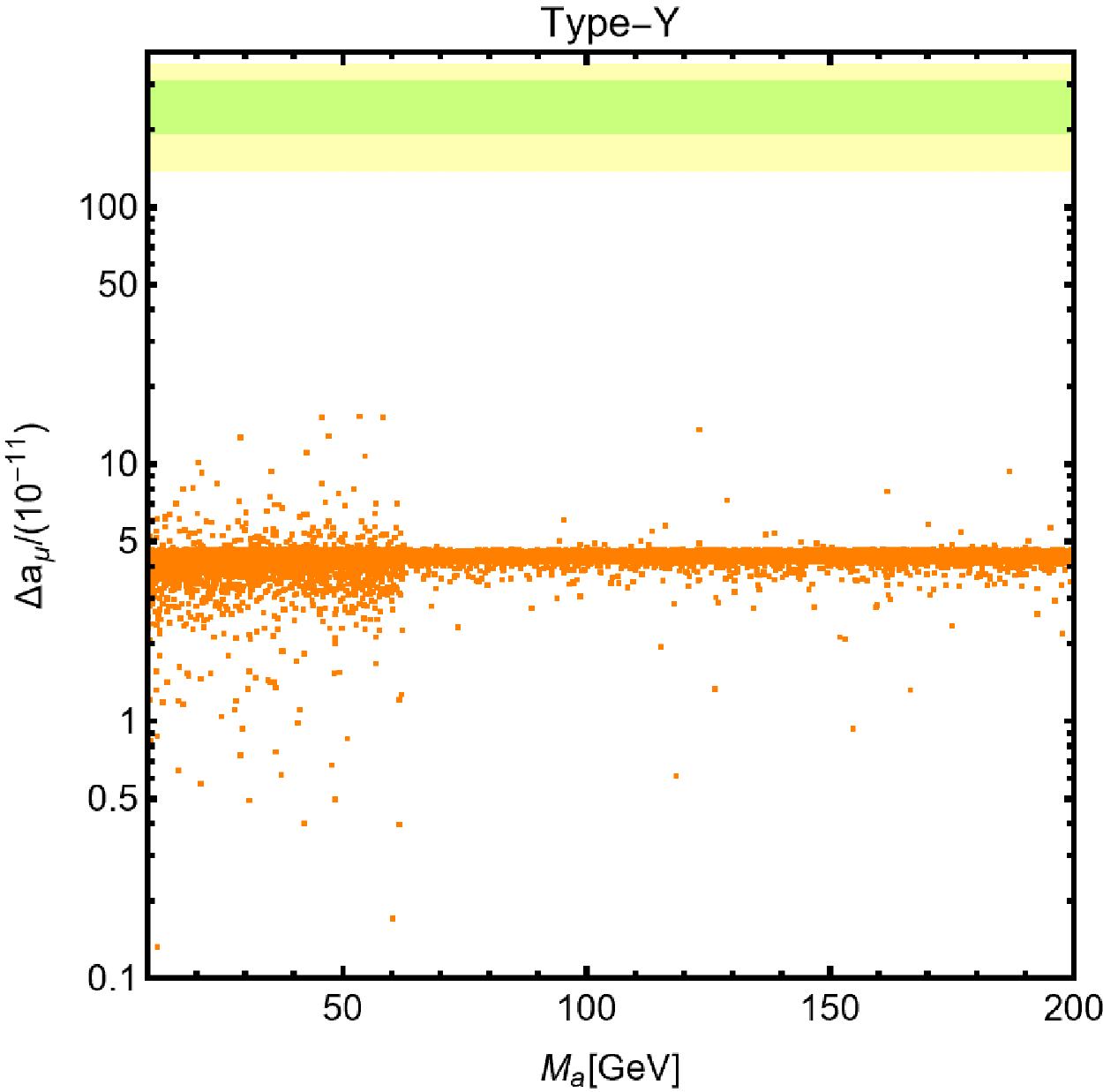}}
 \caption{Values of $\Delta a_\mu$ in units of $10^{-11}$ as a function of $M_a$ for the four 2HD+a model types. The colored points are obtained performing a scan of the model parameters as illustrated in the text.}     
 \label{fig:g-2plot}
\end{figure}

Fig.~\ref{fig:g-2plot} shows the value of $\Delta a^\mu$ as a function of $M_a$ for the model points obtained from the parameter scan illustrated in the previous section but in which, for simplicity,  we have focused on the regime $M_a \leq 200\,\mbox{GeV}$. The four panels of the figure consider individually the four different Yukawa configurations, namely Type I (red points),  II (blue points), X (green points) and  Y (orange points). The results are compared to the $1\sigma$ ($2\sigma$) region of the Fermilab measurement drawn in green (yellow). As can be seen, the Type I and Type Y models, featuring $\tan\beta$ suppressed interactions of the neutral Higgs bosons with the muon, provide a too small contribution to the $(g-2)_\mu$ anomaly, far below the experimental bands.

Hence, a sufficiently large contribution to $a_\mu$ can be achieved only in the Type II case when the mass of $a$ is less than a few 10 GeV, while the contribution in the Type I and  Type Y scenarios is far too small. In the Type X model, on the contrary, the contribution can get close to the experimental sensitivity but falls short to reach the experimental bands. The reason for such a behavior is that in Type II, it is the $b$-quark loop contribution that is enhanced by $\tan\beta$ (both $g_{a bb}$ and $g_{a\mu\mu}$ are proportional  to $\tan\beta$) while in Type X, only the loop involving $\tau$ leptons (which has a lower mass and no color factor compared to $b$-quarks) contributes significantly since only $g_{a\ell \ell } \propto \tan\beta$. As already discussed in Ref.~\cite{Arcadi:2021zdk}, the  $a$ contribution can be made sufficiently large in the Type X case as to explain the $(g-2)_\mu$ excess by invoking extremely large $\tan\beta$ values, $\tan\beta \approx 80$. While they lead to a non-perturbative $b$ Yukawa coupling in the Type II scenario, these $\tan\beta$ values are acceptable in Type X as the coupling $g_{abb}$ is not enhanced and the $\tau$-lepton coupling is still perturbative as a result of the smaller mass. Such extreme assignment of the value of $\tan \beta$ have not been considered in our present study as we assume $\tan\beta\leq 60$ in all cases. For this reason, contrary to Refs.~\cite{Arcadi:2021zdk,Arcadi:2022dmt}, no model points are present in the experimentally favored bands.
  
\subsection{Constraints from direct Higgs searches} 

The most stringent constraints on some of the 2HD+a configurations come from the direct searches at colliders of the additional Higgs bosons compared to the already observed $h$ state. As already mentioned before, bounds were already available on the 2HDM states from negative LEP2 searches at a c.m. energy up to $\sqrt s=209$ GeV \cite{ParticleDataGroup:2020ssz}: $M_A \gsim 90$ GeV in the associated production process $e^+ e^- \to hA$ and $M_{H^\pm} \gsim 80$ GeV in the pair production process $e^+ e^- \to H^+ H^-$. These bounds are independent of the value of $\tan\beta$ and, to a large extent, of the value of the angle $\alpha$ as there is some complementarity between various processes. Added to that, one has the ad-hoc assumption of $M_H > M_h=125$ GeV that we have introduced from the start and which is also favored by LEP2 and present data.\footnote{There is though still the possibility of a light Higgs boson with a mass $M_H \approx 96$ GeV \cite{Biekotter:2019kde} for which there was a slight excess of events at LEP2 and also more recently in CMS. We will not consider this possibility.}  

In the case of a light pseudoscalar $a$ with $M_a \!=\!{\cal O}({\rm a\;few\;10\; GeV})$, there are constraints from LEP1 at high $\tan \beta$ for the Type II and X models via searches of associated production $e^+ e^- \to b\bar b a$ and $e^+ e^- \to \tau^+\tau^- a$ \cite{Djouadi:1990ms,Barate:2003sz}. In addition, besides  constrains from searches of exotic $Z\! \to\! a \gamma$ decays induced by heavy fermion or Higgs boson loops at LEP1 \cite{Barate:2003sz},  there must also be a  constraint from the associated production process  $e^+ e^- \to ha$ at LEP2 when there is significant $Aa$ mixing and one is outside the alignment limit as to make the coupling $g_{haZ} \approx \sin\theta \cos(\beta-\alpha)$ non-zero, as is particularly the case  in the Type I scenario, see Fig.~\ref{fig:h_constr}. 

More significant portions of the parameter space of the 2HD+a model have been probed by direct Higgs searches at the LHC in particular at a c.m. energy of $\sqrt s=13$ TeV with the full collected luminosity of $139\ {\rm fb}^{-1}$ 
\cite{ATLAS:2020zms,CMS:2022rbd,ATLAS:2018rvc,CMS:2019pzc,ATLAS:2018gfm,CMS:2019bfg,ATLAS:2021upq,CMS:2020imj} as will be summarized below.

 \subsubsection{Single production of the heavy H/A and H$^\pm$ states}
 
At the LHC, the heavy 2HDM  neutral bosons $\Phi\!=\!H,A$ can be searched for in several channels, the most important one being their production in the gluon-fusion process as single 
resonances via loops of heavy quarks and their decay into the clean $\tau^+\tau^-$ final states \cite{ATLAS:2020zms,CMS:2022rbd} 
\beq
pp\to gg \to \Phi\!=\! H/A \to \tau^+\tau^- \, .  
\label{eq:Hprod1} 
\eeq
The experimental outcome strongly depends on the type of scenario and on the value of $\tan\beta$. Most of the ATLAS and CMS analyses have been performed in two benchmark scenarios of the MSSM (more precisely, the so-called hMSSM \cite{Djouadi:2013uqa,Djouadi:2015jea} and $M_h^{\rm max}$ \cite{Carena:2013qia} scenarios) which has a Type II configuration with the additional constraint of being close to the decoupling regime that is similar to alignment, i.e with $\alpha \simeq \beta-\frac{\pi} {2}$, making that the $\Phi$ states have similar couplings, with the additional constraint of  being almost degenerate in mass, $M_H \approx M_A$. 

At high $\tan\beta$, that is $\tan\beta \gsim 10$, the dominant contribution in $gg\to H/A$ production is due to loops of $b$-quarks that have enhanced couplings, $g_{\Phi bb} \simeq \tan\beta$. In this case, the process is supplemented by contributions from initiated $b$-quark fusion, $b\bar b \to H/A$ which has a comparable rate. The $\Phi$ states mostly decay into $b$-quarks and $\tau$-lepton pairs with respective branching ratios BR$(\Phi \to b\bar b) \approx 90\%$ and BR$(\Phi \to \tau^+\tau^-) \approx 10\%$ as one also has $g_{\Phi \tau\tau}=\tan\beta$. All other decays are suppressed: the bosonic ones are absent in the alignment limit and the decays into top quark pairs, for $\Phi$ masses above the $t\bar t$ threshold $M_\Phi \gsim 350$ GeV which is favored by constraints from flavor physics, have suppressed rates as  $g_{\Phi tt}= 1/\tan\beta$. 

At low $\tan\beta$, i.e. $\tan \beta \lsim 3$, the $gg\to \Phi$ cross section is mostly generated by loops of top quarks which have couplings that are not strongly suppressed or are even enhanced for $\tan\beta<1$;  the yield of $b\bar b$ fusion becomes negligible in this case.  At the decay end, the only relevant mode would be $\Phi \to t\bar t$ above the favored mass range of $M_\Phi >2m_t$,  with an almost unit branching ratio. One would thus have the partonic process 
\begin{equation} 
gg\! \to \! \Phi \!=\!H/A  \! \to \! t\bar t
\label{eq:Hprod2} 
\end{equation}
for the Higgs signal. One then needs to consider the large QCD background from the process $gg \! \to \! t\bar t$ as well as its interference with the signal as both have the same initial and final states, rendering the interpretation of the searches more problematic \cite{Djouadi:2019cbm}. 

Some of these features are also present at intermediate $\tan\beta$ values, with $3 \lsim \tan\beta \lsim 10$, where the $H/A$ couplings to $b$- and $t$- quarks are comparable and not strong enough. The production rates are not large and there is a competition  between the $b\bar b$ and $t\bar t$ decays modes even for $H/A$ masses above the $2 m_t$ threshold. Any new decay channel into particles that do not have suppressed couplings to the Higgs bosons could be important and even dominating.   This would be the case of, for example,  invisible decays of the $A$ state into pairs of the DM particles, $A\to \chi \chi$, which is not detectable in this particular channel. Concerning the CP-even $H$ state,  there is the possibility of $H \to aa$ decays, if the coupling $g_{Haa}$ is not too small, which would lead to complicated topologies with e.g. $4b,2b2\tau, 4\tau$ final states. 

In the case of the charged Higgs state, besides the LEP bound  $M_{H^\pm} \gsim 80$ GeV, there are searches at LHC (and the Tevatron) of top quark decays $t \to bH^+$ with the subsequent decay $H^- \to \tau \nu$ and eventually $H^- \to c\bar s$, leading to an exclusion of the mass range $M_{H^\pm} \lsim  m_t \approx 170$ GeV for any value
of $\tan\beta$  \cite{ATLAS:2018ntn}. For the larger $H^\pm$ masses that are still allowed by flavor constraints,  the dominant process would be the associated production mechanism 
\beq
gb \to t H^\pm \ \  {\rm with} ~~ H^\pm \to tb, \tau \nu \, , 
\label{eq:Hprod3} 
\eeq
at low $\tan\beta \lsim 1$ or large $\tan \beta \gg 1$ values, for which either the $t$- or the $b$-component of the $g_{H^\pm tb}$ coupling is strong.  In the former case,  $H^\pm$ will decay into $tb$ final states with almost 100\% probability while in the second case, one would have the $tb$ and $\tau\nu$ final states with  branching ratios of BR($H^+ \to tb)\approx 90\%$ and BR($H^+ \to \tau \nu) \approx 10\%$. Both these topologies have been searched for at the LHC but at present,  only loose  constraints have been set at  low $\tan\beta \lsim 1$  and high $\tan\beta \gsim 50$ \cite{ATLAS:2018ntn} which do not compete with the limits from $b\to s\gamma$.  
 
Recently, an updated analysis of these constraints has been performed \cite{Arcadi:2022hve} in the context of the MSSM (and more precisely of the hMSSM in which the value $M_h\!=\!125$ GeV is enforced) which has the following simplified features in the decoupling limit $\alpha \simeq \beta -\frac{\pi} {2}$  (which is reached as soon as $M_A \gsim 500$ GeV and it should be valid in order to cope with the $h$ signal strengths): $i)$ the same Higgs couplings as in the Type II  2HDM in the alignment  limit, $ii)$ only values $1 \lsim \tan\beta \lsim 50$ are allowed, and  $iii)$ the approximate mass degeneracy, $M_H \simeq M_A \simeq M_{H^\pm}$.   Using the full LHC data of 139 fb$^{-1}$ collected at an energy of 13 TeV, and considering the three main search channels of Eqs.~(\ref{eq:Hprod1})--(\ref{eq:Hprod3}), one obtains the excluded area at the 95\% CL of the $[M_A, \tan\beta ]$ plane shown in Fig.~13 of Ref. \cite{Arcadi:2022hve}. 

From this figure, one can see that the $pp \to H/A \to \tau^+ \tau^-$ search is extremely efficient and excludes values of $M_A$ below 1 TeV for the entire  range $\tan\beta \gsim 10$. The exclusion extends to $\tan\beta \approx 5$ for $M_A \approx 700$ GeV and $\tan\beta  \gsim 20 $ for $M_A \approx 1.5$ TeV. The search in the $H/A \to t\bar t$ final state is much less constraining as only values $M_A \lsim 750$ GeV are excluded for $\tan\beta \lsim 2$. The search for the charged Higgs boson in the mode $H^+ \to t \bar b$ excludes masses below $M_{H^\pm} \lsim 700$ GeV both at low $\tan\beta \lsim 2 $ and high $\tan\beta \gsim 40$ values. 

In our 2HD+a context, these limits can be overcome or loosened in the following cases:\smallskip 

-- One can first have a non degenerate mass scenario, $M_H\! \neq\! M_A$, in such a way that in the $pp\!\to \!H/A \!\to \!\tau^+ \tau^-$ process, one looks for two resonances rather than a single one. This lowers the excluded values of $\tan\beta$ by approximately a factor $\sqrt 2$. The mass non-degeneracy would also allow cascade decays, such as $H\! \to \!AZ,  A\!\to\! HZ$ and $H/A \! \to \! H^\pm W^\mp$ depending on the hierarchy of masses,  which lead to lower $\tau\tau$ rates and, hence, looser bounds.\smallskip 

-- The cross section for $A$ production is lowered by a factor $\cos^2\theta$ compared to the MSSM. In addition, as noted previously, one could have the additional  decays $A \to \chi \chi$ as well as $H\to aa$ in the case of the CP-even Higgs (in addition to the cascade decays above) which would lead to less severe bounds, in particular at intermediate $\tan\beta$ values.\smallskip 

-- Finally, in the charged Higgs case, one has the additional decay $H^\pm\! \to \! a W^\pm$  \cite{Borzumati:1998xr} which is always favored by phase-space and which, for the sizable $Aa$ mixing needed for DM issues, can compete with the $H^+ \! \to \! tb$ mode (in addition to the possible $H^\pm \!\to \!W\!+\!H/A$ decays).\smallskip    

All these features would make that the LHC heavy Higgs searches are less constraining in our model compared to the MSSM. Nevertheless, they still exclude a substantial area of the parameter space of the model as will be seen shortly when we move   to our numerical analysis. 
Before that,  let us first adapt all these discussions held for the Type II scenario  to the other 2HD+a model configurations, namely Type I, X and Y.

First, in the Type Y scenario, $\Phi$ production is the same as above since the Higgs couplings to $t$- and $b$-quarks are as in Type II. In turn, as the Higgs couplings to $\tau$-leptons are now $\propto 1/\tan\beta$, only the decay mode $H/A \to b\bar b$ is relevant at high $\tan\beta$ and it is subject to a large QCD background which makes it difficult to probe at the LHC. At low $\tan\beta$ values, the situation is similar as in Type II as only the channel $\Phi \to t\bar t$ is relevant.  
In the case of the $H^\pm$ state, the search channel $gb \to t H^\pm \! \to\! ttb$ is the only relevant one at low $\tan\beta$ but also at high $\tan\beta$ as the coupling $g_{A \tau\tau} \propto 1/\tan\beta$, is suppressed in this case.

\begin{figure}[!h]
    \centering
    \subfloat{\includegraphics[width=0.44\linewidth]{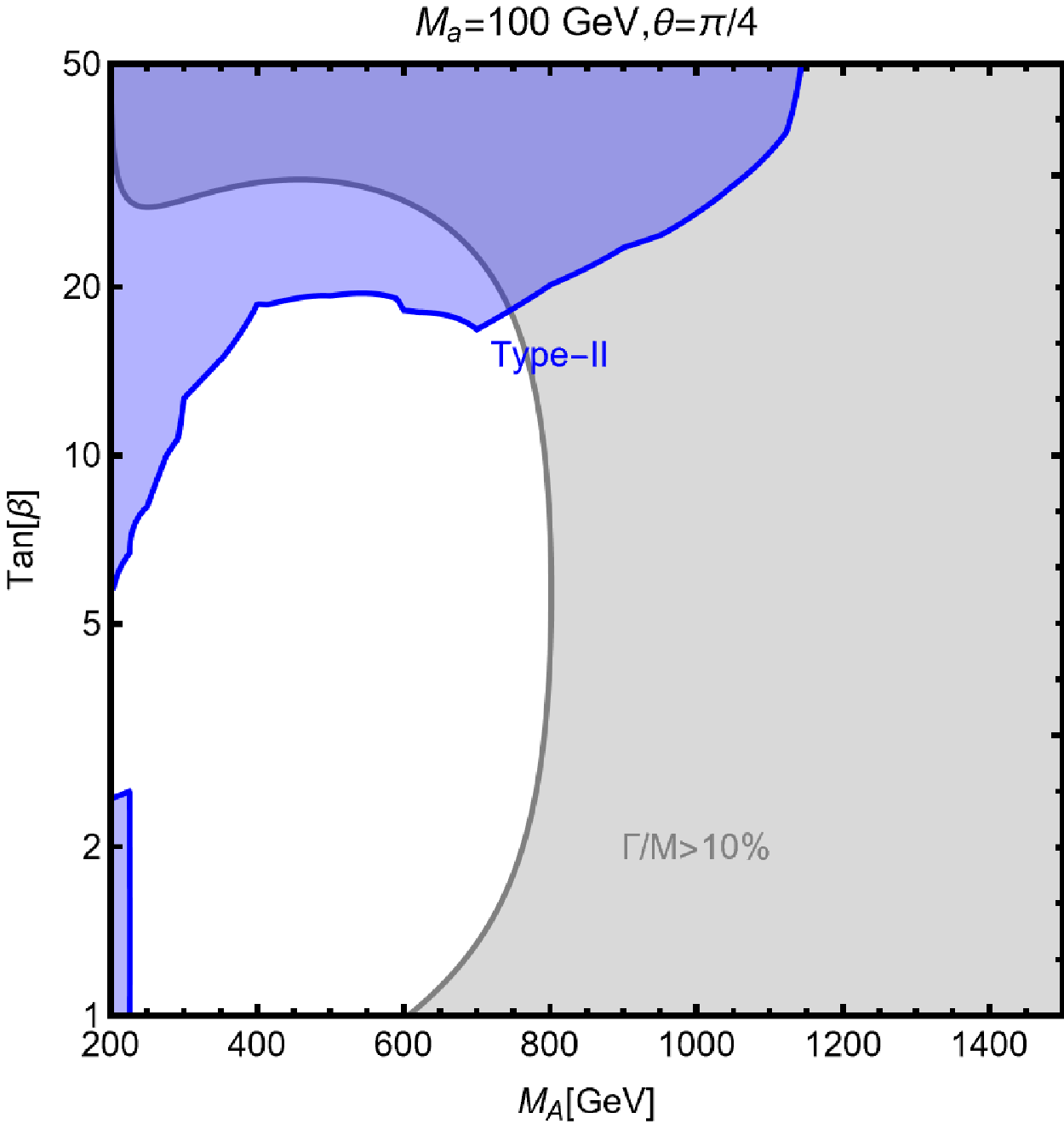}}~~
\subfloat{\includegraphics[width=0.44\linewidth]{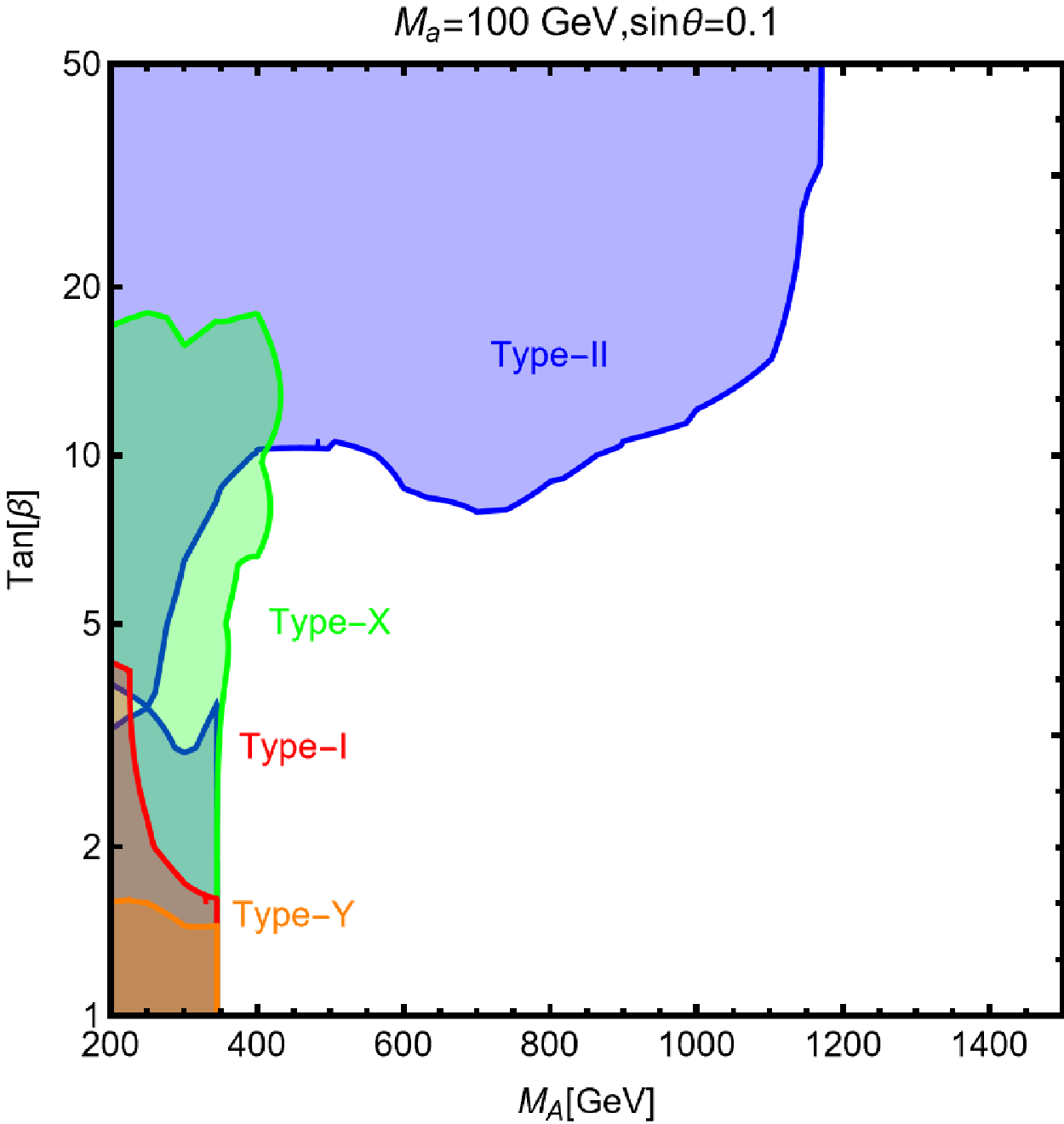}}\\
    \subfloat{\includegraphics[width=0.44\linewidth]{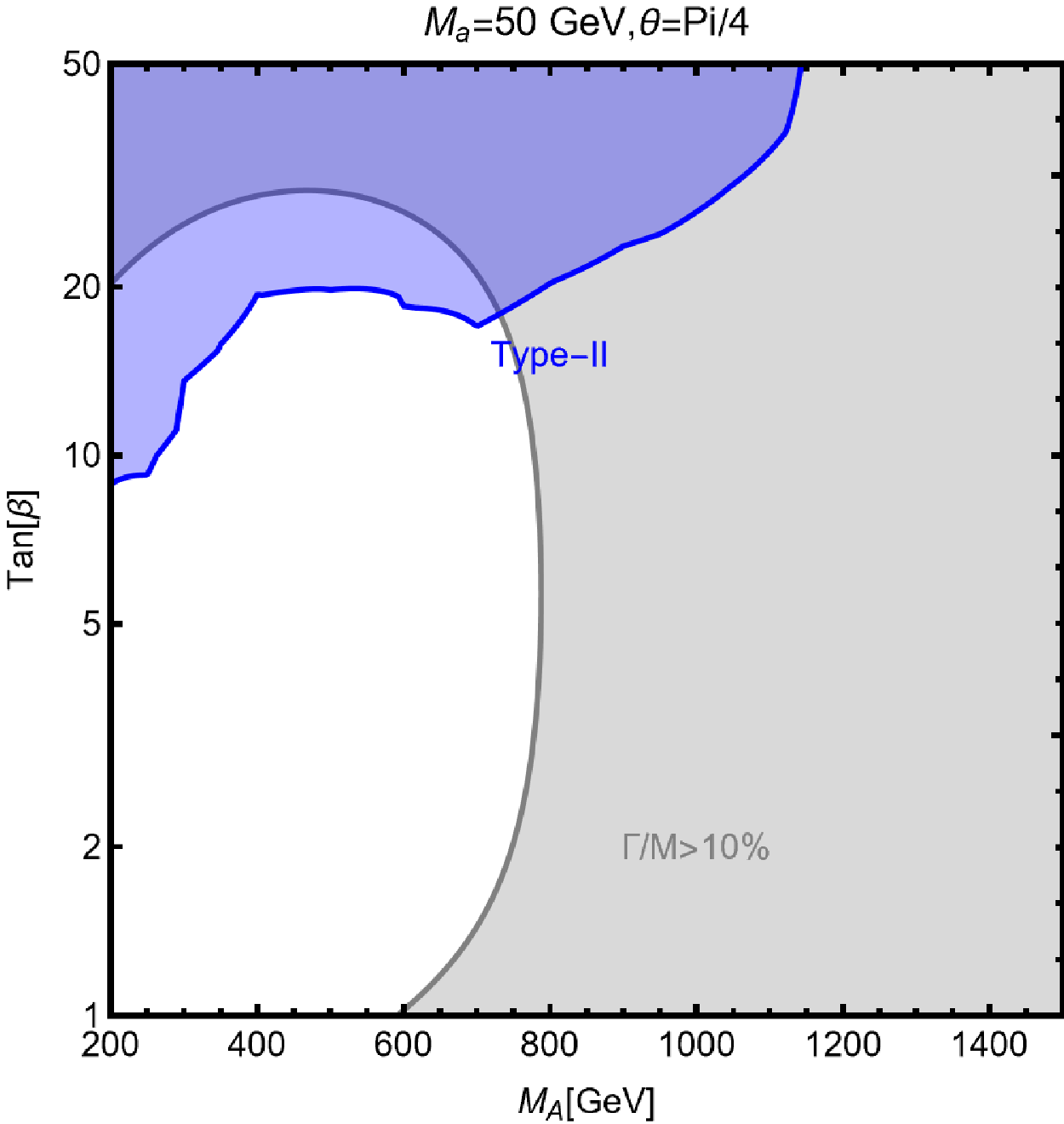}}
    \subfloat{\includegraphics[width=0.44\linewidth]{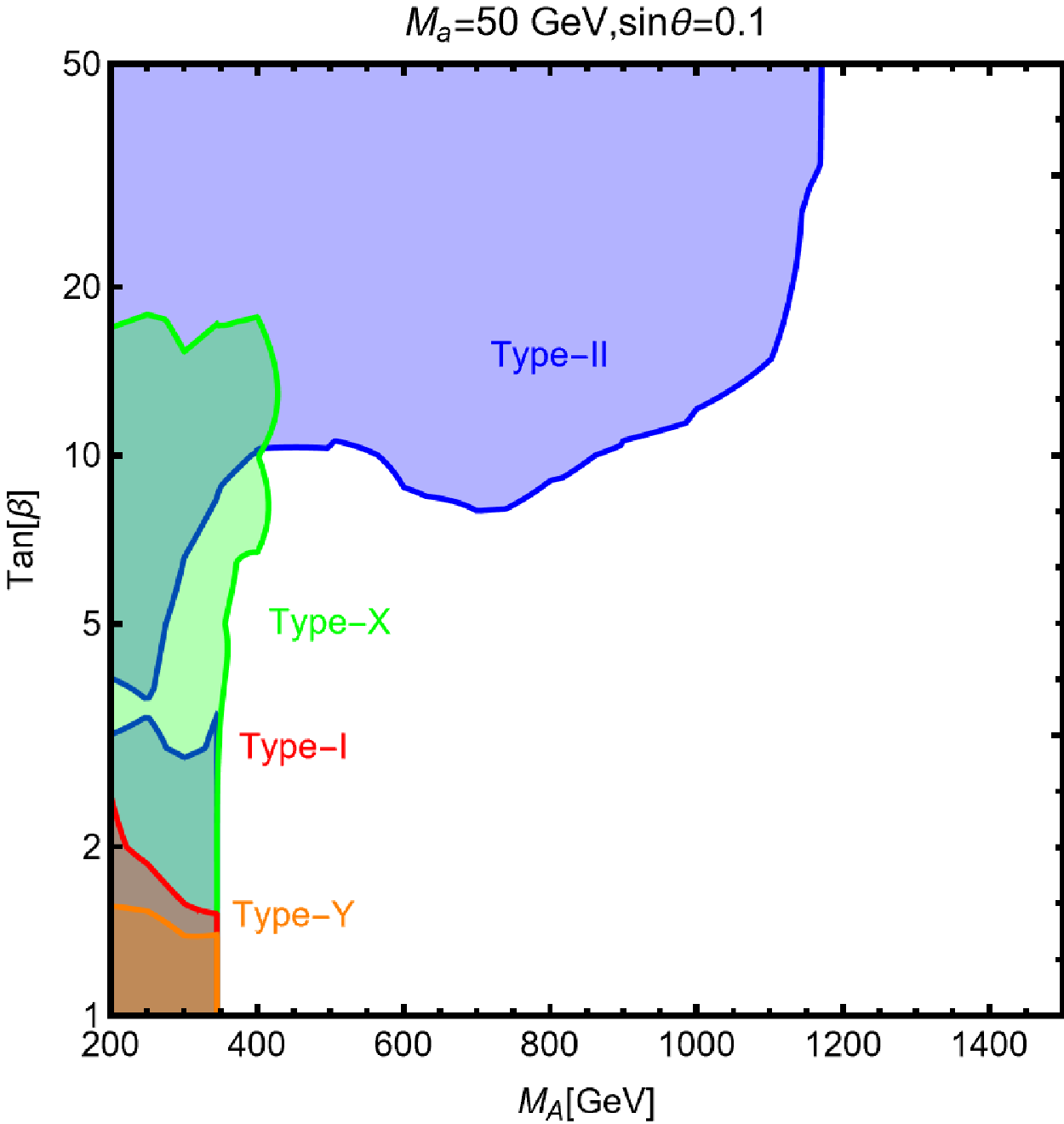}}
    \caption{LHC limits from searches in the $pp\to H/A \to \tau^+\tau^-$ channel in the 
    $[M_A, \tan\beta]$ plane for different assignments of the $(M_a,\sin\theta)$ pair and
    assuming $M_H\!=\!M_A\!=\!M_{H^{\pm}}\!=\! |M|$. The different colored regions correspond to the exclusions for a given Yukawa configuration, namely Type I (red), Type II (blue), Type X (green) and Type Y (orange).  In the panels with $\theta=\frac{\pi}{4}$, the gray regions correspond to the case in which at least one between the $H$ and $A$ states has a total decay width that exceeds $10\,\%$ of its mass.}
    \label{fig:plot_tautau}
\end{figure}

In the Type I scenario, the main process for $H/A$ production is again gluon fusion but it is generated by top quarks loops which give large rates only at low $\tan\beta$ values, $\tan\beta \lsim 3$ when $g_{\Phi tt} \propto 1/\tan\beta$ is strong. In the strict alignment limit, the dominant decay modes for $M_{\Phi} \gsim 350$ GeV are into top quarks with branching ratios of order 100\% unless there are  exotic decays.  For the $H^\pm$ states, the only relevant production process would be again $gb \to tH^\pm$ with a large rate at low $\tan\beta$ where $H^+ \to t\bar b$ decays have a unit branching ratio. 

 Finally in the Type X scenario, neutral Higgs production is as in the  Type I case and is important only at low $\tan\beta$ values, with exclusive $\Phi \to t\bar t$ decays for Higgs masses above the $2m_t$ threshold.  For the $H^\pm$ state, the production is also as in Type I, but at high $\tan\beta$, BR($H^+ \to \tau \nu)$ will be dominant thanks to the enhanced $g_{H^\pm \tau\nu} \propto \tan\beta $ coupling. 

\begin{figure}
    \centering
\subfloat{\includegraphics[width=0.25\linewidth]{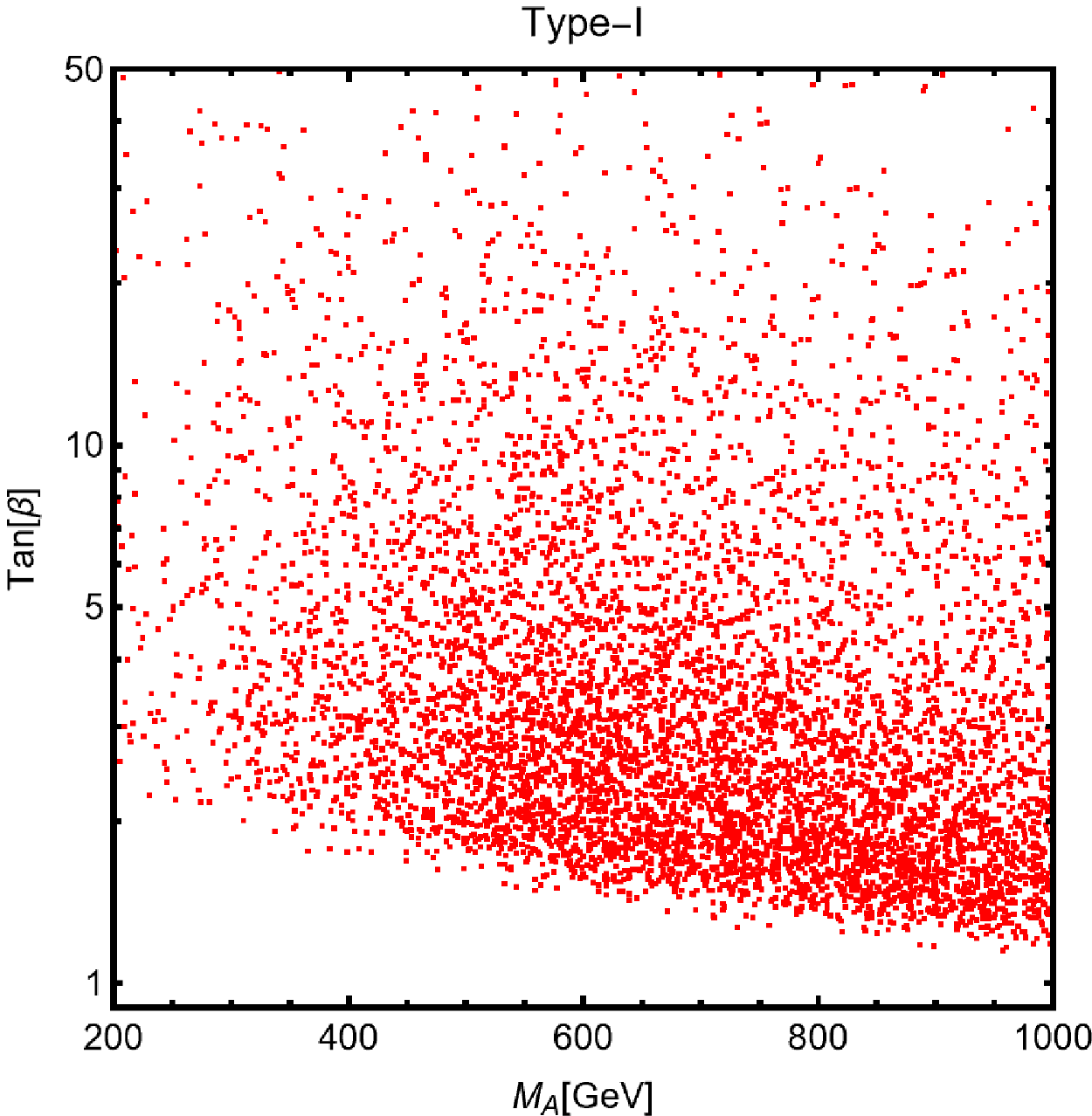}}
\subfloat{\includegraphics[width=0.25\linewidth]{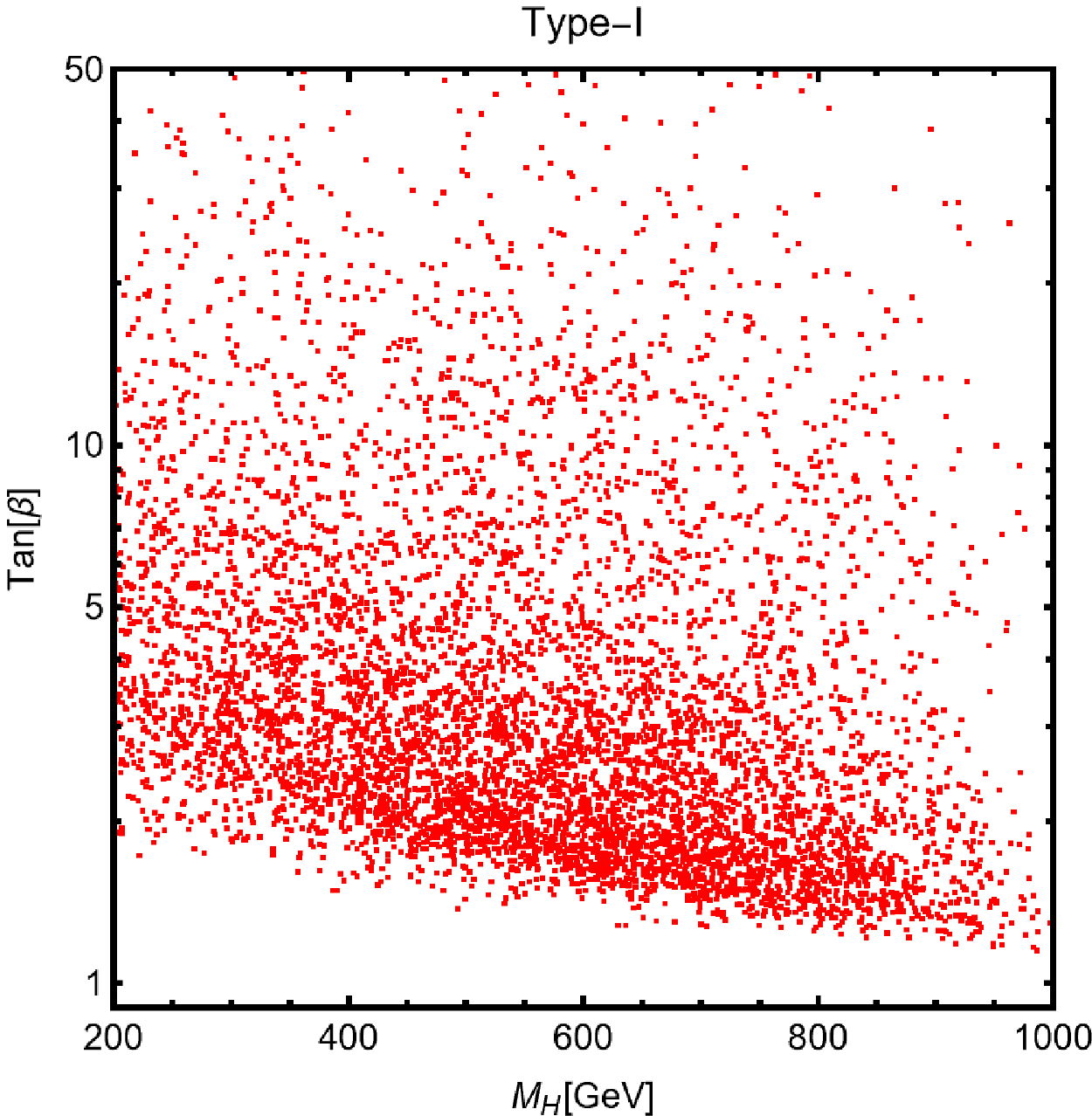}}
\subfloat{\includegraphics[width=0.25\linewidth]{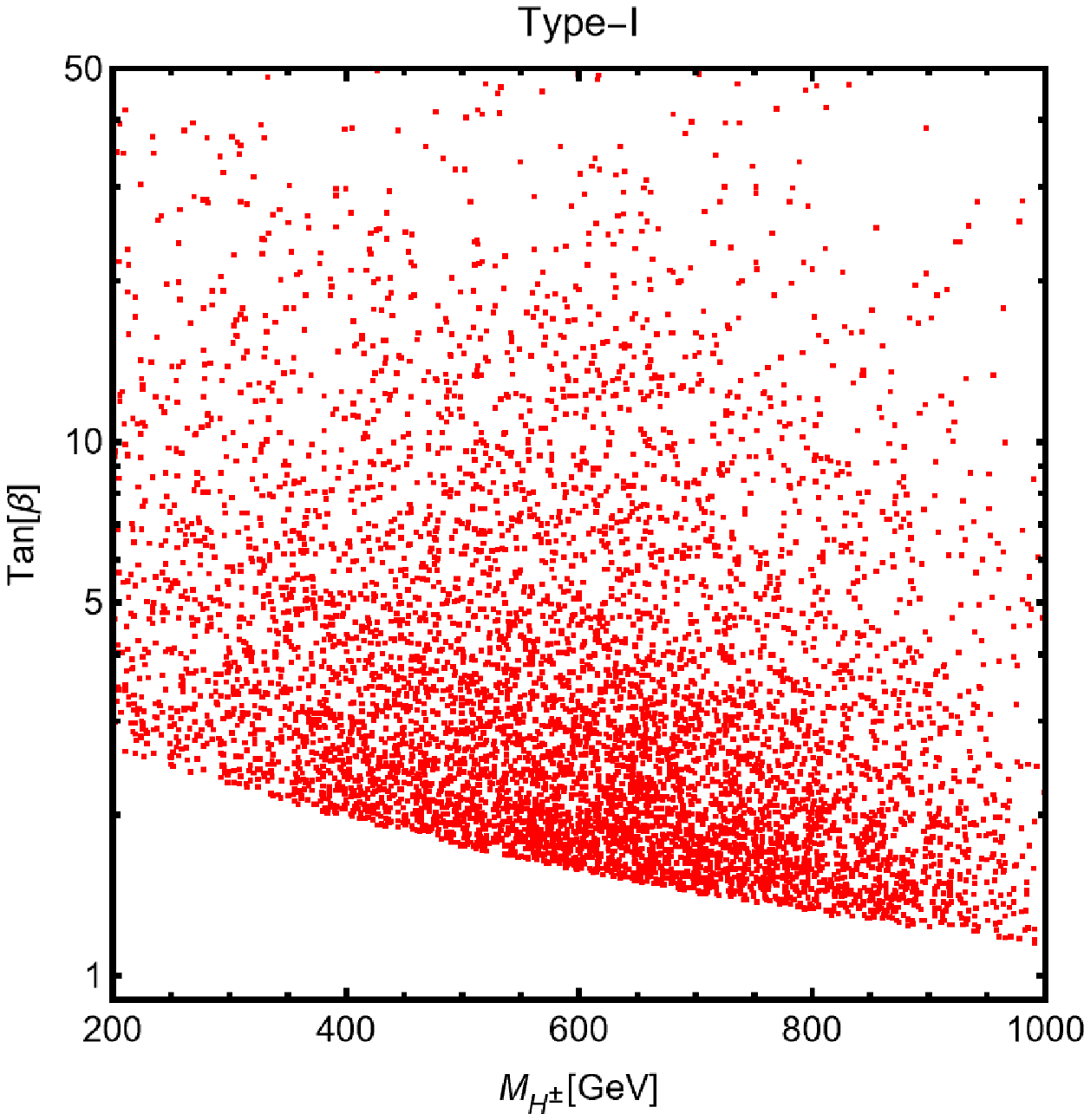}}
\subfloat{\includegraphics[width=0.25\linewidth]{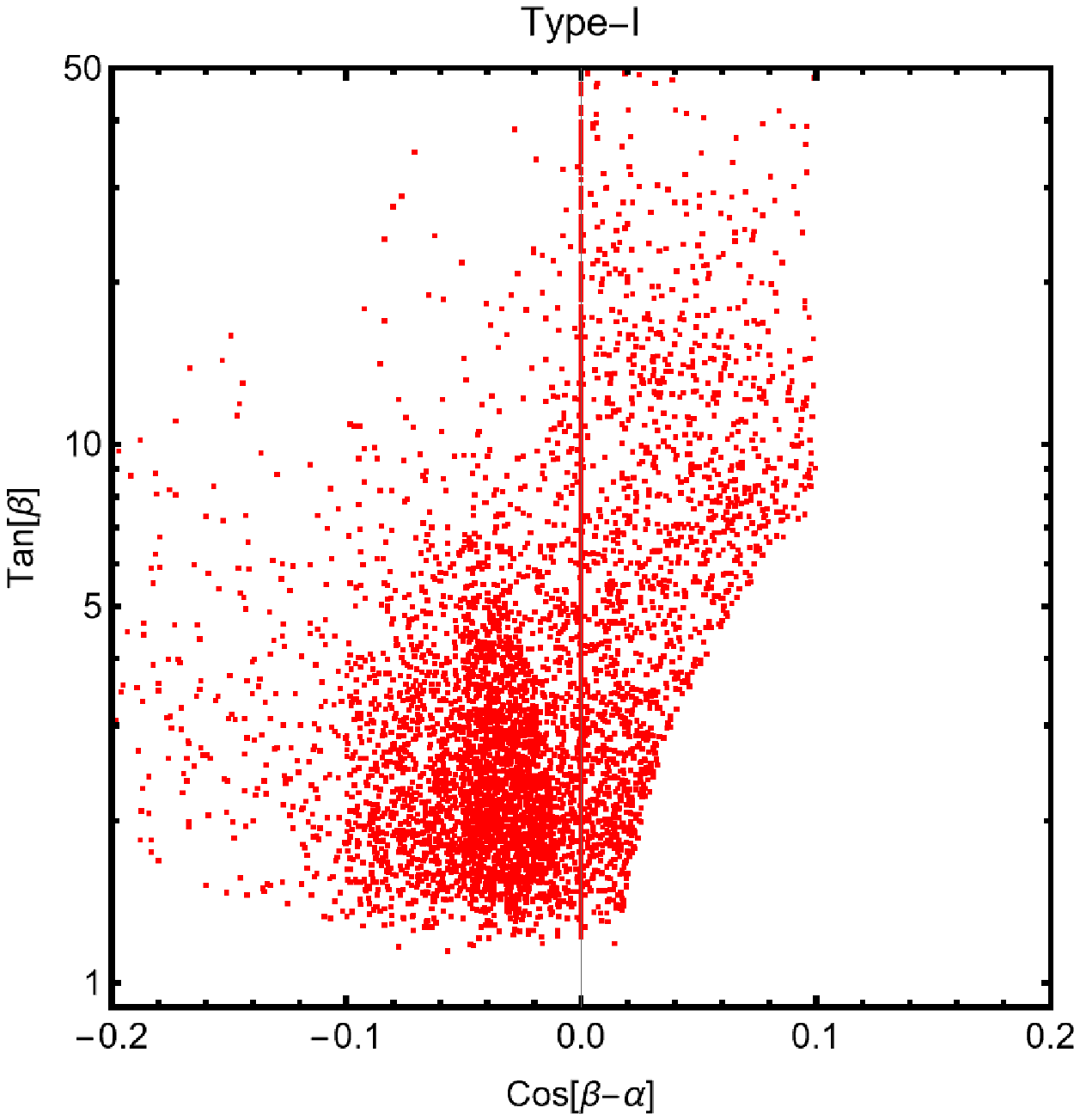}}\\
\subfloat{\includegraphics[width=0.25\linewidth]{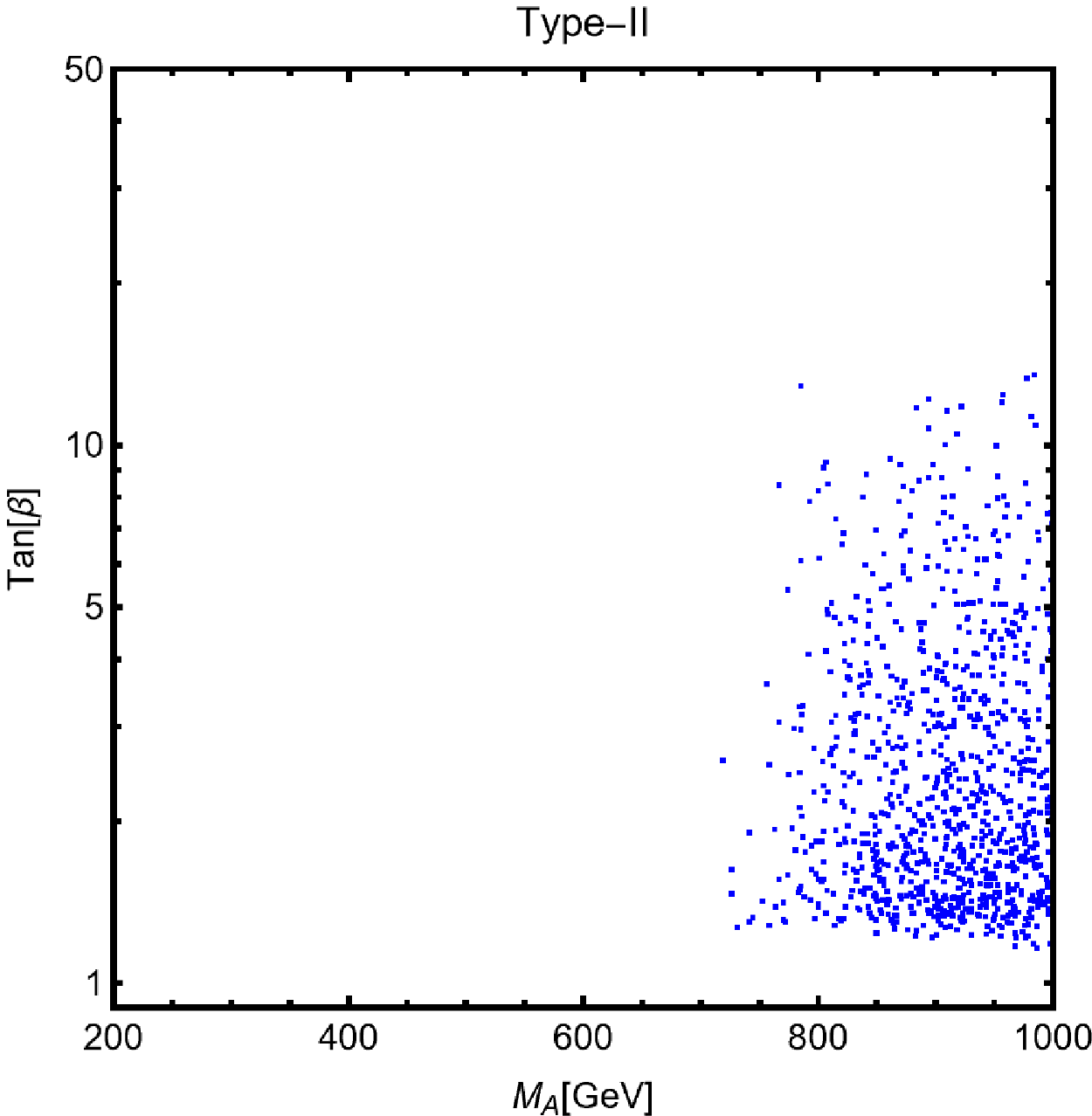}}
\subfloat{\includegraphics[width=0.25\linewidth]{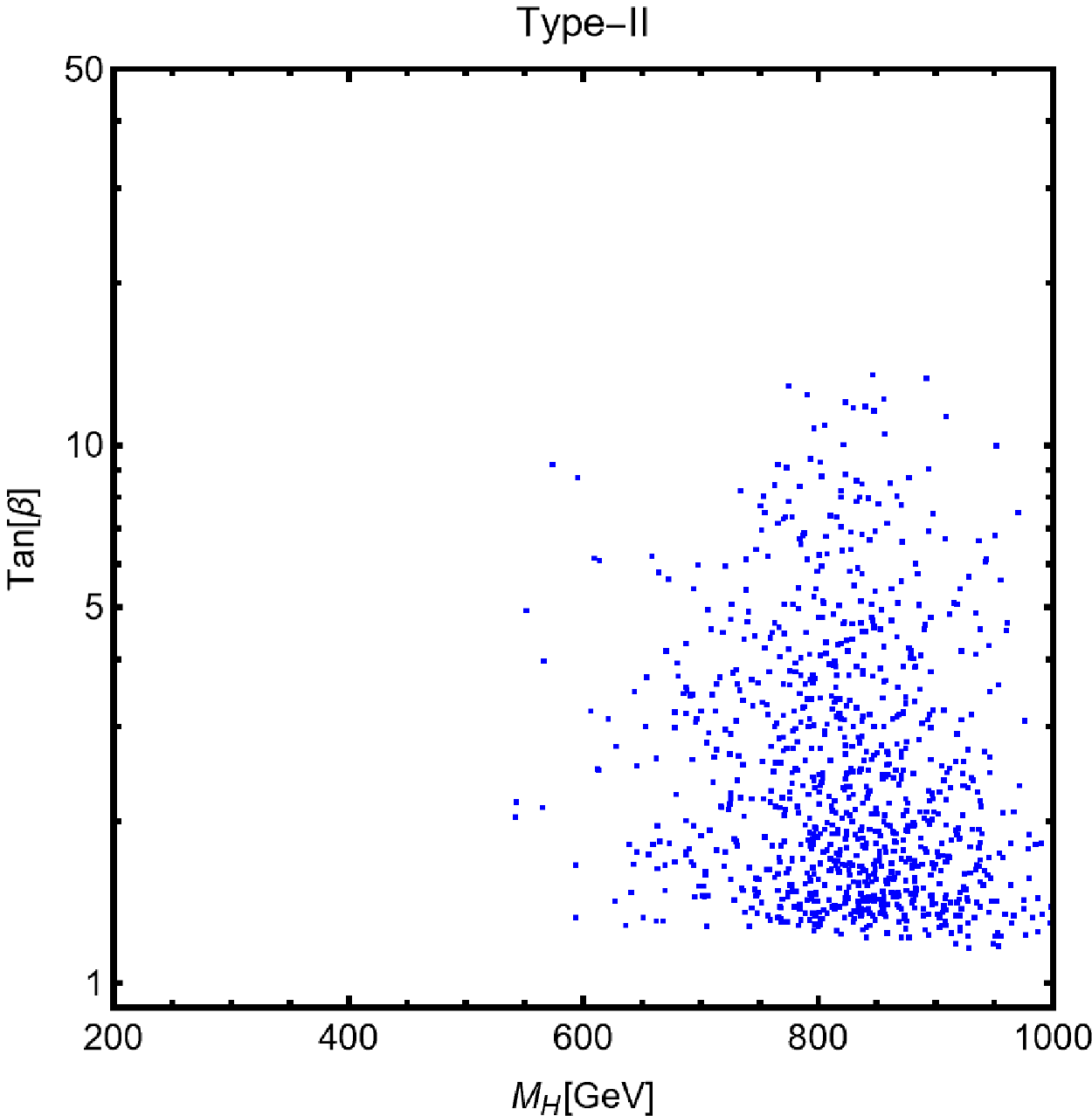}}
\subfloat{\includegraphics[width=0.25\linewidth]{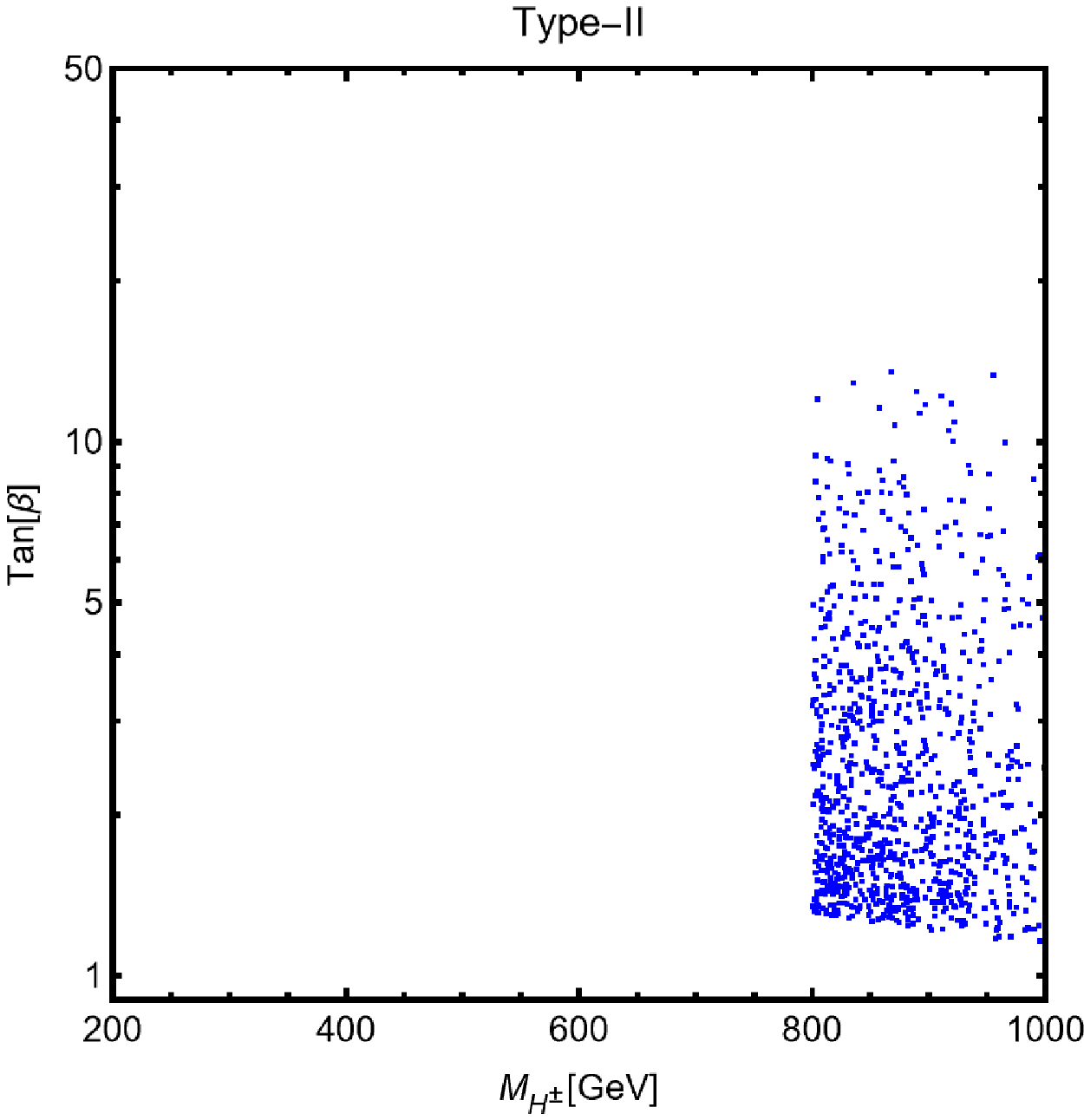}}
\subfloat{\includegraphics[width=0.25\linewidth]{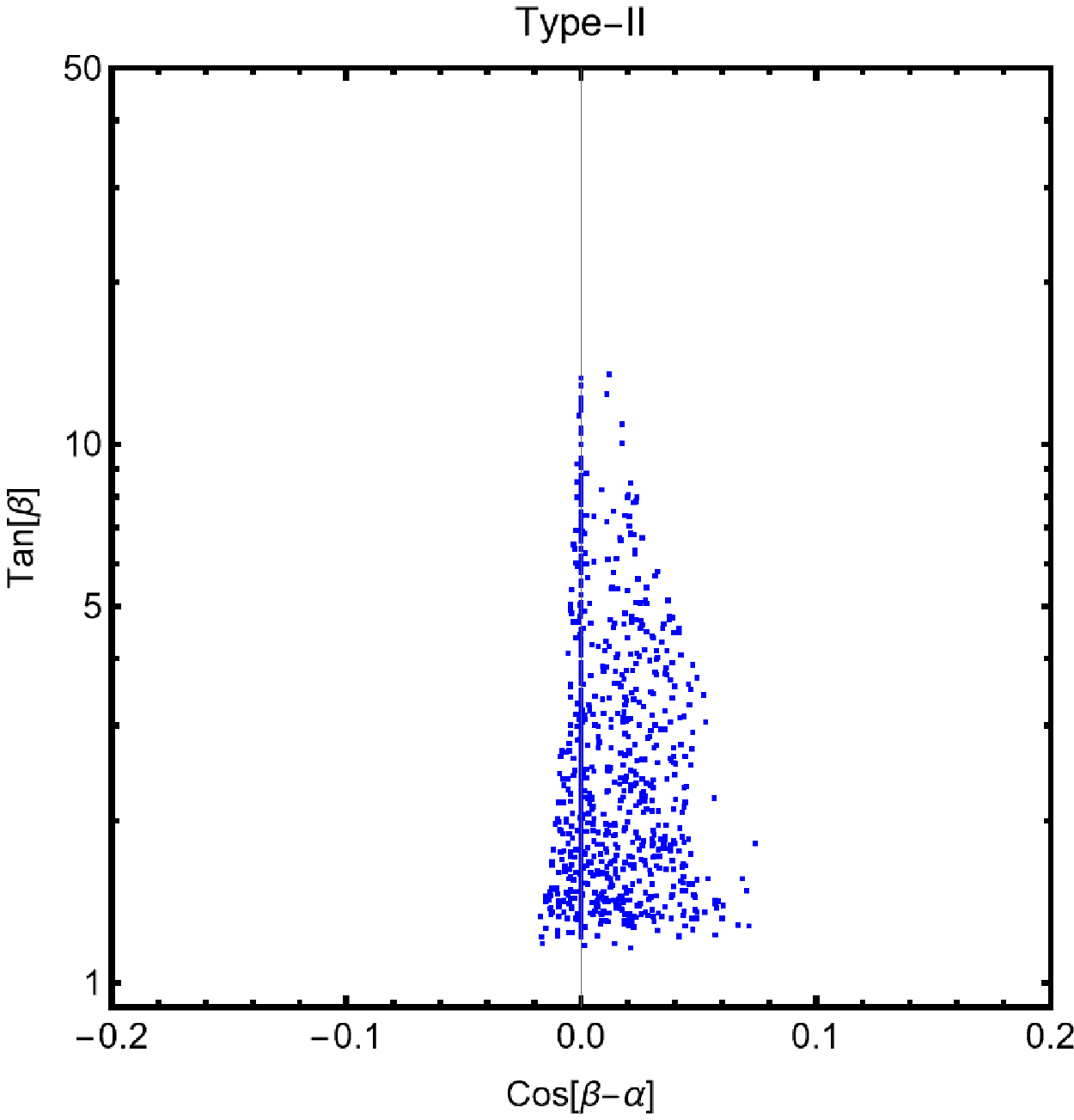}}\\
\subfloat{\includegraphics[width=0.25\linewidth]{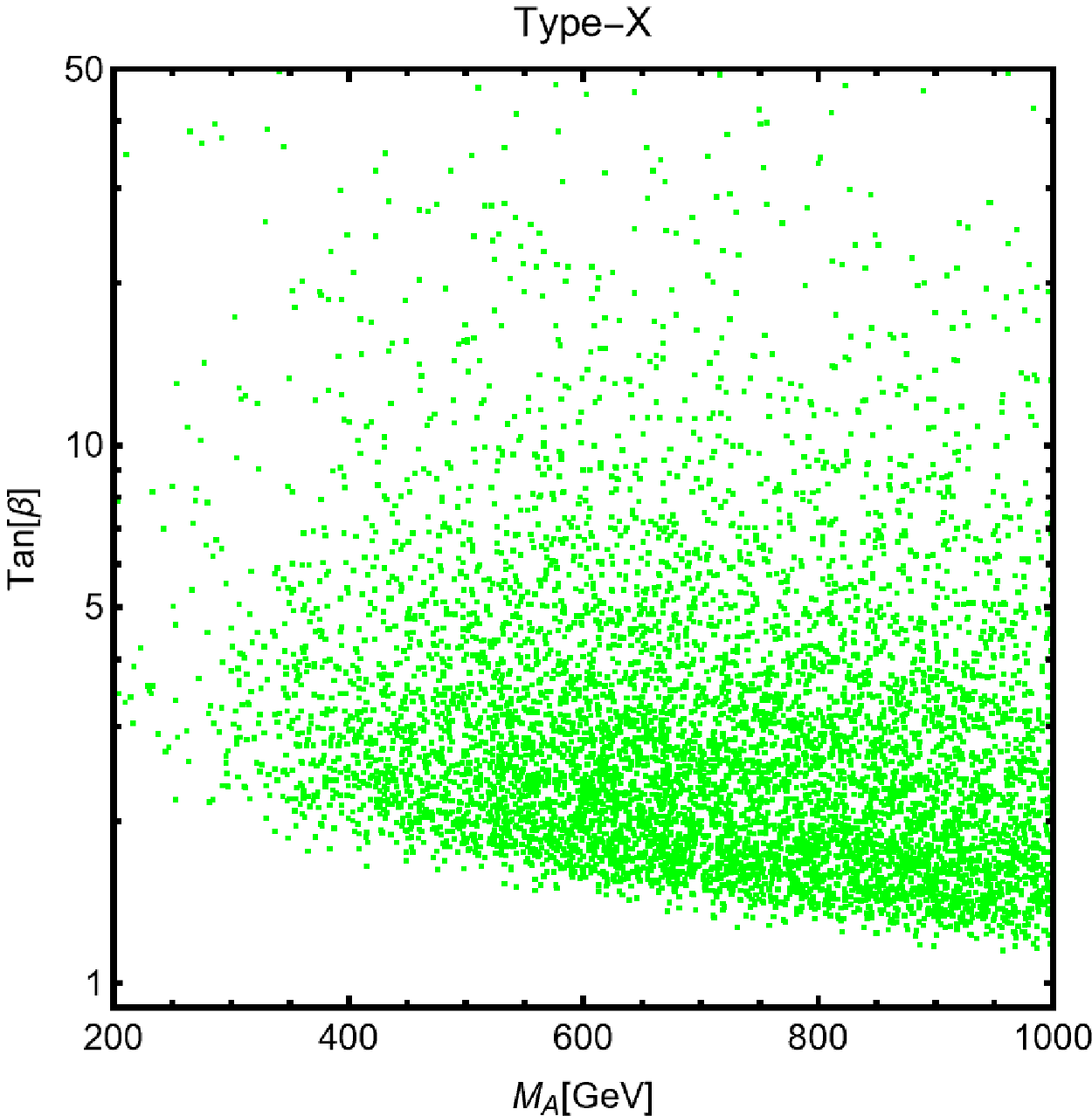}}
\subfloat{\includegraphics[width=0.25\linewidth]{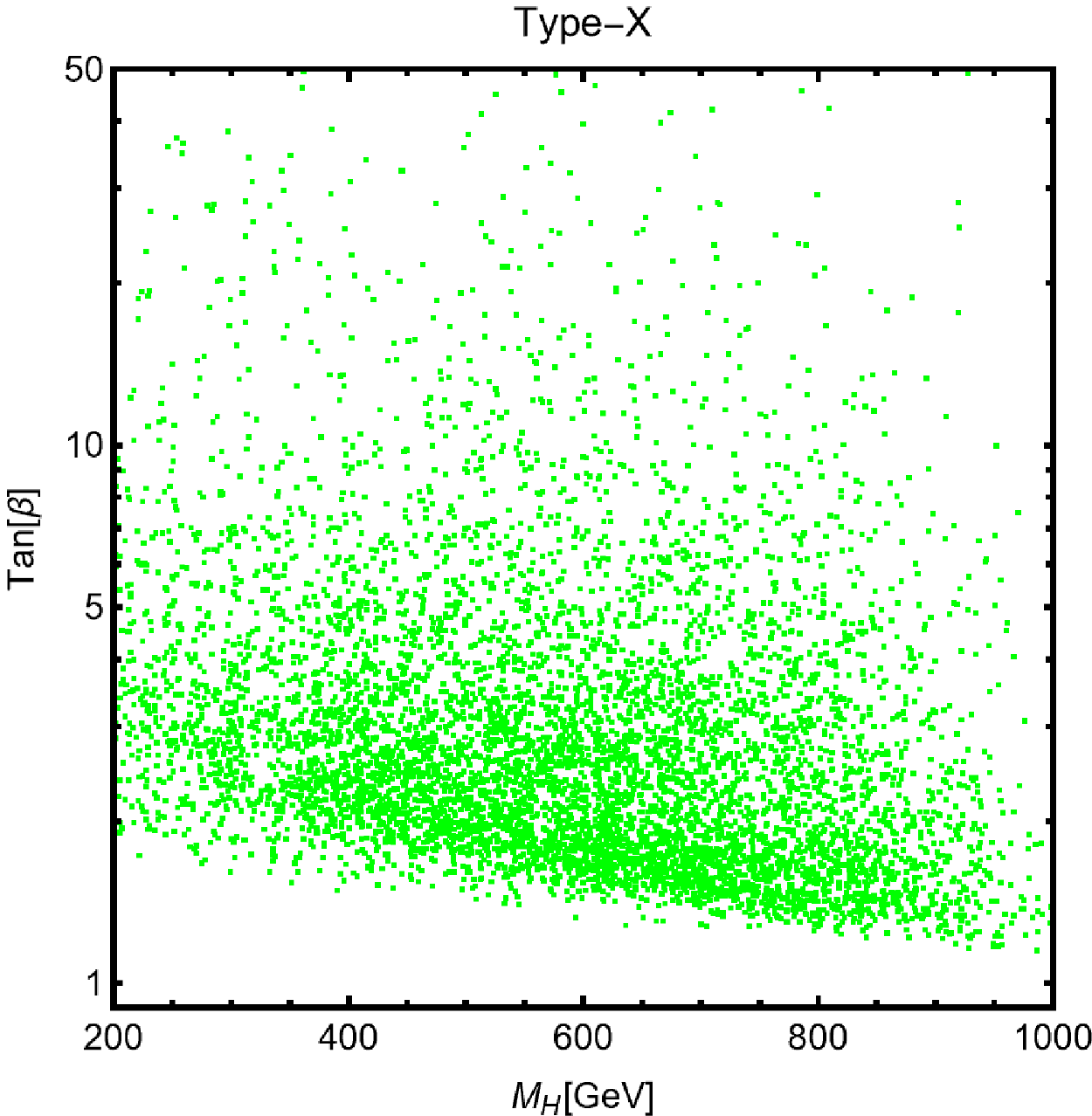}}
\subfloat{\includegraphics[width=0.25\linewidth]{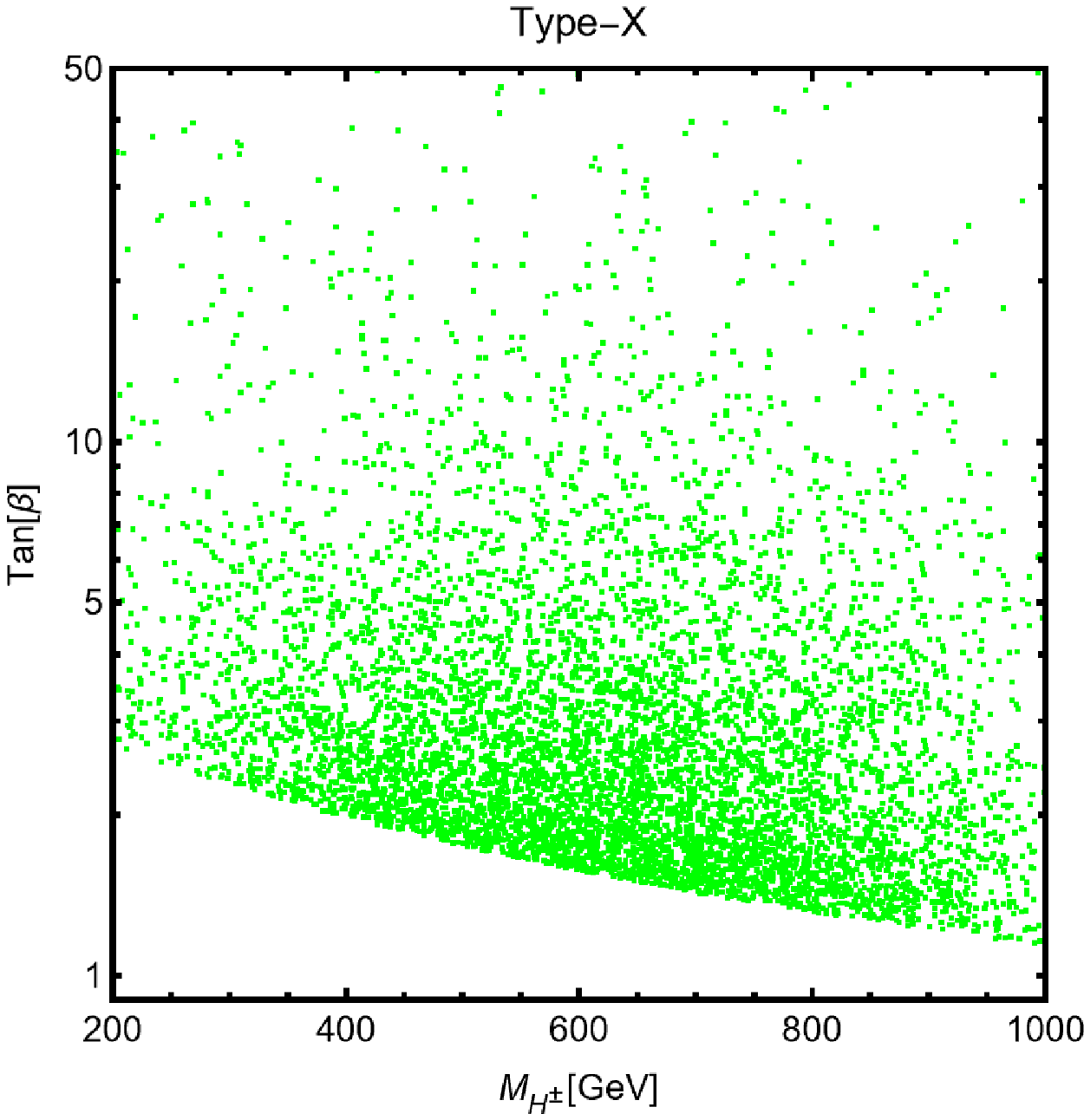}}
\subfloat{\includegraphics[width=0.25\linewidth]{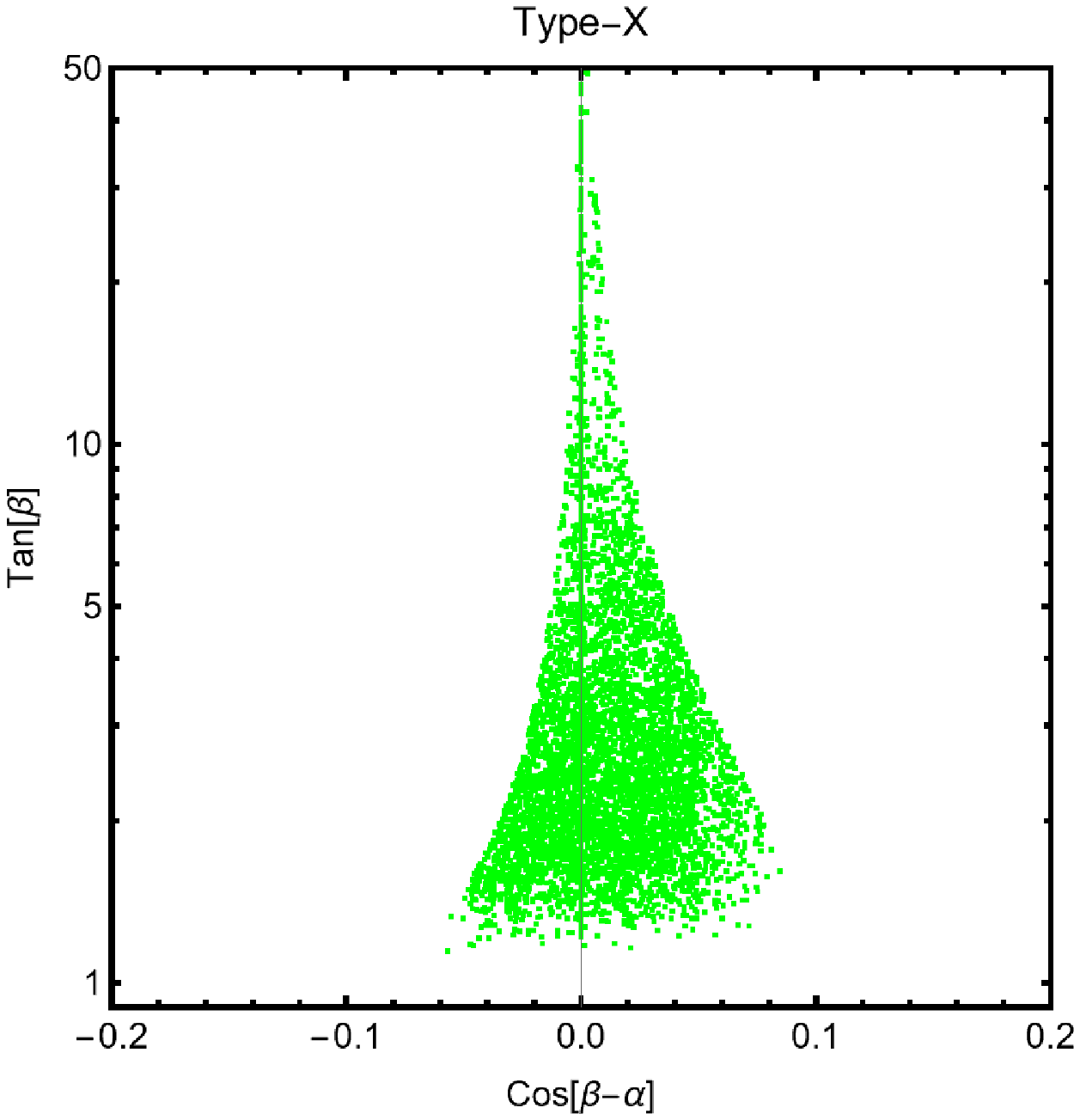}}\\
\subfloat{\includegraphics[width=0.25\linewidth]{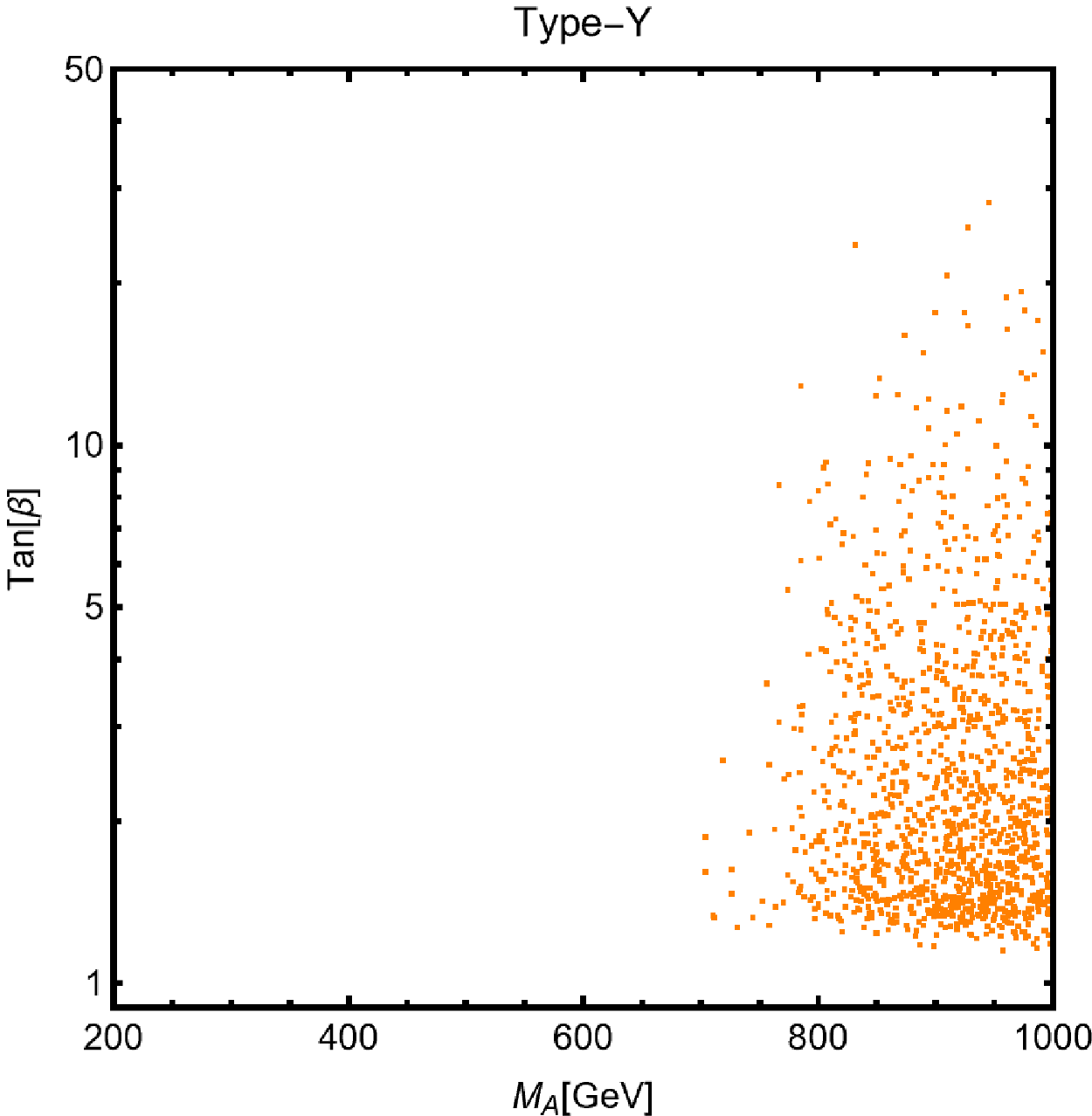}}
\subfloat{\includegraphics[width=0.25\linewidth]{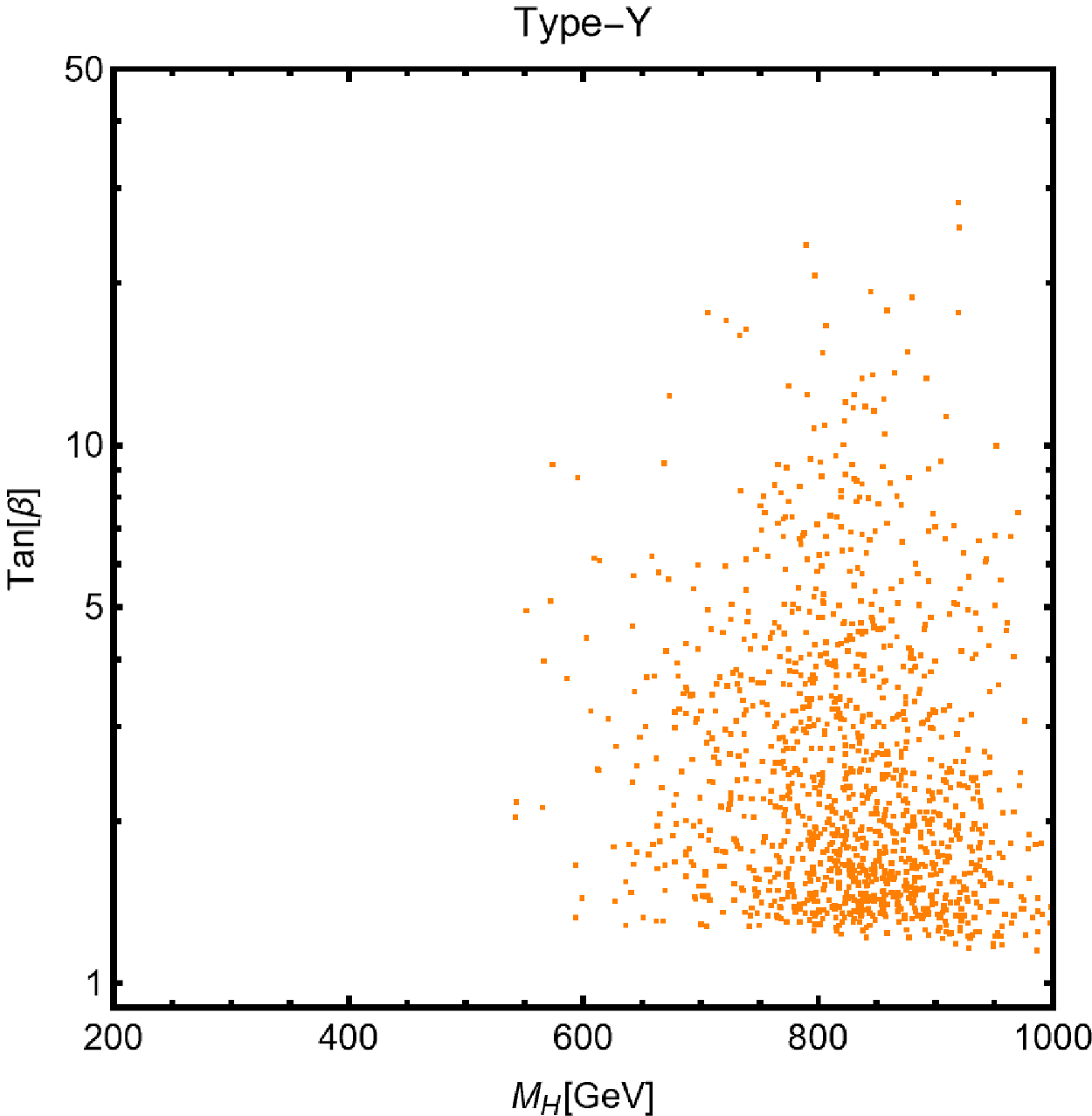}}
\subfloat{\includegraphics[width=0.25\linewidth]{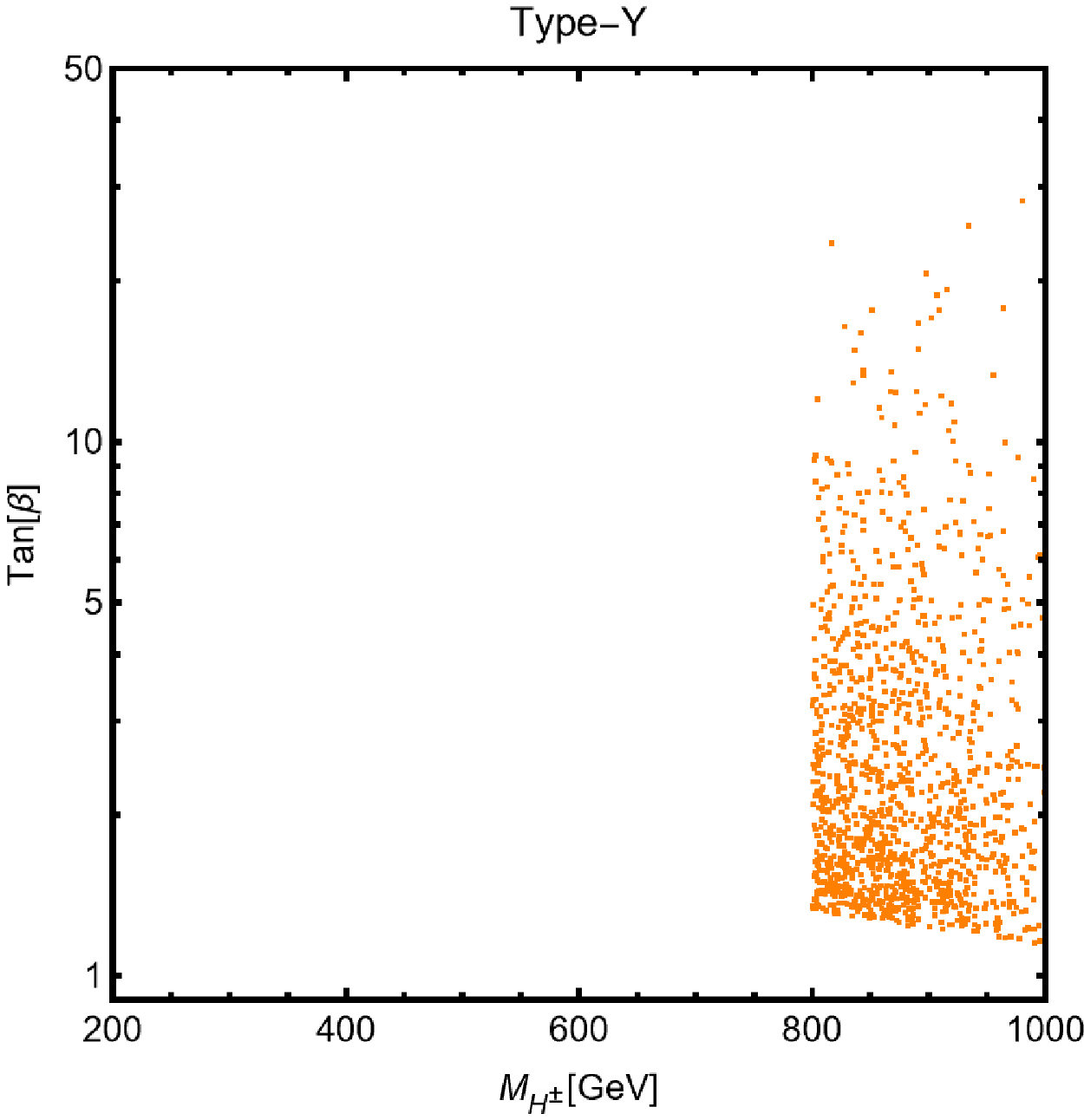}}
\subfloat{\includegraphics[width=0.25\linewidth]{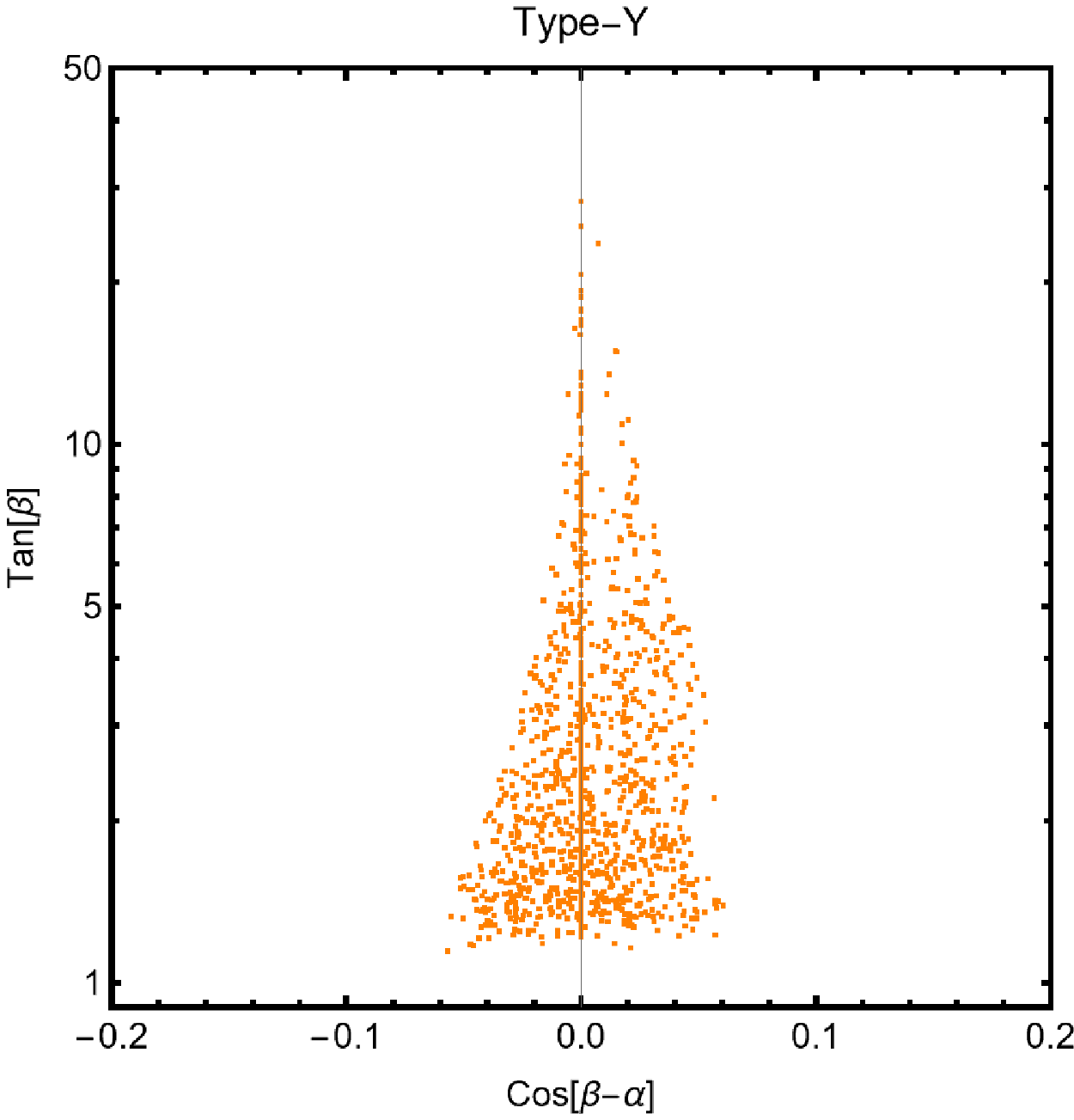}}
 \caption{Model points for the four flavor-preserving Yukawa configurations of the 2HD+a model, which comply with theoretical constraints, the bounds from flavor and Higgs signal strengths, and LHC searches of extra Higgs states (see the main text for details). The points which pass these constraints are shown from left to right, in the $[M_A,\tan\beta]$,$[M_H,\tan\beta]$, $[M_{H^\pm},\tan\beta]$ and $[\cos(\beta-\alpha),\tan\beta]$  planes.}
\label{fig:Hlimits}
\end{figure}

An important remark to be made at this stage is that there is another source of $H,A,H^\pm$ states at the LHC whose rates do not depend on $\tan\beta$ in the alignment, namely production in pairs in the Drell-Yan processes $q\bar q \to H^+H^- , HA$ and $q\bar q' \to HH^\pm, AH^\pm$. These occur through virtual gauge boson exchange and have rates that are unsuppressed by the scalar-vector couplings $g_{HAZ}, g_{H^\pm HW}$ which are maximal in the alignment limit, Eq.~(\ref{HV-couplings}). The production rates are limited only by phase space but as we have already the constraint $M_{H,A,H^\pm} \gsim 500$ GeV in most cases, they should be small and these channels, which have not yet been considered by the experiments, are or should not be very constraining. 

In order to determine LHC limits for the scenario under scrutiny, we have computed the production cross sections of the neutral Higgs bosons using the package SuSHI \cite{Harlander:2012pb,Harlander:2016hcx} in the four flavor-preserving 2HDMs. The results have been then adapted to the corresponding 2HD+a models by applying suitable $\sin^2 \theta, \cos^2 \theta$ factors.

In our scan, the following constraints have been applied to the parameter space as already discussed or at least mentioned previously.

-- The limits from the search in the process  $pp\rightarrow H/A \rightarrow \tau^+ \tau^-$ as given in Ref.~\cite{ATLAS:2020zms}. As already pointed out, this is probably the most constraining search, impacting in particular our 2HD+a model. We have ignored the limits from the processes $gg\to H/A \to t\bar t$ and $gb\to H^-t \to tb, \tau\nu$ as these are sensitive only to small areas of the parameter space that are excluded by the previous channel and by $B$-physics  constraints.  
     
 -- In order to consider also the region outside the alignment limit, which is not realized everywhere in particular in the Type I scenario, we also include the limits from the two search channels $pp \rightarrow A \rightarrow Zh$ \cite{ATLAS:2022enb}  and $pp \rightarrow H \rightarrow ZZ$ \cite{ATLAS:2020tlo}. Both have rates that are suppressed by the factor $g_{HVV}=g_{hAZ}= \cos(\beta-\alpha)$ which vanishes in the alignment limit.

-- To cope with the possibility of a significant mass splitting between the  masses of the heavy neutral Higgs bosons,  $|M_H-M_A|\gsim M_Z$, we also include the search channels  $pp \rightarrow A \rightarrow ZH$ \cite{ATLAS:2020gxx} and  $pp \rightarrow H \rightarrow ZA$ \cite{CMS:2019ogx}. Again, in the alignment limit,  the decays $H\to AZ$ or $A\to HZ$ depending on the mass hierarchy, are not suppressed since $g_{HAZ}\!=\!1$.

-- Finally, and as will be discussed in the next subsection, we will also include the possible decay $H\to Za$ which is always phase-space allowed in our context, since we assume $M_H \gg M_a$. This is done by considering one of the previous channels, $pp \rightarrow H \rightarrow ZA$ \cite{CMS:2019ogx},  formulated in the 2HDM with $M_H \gg M_A$, and adapting it to the case $pp \rightarrow H \rightarrow Za$.

The constraints from the $pp\to H/A\to \tau^+\tau^-$ channel are shown in Fig.~\ref{fig:plot_tautau} in the $[M_A, \tan\beta]$ plane in the four configurations for some choices of the $a$ parameters, $M_a\!=\!100$ and 50 GeV and $\sin\theta=\frac{\sqrt 2}{ 2}$ or $0.1$. In all cases, the equality $M_H=M_{H^\pm}=M_A$ has been assumed.  As can be seen, while the constraint is strong in the Type II case, it is less severe in the other configurations and only 
masses below the $t\bar t$ threshold, $M_A \!<\! 350$ GeV,  are excluded for $\tan\beta$ values of order 20 in Type X and 2 in Type I and Y scenarios. In the left panels, we also show the regions (in gray) in which the total decay width of either the $A$ or $H$ states exceeds 10\% of its mass. While this condition should not be strictly regarded as a constraint, it should however be noted that most of the bounds from resonance searches are given assuming the narrow width approximation. The gray regions thus require a dedicated study which is nevertheless beyond our scope here.
 
The model points passing all constraints are shown in Fig.~\ref{fig:Hlimits}, distinguishing as usual the four different Yukawa configurations. From the left to the right columns, displayed are the viable model points in the bidimensional planes $[M_A,\tan\beta]$, $[M_H,\tan\beta]$, $[M_H^\pm,\tan\beta]$ and finally  $[\cos(\beta-\alpha),\tan\beta]$.
As expected, the Type II case appears to be the most constrained one, allowing only for values of $M_A$ between 700 GeV and 1 TeV and limiting $\tan\beta$ to $2 \lesssim \tan\beta \lesssim 15$. The lower limit on $M_A$ is essentially due to the bound from $b \rightarrow s \gamma$. While the latter strictly applies to the mass of the $H^\pm$ boson, it impacts also the other mass eigenstates since theoretical and electroweak constraints do not allow for arbitrary mass splittings. The low $\tan\beta$ region is ruled out by the constraints on the $B_s \rightarrow \mu^+ \mu^-$ process while high $\tan\beta$ values are disfavored by searches of neutral Higgs decaying into  $\tau^+ \tau^-$. 

As already mentioned, the lower bounds is weaker compared to the hMSSM case as a result of the suppression of the $A$ production cross section as well as its reduced decay branching fraction into $\tau$ pairs. The allowed parameter region is similarly small for the Type Y model. The only difference is the absence of a lower bound on $\tan\beta$, since searches of $\tau^+ \tau^-$ resonances are not effective in this case. The only constrain is represented by the Higgs signal strengths which require to be close to the alignment limit for values  $\tan\beta \gtrsim 10$.  

The Type I and X models are, on the contrary, very weakly affected by collider constraints. The most effective bound is the one on low $\tan\beta$ coming from $B_s \rightarrow \mu^+ \mu^-$. It is also worth noticing that our analysis shows that searches for  $Zh/ZH/ZA/ZZ$ events lead to weaker bounds than the ones obtained by the corresponding searches in the ordinary 2HDM \cite{ATLAS:2022enb,ATLAS:2020gxx,CMS:2019ogx,ATLAS:2020tlo}. This is due to the presence of additional decay channels such as $H \rightarrow aa,  aA$ or $A \rightarrow ha$ which reduce the branching fractions of the considered signals. More dedicated experimental searches for the production of light pseudoscalars from the decays of heavy resonances are needed to efficiently probe the 2HD+a model, besides the Type II.

\subsubsection{Constraints on the light $a$ boson}

Turning to the case of the light pseudoscalar $a$ boson, in addition to the pre-LHC bounds discussed in the beginning of this section, there is first a severe constraint from searches at the LHC in the decay $h \rightarrow aa$  of the SM-like Higgs boson for masses $M_a \lsim 62$ GeV~\cite{Goncalves:2016iyg,Haisch:2018kqx,Tunney:2017yfp}.  The partial decay width involves the $\lambda_{haa}$ coupling of Eq.~(\ref{eq:haa}) and is given by 
\begin{align}
\Gamma(h \rightarrow a a)= \frac{|\lambda_{haa}|^2}{8 \pi M_h} \sqrt{1- 4 M_a^2/M_h^2 } \, . 
\end{align}
 This process has been intensively searched for by the ATLAS and CMS collaborations in various topologies, namely $2b2\mu$, $2b2\tau$, $4b$, $jj\gamma \gamma$, $2\mu 2\tau$ and $4\tau$ and is also constrained by the $h$ invisible branching ratio which can be inferred from the Higgs signal strengths discussed earlier and which was  measured to be BR$(h\! \to \! \mbox{inv})\!<\!0.11$ \cite{ATLAS:2020kdi}.  To evade this constraint, a very small  coupling is required, $\lambda_{haa}/M_h \lsim \mathcal{O}(10^{-3})$. Such a value is achieved by choosing for the  parameters entering Eq.~(\ref{eq:haa}), e.g. $\sin\theta$ and $\lambda_{1P,2P}$, ad-hoc values that lead to an almost vanishing coupling. Hence, one should rely on  blind spots on the coupling and we refer to Ref.~\cite{Arcadi:2021zdk} for a more detailed discussion that involves the possibility of including radiative corrections to $\lambda_{haa}$ that are generated by loops of $b$- and $t$-quarks with enhanced couplings. 

Searches of mono-$Z$ and mono-$h$ signatures, corresponding to $pp \rightarrow Za,ha$ with $a \rightarrow \chi \chi$,  represent a very interesting tool for probing the 2HD+a model; see e.g. Ref.~\cite{Arcadi:2020gge}. The latter searches are, however, effective only if the DM candidate is lighter than $\frac12 M_a$. As will be detailed in the next DM section, we will consider a broader range of values of the DM mass and,  for this reason, we have not included mono-X searches in our analysis.

The most severe constraints on $a$ with a significant mixing  with the heavier $A$ comes from searches of light resonances decaying into muon pairs which have been revived recently \cite{Argyropoulos:2022ezr}
\beq 
p  p \to gg, b\bar b \to a \to \mu^+ \mu^- \, . 
\eeq
As in the case of $H/A$ discussed above, the gluon-fusion process is mediated  by $t$-quark loops at low and $b$-quark loops at high $\tan\beta$ in scenarios like Type II and Y; in the high-$\tan\beta$ case, additional contributions from $b$-quark fusion should also be included. The decay branching ratio BR($a\to \mu^+\mu^-$) is important only in Type II and X scenarios at high $\tan\beta$ when the coupling is $g_{a\ell\ell } = \sin\theta \tan\beta$. Hence, the cross section times branching ratio is significant only in the Type II case with $\tan\beta \gsim 10$ when both production and decay rates are large.    

Two recent searches have been conducted in this channel, one by CMS \cite{CMS:2019buh} and another one by the LHCb collaboration \cite{LHCb:2020ysn}. The latter, which has been interpreted only in the Type Y configuration when setting $\tan\beta= \frac12$ and assuming  $M_a \gsim 10$ GeV, is the strongest. 

\begin{figure}[!ht] 
    \centering
    \subfloat{\includegraphics[width=0.37\linewidth]{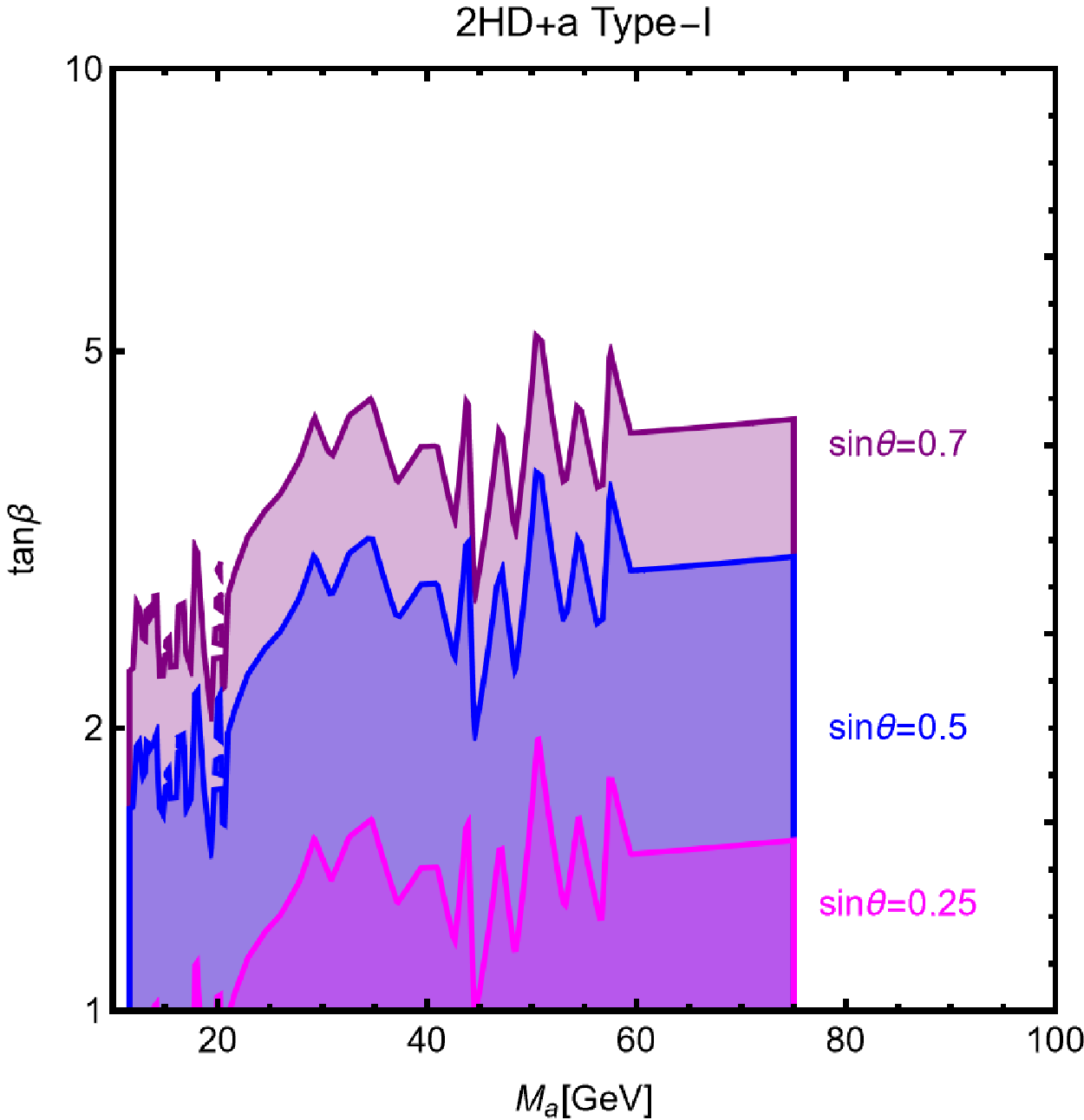}}~~ \subfloat{\includegraphics[width=0.37\linewidth]{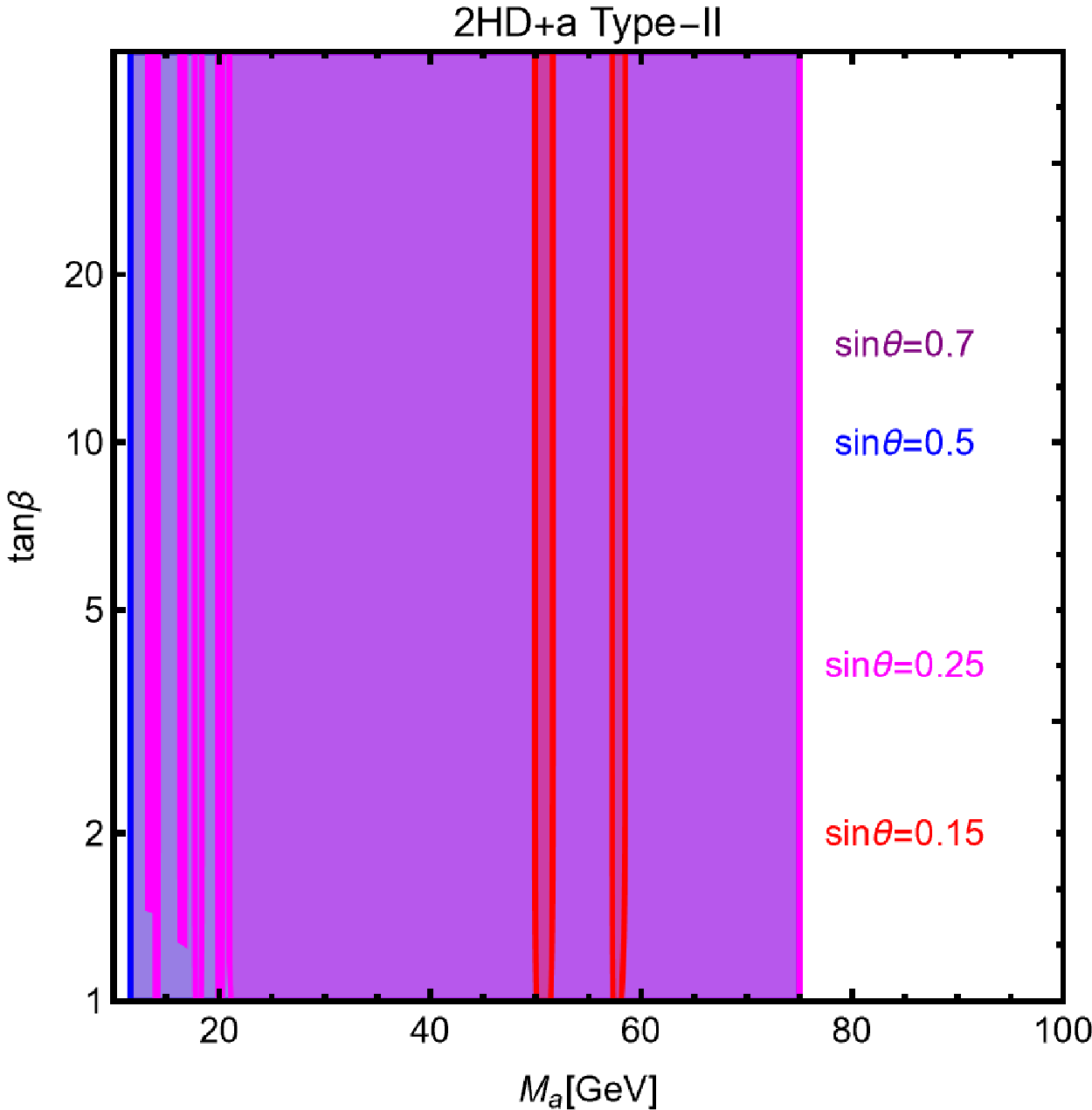}}\\[1mm]  \subfloat{\includegraphics[width=0.37\linewidth]{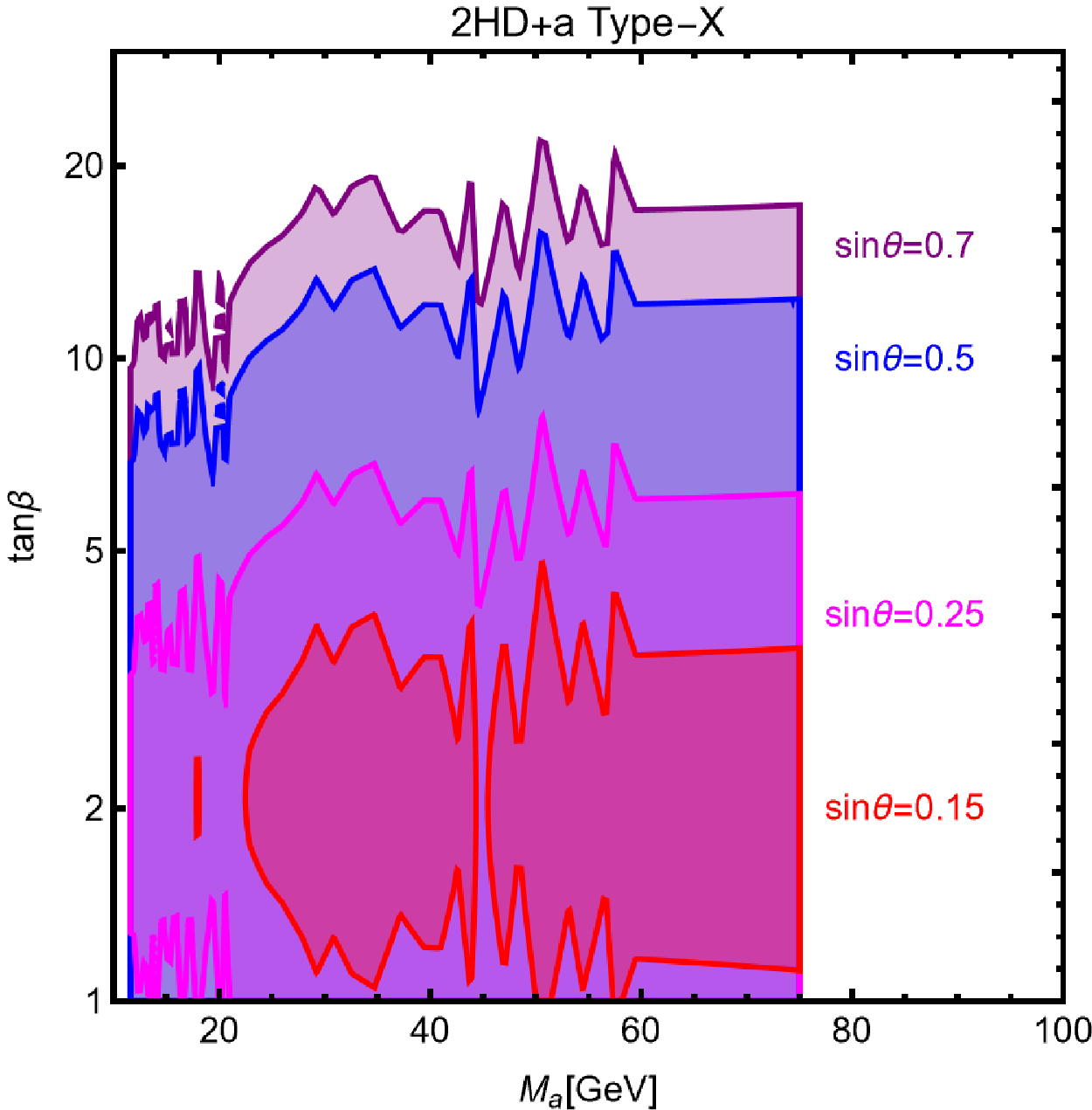}}~~ \subfloat{\includegraphics[width=0.37\linewidth]{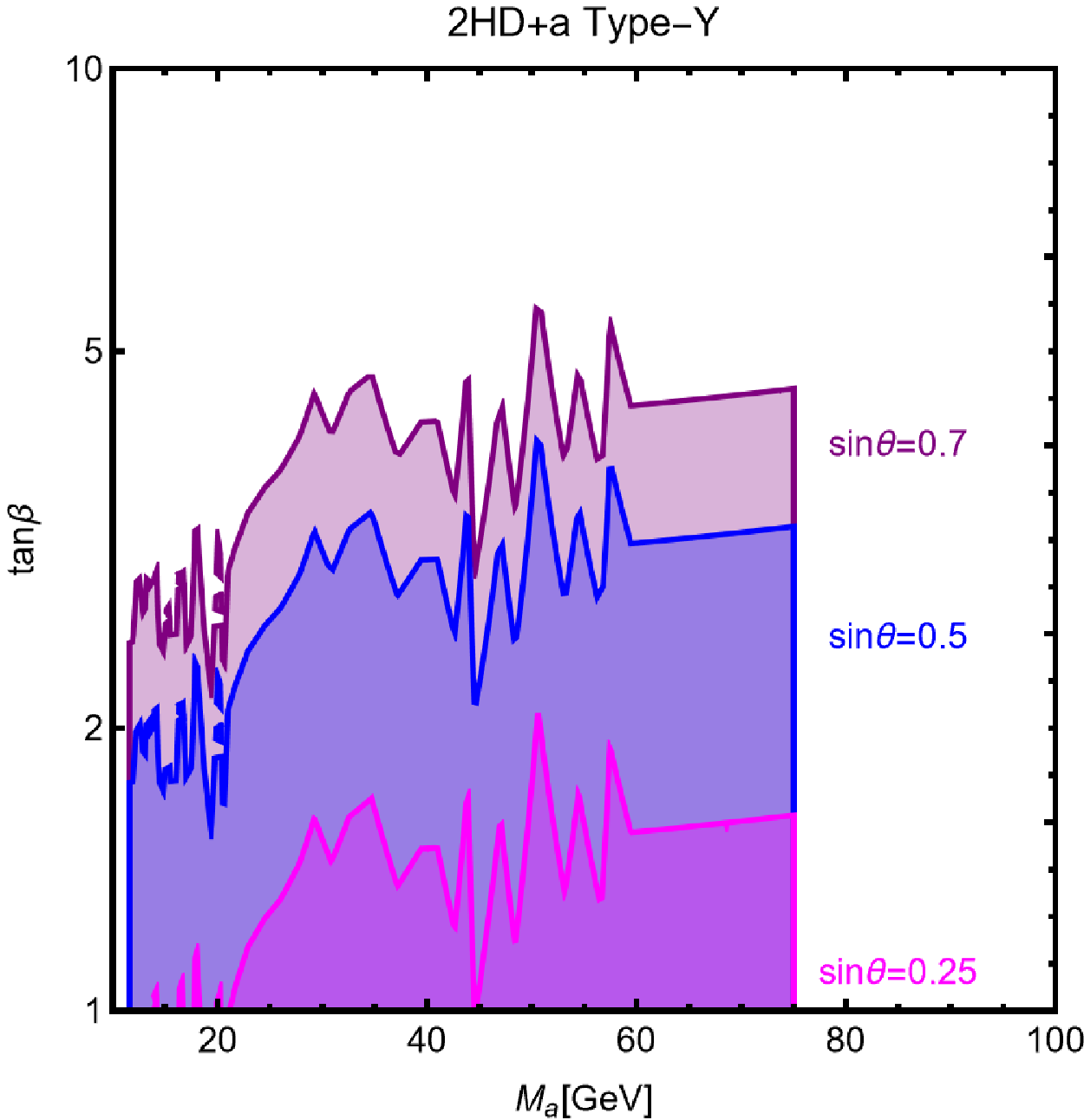}}
\vspace*{-2mm} 
  \caption{Excluded regions in the $[M_a,\tan\beta]$ plane from searches in the process $gg,b\bar b\rightarrow a \rightarrow \mu^+ \mu^-$ in the four types of Yukawa configurations. Each colored region corresponds to different assignments of $\sin\theta$ reported on the panels.}     
 \label{fig:mumu}
 \vspace*{-2mm}
\end{figure}

We have recast the resulting bound of this search in the $[M_a,\sin\theta]$ plane and for larger $a$ masses up to $M_a =\frac12 M_h$. The production rate $\sigma (pp\! \rightarrow \!a)$ has been calculated using the programs HIGLU \cite{Spira:1995mt} and for the decay rate $\mbox{BR}(a \! \rightarrow \! \mu^+ \mu^-)$ using the the program HDECAY \cite{Djouadi:1997yw,Djouadi:2018xqq} and we have compared the obtained result with the corresponding one given by CMS  \cite{CMS:2019buh}.
The excluded regions in the $[M_a, \tan\beta]$ plane and for different assignments of $\sin\theta$ are shown in Fig.~\ref{fig:mumu} in which the four panels correspond to the four Yukawa configurations, namely Type I, II, X and Y.  The mass $M_a$ is varied from 10 to 62 GeV, while the CP-odd mixing angle was assigned the values  $\sin\theta=0.15,0.25,0.5$ and $0.7$.

In agreement with the findings of Refs.~\cite{Arcadi:2021zdk,Arcadi:2022dmt}, as well as in the earlier discussion in Ref.~\cite{Argyropoulos:2022ezr}, the strongest constraint applies on the Type II scenario in which both the production and decay rates are enhanced as $g_{a bb} = g_{a\ell \ell} \propto \tan\beta$. In the absence of additional decays of the $a$ state,  it basically rules out the whole parameter space for which the considered searches have sensitivity. A weaker but still sizable constraint is obtained in the Type X model, as a consequence of the $\tan\beta$ enhancement of the coupling of the $a$ state with muons.  Much weaker are the limits which apply in the Type I and Y models for which one obtains the lower bounds $\tan\beta \gtrsim 2\, (5)$ for $\sin\theta=0.5\, (0.7)$.

Finally, let us note that for a non-zero $Aa$ mixing,  one would also have the possibility of pair production of the $H/A$ and $H^\pm$ states with a light pseudoscalar $a$,   $q\bar q \! \to \! Ha$ and  $q\bar q' \! \to \! H^\pm a$ which occur through virtual vector boson exchange. When $\theta\!=\! \frac{\pi}{2}$, the cross sections are maximal and the processes are favored by  phase space as we expect the $a$ state to be much lighter than the $A$ and $H^\pm$ bosons. They lead to interesting topologies with four fermions in the final states like $H \to b\bar b, t\bar t$ and $H^+ \to tb$ while one should have $a \to \tau\tau, b\bar b$ and even $a\to \mu^+\mu^-$ decays for the light $a$. Nevertheless, except for the $Ha$  case which has been adapted from the 2HDM  CMS search in the channel $pp\to AH$ \cite{CMS:2019ogx} discussed before, these processes have not been explicitly considered by the ATLAS and CMS experiments and there is barely a way to set strong limits. One expects, though, that these limits are not stronger than the ones from the $gg\to a \to \mu^+\mu^-$ process that we discussed here. 

\section{The dark matter and combined constraints}

\subsection{The DM relic density}

In our 2HD+a context, we have introduced a dark matter particle candidate which was assumed to be a Dirac fermion $\chi$ that is isosinglet under the SM gauge group (no substantial change of the results are expected in the case in which the DM were of Majorana type). We also introduced a discrete $Z_2$ symmetry under which the new DM field is odd and transforms as $\chi \to -\chi$ while all other fields are even and transform like $\phi \to +\phi$, making that the $\chi$ particle cannot decay into SM particles and is hence absolutely stable as it should be.  Because it is not charged under the ${\rm SU(2)_L}$ group, $\chi$ has no couplings to gauge bosons and, by virtue of the $Z_2$ symmetry, it couples to  Higgs bosons only in pairs.  

Starting from an initial coupling $i g_\chi a_0 \bar \chi i \gamma^5 \chi$ of the $\chi$ states with the $a_0$ boson (the $\chi$ states do not couple to the 2HDM bosons), and after electroweak symmetry breaking, the DM will interact with the two pseudoscalar Higgs bosons according to the following Lagrangian
\begin{equation}
\mathcal{L}_{\rm DM}=y_\chi \left(\cos\theta a+\sin\theta A\right) \bar \chi i \gamma_5 \chi \, . 
\end{equation}
There are no couplings of the DM fermion to the CP-even Higgs bosons at the tree-level, a feature which will have major consequences as will be discussed shortly.   

The DM fermion will have the correct cosmological relic density, as we will assume the conventional freeze-out mechanism in which the experimentally favored value measured by the Planck collaboration \cite{Planck:2018vyg}
\begin{equation} 
\Omega_{\chi}h^2 = 0.12 \pm 0.0012\, ,  
\end{equation} 
is achieved if the DM thermally averaged pair annihilation cross section has a value in the appropriate range. For the scenario that we are interested in here,  the most relevant DM annihilation channels are the following final states that occur via $a/A$ boson exchange
\begin{equation}
\chi \chi \to a^*, A^* \to \tau^+ \tau^-,\  b\bar b \ {\rm and} \  t \bar t \, , 
\end{equation}
the latter channel occurs only when kinematically accessible, i.e. for $m_\chi \gsim 175$ GeV. The weight of the individual channels depend on the type of Yukawa coupling configuration, namely Type I, II, X and Y as well as on the value of $\tan\beta$.  In addition to annihilation into SM fermion pairs, the following final states could also be relevant
 \begin{equation} 
 \chi \chi \to a^*, A^* \to   h a \, , \ Zh \ {\rm and} \ \ \chi \chi \to aa \, ,  
 \end{equation}
where in the last case, the  $aa$ final state is obtained via $t$-channel exchange of the DM. In the first channel, $a^*$ exchange should be suppressed for $m_\chi \lsim \frac12 M_h$ as the coupling $g_{haa}$ should be very small as to make the decays $h \to aa$ very rare.  The second channel with a $Zh$ final state is  only possible outside the alignment limit when the coupling $g_{hZa}$ is non-zero. 

The numerical determination of the DM relic density is achieved through the implementation of the different 2HD+a scenarios into the package micrOMEGAs \cite{Belanger:2001fz,Belanger:2007zz,Belanger:2008sj}.  For completeness, the annihilation channels into $aA$, $AA$, $ZH$ and $W^{\pm}H^{\mp}$ final states, which  open up only for DM masses above several hundreds GeV, have also been included. 

\begin{figure}
    \centering
\subfloat{\includegraphics[width=0.39\linewidth]{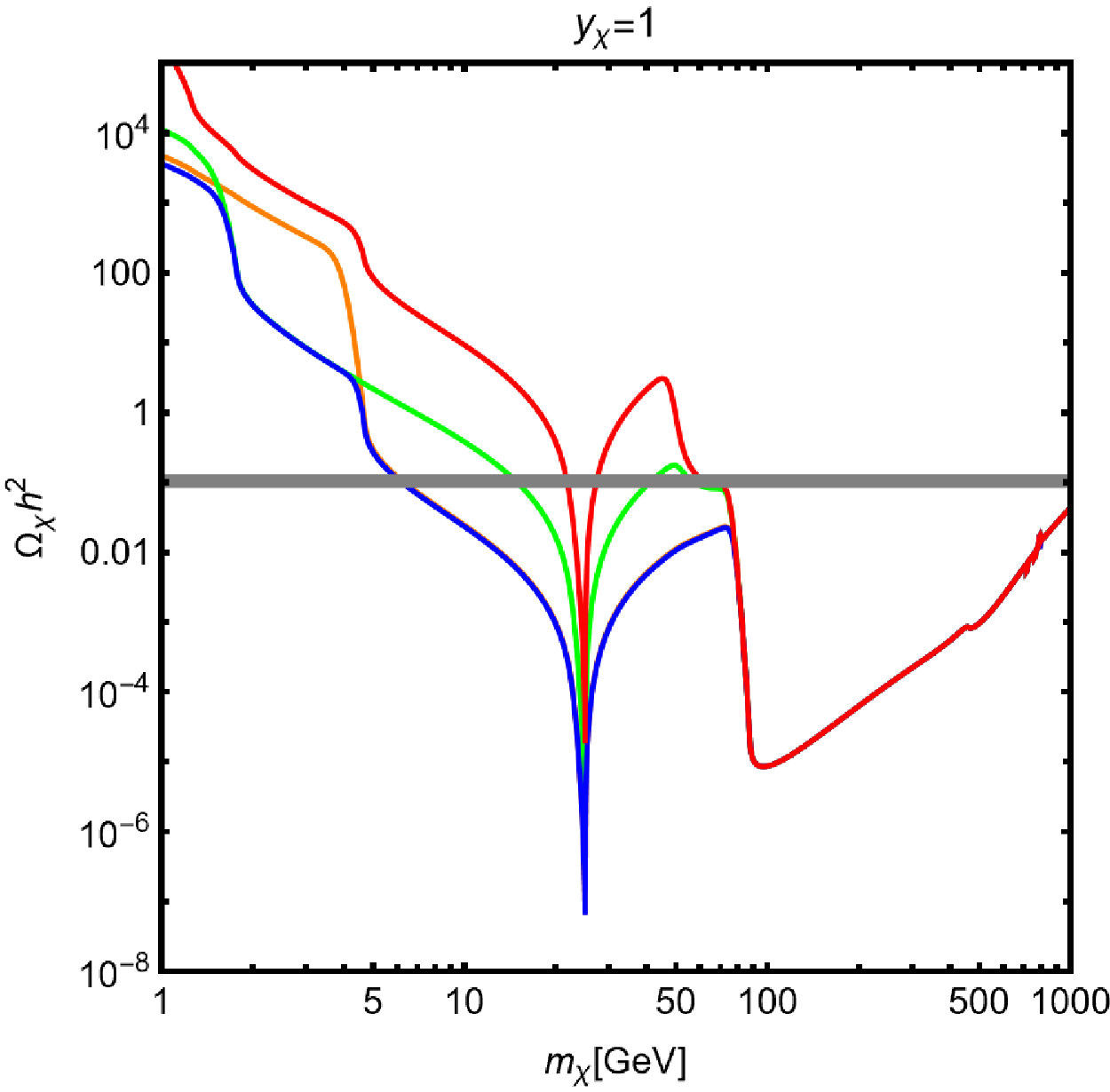}}~~    \subfloat{\includegraphics[width=0.39\linewidth]{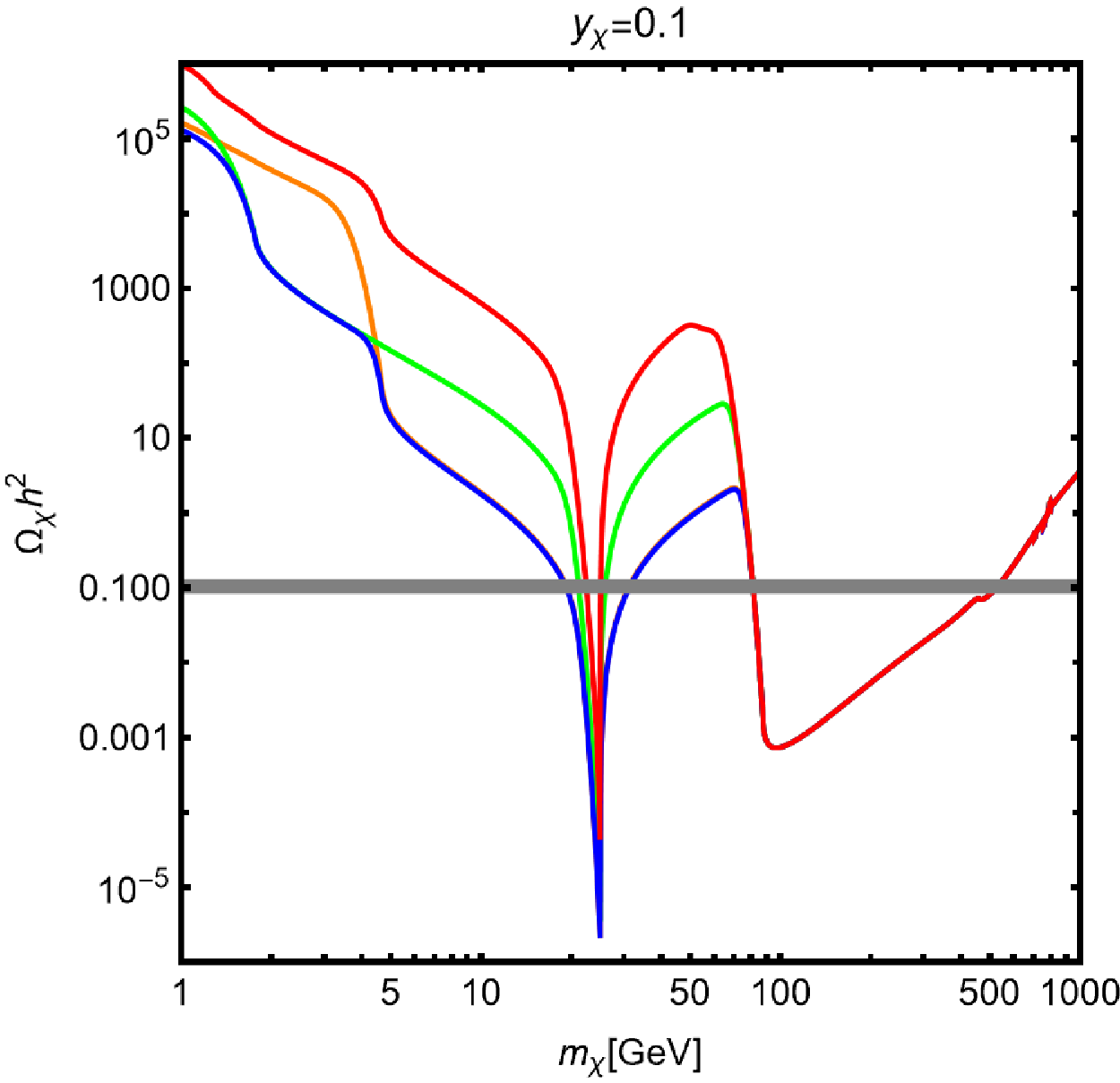}}
\vspace*{-.2mm}
    \caption{The DM relic density as a function of the DM mass for $\tan\beta\!=\!5$, $\cos(\beta-\alpha)\!=\!0$, $\theta\!=\!\frac{\pi}{4}$, $M_a=50\,\mbox{GeV}$ and $M_{H,A,H^\pm}\!=\!800\,\mbox{GeV}$. The different colored lines stand for Type I (red), II (blue), X (green) and Y (orange) and the left (right) panels refer to $y_\chi=1 \, (0.1)$.}
    \label{fig:pOmega}
\end{figure}

A simple illustration of the impact of the relic density constraint is provided by Fig.~\ref{fig:pOmega}. Here, we have considered the same benchmark for all the four Yukawa configurations, namely $\cos(\beta\!-\!\alpha)\!=\!0$, $\tan\beta\!=\!5$, $\theta\!=\! \frac{\pi}{4}$,  $M_H\!=\! M_A\! =\! M_{H^{\pm}}\! =\! M\! =\! 800\,\mbox{GeV}$ and $M_a\! =\! 50\,\mbox{GeV}$ and computed the relic density as a function of the DM mass for two assignments of the coupling $y_\chi$, namely $y_\chi\!=\!1$ (left) and $y_\chi\!=\!0.1$ (right). In each panel, the different colored curves represent the different Yukawa configurations, namely red/blue/green/orange for Type I/II/X/Y, respectively. As can be seen, the DM relic density is sensitive to the different realizations of the 2HD+a, in particular, in the $m_\chi \lesssim M_a$ range. Here, the relic density is mostly due to annihilations into SM pairs via $s$-channel mediation of the $a/A$ states, whose cross section are sensitive to the $\tan\beta$ enhancement/suppression of the Yukawa couplings. Even for $y_\chi\!=\!1$, for the Type I and X models, the correct relic density requires the occurrence of the resonant enhancement of the annihilation cross section for $m_\chi \! \simeq \! \frac12 M_a$. In turn,  the Type II and Y models can have the correct relic density even outside the pole region. As the DM mass increases, the relic density becomes dominated by the $aa$ and, most importantly, $ha$ channels whose cross sections are essentially the same for all the four 2HD+a types.

\subsection{Constraints from direct and indirect detection}

One of the main experimental probes of a weakly interacting and massive DM candidate is represented by direct detection (DD): namely,  the search for the recoil energy deposited in a suitable detector when (elastic) scatterings between the DM particle and the atomic nuclei of the target detector occur. In this regard, the 2HD+a has the very peculiar and interesting property that spin-independent interactions, the ones which are most efficiently probed by present experiments, emerge only at the one-loop level. 

\begin{figure}[!ht]
\vspace*{-5.1cm}
    \centering     \subfloat{\includegraphics[width=1.2\linewidth]{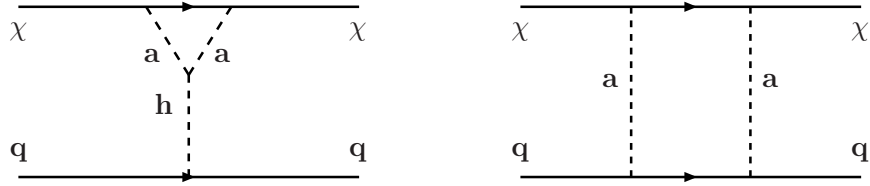}}
\vspace*{-17.4cm}
    \caption{Generic Feynman diagrams for the loop induced scattering of the DM particle on quarks in the 2HD+a model. }
    \label{fig:feynloop}
\vspace*{-2mm}
\end{figure}

The Feynman diagrams responsible for such interactions have two possible topologies shown by the two representative examples given in Fig.~\ref{fig:feynloop}.  The first diagram involves triangle vertices with one CP-even neutral Higgs boson  which is coupled with a the SM quarks and  a pair of pseudoscalar Higgs bosons that couple to the DM state.  The second topology is represented by box-diagrams involving the exchange of two pseudoscalar states between the lines formed by the SM quarks and the fermionic DM candidate.  

We have determined the DM scattering cross section,  adopting the computation performed in  Refs.~\cite{Abe:2018emu,Abe:2019wjw} and slightly refined in Refs.~\cite{Arcadi:2017wqi,Bell:2018zra,Ertas:2019dew} for instance.  We have then compared the results with the strongest exclusion limit as given at the moment by the LZ collaboration \cite{LZ:2022ufs} (which superseded the earlier strong XENON1T limits \cite{XENON:2018voc}; notice that there is also a dedicated study made by the PANDA-X collaboration \cite{PandaX:2022xas}). 

Some of the relevant annihilation channels of DM, in particular the ones into SM fermions pairs via pseudoscalar Higgs exchange, feature an $s$-wave dominated cross section, i.e. the values of the cross section at present times and at freeze-out are very close to each other. Consequently, the viable parameter space for the relic density can be probed by indirect detection (ID) experiments as well that search for the clean products of the annihilation processes. In order to account for indirect detection, we have used the limits from searches of continuous $\gamma$-ray signals determined by the FERMI-LAT experiment in Refs.~\cite{Fermi-LAT:2015att,Fermi-LAT:2015kyq}.  

\begin{figure}[!h]
    \centering
    \subfloat{\includegraphics[width=0.39\linewidth]{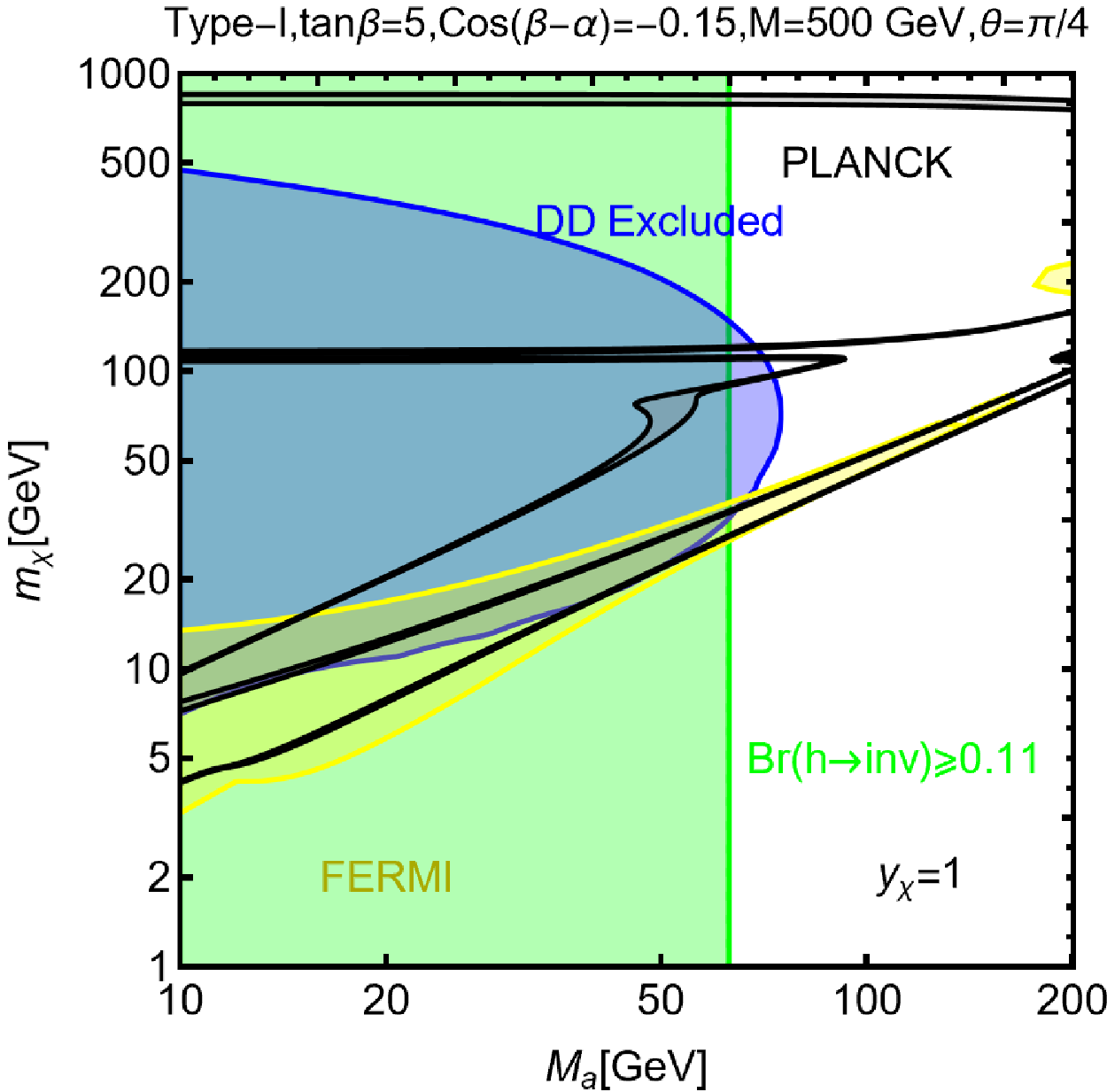}}~~
    \subfloat{\includegraphics[width=0.39\linewidth]{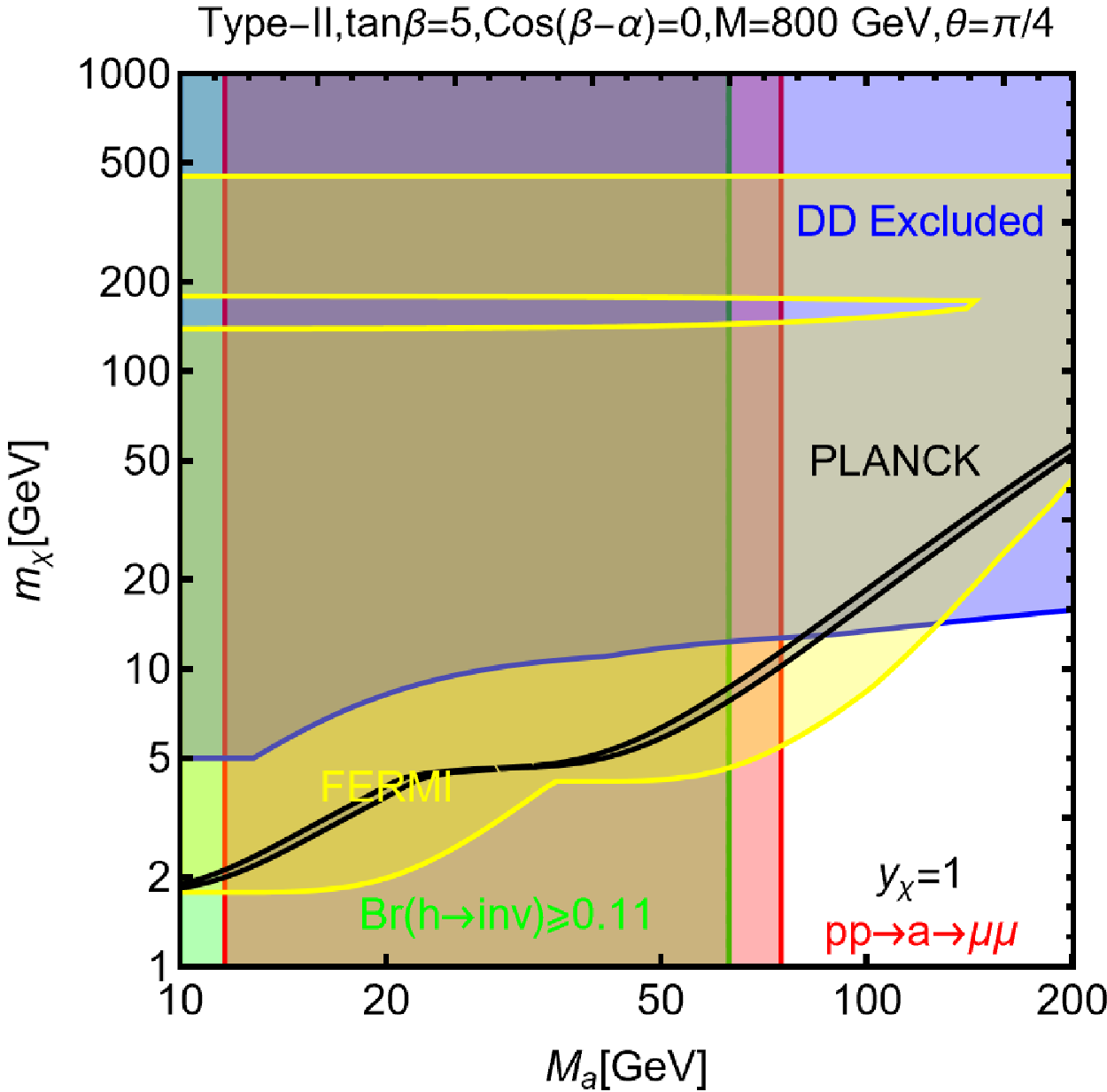}}\\[2mm]
    \subfloat{\includegraphics[width=0.39\linewidth]{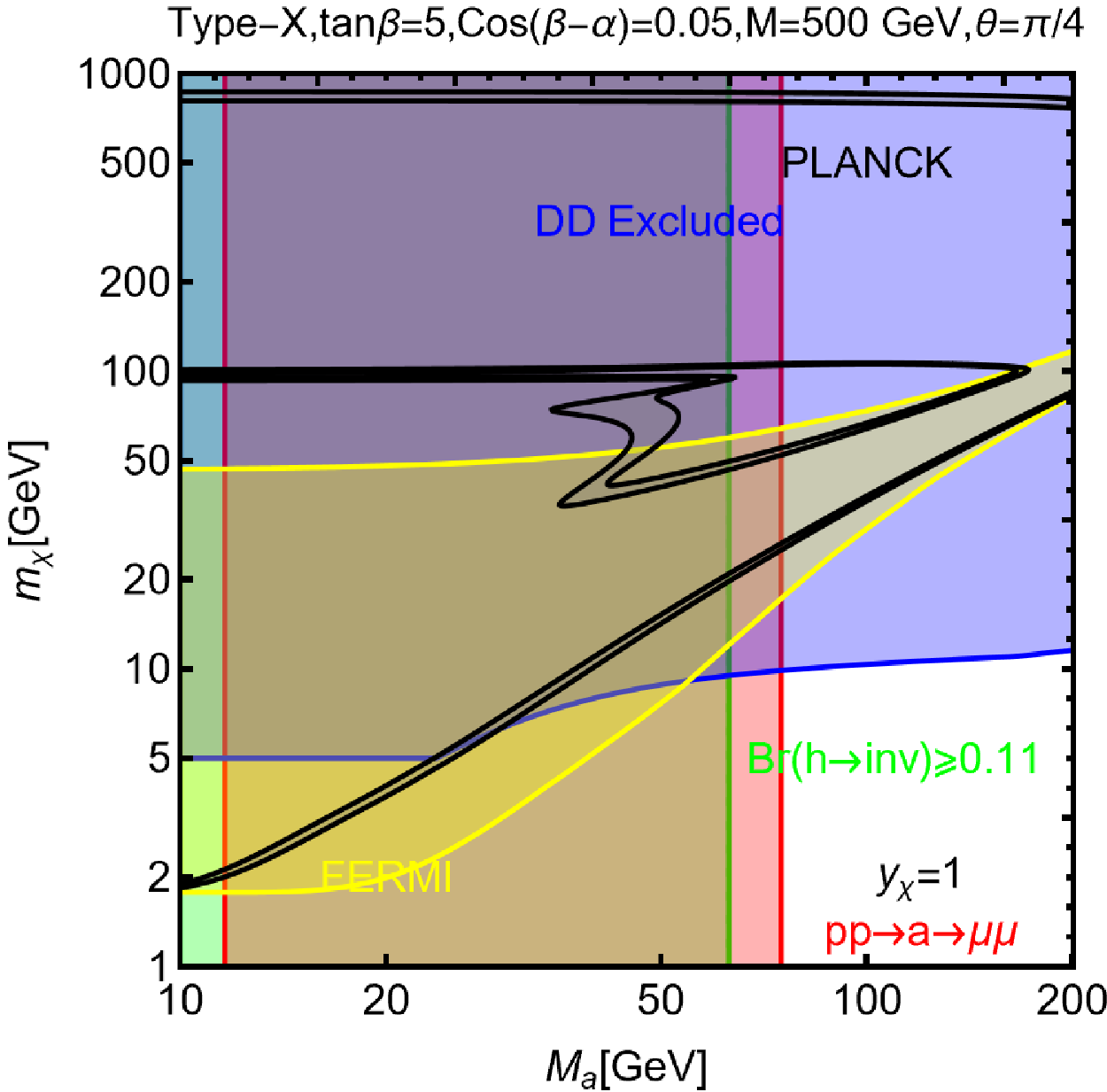}}
    \subfloat{\includegraphics[width=0.39\linewidth]{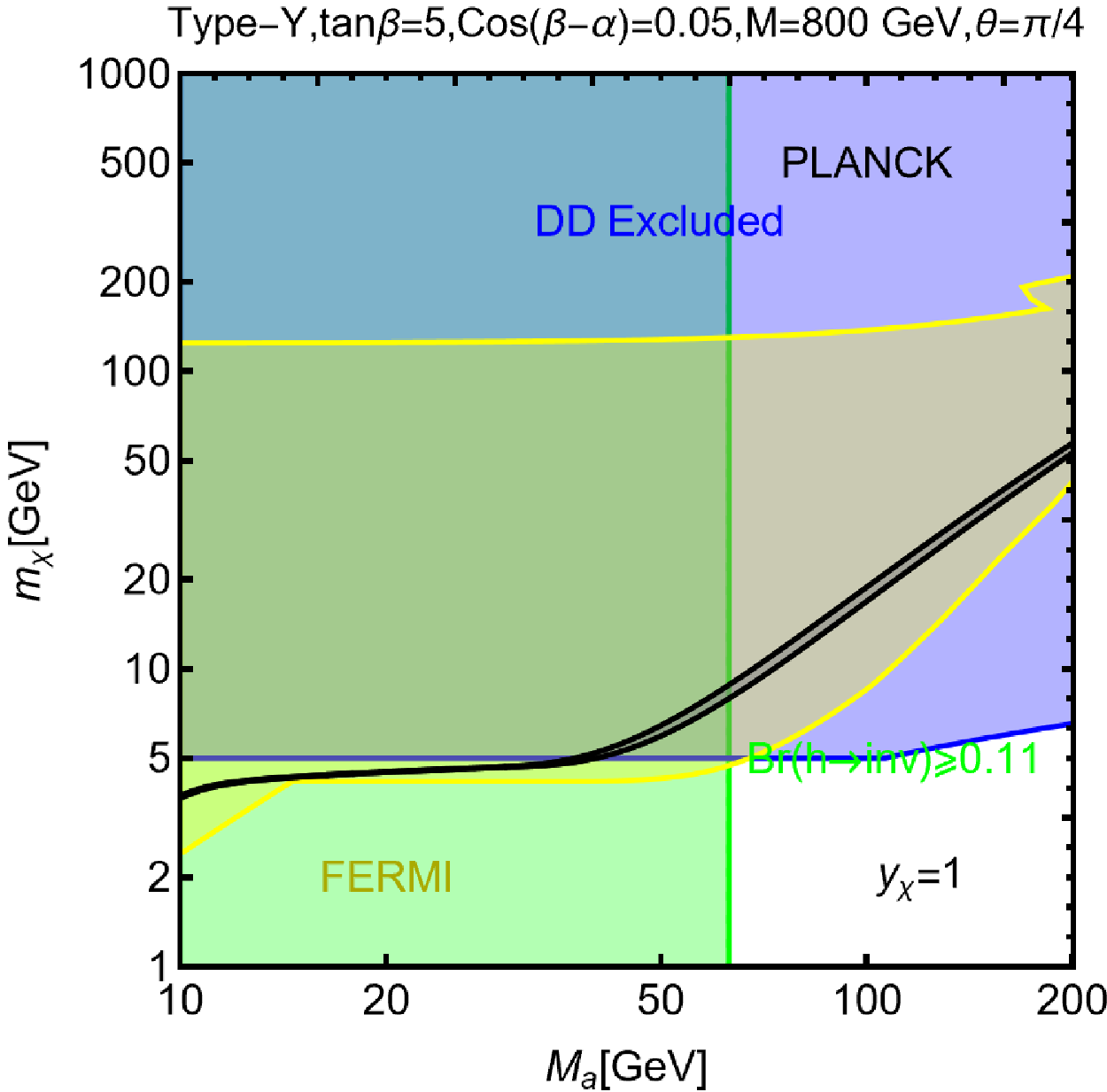}}
    \caption{DM constraints in the $[M_a,m_\chi]$ plane for some benchmark assignments of the 2HD+a parameters in the four configurations of the Yukawa couplings. In each plot, the black isocontours correspond to the correct DM relic density, while the blue (yellow) regions represent the parameters space excluded by DM direct (indirect) detection. For comparison, the region excluded by the invisible width of the 125 GeV Higgs and by searches of light resonances decaying into $\mu^+ \mu^-$ have also been shown, in green and red, respectively.}
    \label{fig:pDM1}
\end{figure}

\begin{figure}[!ht] 
    \centering
    \subfloat{\includegraphics[width=0.39\linewidth]{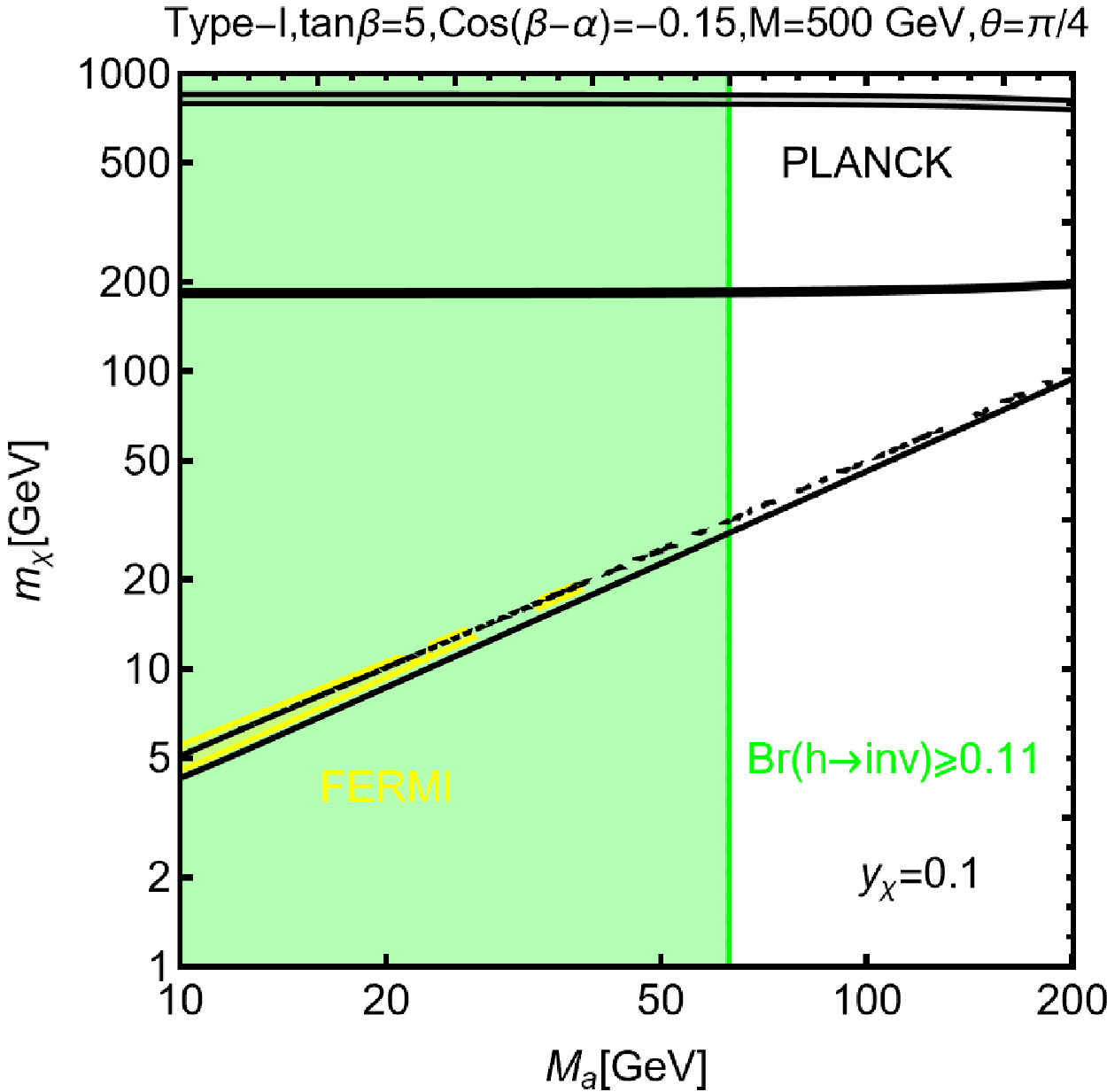}}~~
    \subfloat{\includegraphics[width=0.39\linewidth]{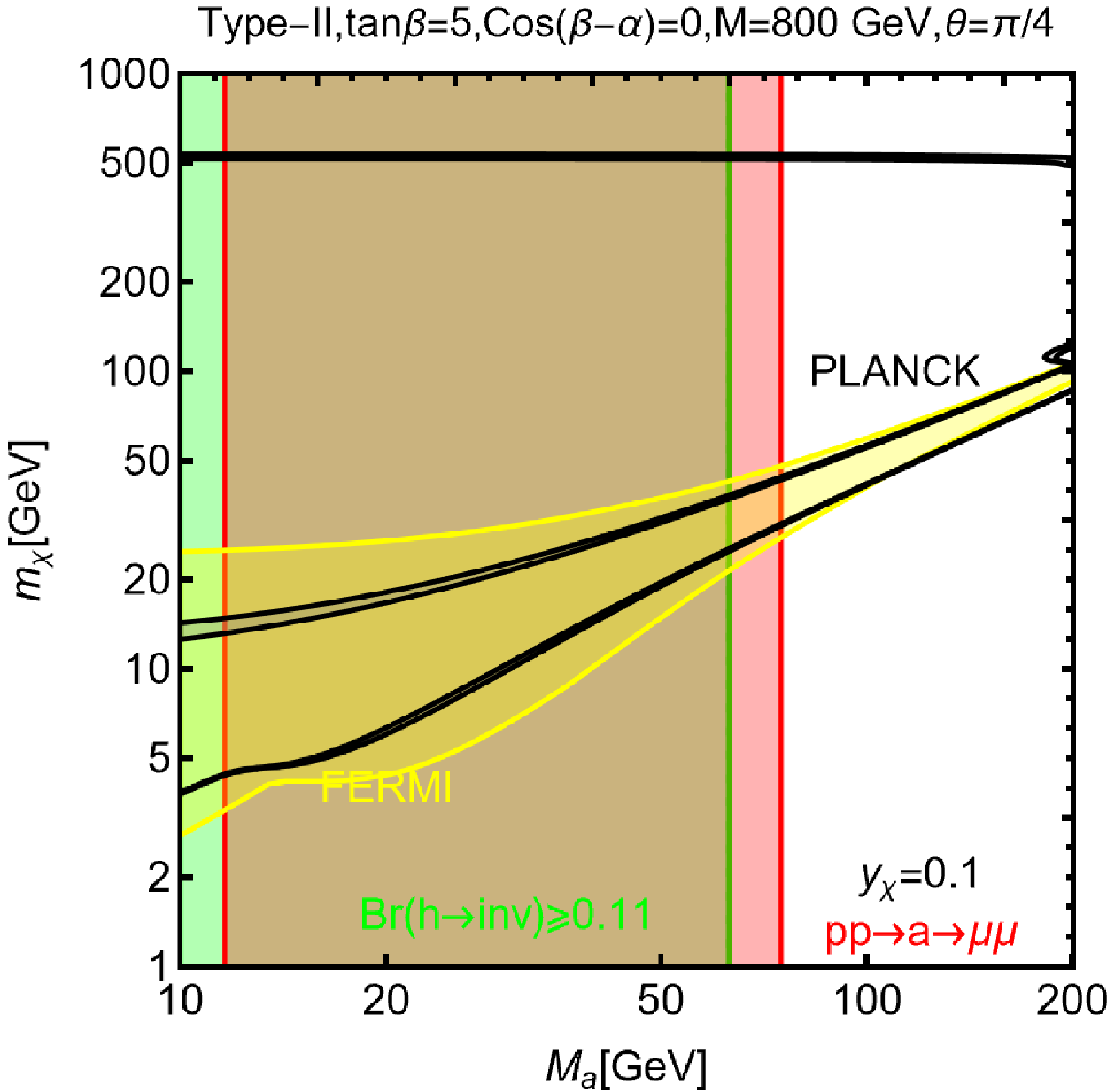}}\\[5mm]
    \subfloat{\includegraphics[width=0.39\linewidth]{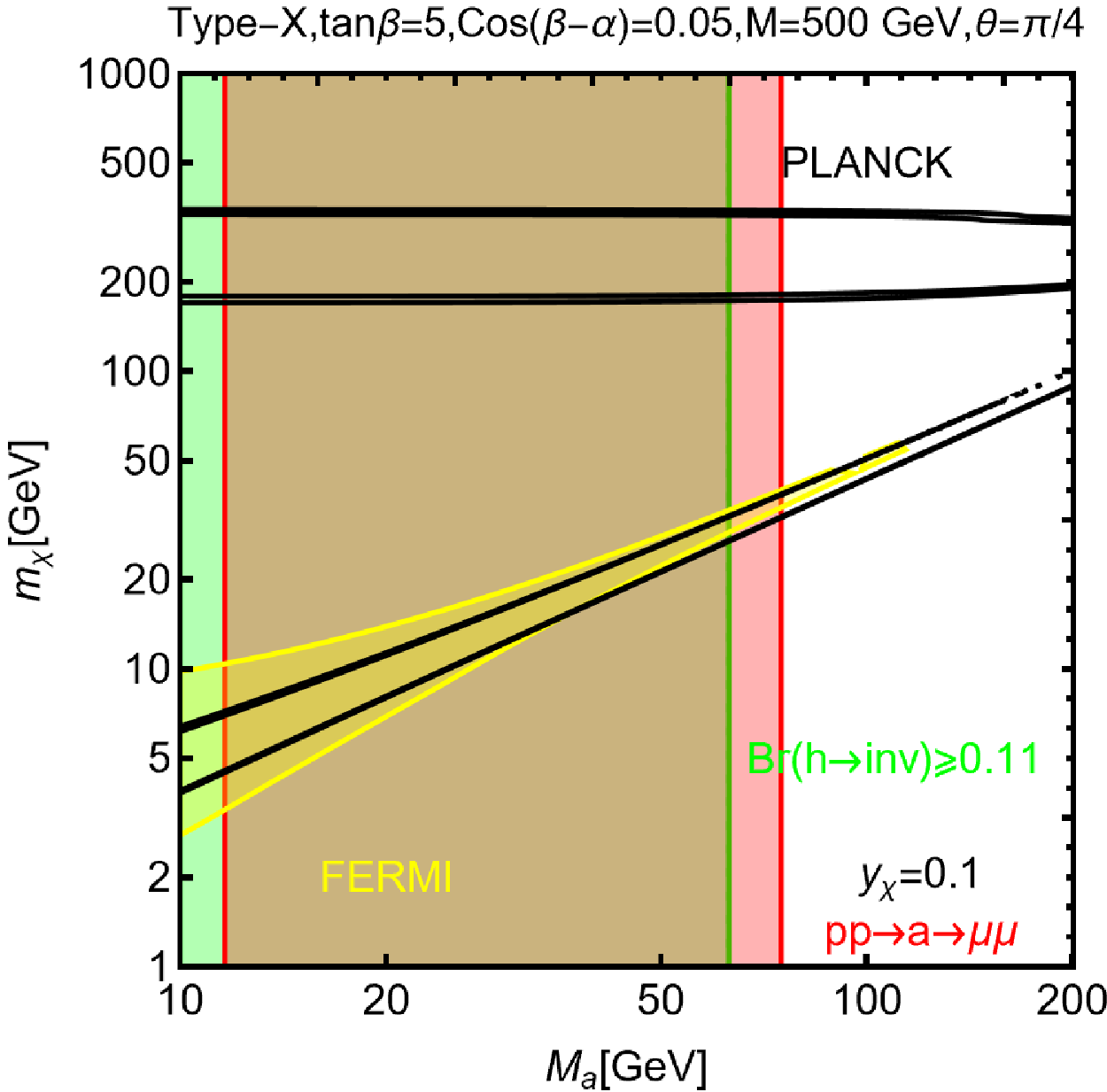}}~~
    \subfloat{\includegraphics[width=0.39\linewidth]{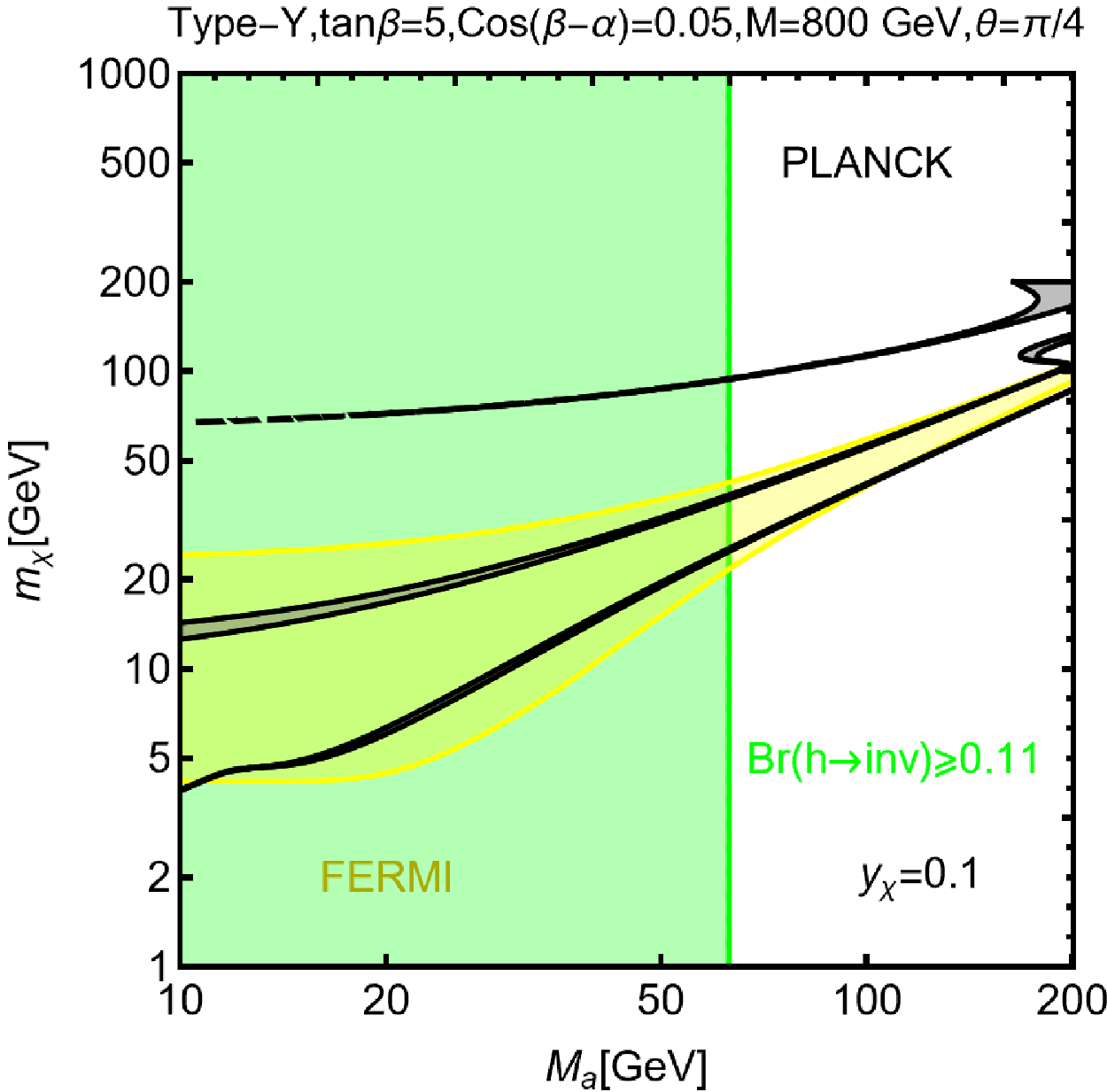}}
    \caption{DM constraints in the $[M_a,m_\chi]$ plane for the four types of the 2HD+a model. Everything is the same 
    as in Fig.~\ref{fig:pDM1} but taking $y_\chi=0.1$ instead of $y_\chi=1$.}
    \label{fig:pDM01}
\end{figure}

Before discussing our main results, obtained via  scans of the parameter space, we provide in Figs.~\ref{fig:pDM1} and Fig.~\ref{fig:pDM01}, two simplified illustrations of the impact of the DM constraints with some other relevant bounds, namely the one from the invisible width of the $h$ boson which accounts for possible $h\to aa$ decays and light $a$ boson searches in the $pp\! \rightarrow \! a\! \to \! \mu^+ \mu^-$ mode. The two figures illustrate the limits in the $[M_a,m_\chi]$ plane for all the four flavor-preserving Yukawa configurations for some fixed assignments of the parameters $\cos(\beta-\alpha)$, $\tan\beta$ and $\theta$. For simplicity, we have assumed mass degeneracy for the heavy 2HDM states, $M_H\!=\!M_A\!=\!M_{H^{\pm}} \!=\! |M|$. The two figures differ only in the assignment of the $a\chi \chi$ coupling, which has been taken to be large $y_\chi=1$ in Fig.~\ref{fig:pDM1} and small $y_\chi=0.1$ in Fig.~\ref{fig:pDM01}. In each plot, the correct DM relic density is achieved along the black isocontours while the blue, yellow, green and red regions are excluded, respectively, by direct detection by LZ, indirect detection by FERMI-LAT, the invisible branching ratio of the SM-like Higgs boson and LHC searches of light resonances decaying into $\mu^+ \mu^-$ final states.

In summary, the viable parameter space for a given benchmark corresponds to the case in which the black isocontour of the relic density lies outside all the colored regions. As it can be seen from Fig.~\ref{fig:pDM1}, despite of their radiative origin, DM spin-independent interactions can be strongly constrained thanks to the high sensitivity reached by current generation of multi-ton detectors. On the other hand, the scattering cross section is strongly sensitive to the DM coupling, as the spin-independent cross section of the DM on protons behaves as  $\sigma_{\chi p}^{\rm SI} \propto y_{\chi}^4$. Moving from the $y_\chi=1$ to the  $y_\chi=0.1$ case renders the DM direct detection limits irrelevant while it is still possible to achieve a correct relic density. 

One also notices that the region $M_a \lesssim \frac12 M_h$ is completely ruled out by the bound  imposed on the invisible branching ratio  of the 125 GeV Higgs, BR$(h \to {\rm inv}) \leq 0.11$.   As already pointed out, the latter includes also the decay channel $h\to aa$ which can be evaded only by imposing specific conditions on the parameter of the scalar potential that leads to a very tiny $\lambda_{haa}$ coupling. This has not been enforced in  the benchmark considered in Fig.~\ref{fig:pDM01}.

\subsection{Combined constraints}

\begin{figure}[!h]
\vspace*{-4mm}
    \centering    \subfloat{\includegraphics[width=0.29\linewidth]{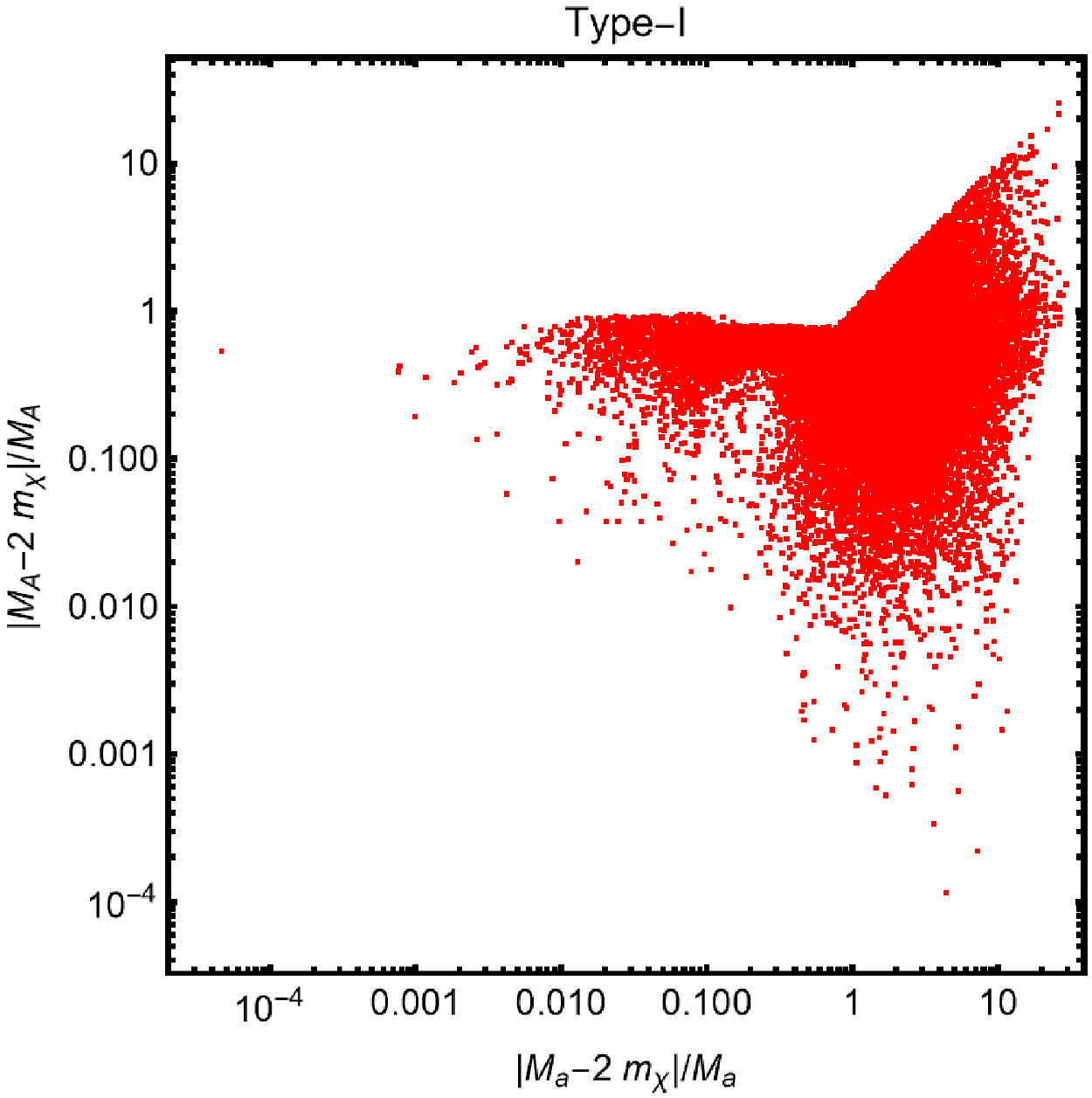}}~~~
    \subfloat{\includegraphics[width=0.29\linewidth]{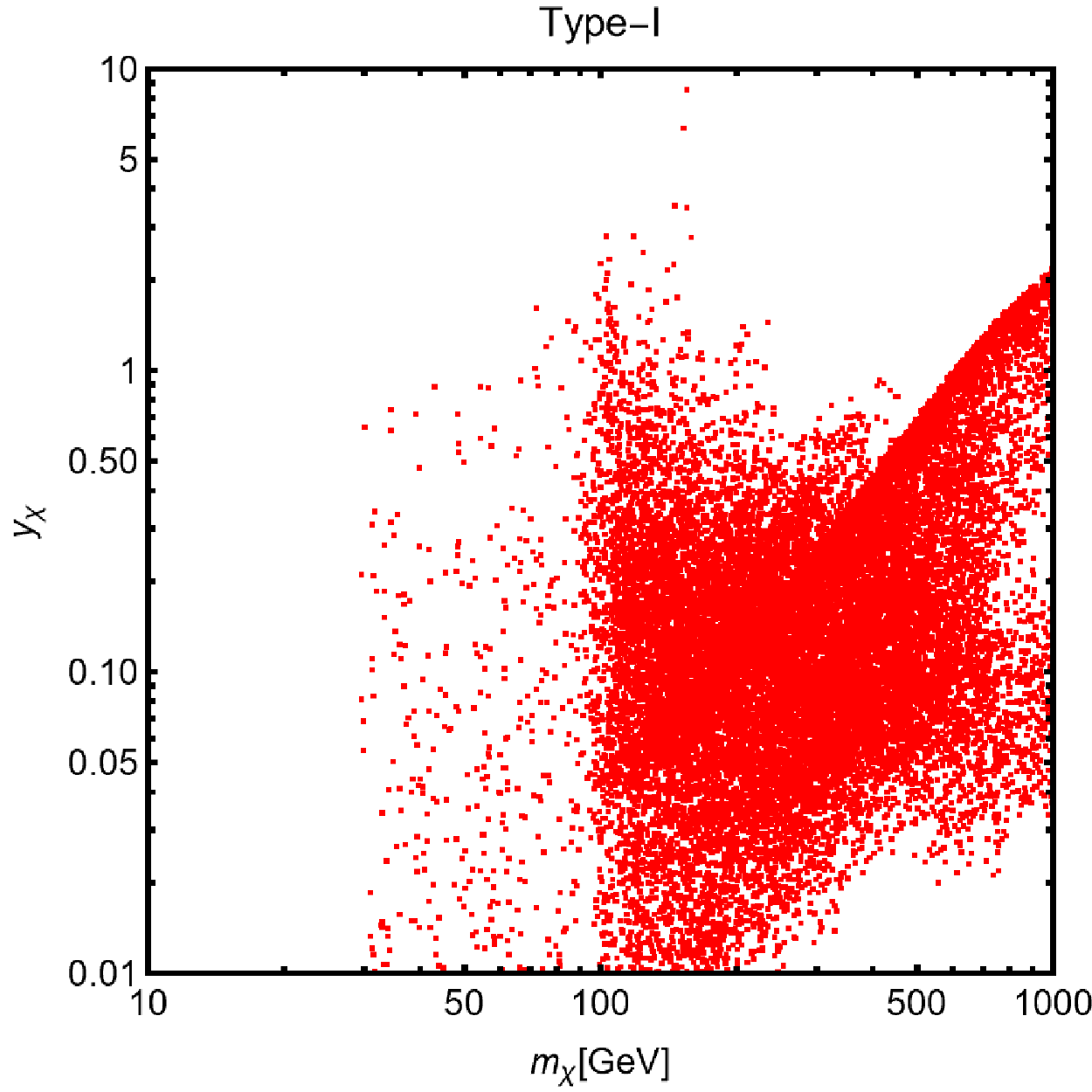}}\\
    \subfloat{\includegraphics[width=0.29\linewidth]{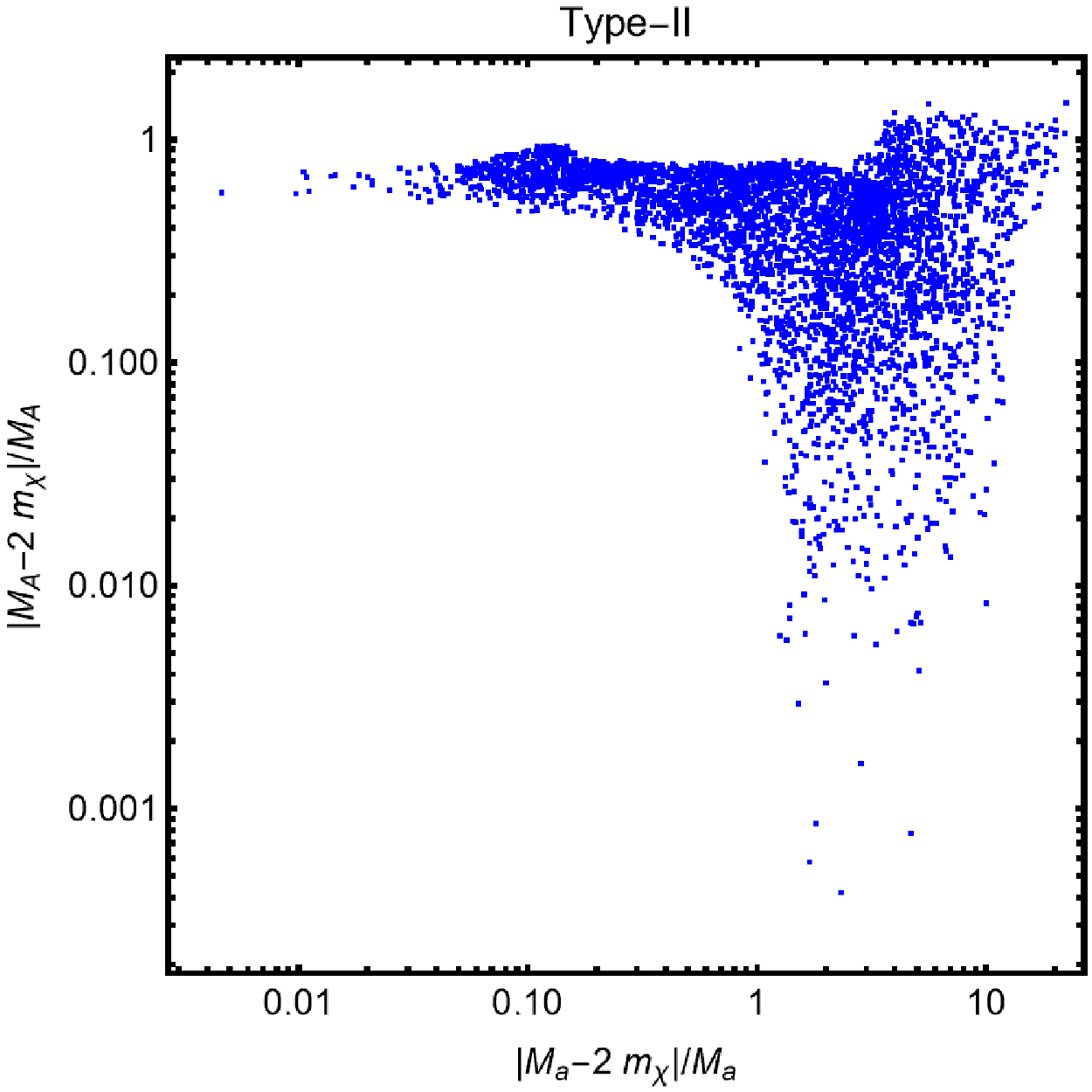}}~~~
    \subfloat{\includegraphics[width=0.29\linewidth]{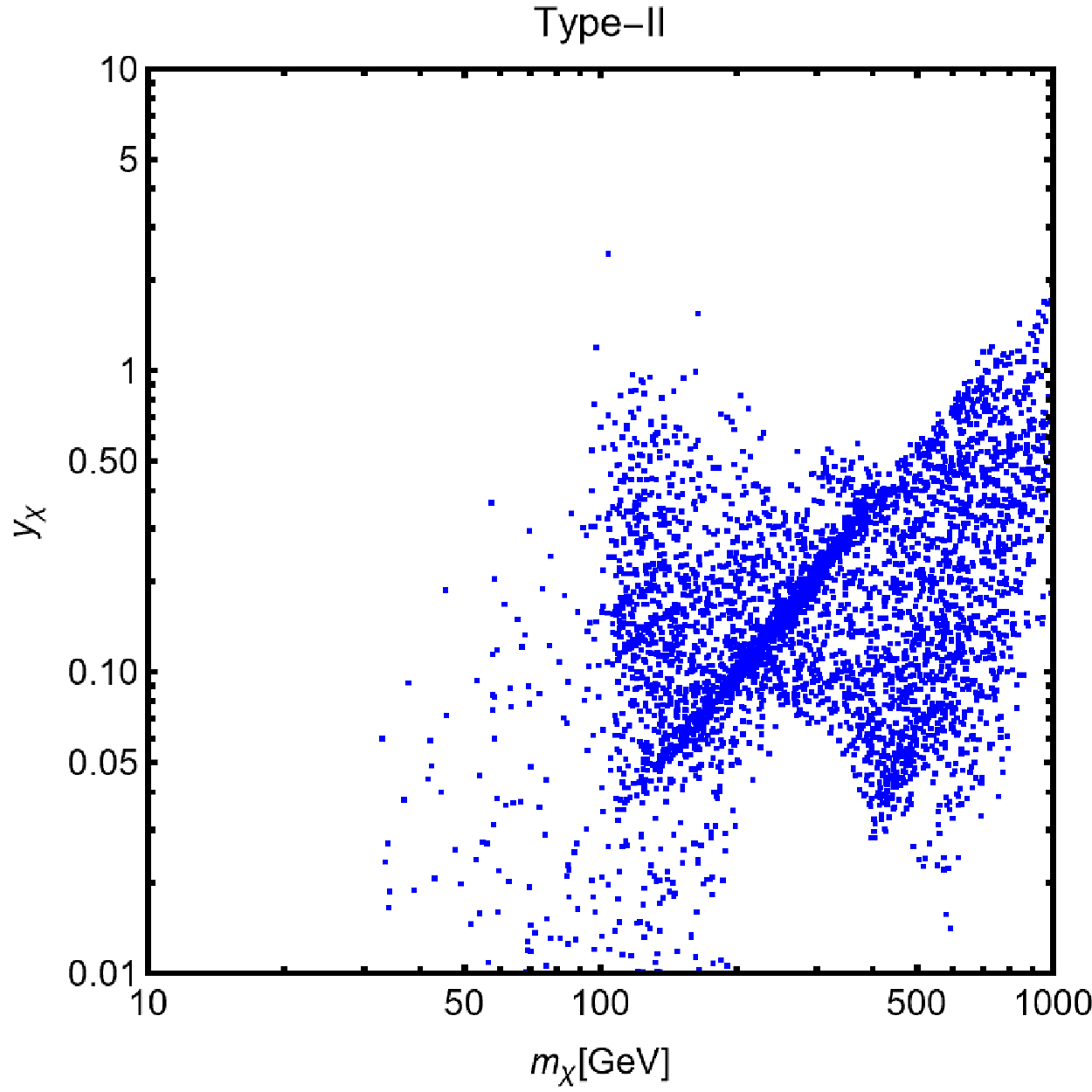}}\\
    \subfloat{\includegraphics[width=0.29\linewidth]{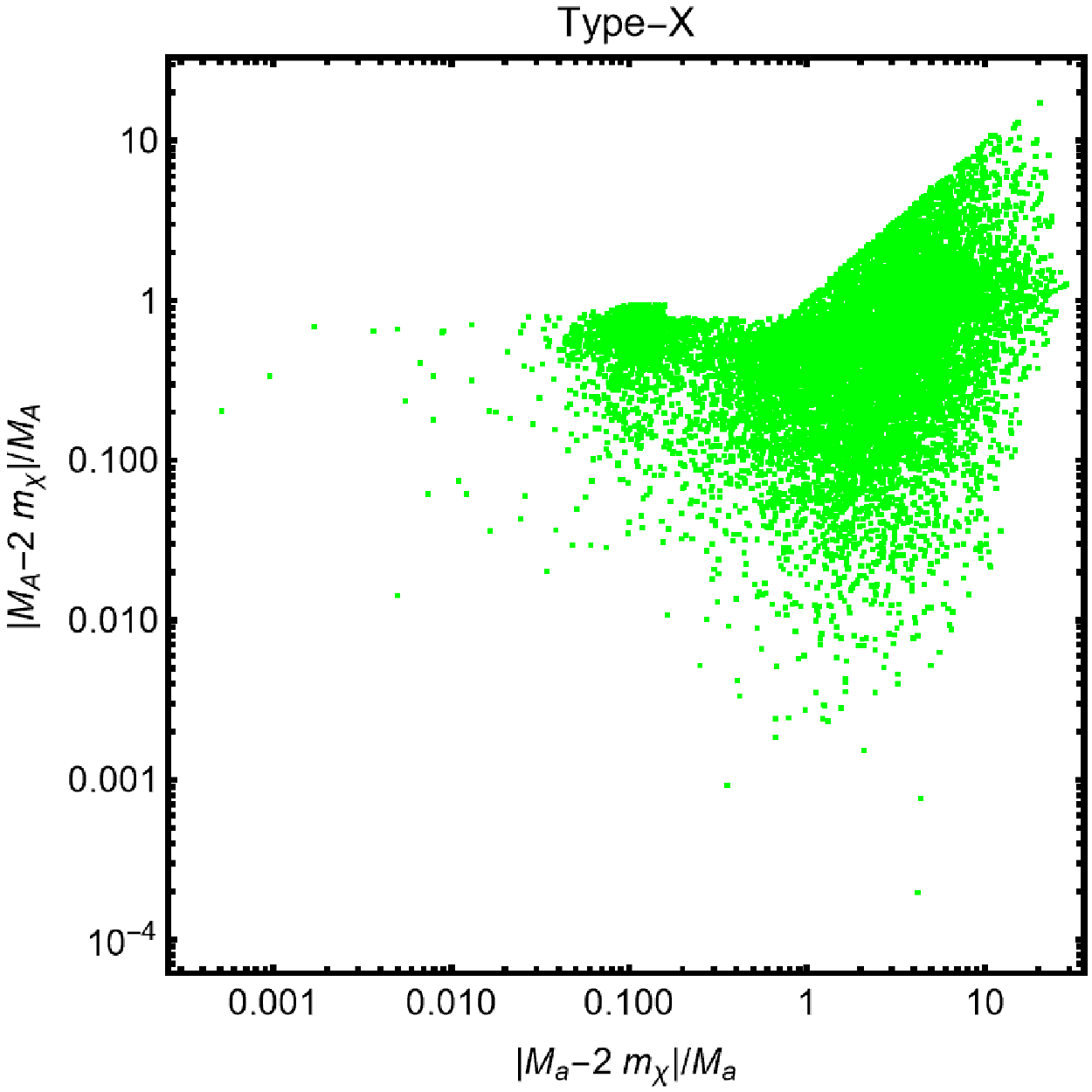}}~~~
    \subfloat{\includegraphics[width=0.29\linewidth]{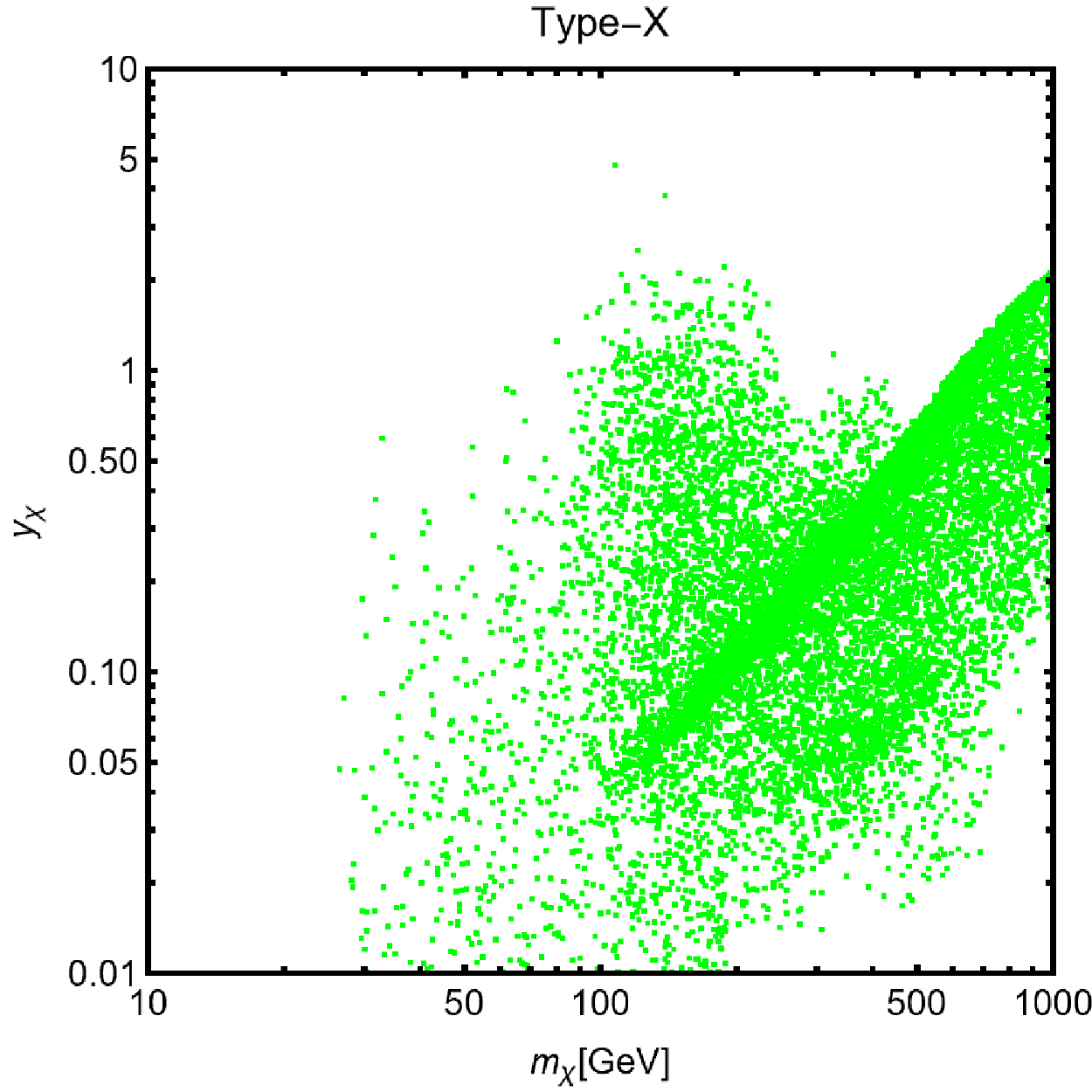}}\\
    \subfloat{\includegraphics[width=0.29\linewidth]{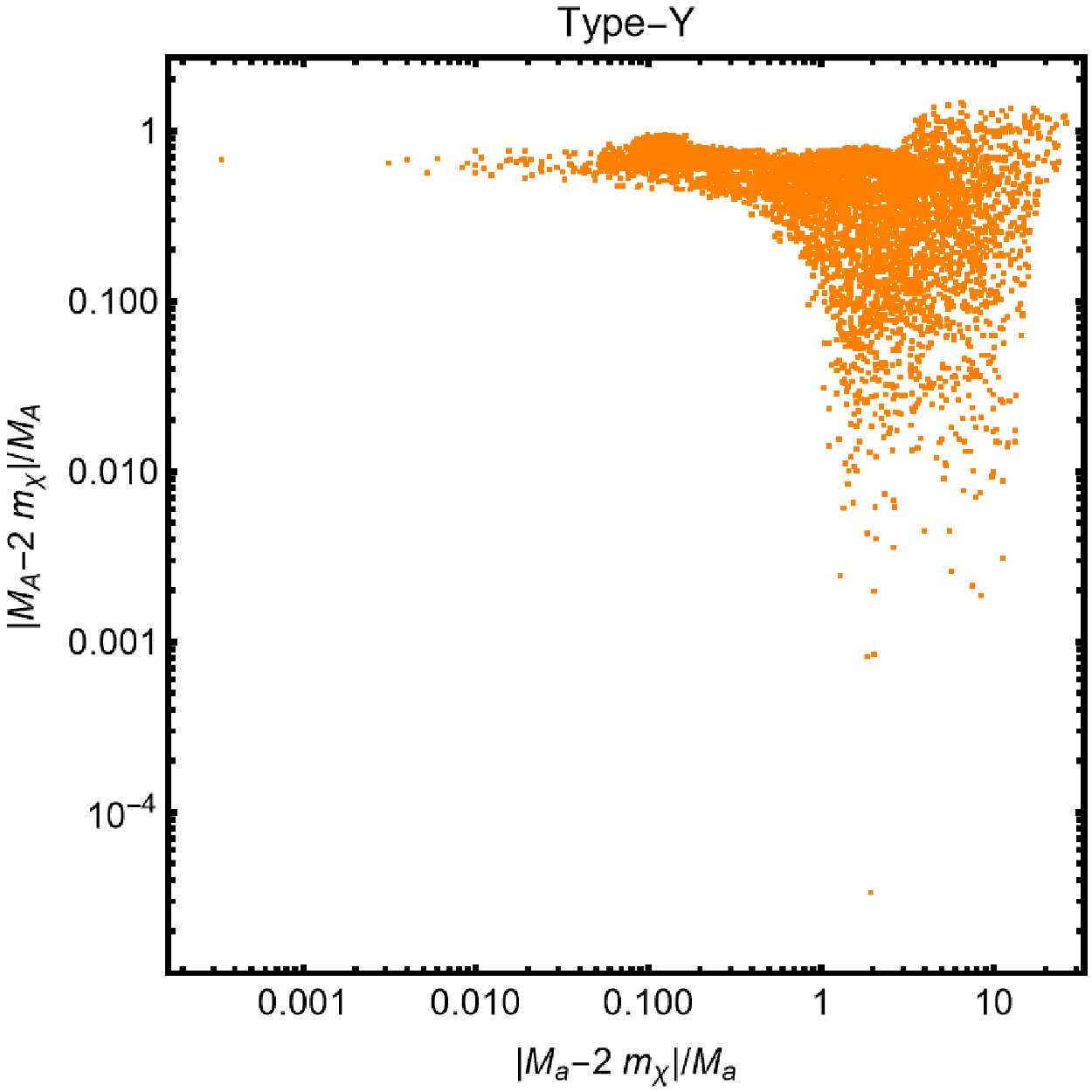}}~~~
    \subfloat{\includegraphics[width=0.29\linewidth]{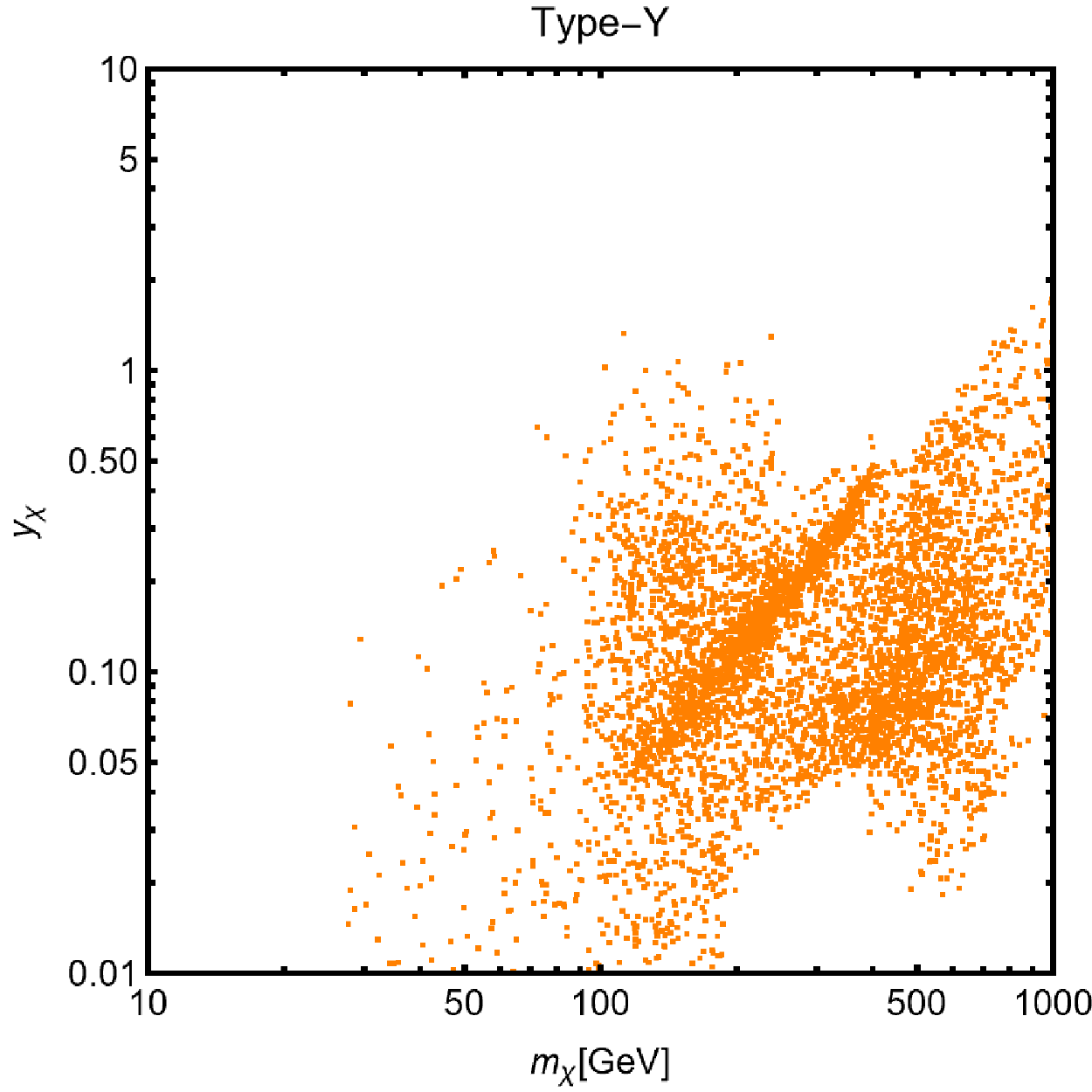}}
    \vspace*{-3mm}
    \caption{Outcome of the parameter scan including DM parameters and constraints in the four types of scenarios  (see main text for details). Each plot contains the model points complying with the correct relic density, a spin-independent cross section below the LZ direct limit, and an annihilation cross section complying with indirect constraints. Limits from $B$-physics and LHC searches (mostly $H/A\! \rightarrow\! \tau^+ \tau^-$ and $a \!\rightarrow \!\mu^+ \mu^-$) are accounted for.}
    \label{fig:DMplot1}
\vspace*{-2mm}
\end{figure}

We have now all the ingredients to assess in a more systematic manner the impact of the DM constraints and to combine them with the collider ones. To achieve this task, we have  conducted an analogous parameter scan as the one considered in the previous sections.  Some simplifying assumptions, to reduce the dimensionality of the parameter space, have been considered though: namely $M_H\!=\!M_A\!=\!M_{H^\pm}$ and $\lambda_{1P}\!=\!\lambda_{2P}\!=\!3$. For what concerns the DM parameters, i.e. its mass $m_\chi$ and its coupling $y_\chi$, we have considered the following range of variation
\begin{equation}
    m_\chi \in \left[1,1000\right]\,\mbox{GeV}\,\,\,\,\mbox{and}\,\,\,\,y_\chi \in [10^{-2},10].
\end{equation}
A first result of such a parameter scan is shown in Fig.~\ref{fig:DMplot1} in which each panel reports the model points that satisfy the correct DM cosmological relic density and evade the bounds from DM direct as well as indirect detection. Furthermore, we have applied to the Type II and Y models, the lower bound $M_{H^{\pm}}>800\,\mbox{GeV}$ from the $b\rightarrow s\gamma$ constraint, and to all models, the LHC bounds from searches of heavy resonances decaying into $\tau^+ \tau^-$ and of light resonances decaying into $\mu^+ \mu^-$.

The left column of the figure displays the model points in the $\left[\frac{|M_a-2 m_\chi|}{M_a},\frac{|M_A-2 m_\chi|}{M_A}\right]$ plane, while the right column illustrates the results in the $\left[m_\chi,y_\chi\right]$ plane. The distribution of model points in the first column of the plot is mostly sensitive to the relic density constraint. In agreement with previous findings, one gets very similar results for the four Yukawa configurations. In all cases, the distribution of model points have tails covering the regions in which either $\frac{|M_a -2 m_\chi|}{M_a}\ll 1$ or $\frac{|M_A-2 m_\chi|}{M_A}\ll 1$. This corresponds to the $s$-channel resonance regions $m_\chi \simeq \frac12 M_{a}$ or $\frac12 M_{A}$ for which the correct relic density can be achieved also for very small values of the DM coupling $y_\chi$. 

Besides the $M_{a}$ and $M_A$ $s$-channel poles, the other favored regions of the parameter space correspond to the case in which the DM is heavier than one or both the pseudoscalar Higgs bosons. As already pointed out, in this regime, the relic density constraint has as analogous impact in the four Yukawa configurations. The most notable difference is in the value of ${|M_A-2 m_\chi|}/{M_A}$ which does not exceed ${\cal O}(1)$ for the Type II and Y scenarios. This is due to the strong bounds from the LHC searches and $B$-physics observables together with the chosen ranges for the scanned parameters. 

The impact of constraints from direct and indirect DM searches can be more clearly appreciated by looking to the $[m_\chi,y_\chi]$ planes. As already mentioned, to evade direct detection constraints, one needs to require $y_\chi \lesssim 1$. DM indirect detection, instead, rules out most of the viable parameters space for $m_\chi \lesssim 100\,\mbox{GeV}$. The residual points for light DM masses still present in Fig.\ref{fig:DMplot1} correspond to DM annihilation in the $m_\chi \simeq \frac12 M_a$ pole. This is because there is not any longer matching between the DM annihilation cross section at thermal freeze-out and present times in the case of an $s$-channel resonant enhancement \cite{Griest:1990kh}.

\begin{figure}[!h]
\vspace*{-4mm}
    \centering
    \subfloat{\includegraphics[width=0.29\linewidth]{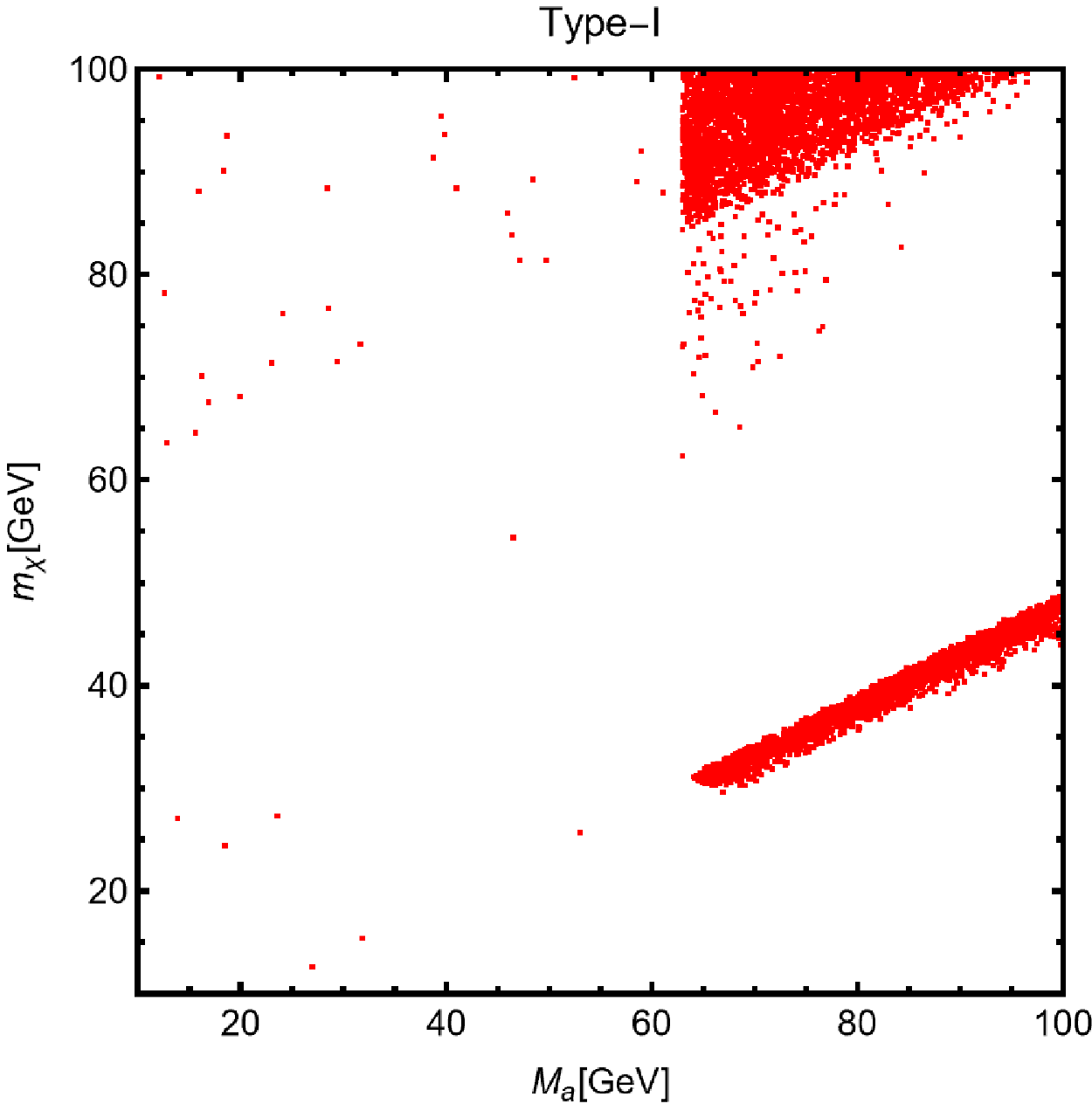}}
    \subfloat{\includegraphics[width=0.29\linewidth]{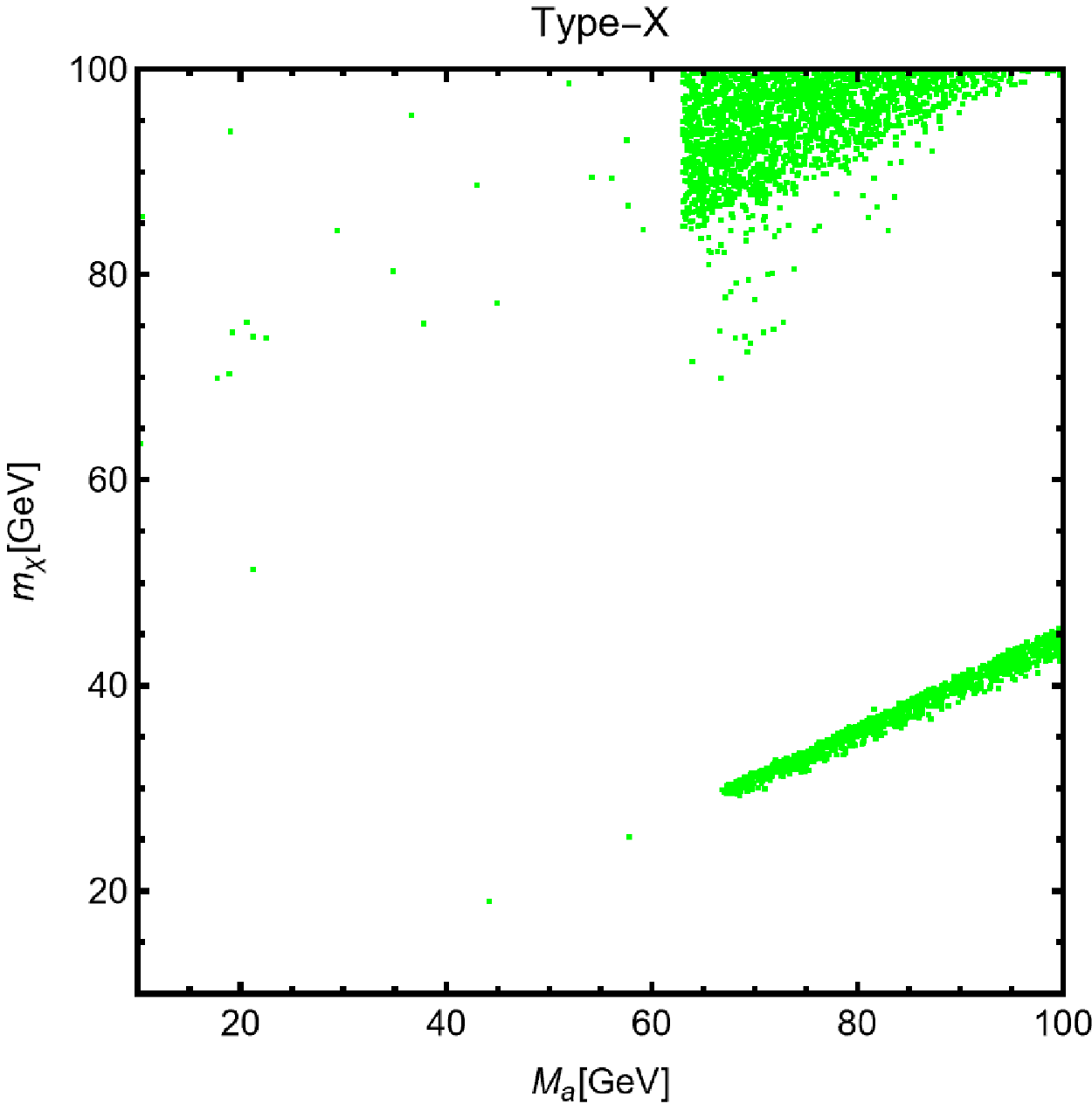}}
    \subfloat{\includegraphics[width=0.29\linewidth]{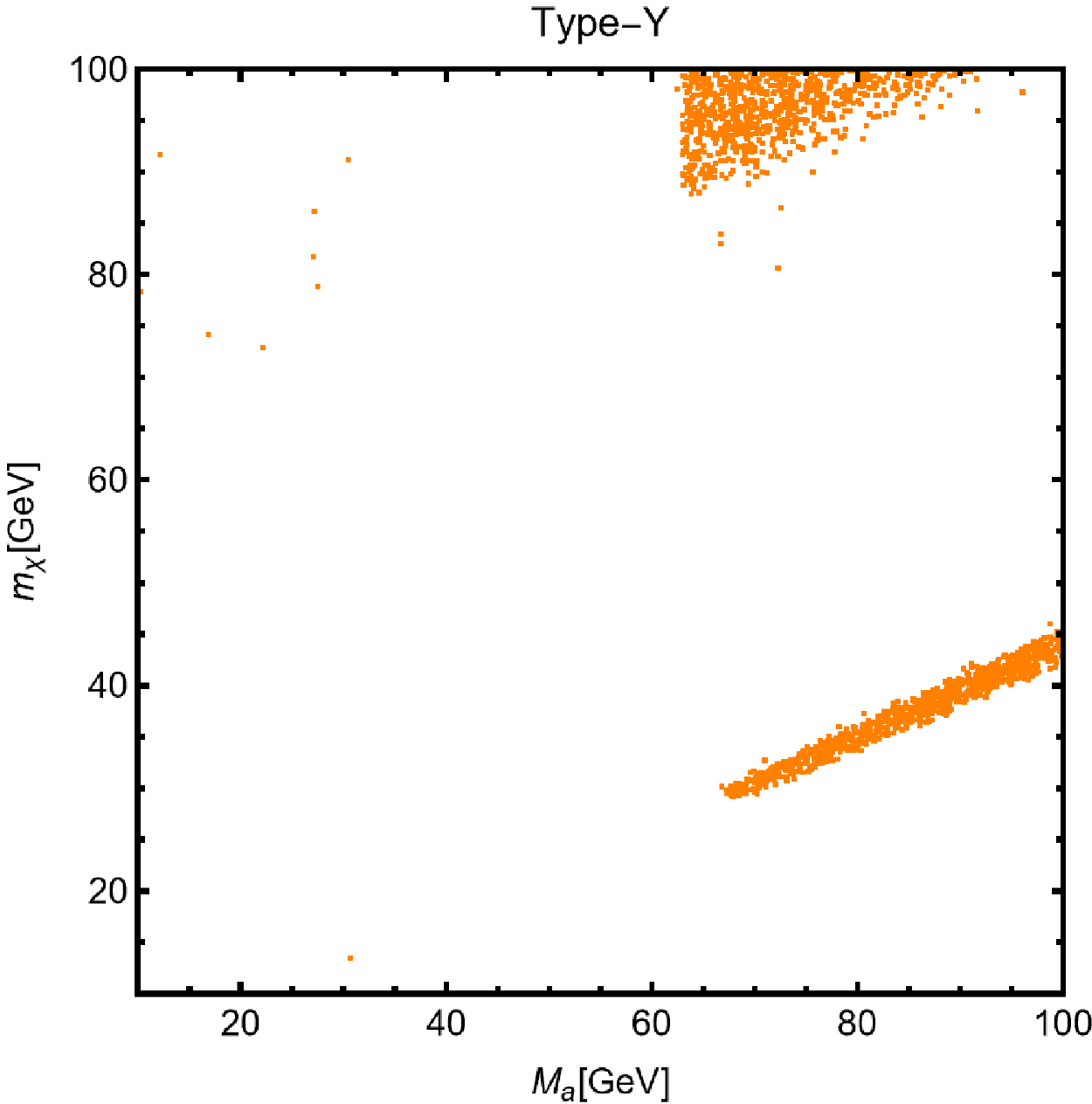}}
    \caption{Accepted model points  in the $[M_a, m_\chi]$ plane of a parameter scan focused on the light $a$ region. The figure contains only three panels, relative to the Type I, Type X and Type Y configurations as the Type II 2HD+a scenario is almost entirely ruled out by LHC searches of light resonances.}
    \label{fig:plotDM2}
\end{figure}

Figure \ref{fig:plotDM2} shows a further illustration of the DM constraints focusing on the $M_a < 100\,\mbox{GeV}$ region. We have repeated the previous parameter scan and selected the viable model points in the $[M_a,m_\chi]$ plane, by considering this low $M_a$ range and keeping fixed the masses $M_H=M_A=M_{H^{\pm}} $ to $|M|=500$ GeV in the case of the Type I and Type X scenarios and to $|M|=800$ GeV in the Type Y case. The Type II model has been not included in this analysis since most of the low $M_a$ region is already ruled out by searches of light resonances decaying into muon pairs as it has been discussed earlier.

As can be seen from the figure, the distributions of the model points are rather similar for the three Yukawa configurations. First, one notices the almost sharp cut of the viable parameter space for $M_a \lesssim 60\,\mbox{GeV}$ which is essentially due to the bound on the invisible $h$ boson width which can be evaded only by choosing fine-tuned blind spot configurations for the model parameters to suppress or forbid $h \to aa$ decays. Low values of $M_a$ are also subject to the bounds from searches of light resonances, for instance decaying into muon pairs. For this reason, the Type I scenario features more viable model points for $M_a \lesssim 60\,\mbox{GeV}$ since it is the least subject to the latter bounds. 

Moving to the range $M_a \gtrsim 60\,\mbox{GeV}$, the viable model points occupy two very specific regions, the pole $m_\chi \simeq \frac12 M_a$ region and the $m_\chi \geq M_a$ area. This outcome is mostly due to the constraints from DM indirect detection. A DM state lighter than ${\cal O}(100\,\mbox{GeV})$ and annihilating into SM fermion pairs is generally strongly disfavored. In the 2HD+a model, this problem can be circumvented by being either in the pole region,  as a consequence of the fact that there is not exact matching between the DM annihilation cross section at the time of thermal freeze-out and at present times, or in the $m_\chi > M_a$ regime such that the $\chi \chi \rightarrow aa$ process is kinematically allowed. This process, indeed, features a $p$-wave dominated cross section for which indirect detection constraints are irrelevant.

\begin{figure}[!h]
\vspace*{-2mm}
    \centering
    \subfloat{\includegraphics[width=0.29\linewidth]{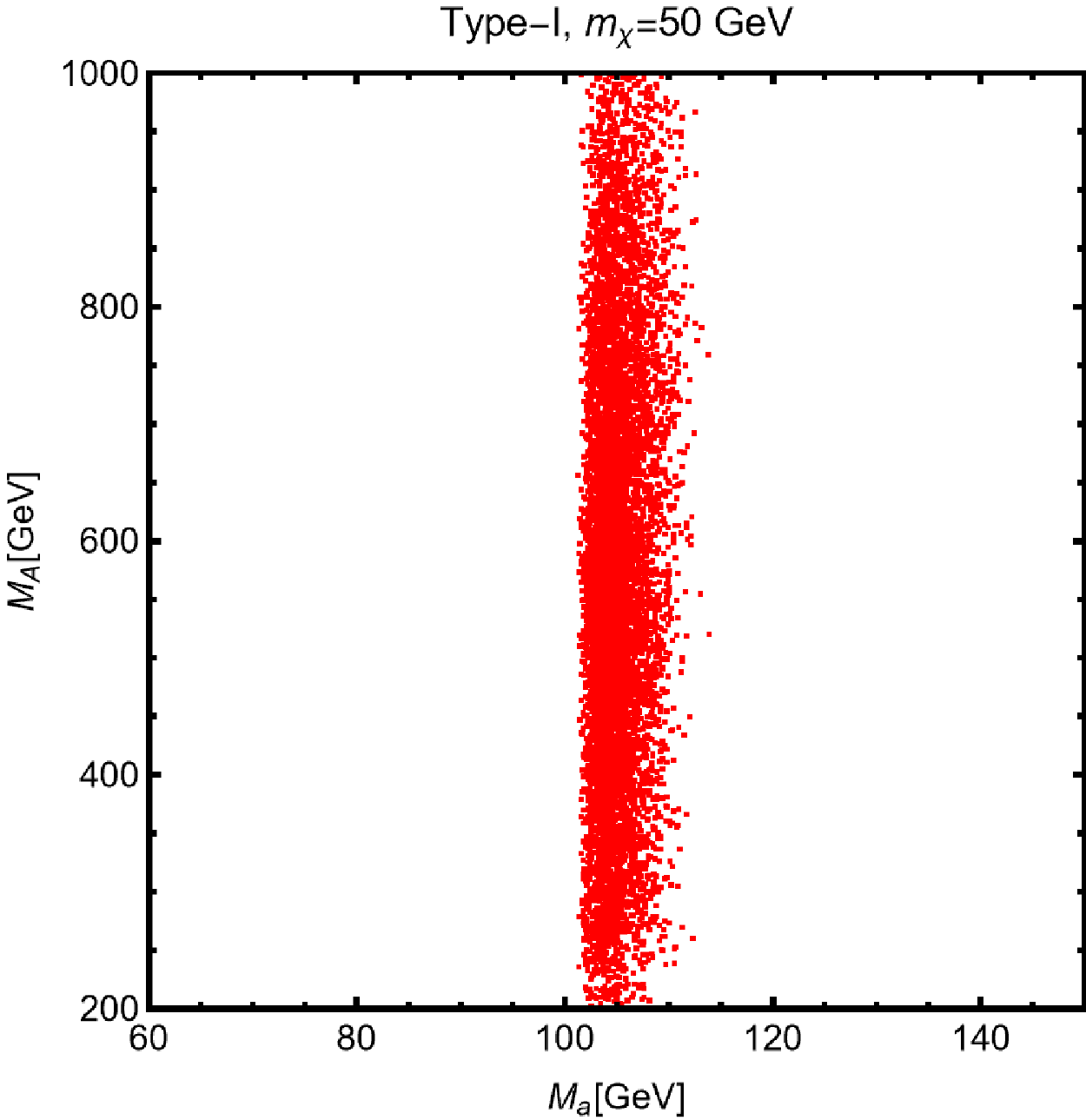}}~
    \subfloat{\includegraphics[width=0.29\linewidth]{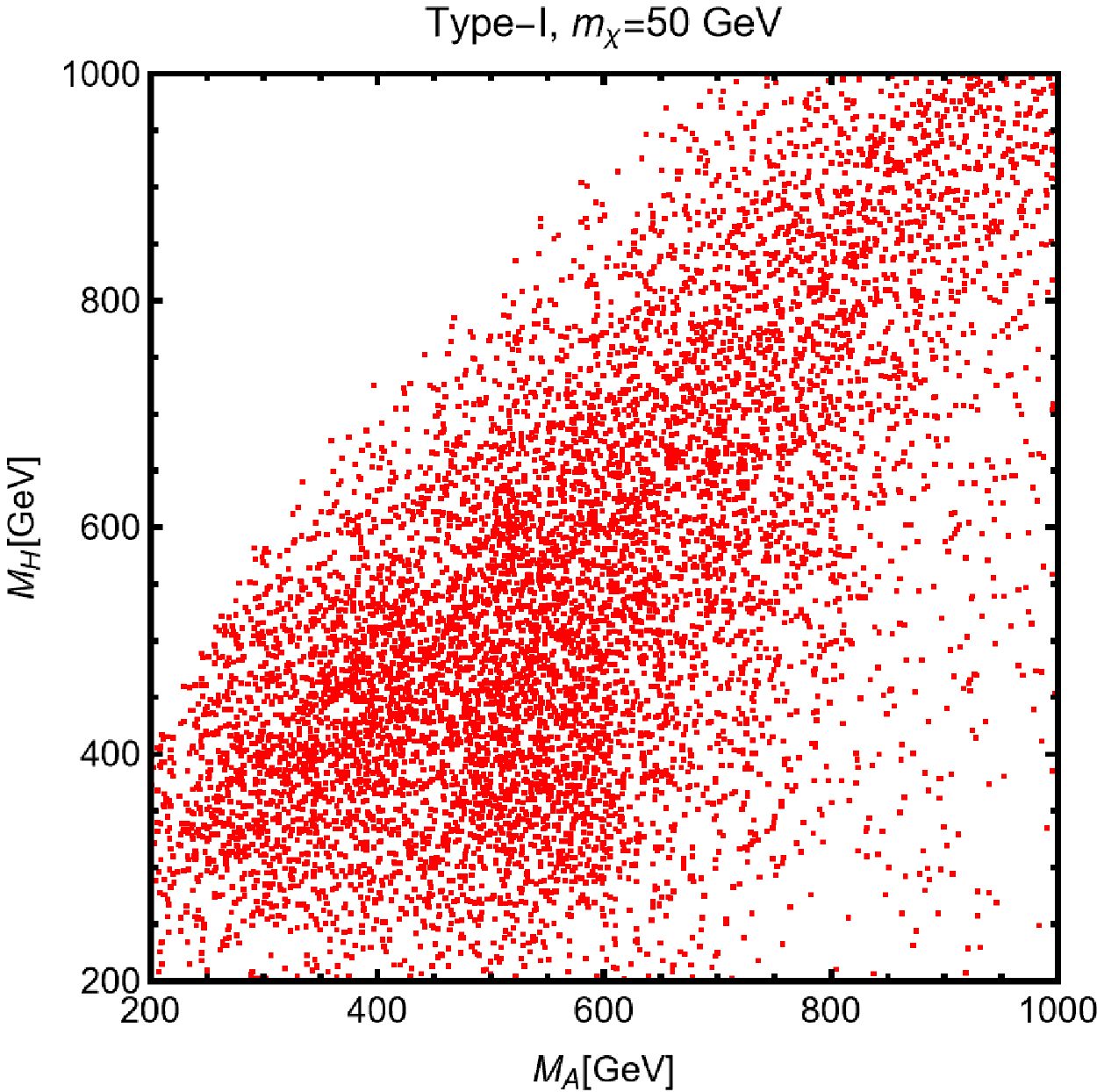}}~
    \subfloat{\includegraphics[width=0.29\linewidth]{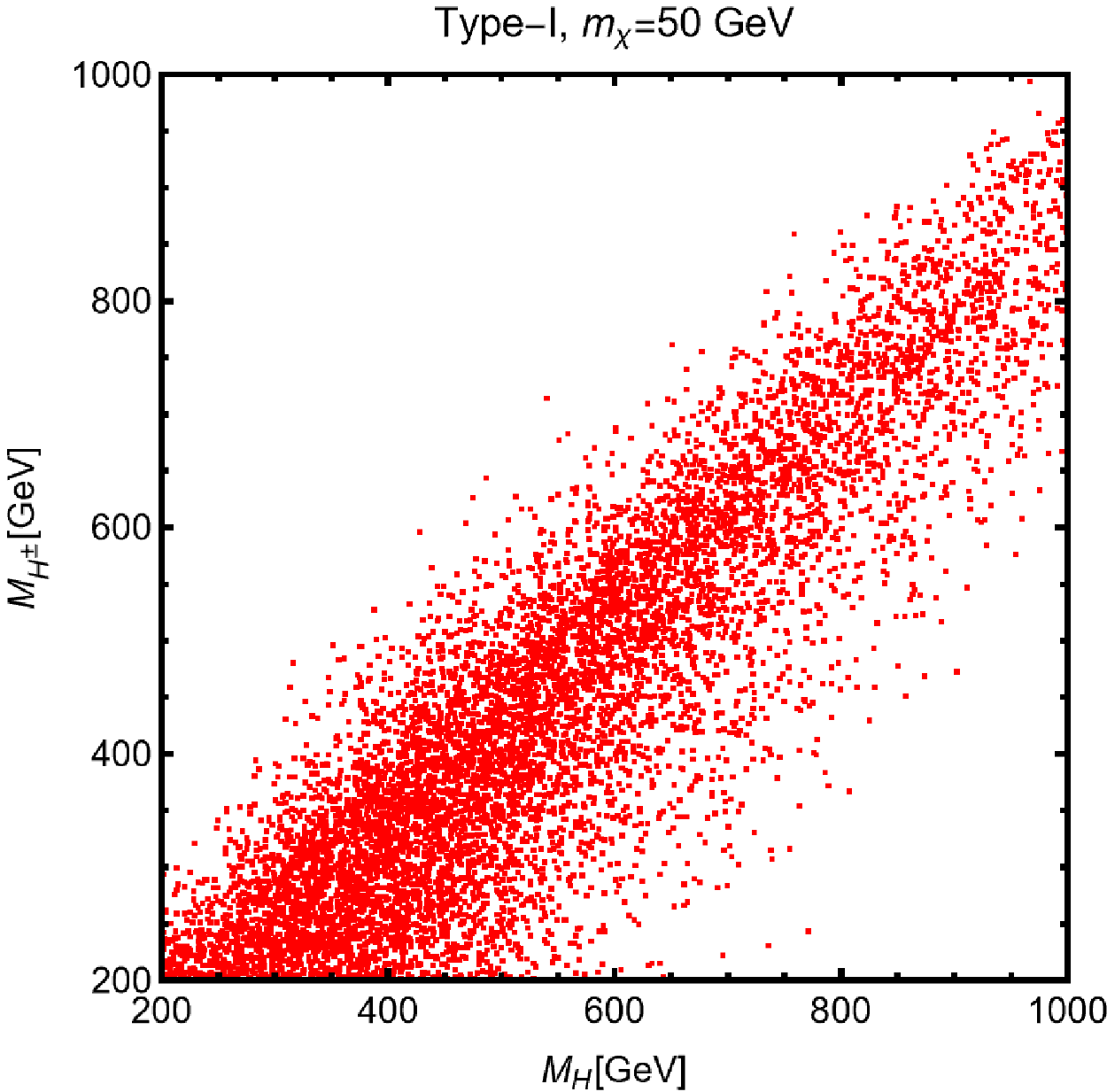}}\\
     \subfloat{\includegraphics[width=0.29\linewidth]{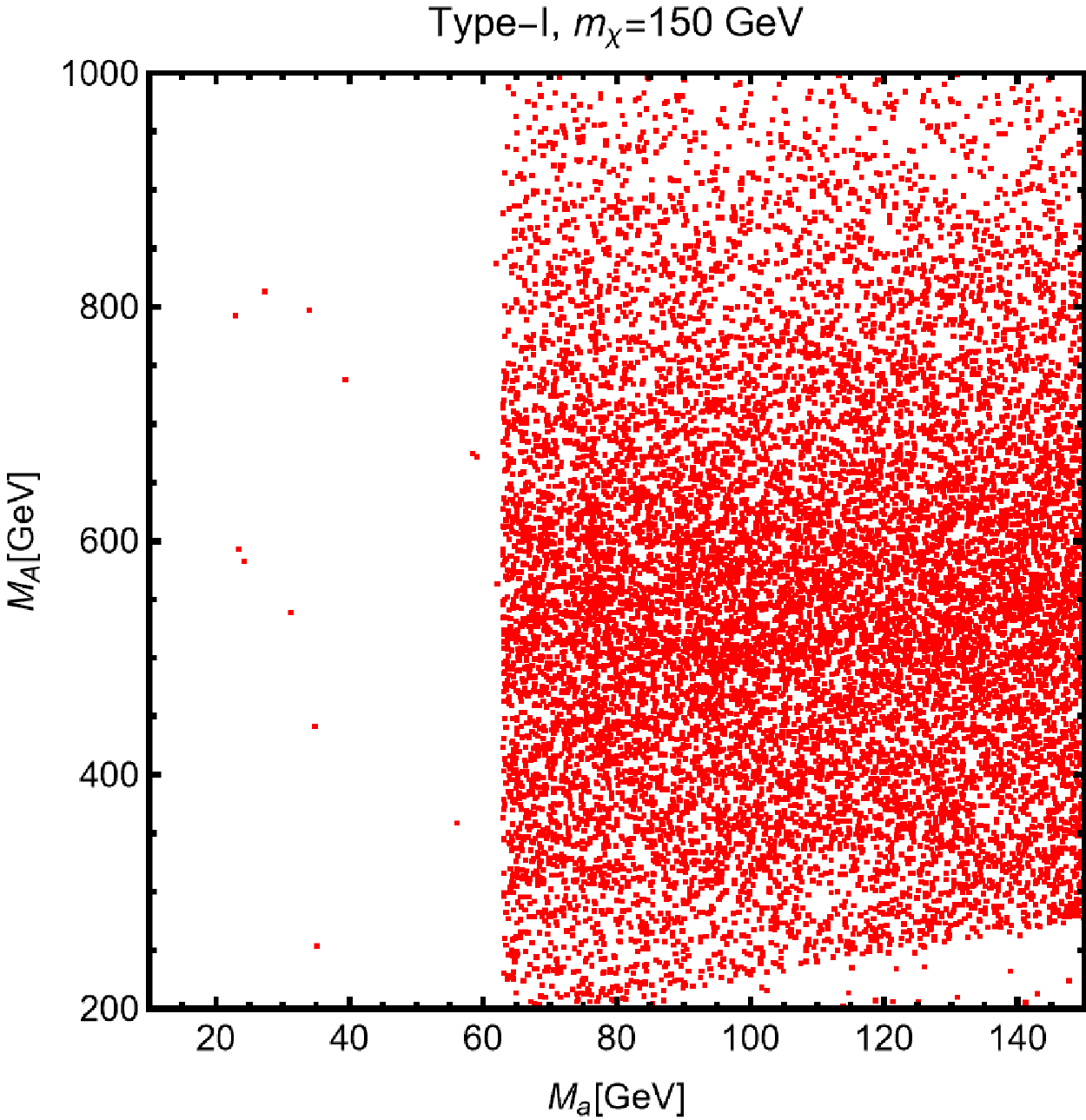}}~
    \subfloat{\includegraphics[width=0.29\linewidth]{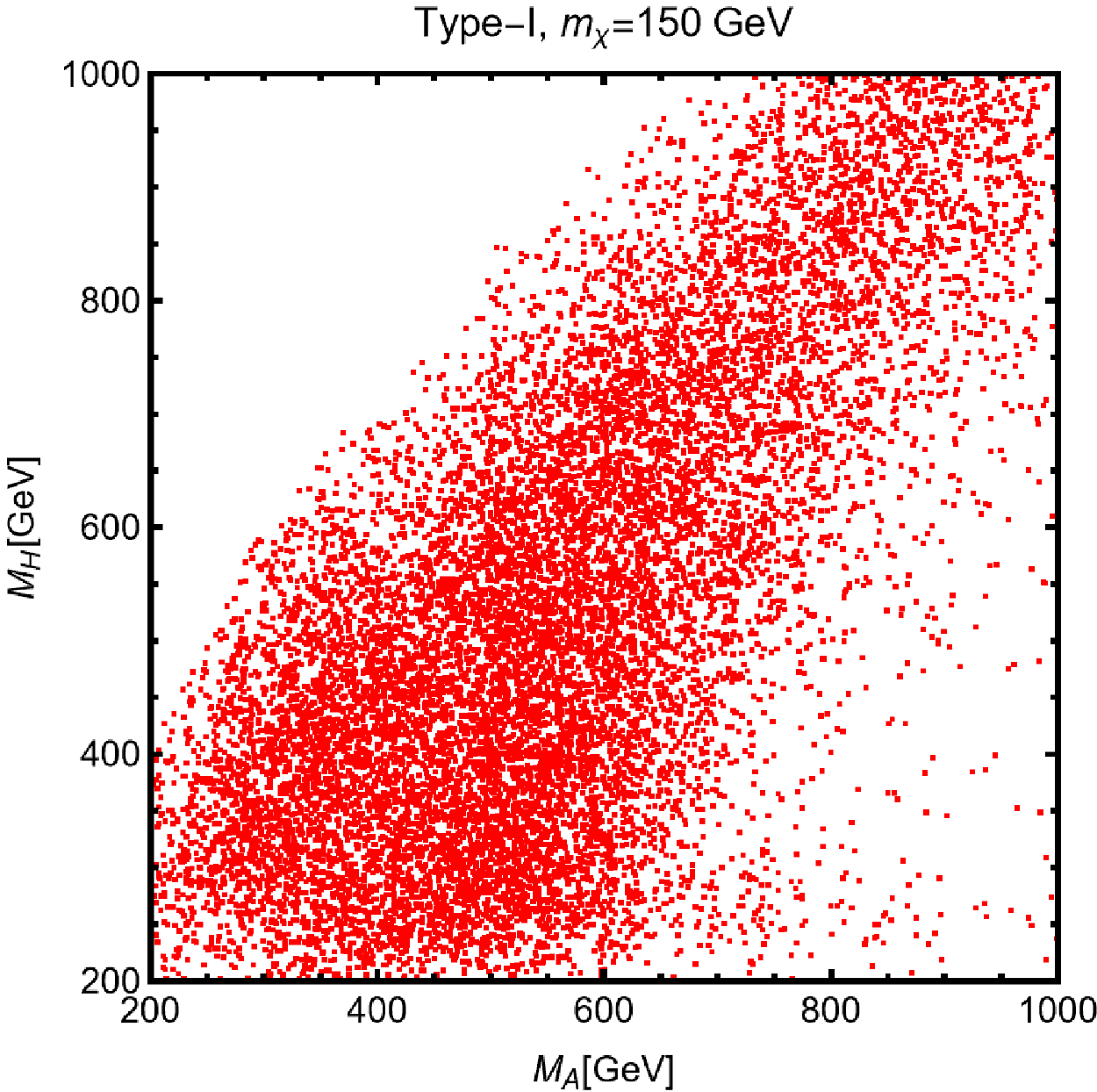}}~
    \subfloat{\includegraphics[width=0.29\linewidth]{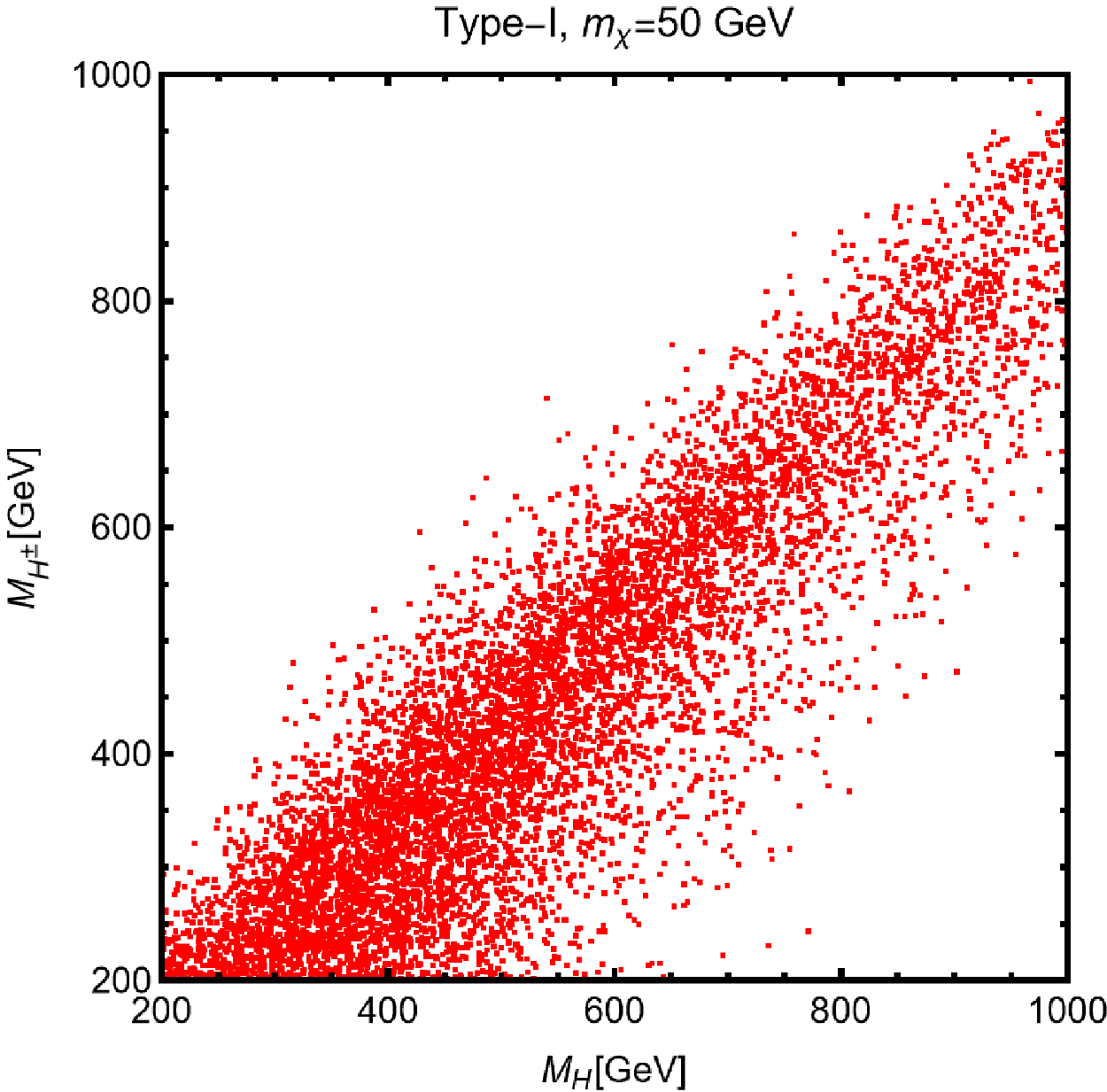}}
    \caption{Outcome of a parameter scan in which all the model parameters are varied but the DM mass, which has been fixed to two values, namely $m_\chi=50$ GeV and $m_\chi=150$ GeV. All the points shown in the panels comply with constraints from DM phenomenology, LHC searches and theoretical constraints. The Type I configuration has been assumed for the Yukawa couplings of the additional Higgs bosons.}
    \label{fig:plotDM3}
\end{figure}

\begin{figure}[!h]
\vspace*{-4mm}
    \centering
    \subfloat{\includegraphics[width=0.29\linewidth]{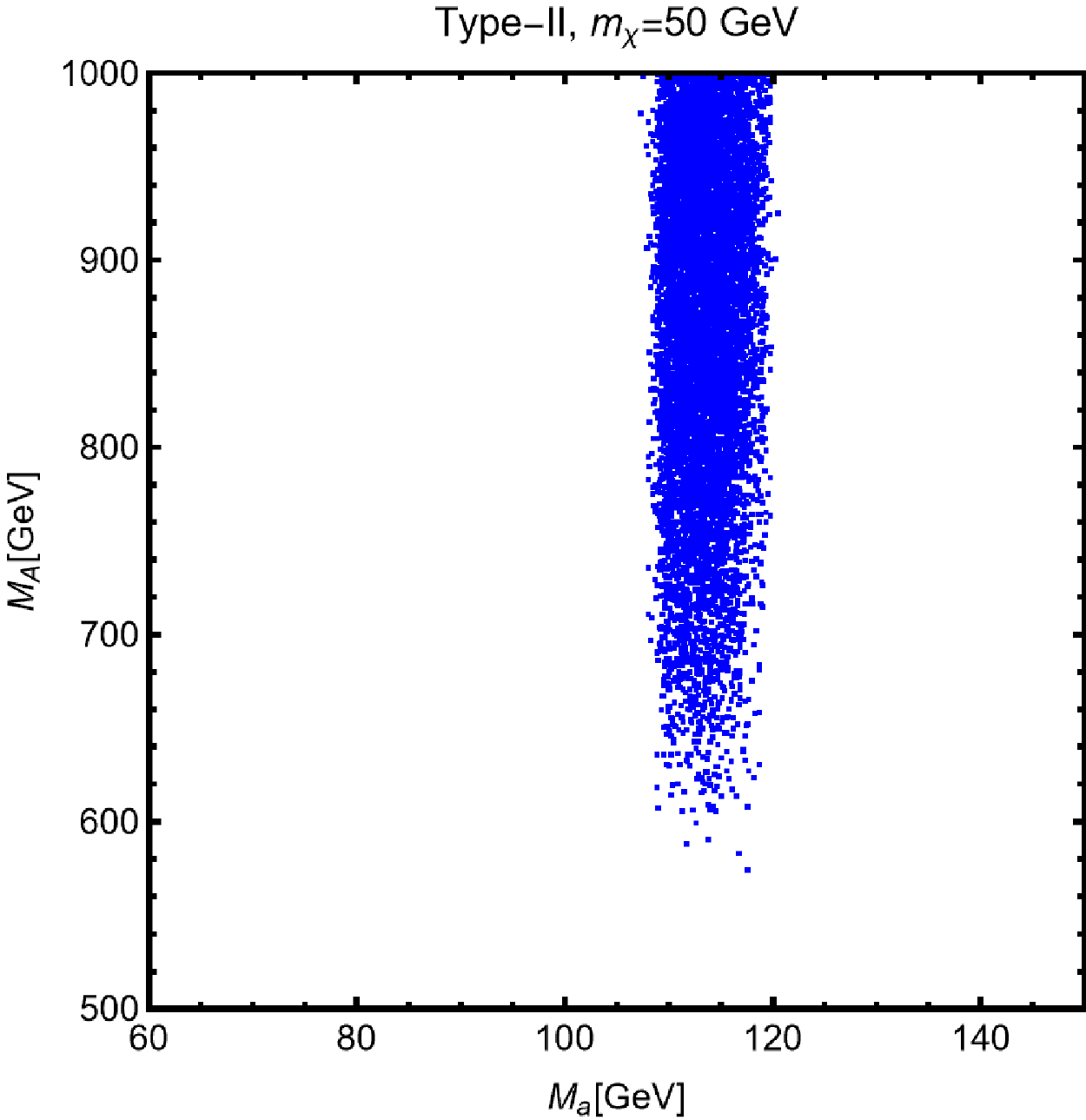}}~
    \subfloat{\includegraphics[width=0.29\linewidth]{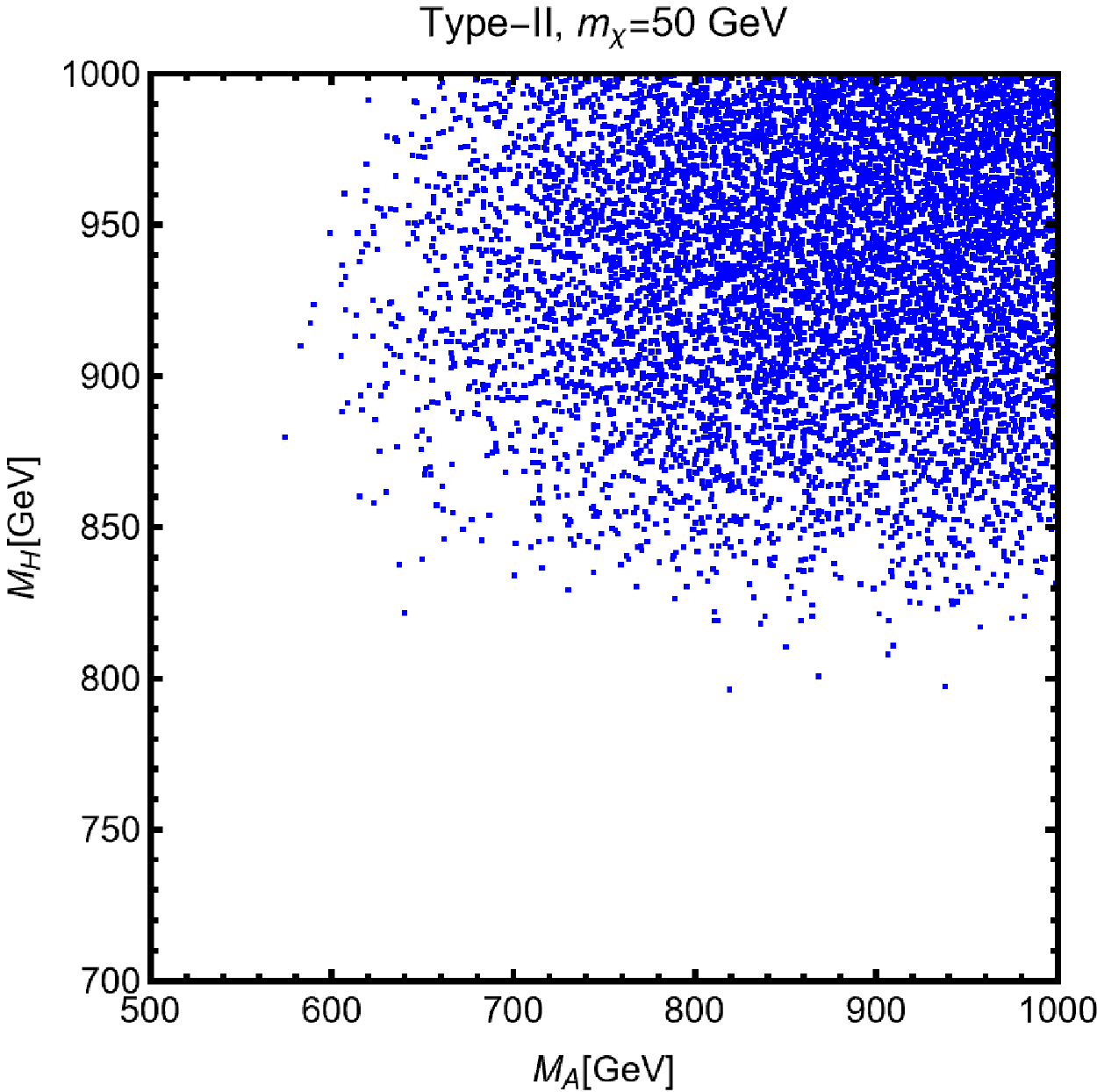}}~
    \subfloat{\includegraphics[width=0.29\linewidth]{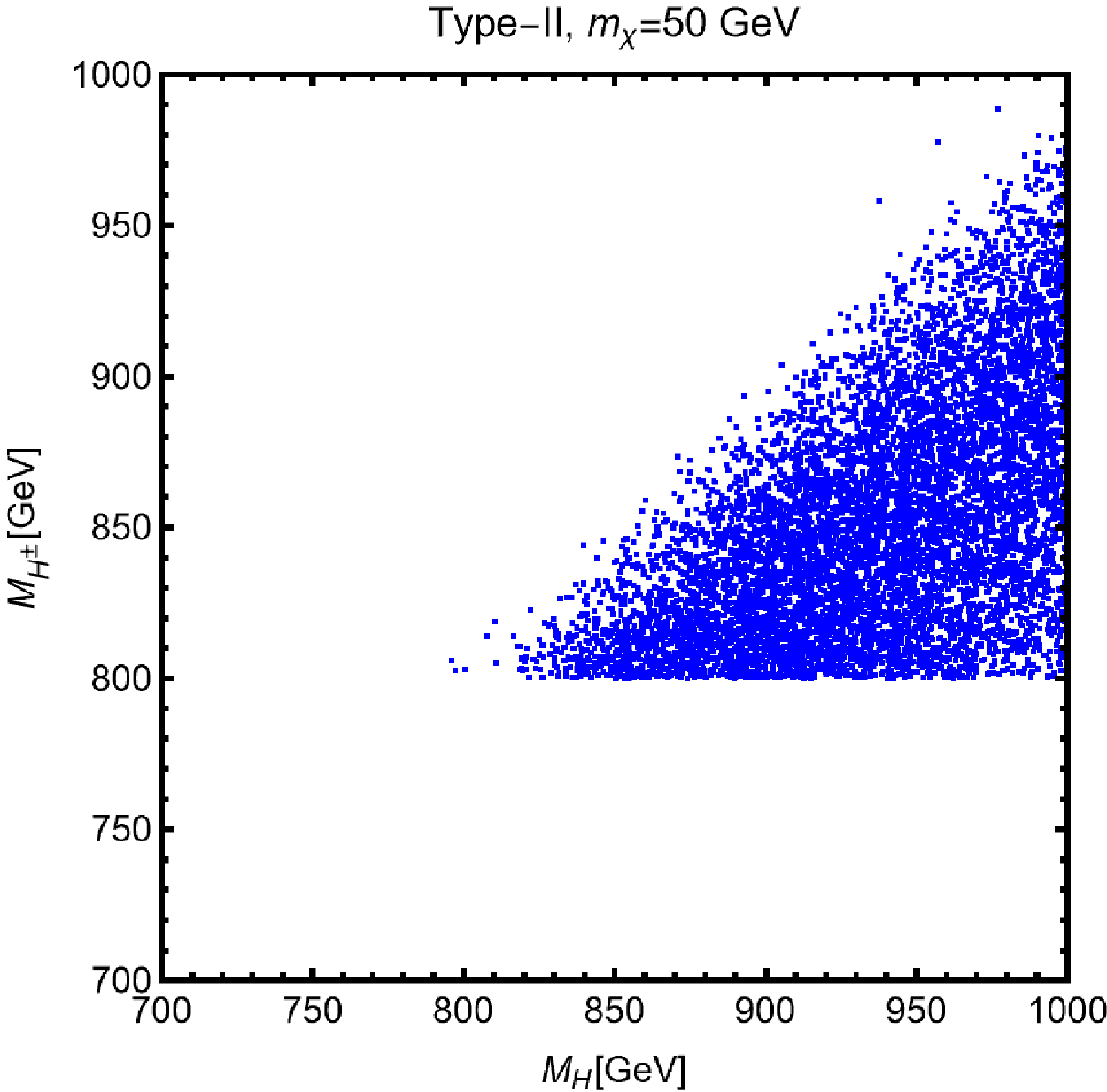}}\\
     \subfloat{\includegraphics[width=0.29\linewidth]{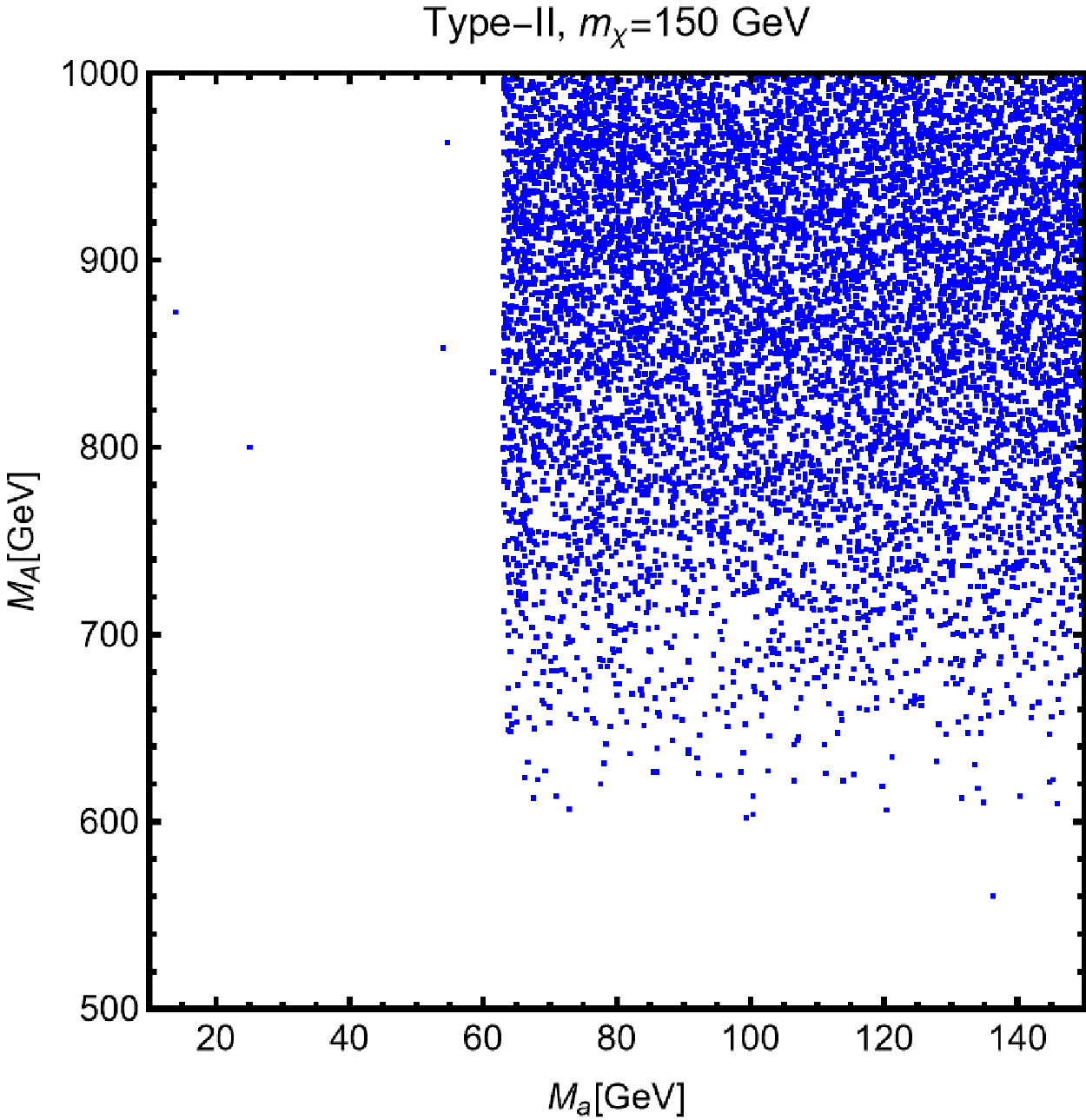}}~
    \subfloat{\includegraphics[width=0.29\linewidth]{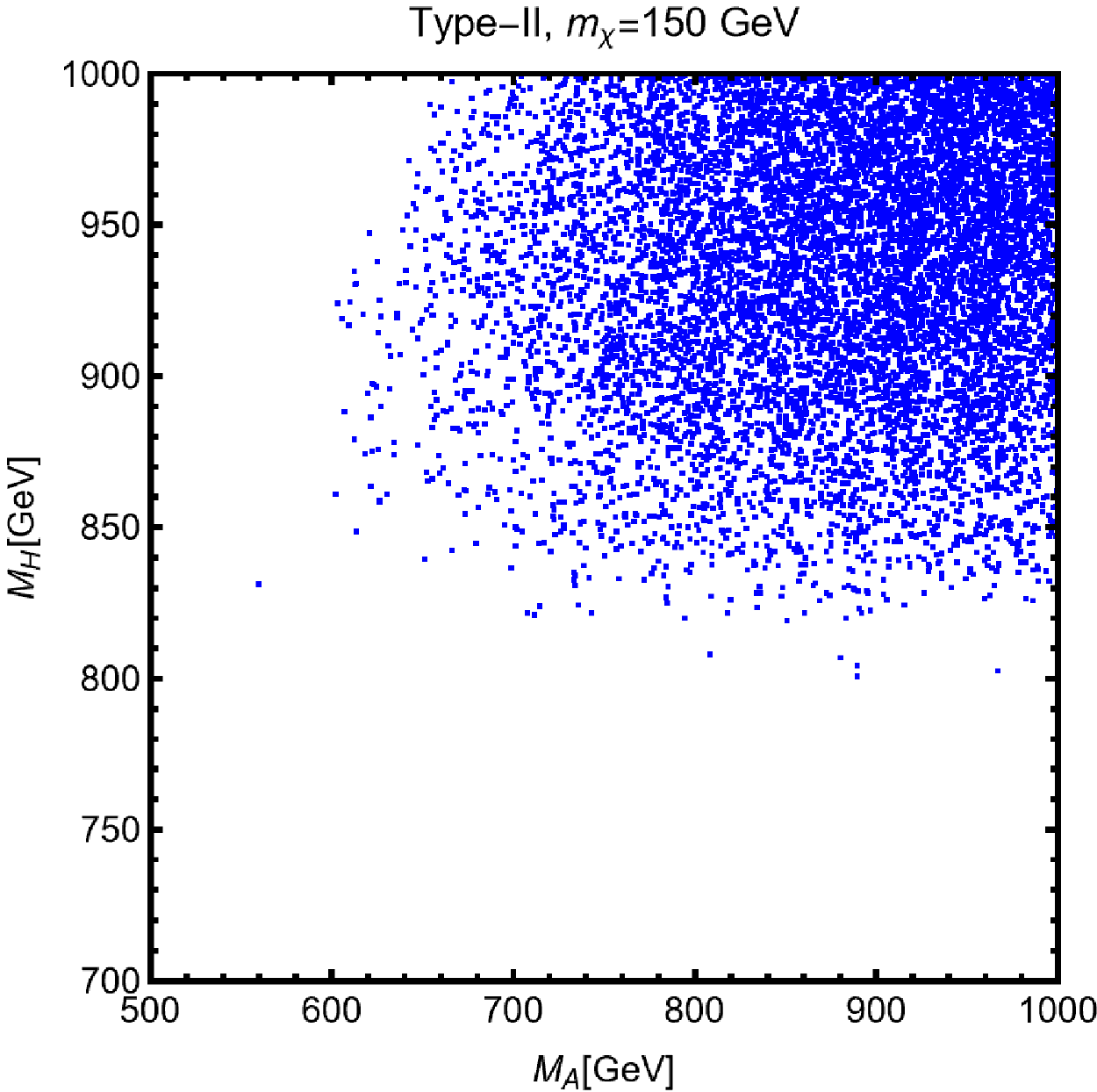}}~
    \subfloat{\includegraphics[width=0.29\linewidth]{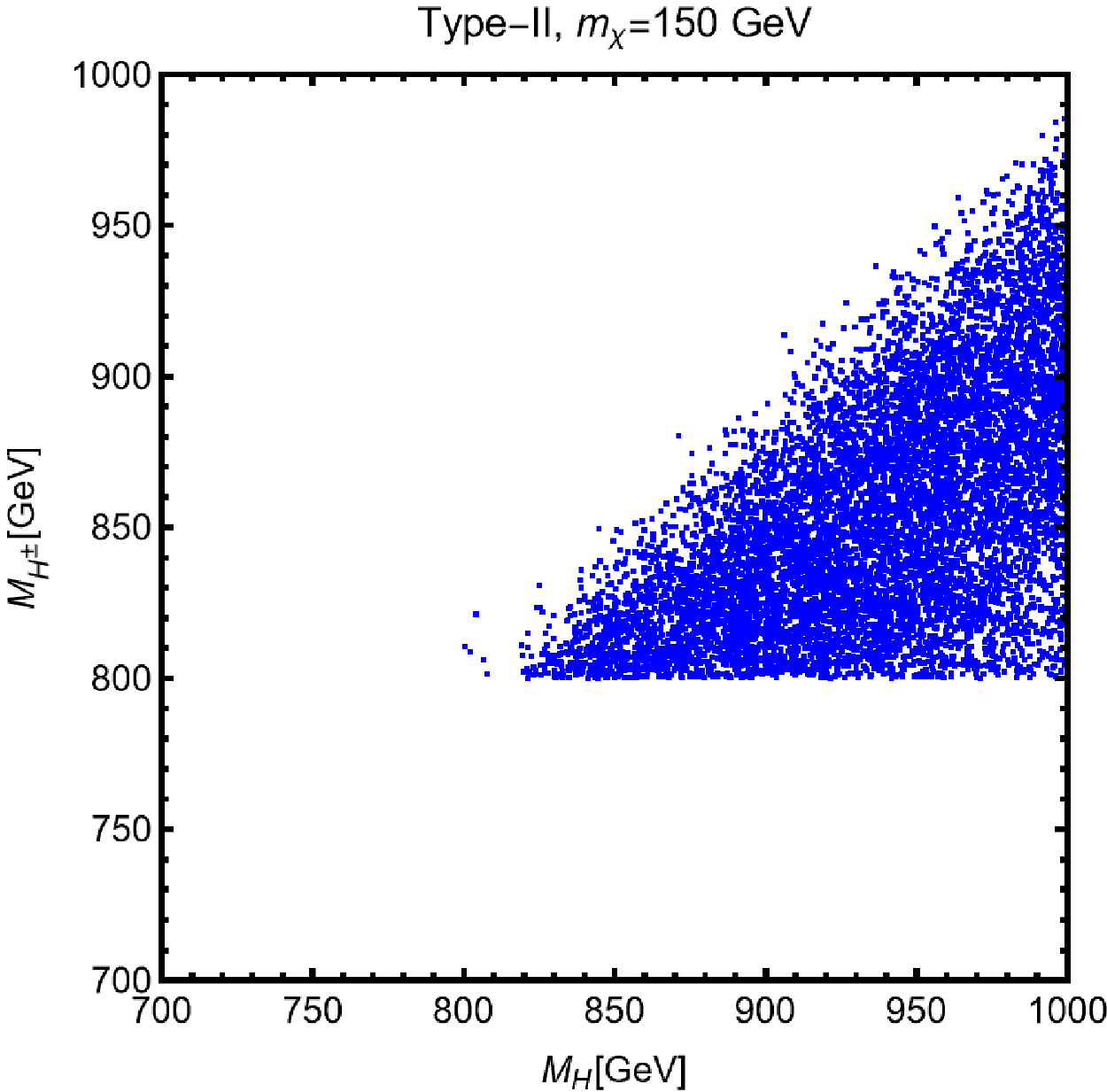}}
    \caption{The same as in Fig. \ref{fig:plotDM3} but considering the Type II scenario. Note the different scales on the $x$ and $y$ axes for the heavy Higgs boson masses  compared to the previous case.}
    \label{fig:plotDM4}
\end{figure}

As already pointed out before, Figs. \ref{fig:DMplot1} and \ref{fig:plotDM2} show the results of parameter scans performed under rather  simplifying assumptions, namely degenerate masses for the additional Higgs bosons except for the $a$ state, and fixed values of the couplings  $\lambda_{1P},\lambda_{2P}$ and $\lambda_3$. To understand how the latter parameters affect DM phenomenology, we have conducted further parameter scans fixing the DM mass to two values, namely $m_\chi=50$ GeV and $m_\chi=150$ GeV, and assuming non degenerate masses for the heavy Higgs bosons while varying freely the quartic couplings of the scalar potential. 
Without loss of generality, we have restricted ourselves to the Type I and Type II configurations, the results of which are given in Figs. \ref{fig:plotDM3} and \ref{fig:plotDM4}, respectively.  


Each figure shows two series of panels, corresponding to the two chosen values of the DM mass. Each one of these  series shows the model points which comply with all the constraints considered in this work in the $[M_a,M_A], [M_H,M_A]$ and $ [M_H,M_{H^{\pm}}]$ planes (results in the 
$[M_A,M_{H^{\pm}}]$ plane are more or less similar to those obtained in the $[M_H,M_{H^{\pm}}]$ plane). By looking at the distribution of the model points, we notices that the DM constraints, namely having the correct relic density and complying with the bounds from direct and indirect detection, do not substantially modify the allowed parameter space regions with respect to the LHC and theoretical bounds discussed in the previous sections. The only exception concerns the mass of the light pseudoscalar boson $a$. Indeed, for $m_\chi< M_a \lesssim 100\,\mbox{GeV}$, the correct DM relic density is achieved while being compatible with the other constraints, only in the pole $m_\chi \sim \frac12 M_a$ region. In the case $m_\chi=50\,\mbox{GeV}$, the mass of $a$ is consequently constrained to lie around $100\,\mbox{GeV}$.

This ends the discussion on the combined collider and astroparticle physics constraints on the 2HD+a parameter space. From this, one concludes that the constraints are rather strong, in particular in the Type II scenario. They nevertheless leave significant regions in which the model is still viable, in particular if it does not have to explain the anomalous $(g-2)_\mu$ result which requires too light $a$ states with too strong couplings to isospin $-\frac12$ fermions. Part of these regions could nevertheless be challenged in the near future by more sensitive LHC and DM direct detection searches.

\clearpage 

\section{Phase transitions and gravitational waves}

A promising way to probe the 2HD+a model is through the detection of the stochastic gravitational-wave background. These gravitational waves, originating from the electroweak first-order cosmic phase transitions, propagate freely, being only redshifted by the expansion of the Universe. They can potentially be detected by future space-based GW detectors such as LISA, BBO or DECIGO. The aim of this section is to compute the GW signal from the electroweak  first order phase transition (FOPT) within the 2HD+a model.

Since the fermionic contributions are not significant for the treatment of the thermal phase transitions,\footnote{Indeed, a high-temperature expansion of the thermal function further defined in Eq.~(\ref{eq:thermal_func}) would show that, unlike for bosons, the function for fermions is lacking of a cubic term, a crucial ingredient to generate a barrier in the effective potential.} it turns out that it does not matter whether we consider the Type I,  II,  X or  Y configuration for our 2HD+a model and the GW signal will be independent of this choice. When computing the stochastic gravitational-wave background, we therefore arbitrarily consider the Type II configuration. We have  nevertheless explicitly checked that we obtain very similar results in the Type I scenario; the Type X and Y scenarios will give exactly the same results as in Type II and I, respectively, as the only difference comes from the different coupling of the $\tau$-lepton which plays a negligible role in this context. At the end of our discussion, we consider the 2HDM limit of our model to underline the impact of the parameters related to the pseudoscalar $a$ boson, namely its mass, mixing and couplings.

\subsection{One-loop thermal effective potential}

In order to study the electroweak phase transition, both zero-temperature quantum corrections and thermal effects must be incorporated into the full effective potential needed for the analysis of phase transitions. However, let us first revisit the tree-level potential.

We consider phase transitions occurring in the field space ($h_0, H_0$), 
where $h_0$ and $H_0$ are the CP-even components of $\Phi_1$ and $\Phi_2$ respectively, which are defined in the gauge basis as
\begin{equation}
 \Phi_1 = \frac{1}{\sqrt{2}}\left(
\begin{array}{c} G_1 + i G_2 \\ h_0 +v_1 + i G_0 \end{array} \right),\quad \Phi_2 = \frac{1}{\sqrt{2}}\left(
\begin{array}{c} H_{c_1} + i H_{c_2} \\ H_0+v_2 + i A_n \end{array} \right). 
\end{equation}
The tree-level potential~(\ref{eq:V2HDa}) in terms of classical background fields thus reduces to
\begin{align}
V_0 &=  \frac{m_{11}^2}{2}h_0^2 + \frac{m_{22}^2}{2}H_0^2 - m_{12}^2 h_0 H_0 + \frac{\lambda_1}{8}h_0^4 + \frac{\lambda_2}{8}H_0^4 +\frac{\lambda_3+\lambda_4+\lambda_5}{4}h_0^2H_0^2.
\label{eq:V0}
\end{align}
The Hessian matrix of~ the tree-level potential (\ref{eq:V2HDa}) is a $9\times 9$ matrix (four degrees of freedom or dof from $\Phi_1$, four from $\Phi_2$ and one from $a_0$, the CP-odd light pseudoscalar). Then, considering only the fields $h_0, H_0$,  we obtain a block-diagonal matrix, with four blocks. The first one is the $2\times 2$ matrix $m_S^2$ of the CP-even states, the second one is the $3\times 3$ matrix $m_P^2$ of the CP-odd states, while the two last ones are the $2\times 2$ matrix $m_C^2$ of the charged states.

One-loop quantum corrections are encoded in the Coleman-Weinberg potential~\cite{Coleman:1973jx}
\begin{equation}
V_{\rm CW} = \frac{1}{64\pi^2}\sum_i n_i m_i^4 \left( \ln\frac{m_i^2}{\mu^2}-c_i \right),
\label{eq:Vcw}
\end{equation}
with $i\in\{t, b, \chi, W^\pm_T, W^\pm_L, Z_T, Z_L, \gamma_L \}$ and also runs over the states from the scalar potential~(\ref{eq:V2HDa}). The degrees of freedom are encoded in $n_i$, where $n_t=n_b = -12, n_\chi=-2, n_{W^\pm_T}=4, n_{W^\pm_L}=n_{Z_T}=2, n_{Z_L}=n_{\gamma_L}=1$, and the dof for each of the scalar neutral states is 1 and 2 for the charged states. The renormalization scale $\mu$ is set to the vev $v = \sqrt{v_1^2 + v_2^2}$. The value of constant $c_i$ arising from dimensional regularization in the $\overline{\text{MS}}$ scheme is 3/2 for fermions, scalars, longitudinal vector bosons and 1/2 for transverse vector bosons. Finally, $m_i^2\equiv m_i^2(h_0, H_0)$ corresponds to the eigenvalues of the field-dependent mass matrix.

In the Type II or Y models, which differ only by the contributions of the $\tau$-lepton that we ignore as its effects are far too small  because of its very small mass, the field-dependent masses for the SM states and the DM candidate $\chi$ are
\begin{align}
    m^2_t = \frac{y_t^2}{2\sin^2\beta}H_0^2,\quad m^2_b = \frac{y_b^2}{2\cos^2\beta}h_0^2, \quad M^2_\chi = m_\chi^2+g_\chi^2 a_0^2,\\
     M^2_{W} = \frac{g_2^2}{4}(h_0^2+H_0^2),\quad M^2_Z = \frac{g_1^2+g_2^2}{4}(h_0^2+H_0^2), \quad m^2_\gamma=0,
\end{align}
with $y_t, y_b, g_1$ and $g_2$ the top Yukawa coupling, the bottom Yukawa coupling, the $U(1)_Y$ and $SU(2)_L$ gauge couplings respectively. Since we are considering $a_0 = 0$ (see Section~\ref{sec:pt_and_gws}), the $M_\chi^2$ term in the effective potential only contributes to the cosmological constant. On the other hand, in the Type-I or -X model, the bottom field dependent mass is given by $m^2_b = (y_b^2)/(2\sin^2\beta) \, H_0^2$.  The eigenvalues of  $m^2_S$ and $m^2_P$ yield the field-dependent masses for the neutral states, while those of $m^2_C$ yields the field-dependent masses of the charged states.

In order to compensate the shift from $V_{\rm CW}$ to the vevs, masses and mixing in the  electroweak  vacuum, we consider the counter-terms
\begin{equation}
V_{\rm CT} = \delta m_{11}^2 h_0^2 + \delta m_{22}^2 H_0^2 + \delta m_{12}^2 h_0 H_0 + \delta \lambda_1 h_0^4 + \delta \lambda_2 H_0^4, 
\end{equation}
where these (finite) counter-terms satisfy the following renormalization conditions\footnote{Divergences arising from Goldstone contributions is treated with the method described in Ref.~\cite{Cline:2011mm}.} in the  electroweak  vacuum ($v_1, v_2$):
\begin{align}
    \partial_{h_0}\left(V_{\rm CW} + V_{\rm CT}\right)\Big\vert_{(v_1, v_2)} &= 0, \quad \partial_{H_0}\left(V_{\rm CW} + V_{\rm CT}\right)\Big\vert_{(v_1, v_2)} = 0, \\
    \partial^2_{h_0^2}\left(V_{\rm CW} + V_{\rm CT}\right)\Big\vert_{(v_1, v_2)} &= 0, \quad \partial^2_{H_0^2}\left(V_{\rm CW} + V_{\rm CT}\right)\Big\vert_{(v_1, v_2)} = 0,\\
    \partial^2_{h_0 H_0}\left(V_{\rm CW} + V_{\rm CT}\right)\Big\vert_{(v_1, v_2)} &= 0,\quad 
\end{align}
and are given by
\begin{align}
\delta\lambda_1 &= \frac{1}{8 v_1^3}\left(\partial_{h_0} V_{\rm CW} - \partial^2_{h_0^2} V_{\rm CW} v_1 -\partial^2_{h_0 H_0} V_{\rm CW} v_2\right)\Bigg\vert_{(v_1, v_2)}, \\
\delta\lambda_2 &= \frac{1}{8v_2^3}\left(\partial_{H_0} V_{\rm CW} - \partial^2_{H_0^2} V_{\rm CW} v_2 -\partial^2_{h_0 H_0} V_{\rm CW} v_1\right)\Bigg\vert_{(v_1, v_2)},\\
\delta m_{11}^2 &= \frac{1}{4v_1}\left(-3 \partial_{h_0} V_{\rm CW} + \partial^2_{h_0^2} V_{\rm CW} v_1 + 3 \partial^2_{h_0 H_0} V_{\rm CW}v_2\right)\Bigg\vert_{(v_1, v_2)}, \\
\delta m_{22}^2 &= \frac{1}{4v_2}\left(-3 \partial_{H_0} V_{\rm CW} + \partial^2_{H_0^2} V_{\rm CW} v_2 + 3 \partial^2_{h_0 H_0} V_{\rm CW}v_1\right)\Bigg\vert_{(v_1, v_2)},\\
\delta m_{12}^2 &= -\partial^2_{h_0 H_0} V_{\rm CW}\Big\vert_{(v_1, v_2)}.
\label{eq:dma0} 
\end{align}
Finally, one must consider thermal effects since the phase transition occurs in the early Universe and thus at very high temperature. These thermal corrections are given by~\cite{Dolan:1973qd}
\begin{equation}
    V_T =\frac{T^4}{2\pi^4}\sum_i n_i J\left(\frac{m_i^2}{T^2}\right),
    \label{eq:VT}
\end{equation}
where the thermal function is defined as~\cite{Dolan:1973qd}
\begin{equation}
    J(y^2) = \int_0^\infty dx~x^2\ln\left(1 + (-1)^B e^{-\sqrt{x^2+y^2}}\right),
    \label{eq:thermal_func}
\end{equation}
with $B=1(0)$ for bosons (fermions).

In order to avoid infrared divergences from the zero Matsubara modes, one resums the daisy diagrams, which amounts to a shift in the mass parameter $m^2$ with a leading-order thermal contribution in the propagator: $m^2\rightarrow m^2 + cT^2$, with $c$ a constant depending on dimensionless couplings. This thermal mass resummation is made in the gauge basis and only then mass matrices are diagonalized to obtain the thermal field-dependent eigenvalues $m^2_i(h_0, H_0, T)$.

In addition the usual Debye mass $cT^2$ for the SM content, the Debye mass peculiar to the Type II or Y 2HD+a model are given by
\begin{align}
    c_1 &=  (g_1^2 + 3 g_2^2+4y_b^2/\cos^2\beta)/16 + (3\lambda_1+2 \lambda_3 + \lambda_4 + \lambda_{1P})/12,\label{eq:c1}\\
    c_2 &= (g_1^2 + 3 g_2^2 + 4y_t^2/\sin^2\beta)/16 + (3\lambda_2+2 \lambda_3 + \lambda_4 + \lambda_{2P})/12,
\end{align}
and where in the case of Type I or X 2HD+a model, $\cos\beta$ is replaced with $\sin\beta$ in Eq.~(\ref{eq:c1}).

The resulting one-loop thermal effective potential is then given by
\begin{equation}
V_{\rm eff}(h_0, H_0, T) = V_0 + V_{\rm CW} + V_{\rm CT} + V_T.    
\end{equation}

\subsection{Key parameters for phase transitions}

Initially, before the  electroweak  phase transition, the Universe is in the symmetric phase. As the temperature decreases, there appears a new minimum in the scalar potential -- a new (broken) phase. With decreasing temperature, this minimum eventually becomes deeper that the one in the symmetric phase, thus making it metastable. This metastable or false vacuum eventually decays into the stable or true vacuum. The cosmic first-order phase transition occurs through the nucleation of a bubbles of true vacuum, which expand and collide with each other, converting the symmetric phase into the broken phase. The decay rate of the false vacuum or the bubble nucleation rate per time per volume $\Gamma$ is given by~\cite{Linde:1981zj}
\begin{equation}
\Gamma \sim T^4 e^{-S/T},
\label{eq:bubble_rate}
\end{equation}
where S is the three-dimensional Euclidean action minimized by the \textit{bounce} or O(3) critical bubble.

The nucleation temperature $T_n$ is defined such that the number of nucleated bubbles per Hubble time per Hubble volume is unity: $\Gamma H^{-4} \sim O(1)$ with $H$ the Hubble parameter. The latter is expressed in a radiation-dominated Universe as
\begin{equation}
    H^2 = \frac{8\pi\rho_{\text{rad}}}{3 M^2_p},
\end{equation}
where $\rho_{\text{rad}}=\pi^2/30 g_* T^4$ is the energy density of the plasma in the false vacuum, with $g_*$ the effective number of relativistic degrees of freedom at $T$ and where $M_p$ is the Planck mass.

Considering the electroweak scale, one has $T\sim O(100)$ GeV and $g_*\sim O(100)$. Therefore, using Eq.~(\ref{eq:bubble_rate}), one obtains $S/T\sim 140$ at the nucleation temperature $T_n$.

The strength of a first-order phase transition is given by~\cite{Espinosa:2010hh}:
\begin{equation}
\label{eq:alpha}
\alpha \equiv\frac{\Delta\epsilon}{\rho_{\text{rad}}}\Big|_{T=T_*},   \quad \Delta \epsilon \equiv \epsilon \big |_{\text{false vacuum}} - \epsilon \big |_{\text{true vacuum}}
\end{equation}
with $\epsilon = V_{\rm eff} - \frac{T}{4}\frac{\partial V_{\rm eff}}{\partial T}$, the vacuum energy.

Finally, the inverse time duration $\beta$ of the PT is defined as~\cite{Grojean:2006bp}:
\begin{equation}
    \label{eq:beta}
    \frac{\beta}{H_n} = T_n\frac{d(S/T)}{dT}\Big |_{T_n},
\end{equation}
where the parameters are evaluated at the nucleation temperature $T_n$.

\subsection{Predictions for gravitational-wave signals}

A single bubble of true vacuum alone cannot be responsible for the generation of gravitational waves because of its spherical symmetry (zero quadrupole moment). A stochastic gravitational-wave background, however, is possible when at least two bubbles collide with each other. The resulting gravitational power spectrum $h^2\Omega_{\text{GW}}$ mainly comes from three contributions\footnote{A new contribution, from feebly interacting particles, has been recently studied in Ref.~\cite{Jinno:2022fom}. This contribution is more appropriate for phase transitions in the dark sector, therefore we omit it in our analysis.}~\cite{Caprini:2015zlo}: $h^2\Omega_{\text{GW}}\simeq h^2\Omega_{\text{col}} + h^2\Omega_{\text{sw}} + h^2\Omega_{\text{turb}}$.

The contribution from bubble collisions is given in the envelope approximation by~\cite{Huber:2008hg}
\begin{equation}
 h^2\Omega_{\text{col}}(f) = h^2\Omega_{\text{col}}^{\text{peak}} S_\text{col}(f),
\end{equation}
with
\begin{align}
& h^2\Omega_{\text{col}}^\text{peak} = 1.67\times 10^{-5}\left(\frac{H_n}{\beta}\right)^2\left(\frac{\kappa_\text{col}\alpha}{1+\alpha}\right)^2\left(\frac{100}{g_n}\right)^{1/3}\left(\frac{0.11 v_w^3}{0.42+v_w^2}\right), \\
& S_\text{col} = \frac{3.8\left(f/f_\text{col}\right)^{2.8}}{1+2.8\left(f/f_\text{col}\right)^{3.8}},
\end{align}
where $\kappa_\text{col}$ is the efficiency factor for the conversion of the vacuum energy into the gradient energy of the scalar field, $v_w$ is the bubble-wall speed in the rest frame of the plasma far away from the bubble~\cite{Caprini:2015zlo}, $f_\text{col}$ is the frequency at the peak of the power spectrum, $h^2\Omega_{\text{col}}^\text{peak}$ and $S_\text{col}$ is the spectral shape of the GW spectrum $h^2\Omega_{\text{col}}$.

The overlap of sound waves yield a contribution given by~\cite{Hindmarsh:2017gnf, Caprini:2019egz, Schmitz:2020rag}
\begin{equation}
 h^2\Omega_{\text{sw}}(f) = h^2\Omega_{\text{sw}}^{\text{peak}} S_\text{sw}(f),
\end{equation}
with
\begin{align}
& h^2\Omega_{\text{sw}}^\text{peak} = 1.23\times 10^{-6}\left(\frac{H_n}{\beta}\right)\left(\frac{\kappa_\text{sw}\alpha}{1+\alpha}\right)^2\left(\frac{100}{g_n}\right)^{1/3}v_w\Upsilon, \\
& S_\text{sw} = \left(\frac{f}{f_\text{sw}}\right)^3\left(\frac{7}{4+3\left(f/f_\text{sw}\right)^{2}}\right)^{7/2},
\end{align}
where $\kappa_\text{sw}$ is the efficiency factor for the conversion of the vacuum energy into the bulk motion of the plasma, $f_\text{sw}$ is the sound-wave peak frequency and $S_\text{sw}$ is the spectral shape of the GW spectrum $h^2\Omega_{\text{sw}}$. The suppression factor, accounting for the finite lifetime $\tau_\text{sw}$ of the sound waves, is defined in a radiation-dominated Universe as~\cite{Guo:2020grp, Ellis:2019oqb} 
\begin{equation}
\Upsilon=1-\frac{1}{\sqrt{2\tau_\text{sw}H_n + 1}}.
\end{equation}

The MHD-turbulence contribution is given by~\cite{Caprini:2015zlo,Schmitz:2020syl}
\begin{equation}
 h^2\Omega_{\text{turb}}(f) = h^2\Omega_{\text{turb}}^{\text{peak}} S_\text{turb}(f),
\end{equation}
with
\begin{align}
& h^2\Omega_{\text{turb}}^\text{peak} = 3.35\times 10^{-4}\left(\frac{H_n}{\beta}\right)\left(\frac{\kappa_\text{turb}\alpha}{1+\alpha}\right)^{3/2}\left(\frac{100}{g_n}\right)^{1/3}v_w \frac{1}{N_\text{turb}}, \\
&S_\text{turb} = \frac{\left(f/f_\text{turb}\right)^3}{\left[1+\left(f/f_\text{turb}\right)\right]^{11/3}} \frac{N_\text{turb}}{1+8\pi f/h_n},\\
& N_\text{turb} = 2^{11/3}\left(1+8\pi f_\text{turb}/h_n\right),
\end{align}
where $\kappa_\text{turb}$ is the efficiency factor for the conversion of the vacuum energy into turbulent flows, $f_\text{turb}$ is the MHD-turbulence peak frequency, $S_\text{turb}$ is the spectral shape of the GW spectrum $h^2\Omega_{\text{turb}}$, $N_\text{turb}$ is a normalization factor such that $S_\text{turb}(f=f_\text{turb})=1$ and $h_n$ is the value of the Hubble rate at $T_n$, red-shifted to today.

\subsection{Phase transitions and GW signals in the plane of $h_0$ and $H_0$}
\label{sec:pt_and_gws}
Considering phase transitions that only occur in the plane ($h_0$, $H_0$), we set $a_0$ to zero. This configuration remains different from the 2HDM model because $a_0$ and its associated couplings intervene through loops via Eqs.~(\ref{eq:Vcw}) and~(\ref{eq:VT}).

In the following we consider the Type II model in the alignment limit $\alpha = \beta -\pi/2$. As already mentioned in the beginning of this section, the impact of the fermionic sector in this context is rather modest and we have checked that the results are the same in the Type I  scenario (as well as in the Type X and Y cases as we neglect the impact of the $\tau$ lepton).  We perform a random sampling in the following parameter space:
\begin{align}
    & M_H, M_A, M_{H^\pm} \in [500, 1250] \text{~GeV}, \quad M_a \in [10, 200] \text{~GeV}, \nonumber\\
    & \tan\beta \in [0.1, 50], \quad \sin\theta \in [\sqrt{2}/2, 1],      \nonumber\\
   & \lambda_a \in [0, 4\pi], \quad \lambda_{1P}, \lambda_{2P} \in [-\pi, 4\pi],
    \label{eq:param_space}
\end{align}
where we have fixed the additional parameter $m_{12}$ to $m_{12}^2 = \frac12 M_H^2 \sin(2\beta)$ ($\vert M\vert = M_H$). The resulting points satisfy the constraint of perturbative unitarity Eq.~(\ref{eq:unitarity}), stability of the potential~Eq.~(\ref{eq:boundedbelow}) and allow a  moderate mass splitting between the heavy scalar particles $\vert M_i-M_j\vert\leq500$ GeV ($i,j \in \{H, A, H^\pm\}$).

We then scan this region of the parameter space with the package \texttt{CosmoTransitions}~\cite{Wainwright:2011kj} and only consider strong first-order phase transition (FOPT), that is with $v_n/T_n \geq 1$~\cite{Postma:2020toi}, with $v_n\equiv \sqrt{(\langle h_0\rangle_\text{sym.}-\langle h_0\rangle_\text{brok.})^2+(\langle H_0\rangle_\text{sym.}-\langle H_0\rangle_\text{brok.})^2}$ evaluated at the nucleation temperature $T_n$, in order to avoid any ambiguity about the kind of the phase transition~\cite{Kajantie:1996mn,Biekotter:2022kgf}. Finally, regarding the GW spectrum, we consider that the bubbles runaway and consider $v_w=1$ for the velocity of the bubble wall.

\begin{figure}[!ht]
    \centering
    \includegraphics[width=0.33\linewidth]{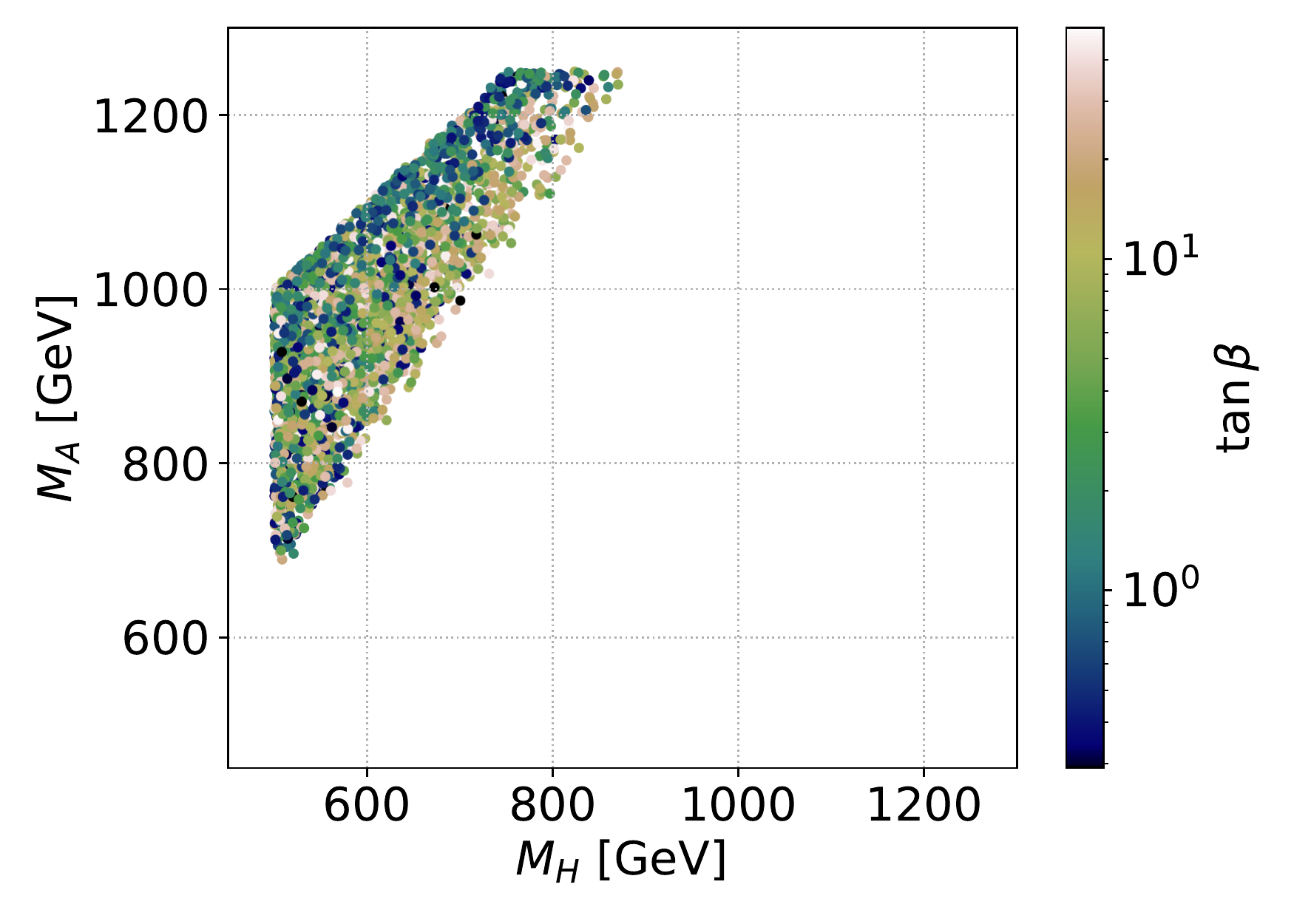}\!\includegraphics[width=0.33\linewidth]{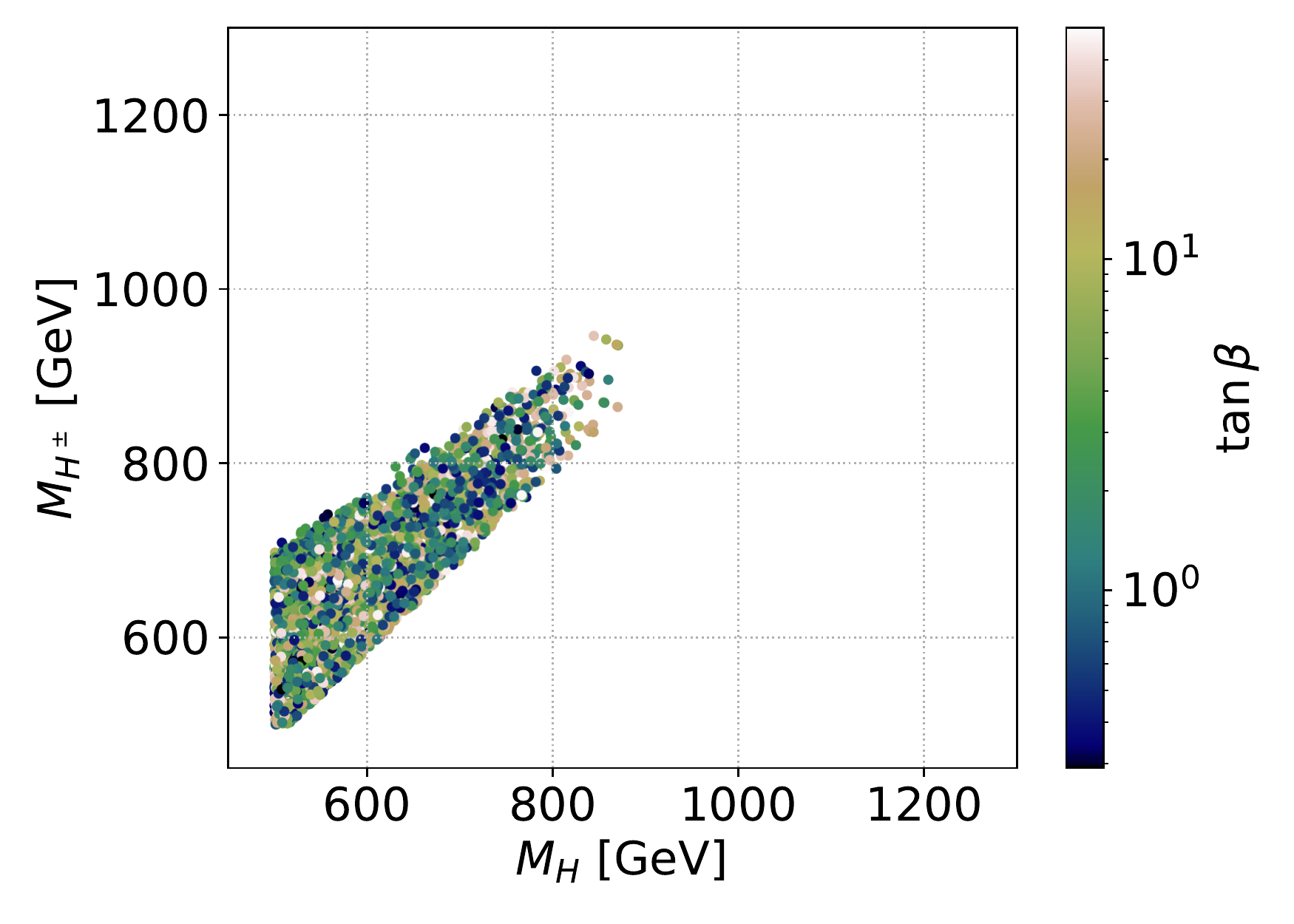}\!\includegraphics[width=0.33\linewidth]{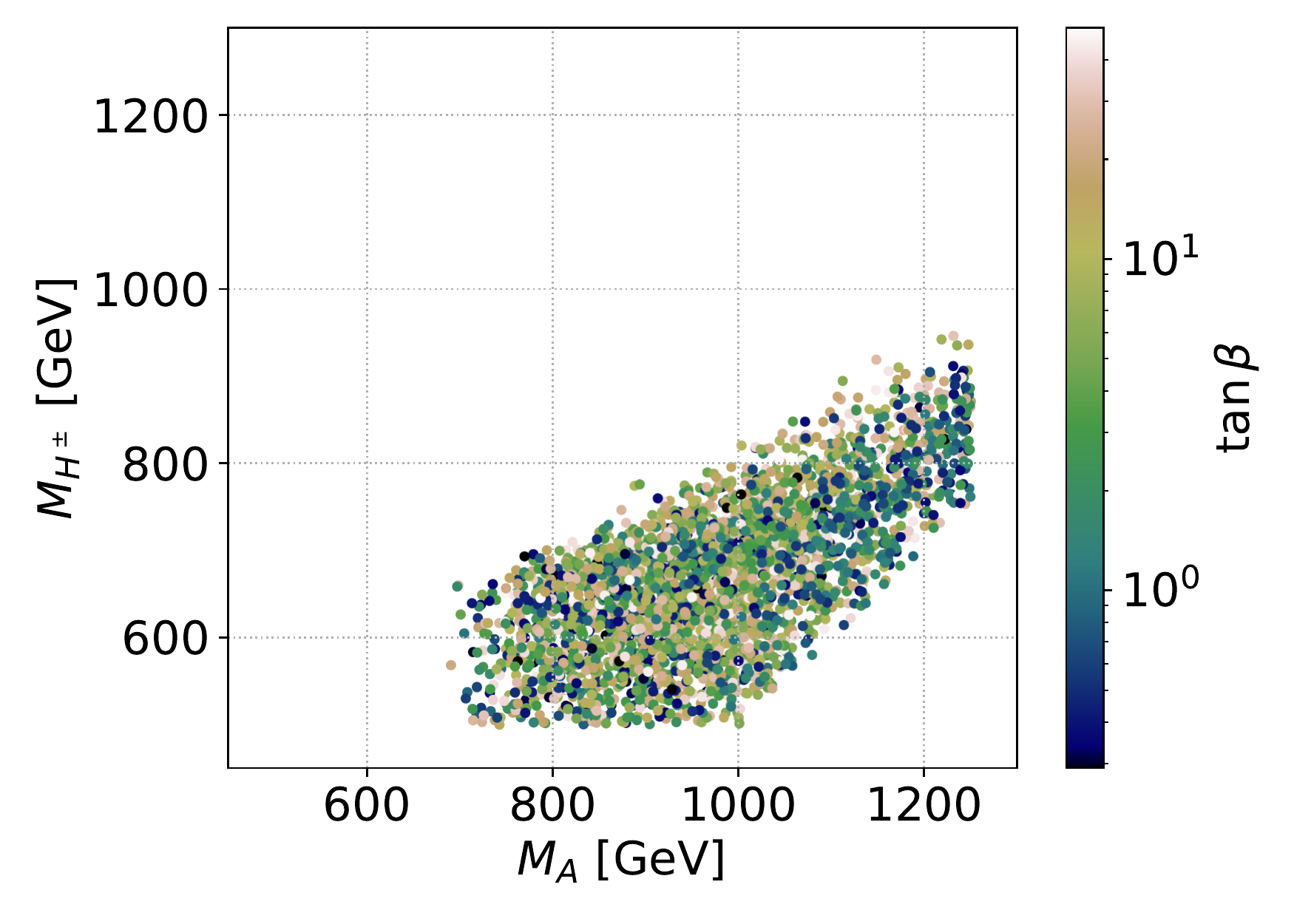}
    \caption{Points, from the scanned parameter space, giving rise to strong FOPTs that generate a GW signal. Left: $M_A$ vs $M_H$. Middle: $M_{H^\pm}$ vs $M_H$. Right: $M_{H^\pm}$ vs $M_A$. The color code indicates values of $\tan \beta$. Shades of blue characterize points with $0.1\leq\tan\beta<1$, shades of green characterize points with $1\leq\tan\beta<10$, while yellow to white shows points with $10\leq\tan\beta<50$. }
    \label{fig:masses_general}
\end{figure}

The points that give rise to strong FOPT are shown in a series of two-dimensional projected spaces. Fig.~\ref{fig:masses_general} shows these points with the value of the mass of the heavy scalar particles, color-coded by $\tan \beta$. One can see from the three panels that the largest value is found for $M_A$, which reaches the upper limit of 1250 GeV in Eq.~(\ref{eq:param_space}). As for $M_H$ and $M_{H^\pm}$, their values remain below 1 TeV. 

\begin{figure}[!ht]
    \centering
    \includegraphics[width=0.33\linewidth]{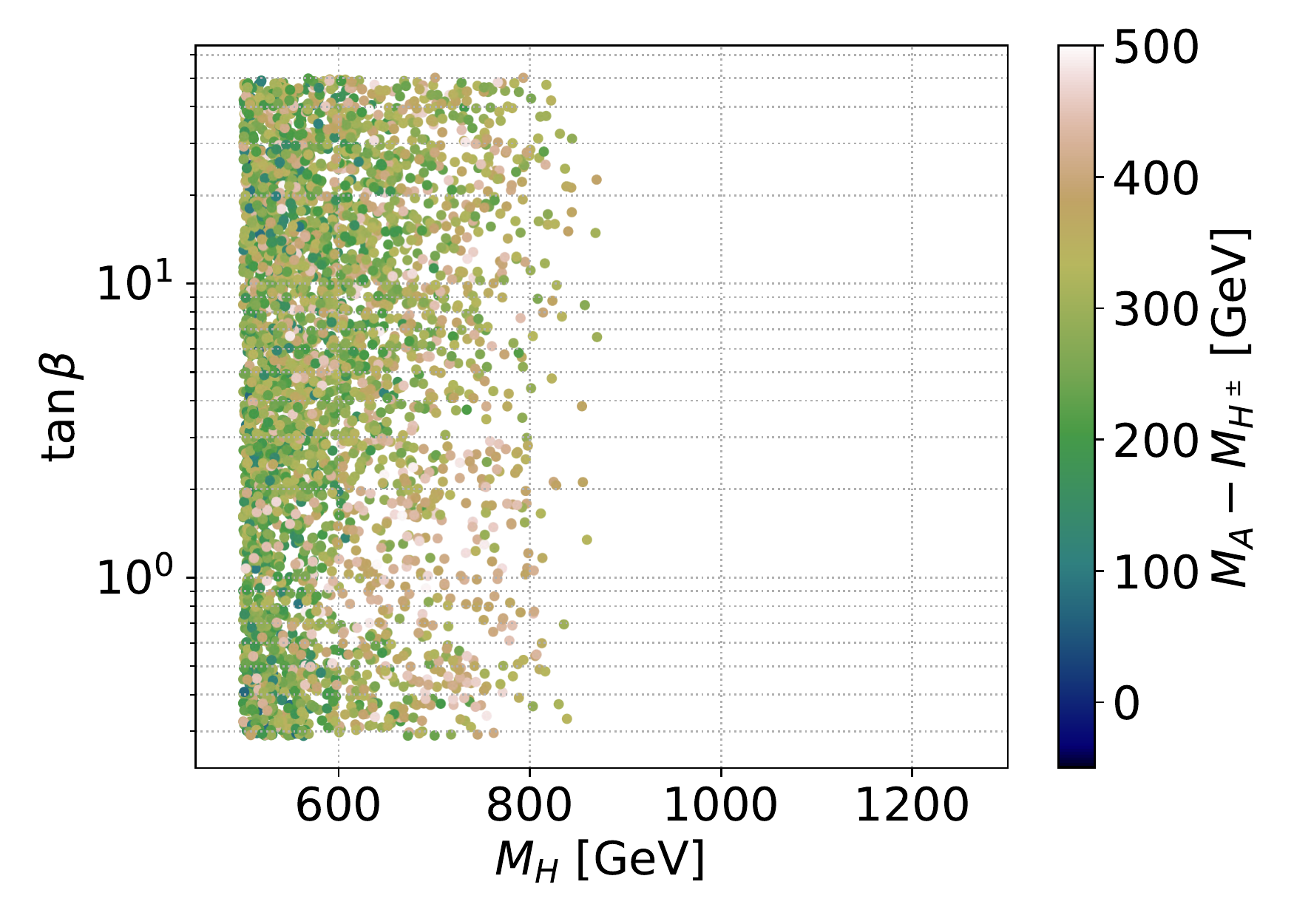}\!\includegraphics[width=0.33\linewidth]{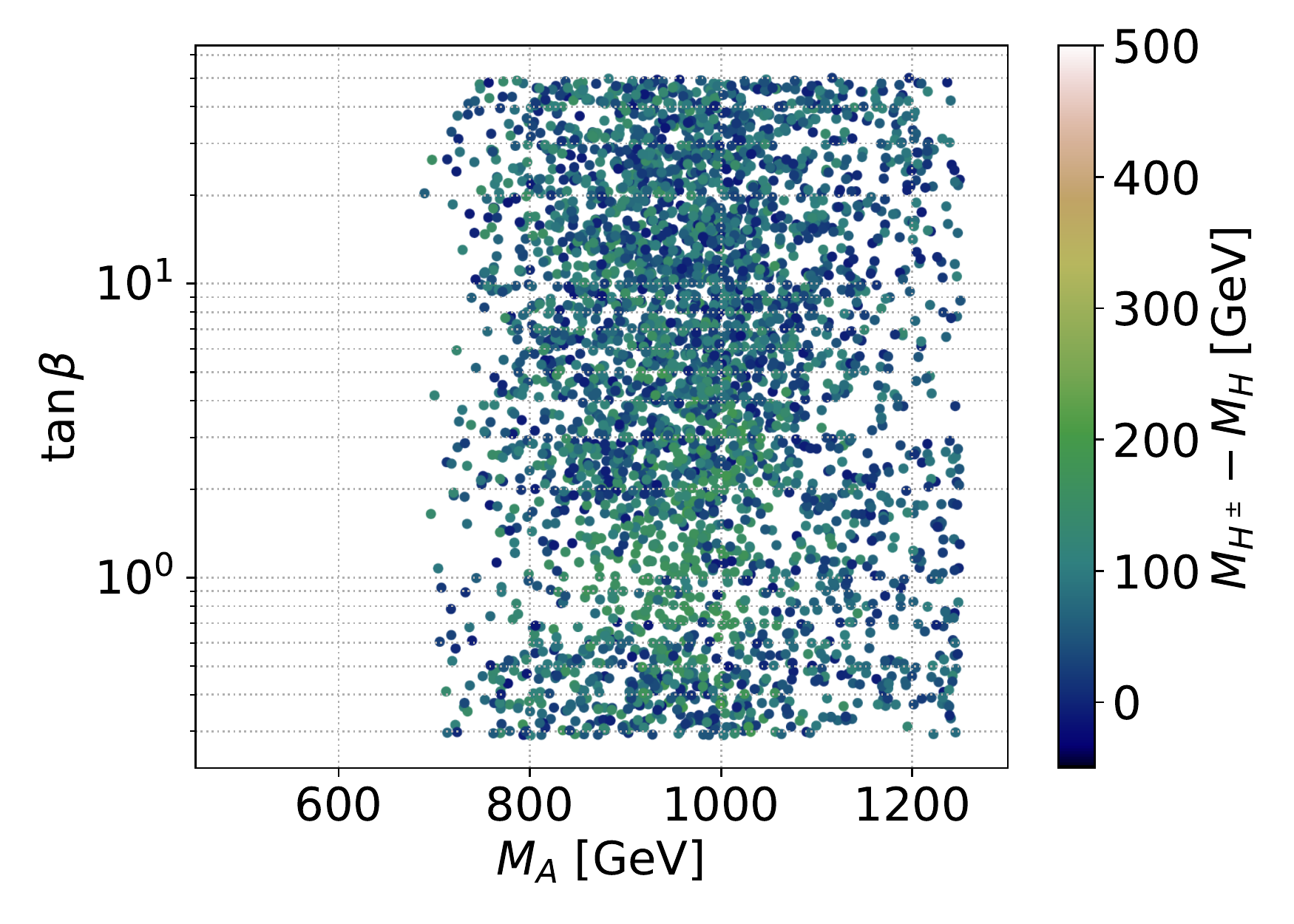}\!\includegraphics[width=0.33\linewidth]{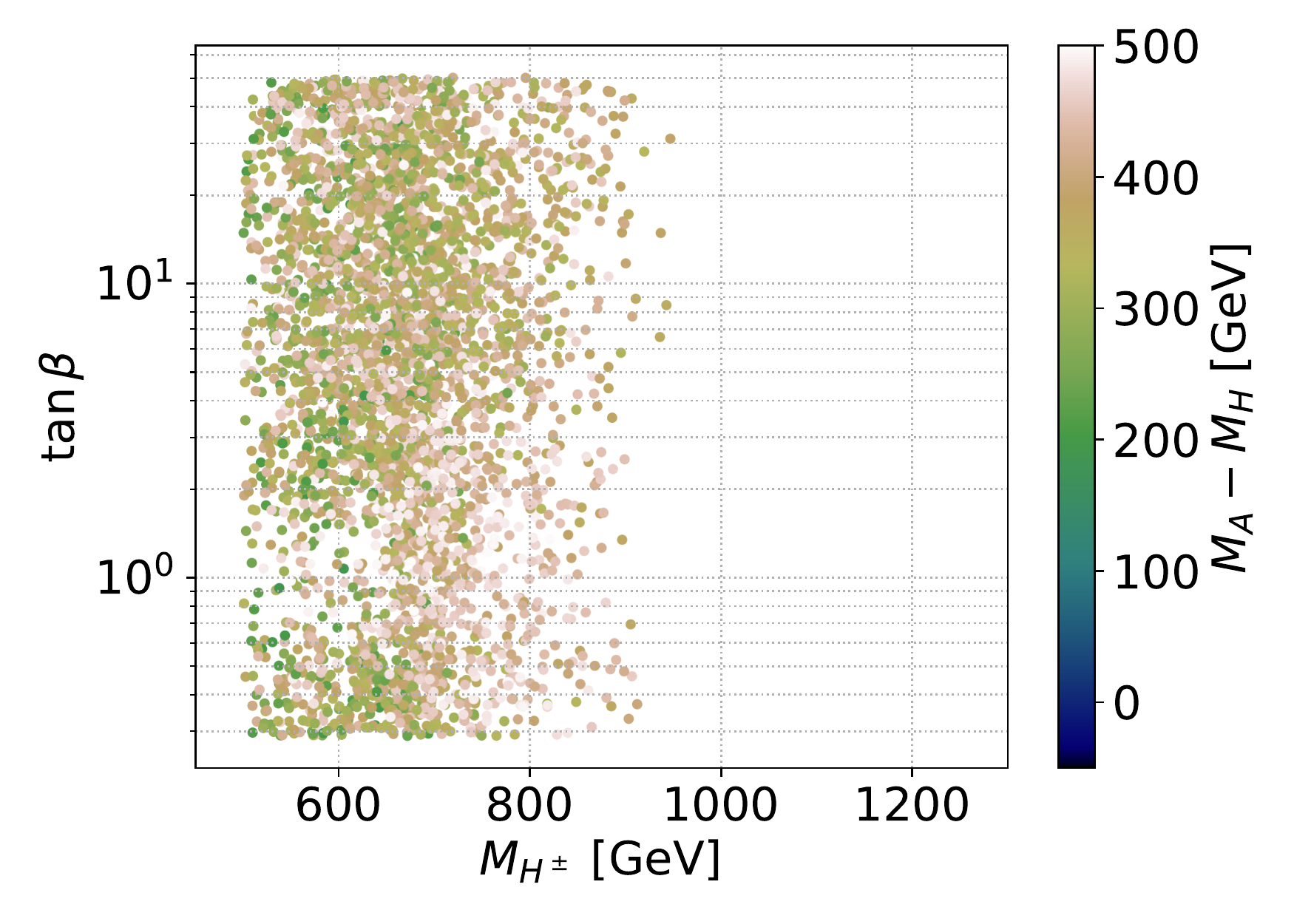}
    \caption{Parameter space with strong FOPTs. Left: $\tan\beta$ vs $M_H$. Middle: $\tan\beta$ vs $M_A$. Right: $\tan\beta$ vs $M_H^{\pm}$. The color code quantifies the mass splitting between two of the three heavy scalar states. Only the middle panel presents a zero mass splitting.}
    \label{fig:m_tB_general}
\end{figure}

In Fig.~\ref{fig:m_tB_general}, the same parameters as in the previous figure are involved but this time, the color code measures the mass splitting between two of the three heavy scalar states, the third one being on the abscissa. In this plot, one more clearly sees the allowed range of value for the masses. In the middle panel one can see that a zero mass splitting between $M_{H^\pm}-M_H$ is allowed, while it is not the case for the two other combinations.

\begin{figure}[!ht]
    \centering
    \includegraphics[width=0.33\linewidth]{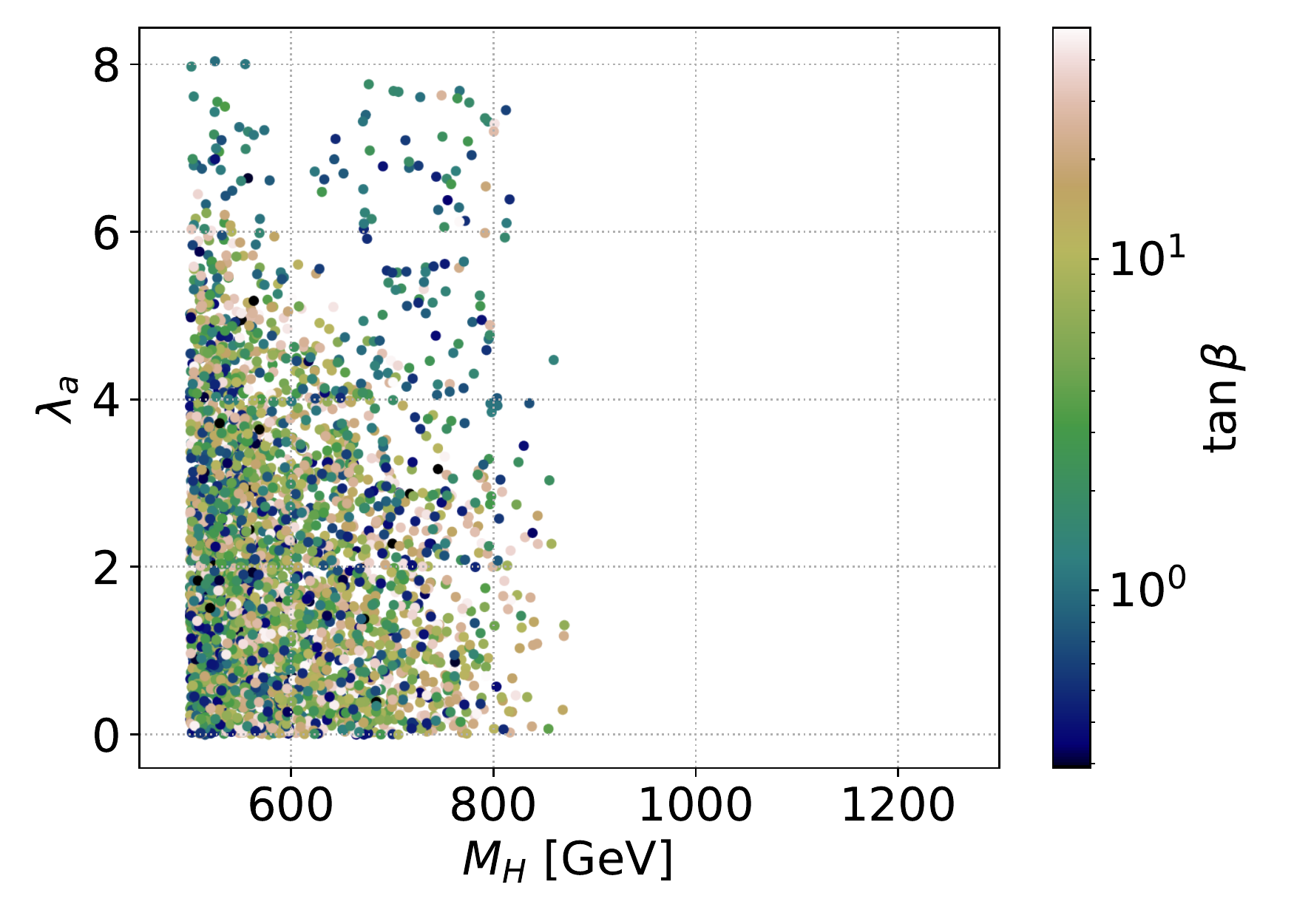}\includegraphics[width=0.33\linewidth]{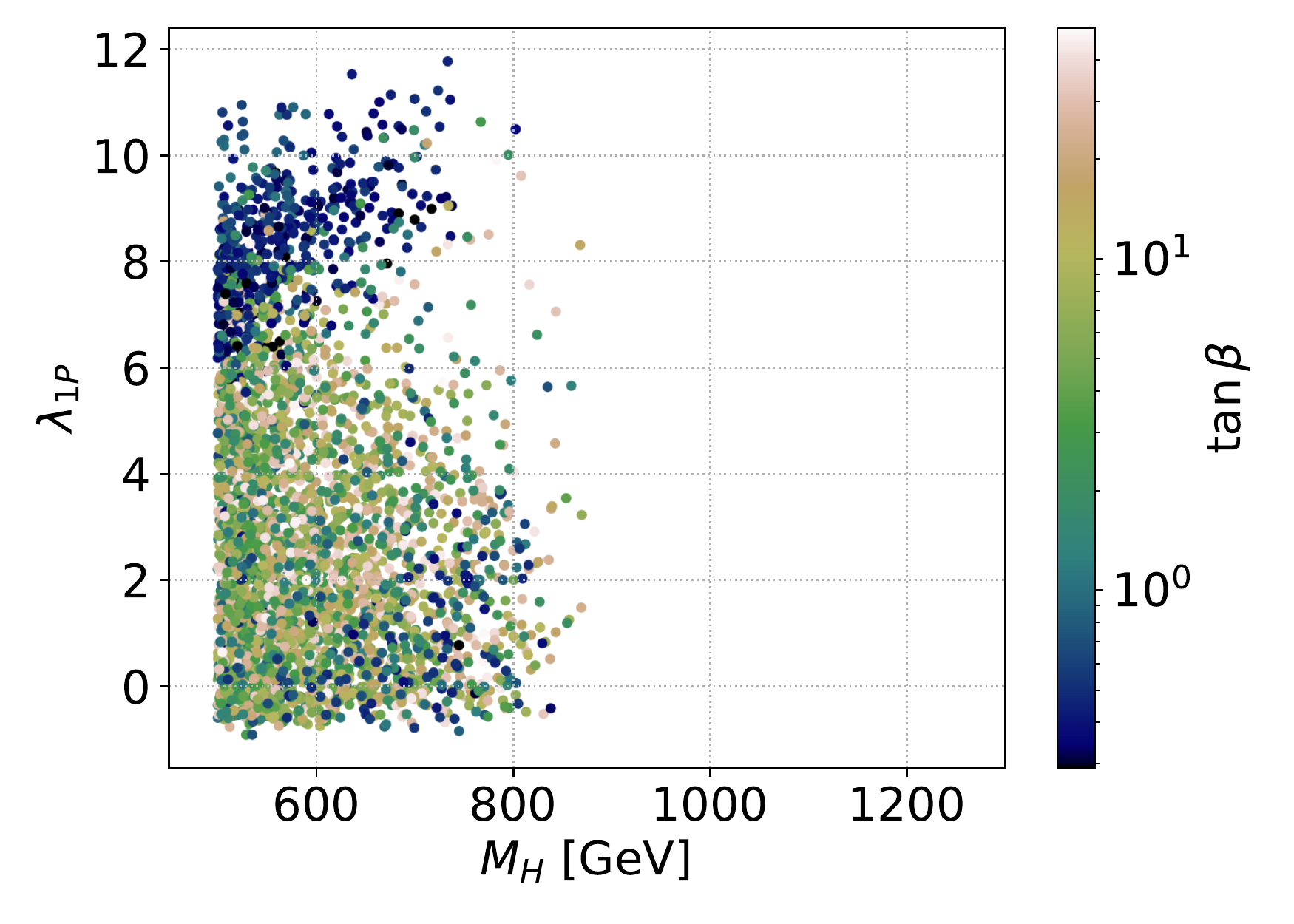}\includegraphics[width=0.33\linewidth]{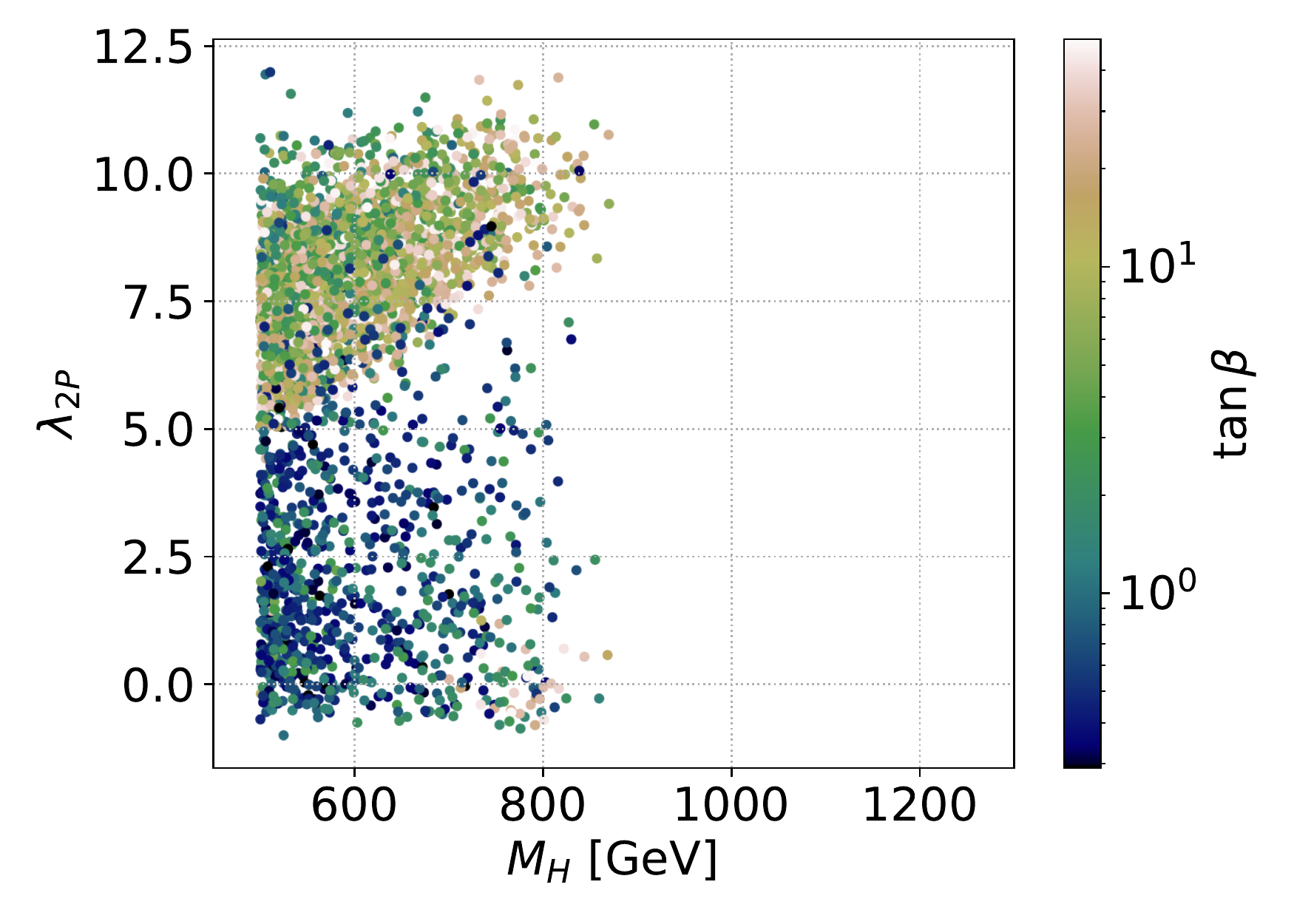}
    \caption{Parameter space with strong FOPTs. Left: $\lambda_a$ vs $M_H$. Middle: $\lambda_{1P}$ vs $M_H$. Right: $\lambda_{2P}$ vs $M_A$. The color code indicates values of $\tan \beta$.}
    \label{fig:lambda_general}
\end{figure}

For each panel of Fig.~\ref{fig:lambda_general}, we show on the ordinate one of the quartic couplings associated to the light pseudoscalar. Similarly to Fig.~\ref{fig:masses_general}, the color code represents the value of $\tan\beta$. This color code clearly indicates two regions in both the middle and right panel. In the middle panel, it shows that large values of $\lambda_{1P}$ are found with small values of $\tan\beta$ and vice-versa. On the other hand, in the right panel $\lambda_{2P}$ and $\tan\beta$ are positively correlated: when $\lambda_{2P}$ is small, $\tan\beta$ is small, and vice-versa. Regarding the phase-transition parameters, the left panel of Fig.~\ref{fig:GW_vs_f_vs_tanBeta_general} shows the usual correlation between $\beta/H$ and $\alpha$: the slower the phase transition, the stronger it is. The color code represents the value of $\tan\beta$.

\begin{figure}[!ht]
    \centering
    \includegraphics[width=0.33\linewidth]{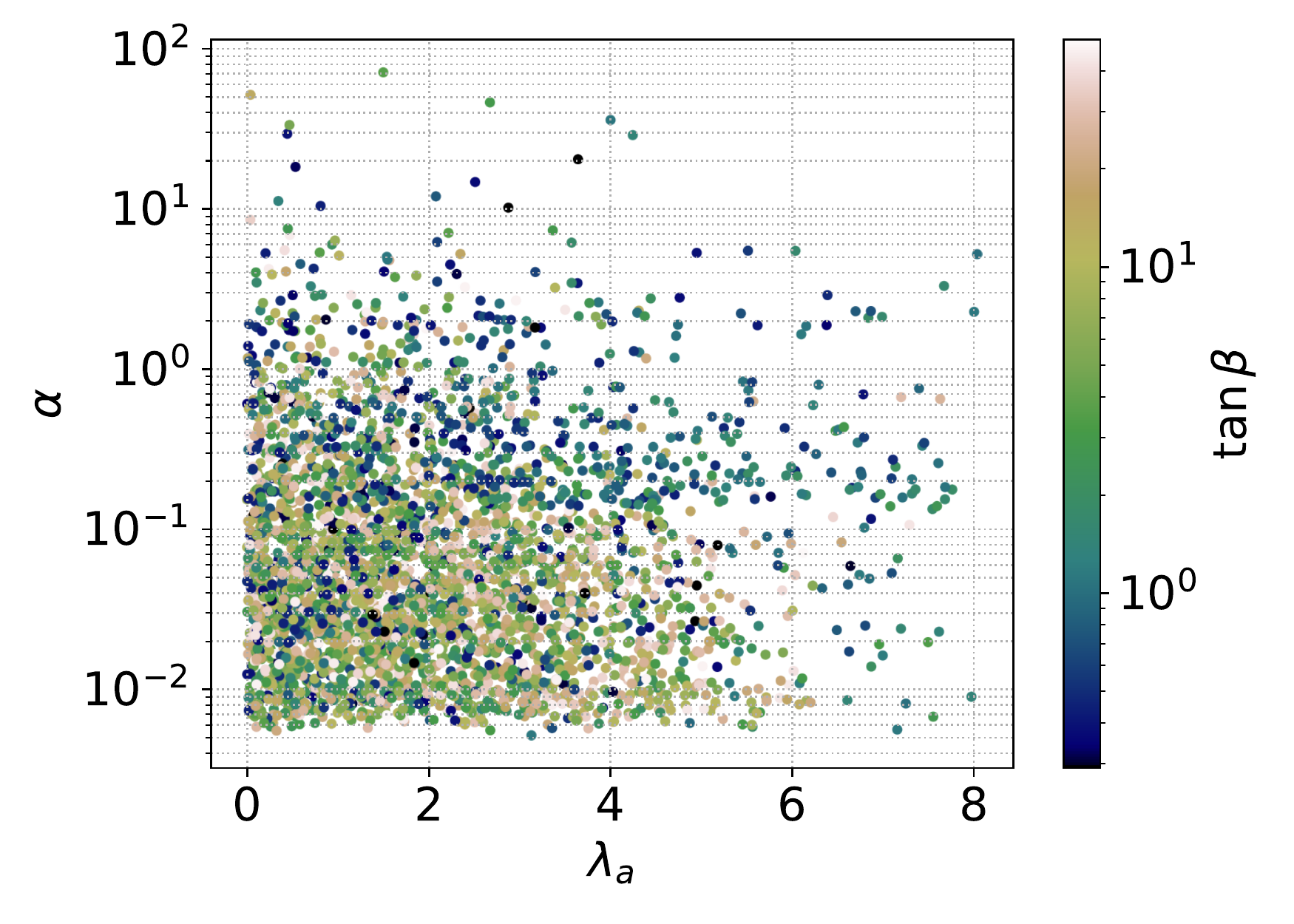}\includegraphics[width=0.33\linewidth]{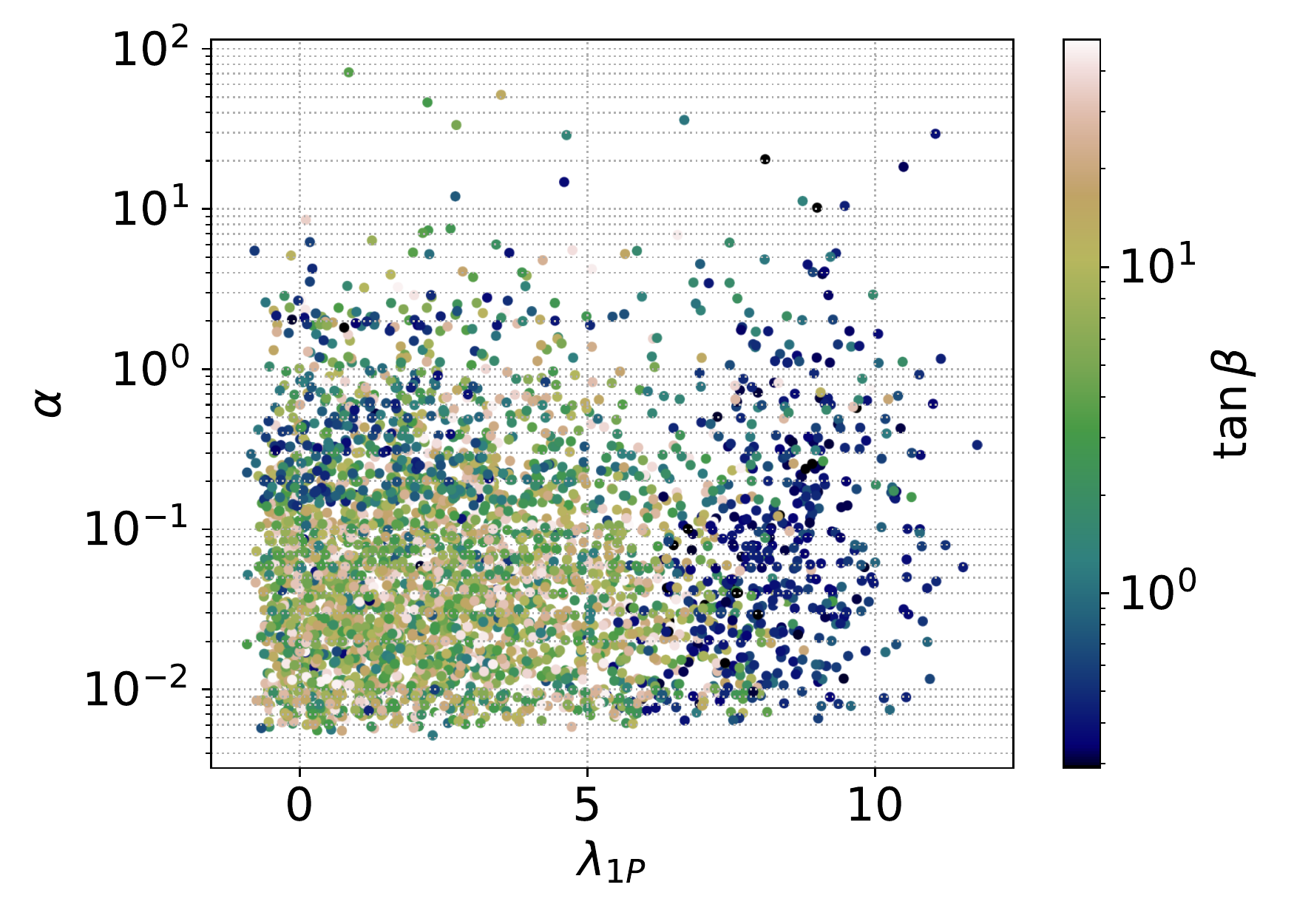}\includegraphics[width=0.33\linewidth]{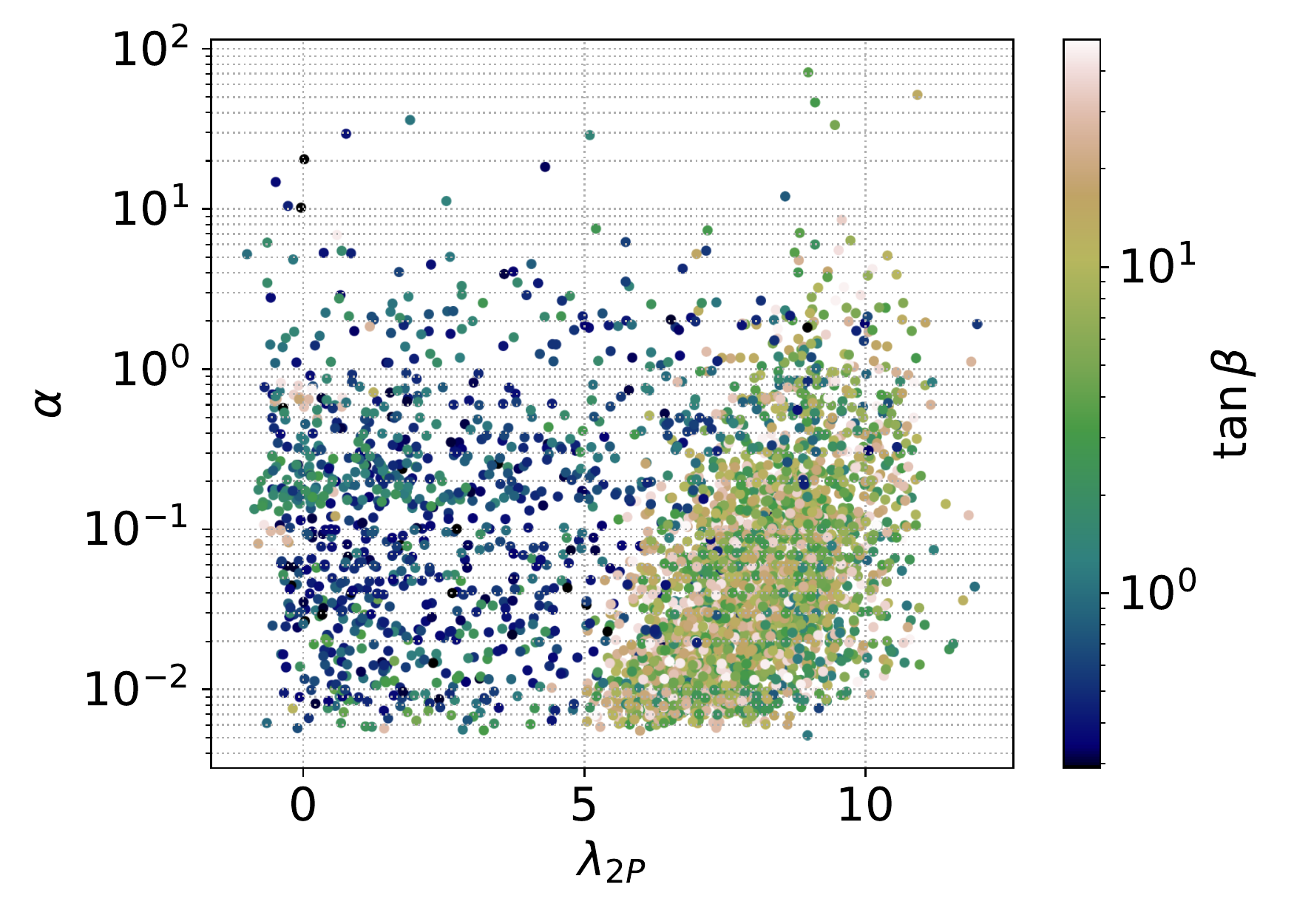}
    \caption{Parameter space with strong FOPTs. Left: $\alpha$ vs $\lambda_a$. Middle: $\alpha$ vs $\lambda_{1P}$. Right: $\alpha$ vs $\lambda_{2P}$. The color code represents the value of $\tan\beta$.}
    \label{fig:lambda_alpha_general}
\end{figure}

\begin{figure}[!ht]
    \centering
    \includegraphics[width=0.49\linewidth]{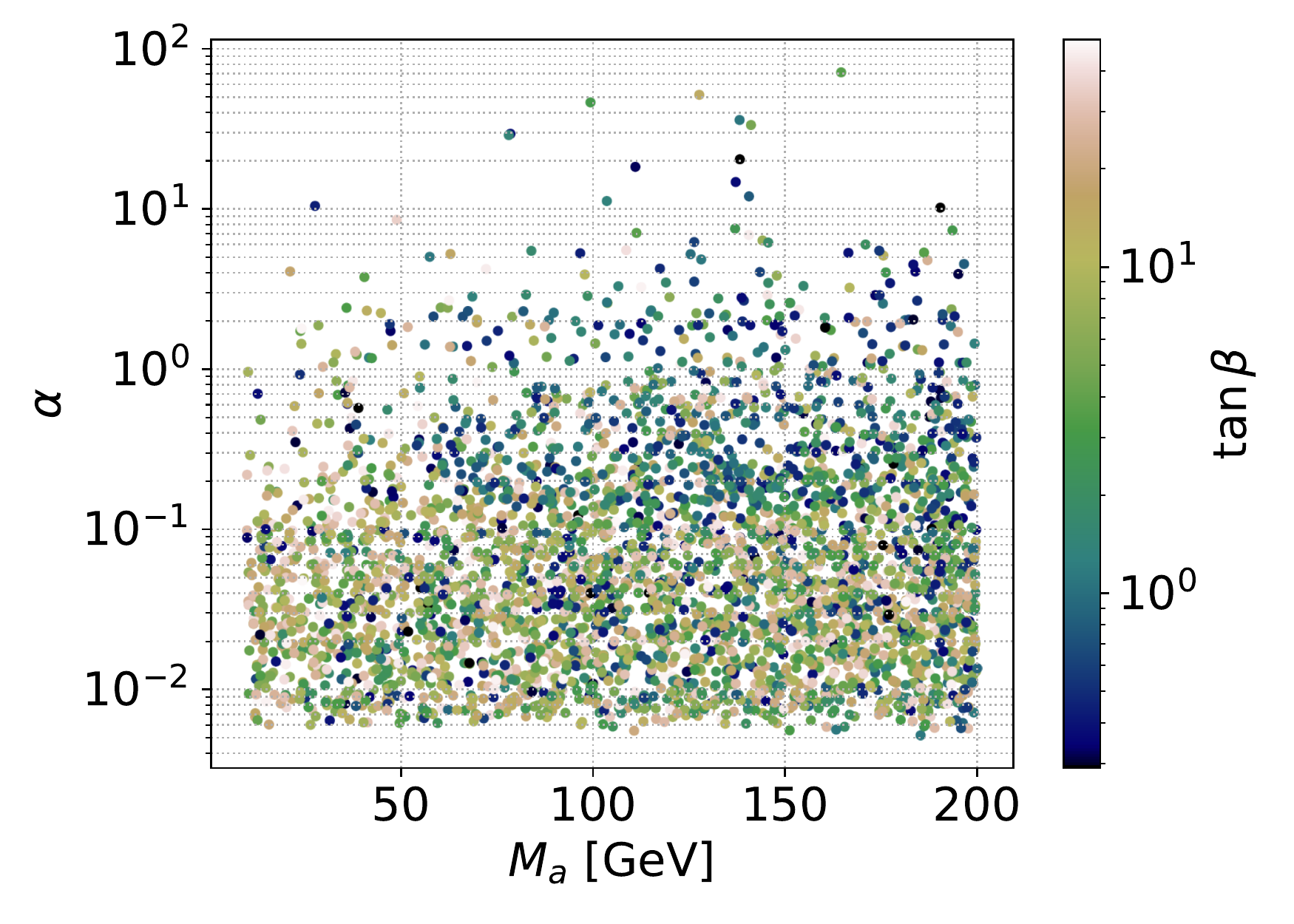}
    \includegraphics[width=0.49\linewidth]{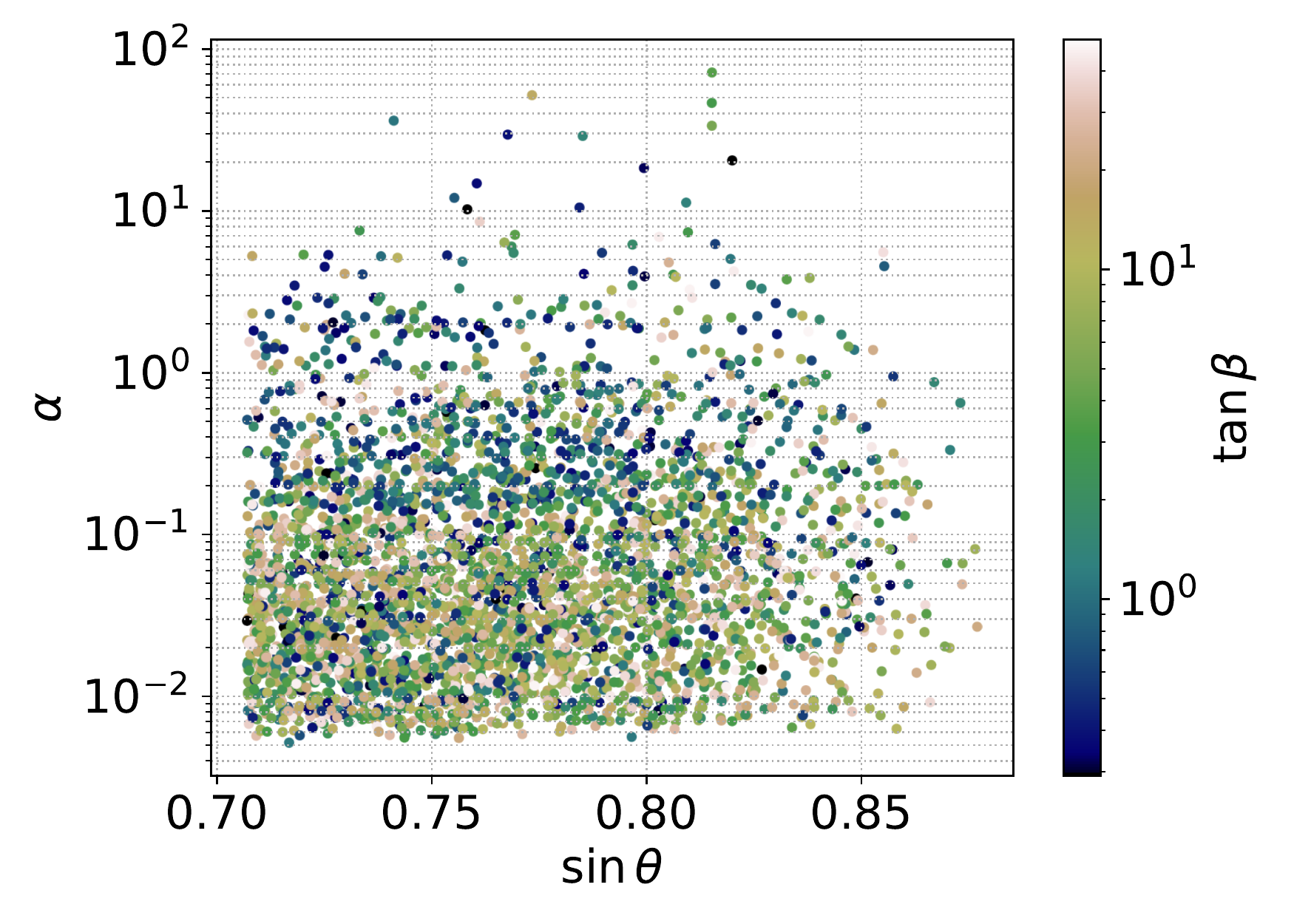}
    \caption{Parameter space with strong FOPTs. Left: $\alpha$ vs $M_a$. Right: $\alpha$ vs $\sin\theta$. The color code indicates the value of $\tan\beta$.}
    \label{fig:ma_alpha_general}
\end{figure}

Let us now investigate the impact of the parameters related to the light pseudoscalar on the strength of the phase transition $\alpha$. Fig.~\ref{fig:lambda_alpha_general} shows that the maximal value for the pseudoscalar self-coupling $\lambda_a$ is smaller than the maximal value for the portal couplings $\lambda_{1P}$ or $\lambda_{2P}$, and it seems easier to obtain SFOPT for smaller $\lambda_a$, as beyond $\lambda_a=5$, the plot is less populated. The range of values for both $\lambda_{1P}$ and $\lambda_{2P}$ are quite similar. However contrary to $\lambda_a$, moderately small negative values of these portal couplings are allowed to give rise to  strong FOPTs. In Fig.~\ref{fig:ma_alpha_general}, one can see  in the left panel that increasing $M_a$ can slowly raise the value of the strength of phase transition $\alpha$. In the right panel, a not too large mixing angle is favored for strong FOPTs.

The right panel of Fig.~\ref{fig:GW_vs_f_vs_tanBeta_general} shows the peak amplitude of the GW signal as a function of the peak frequency. The power-law integrated sensitivity curves for the GW detectors are constructed for an observation time of four years for LISA (solid line), BBO (dashed line) and DECIGO (dotted line) and GWs are considered to be detectable if the signal-to-noise ratio is above 10~\cite{Azatov:2019png}. This figure shows points yielding a signal strong enough to be potentially detected by LISA, BBO or DECIGO.

\begin{figure}[!ht]
    \centering\includegraphics[width=0.49\linewidth]{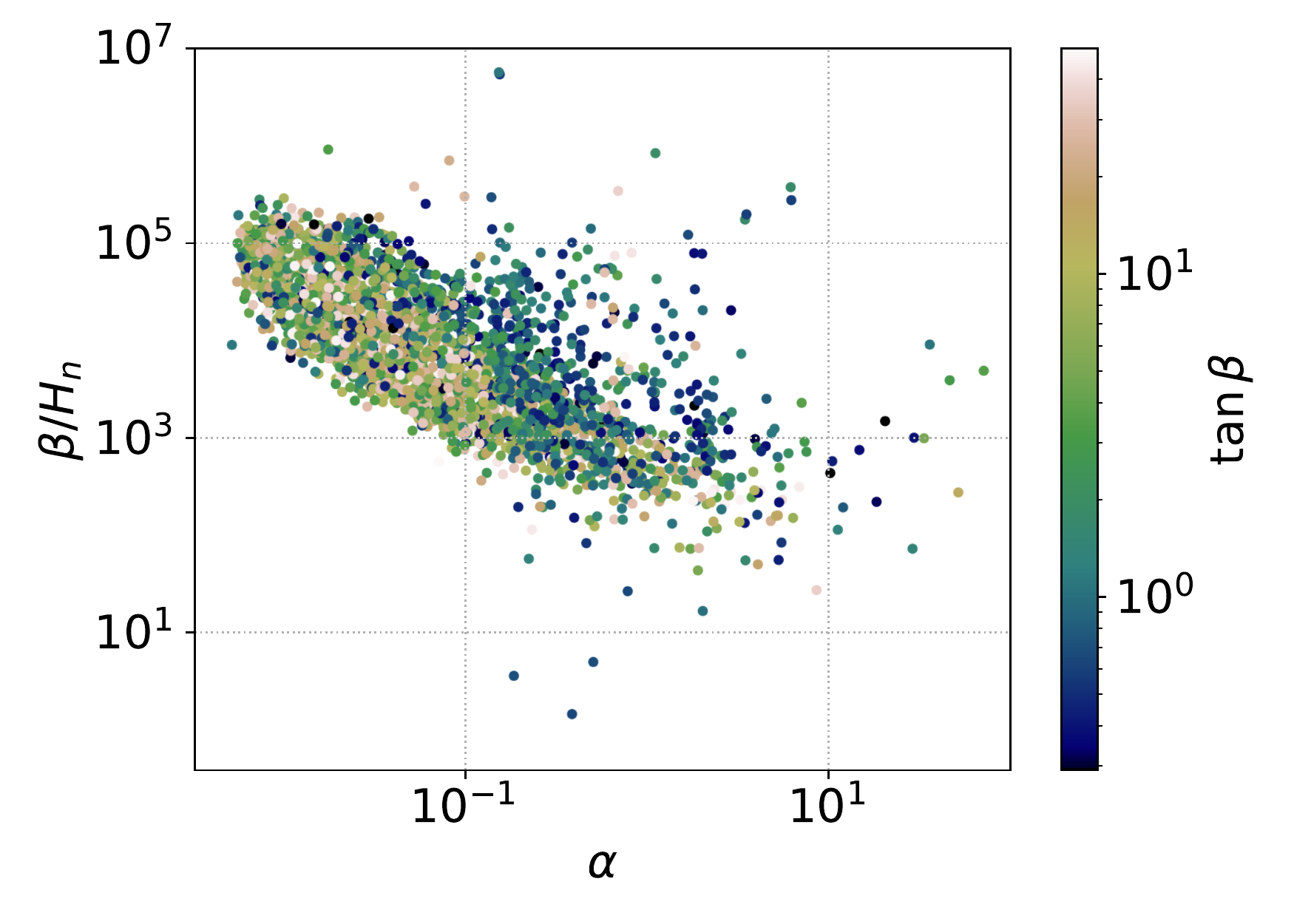}
    \includegraphics[width=0.49\linewidth]
    {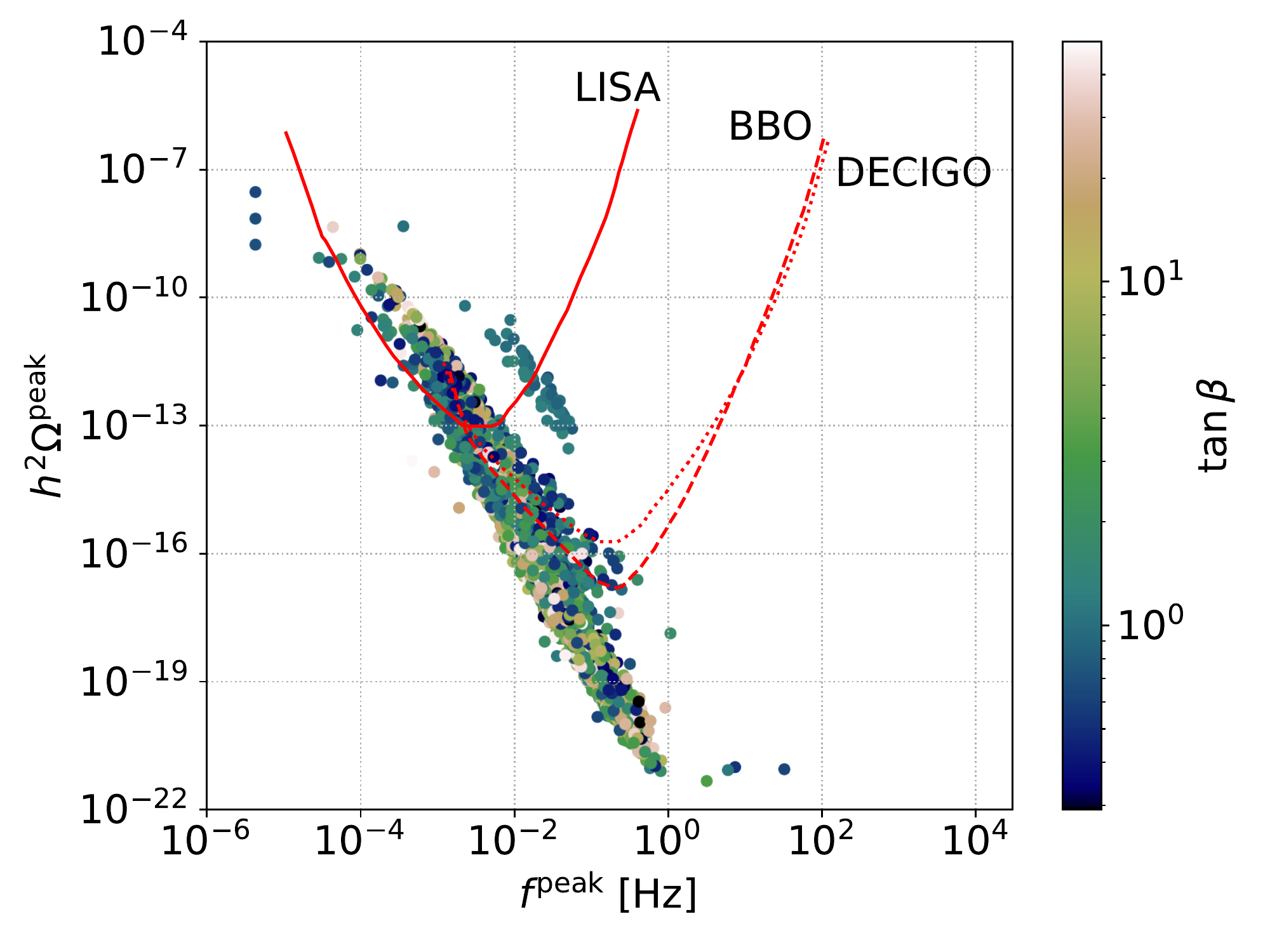}
    \caption{Parameter space with strong FOPTs. Left: $\beta/H_n$ vs $\alpha$. Right: Peak amplitude of the GW signal as a function of peak frequency. Also shown are the sensitivity curves of future detectors LISA, BBO and DECIGO.}
    \label{fig:GW_vs_f_vs_tanBeta_general}
\end{figure}

\subsubsection{Comparison with the 2HDM}

\begin{figure}[!ht]
    \centering
    \includegraphics[width=0.33\linewidth]{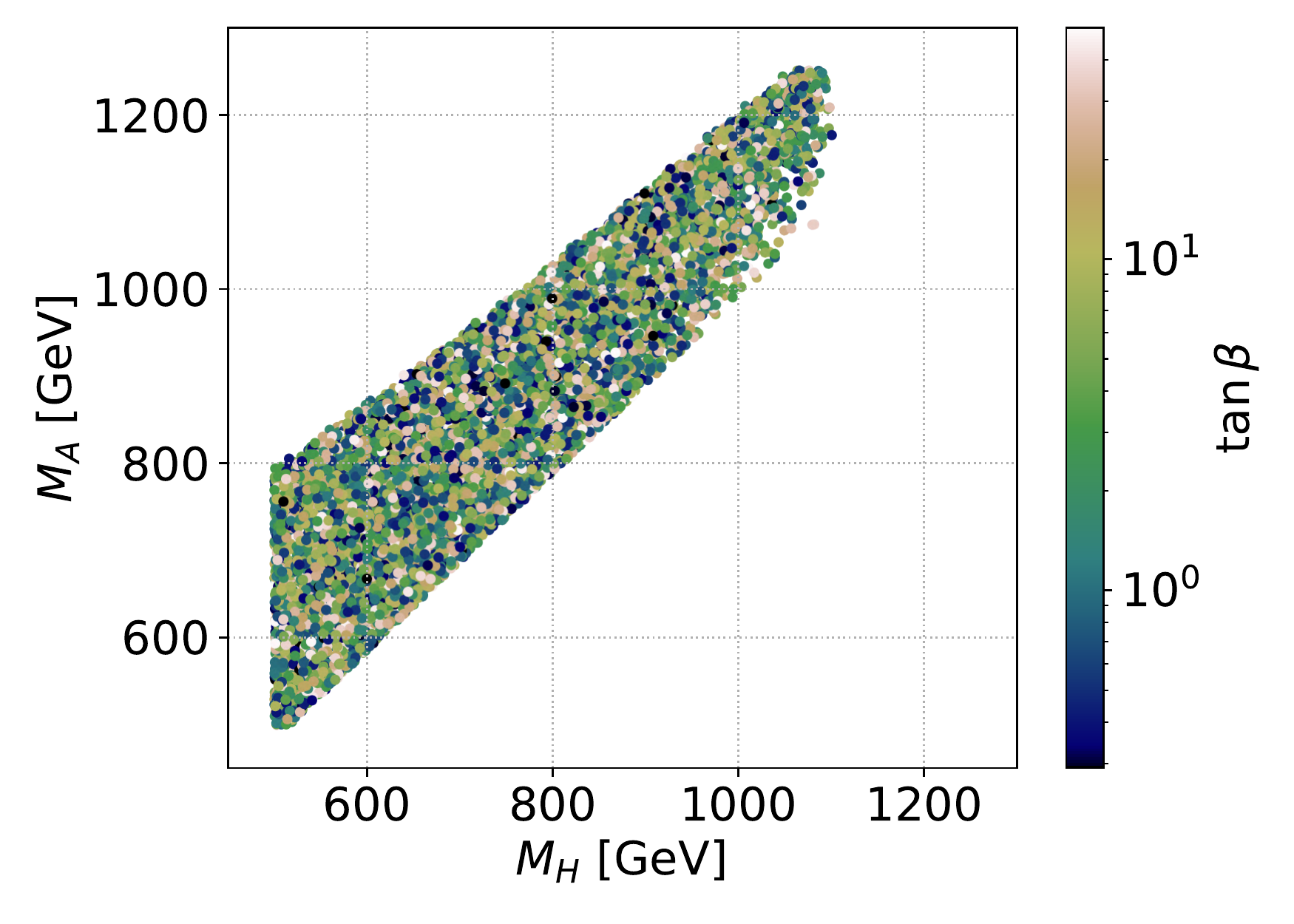}\includegraphics[width=0.33\linewidth]{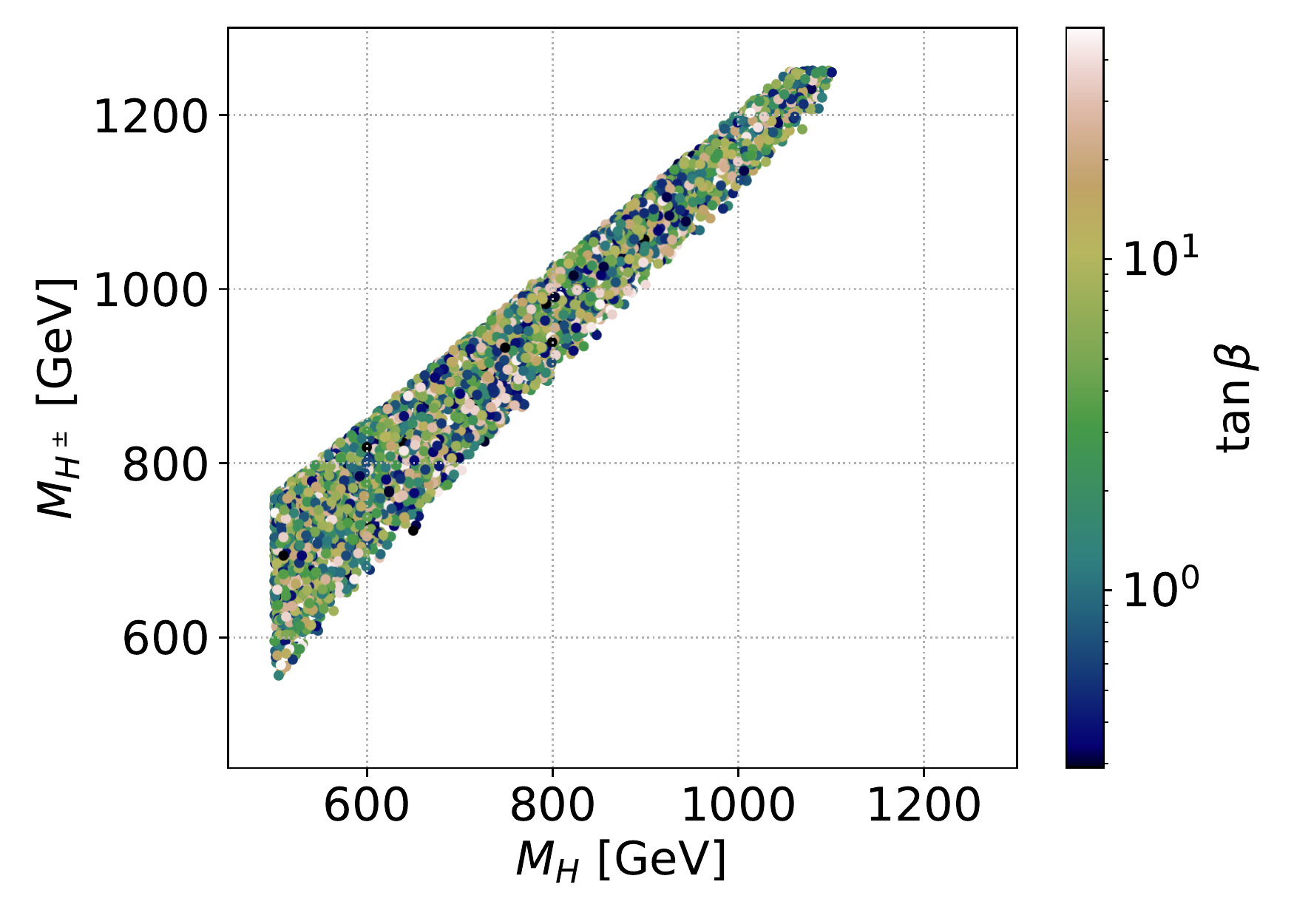}\includegraphics[width=0.33\linewidth]{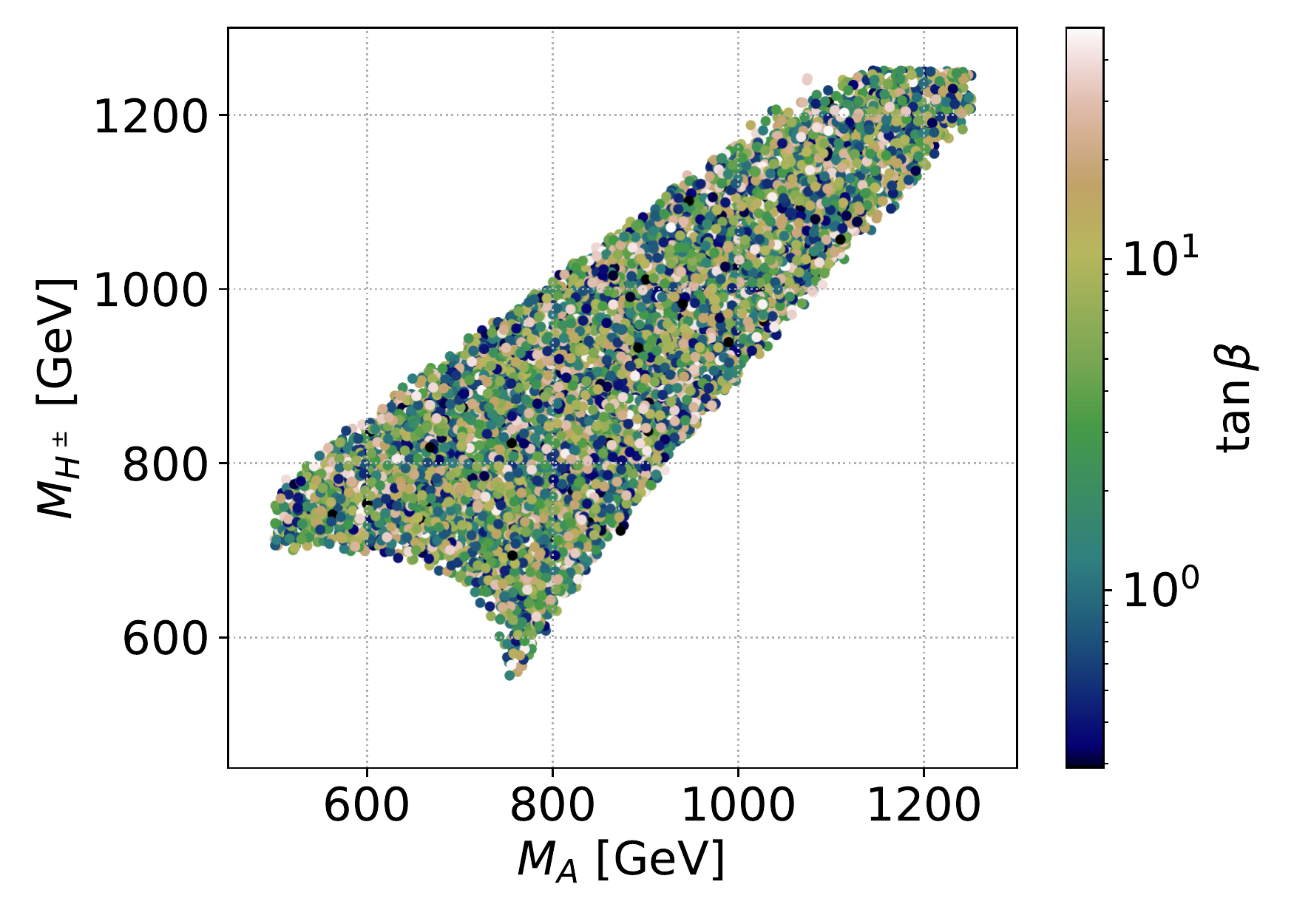}
    \caption{Parameter space with strong FOPTs for pure 2HDM. Left: $M_A$ vs $M_H$. Middle: $M_{H^\pm}$ vs $M_H$. Right: $M_{H^\pm}$ vs $M_A$. The color code indicates values of $\tan \beta$.}
    \label{fig:masses_general-2HDM_typeII}
\end{figure}

\begin{figure}[!ht]
    \centering
    \includegraphics[width=0.33\linewidth]{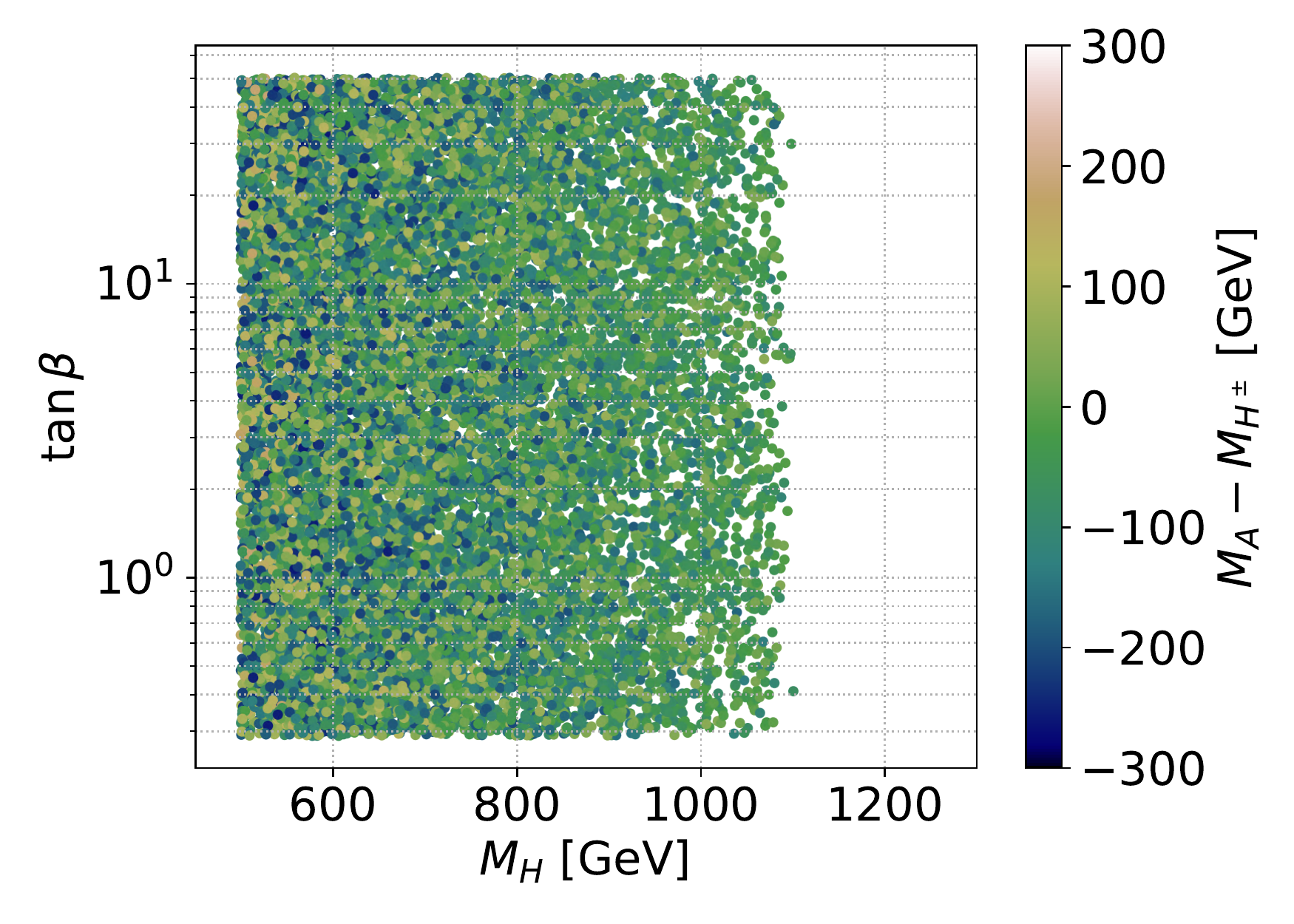}\includegraphics[width=0.33\linewidth]{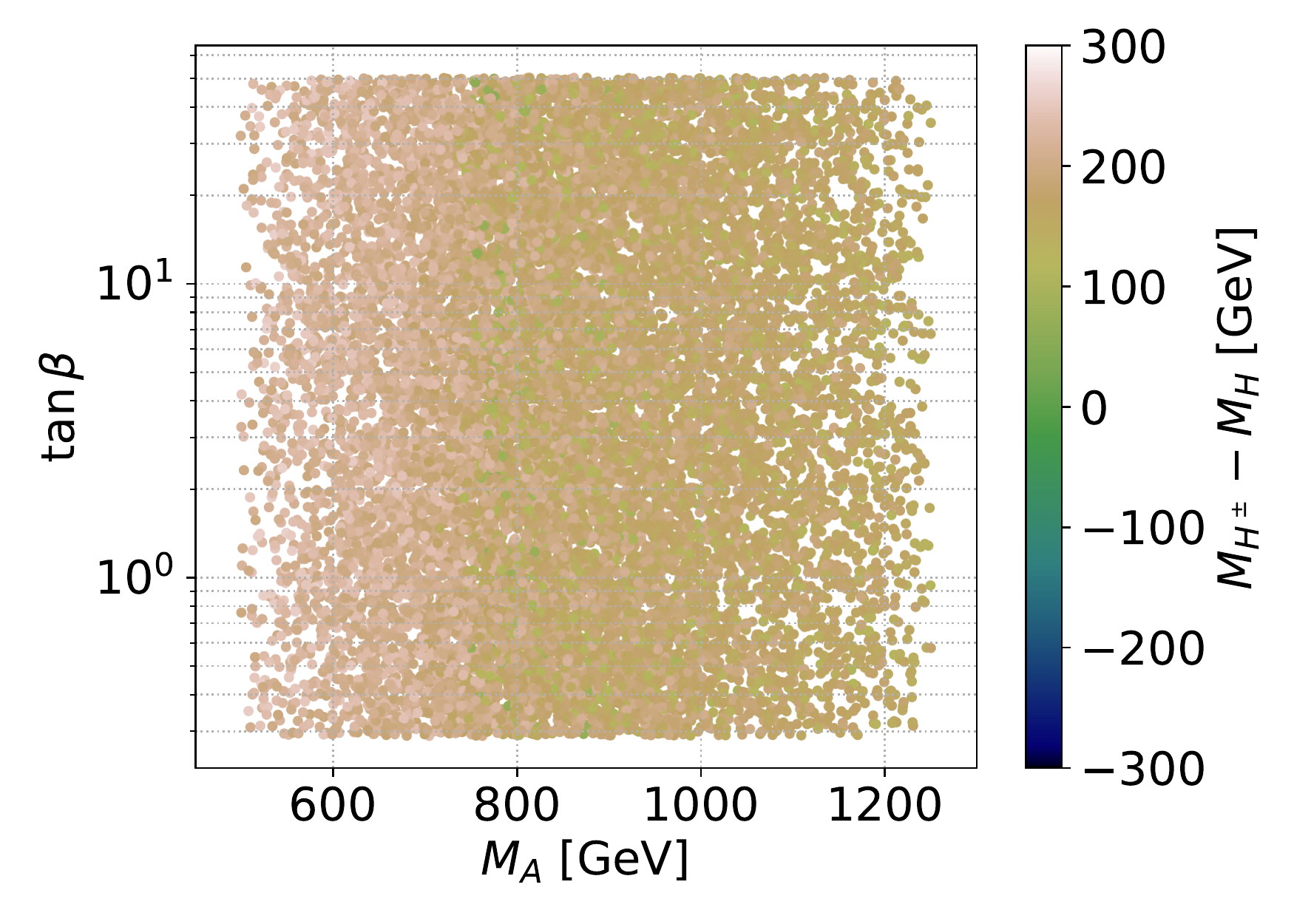}\includegraphics[width=0.33\linewidth]{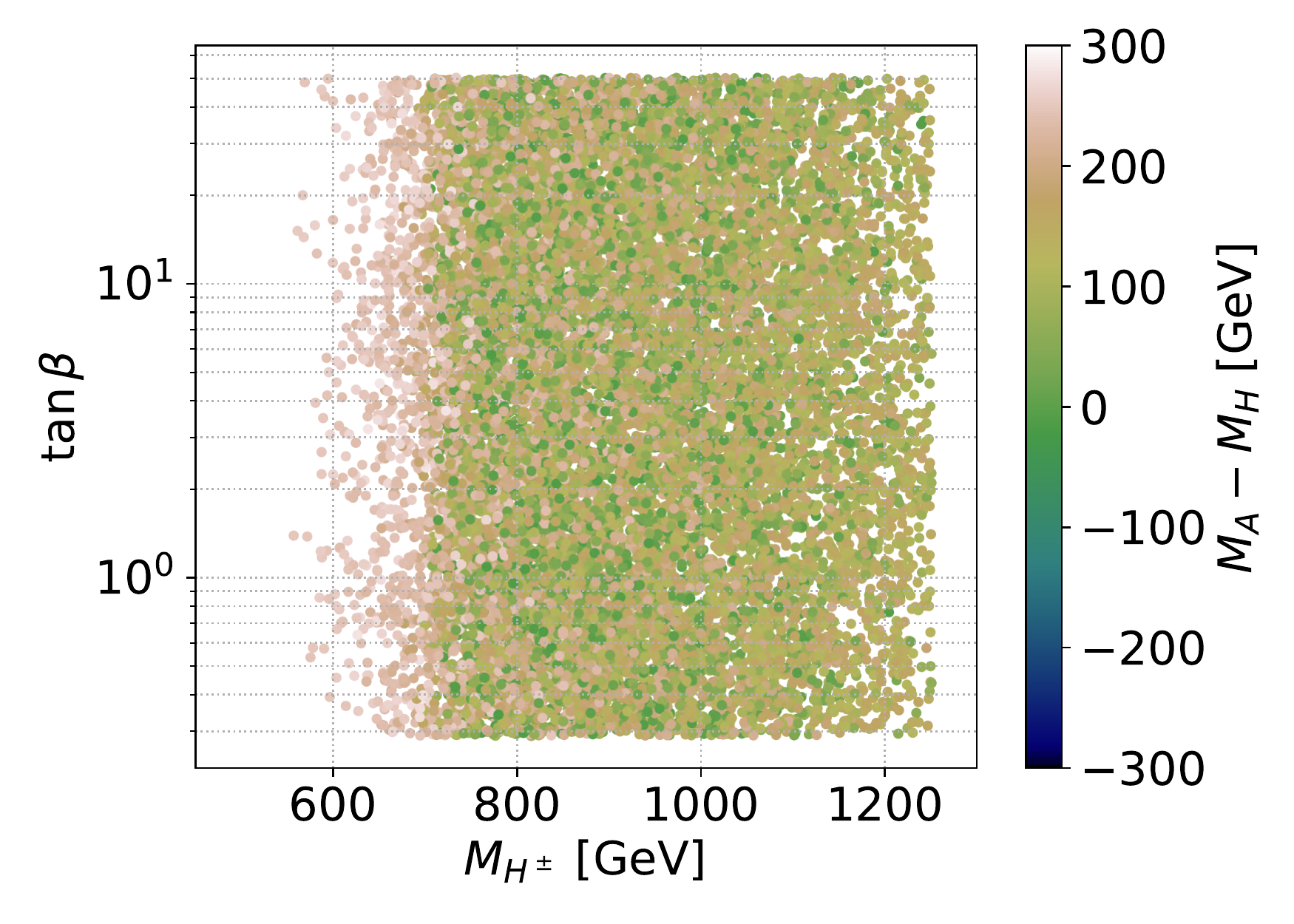}
    \caption{Parameter space with strong FOPTs for pure 2HDM. Left: $\tan\beta$ vs $M_H$. Middle: $\tan\beta$ vs $M_A$. Right: $\tan\beta$ vs $M_H^\pm$. The color code quantifies the mass splitting between two of the three heavy scalar states. Both the left and right panel present a zero mass splitting.}
    \label{fig:m_tB_general-2HDM_typeII}
\end{figure}

In this section, we consider the 2HDM limit, which means $M_a=\sin\theta=\lambda_a=\lambda_{1P}=\lambda_{2P}=0$, to put in evidence the impact of these parameters in the 2HD+a model. We again consider the parameter space~(\ref{eq:param_space}), constrained by perturbative unitarity and the requirement of a potential bounded from below. These constraints can be found in the 2HDM review~\cite{Branco:2011iw}. In Fig.~\ref{fig:masses_general-2HDM_typeII}, we see that the allowed splitting between $M_A$ and $M_H$ is larger than in the case of 2HD+a in Fig.~\ref{fig:masses_general}. Moreover, all the three masses $M_H, M_A$ and $M_{H^\pm}$ can produce strong FOPTs also for higher values, i.e. beyond 1 TeV. While for the case of strong FOPTs in 2HD+a, we find that only the mass splitting $M_{H^\pm} - M_H$ can be zero (Fig.~\ref{fig:m_tB_general}), then in the 2HDM limit, we only find zero splitting in $M_A - M_{H^\pm}$ and $M_A - M_{H^\pm}$, see Fig.~\ref{fig:m_tB_general-2HDM_typeII}.

\begin{figure}[!ht]
    \centering
    \includegraphics[width=0.49\linewidth]{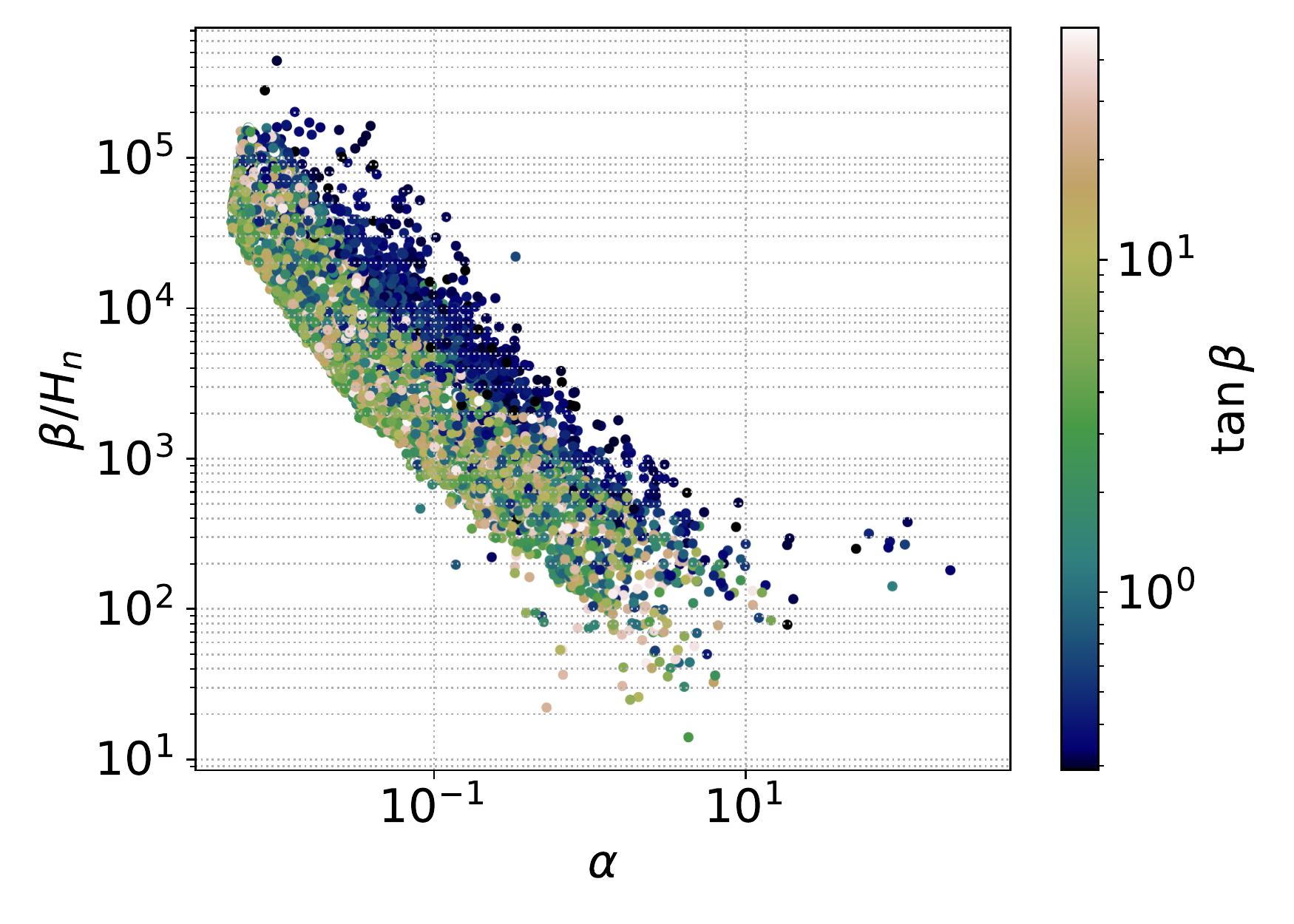}
    \includegraphics[width=0.49\linewidth]{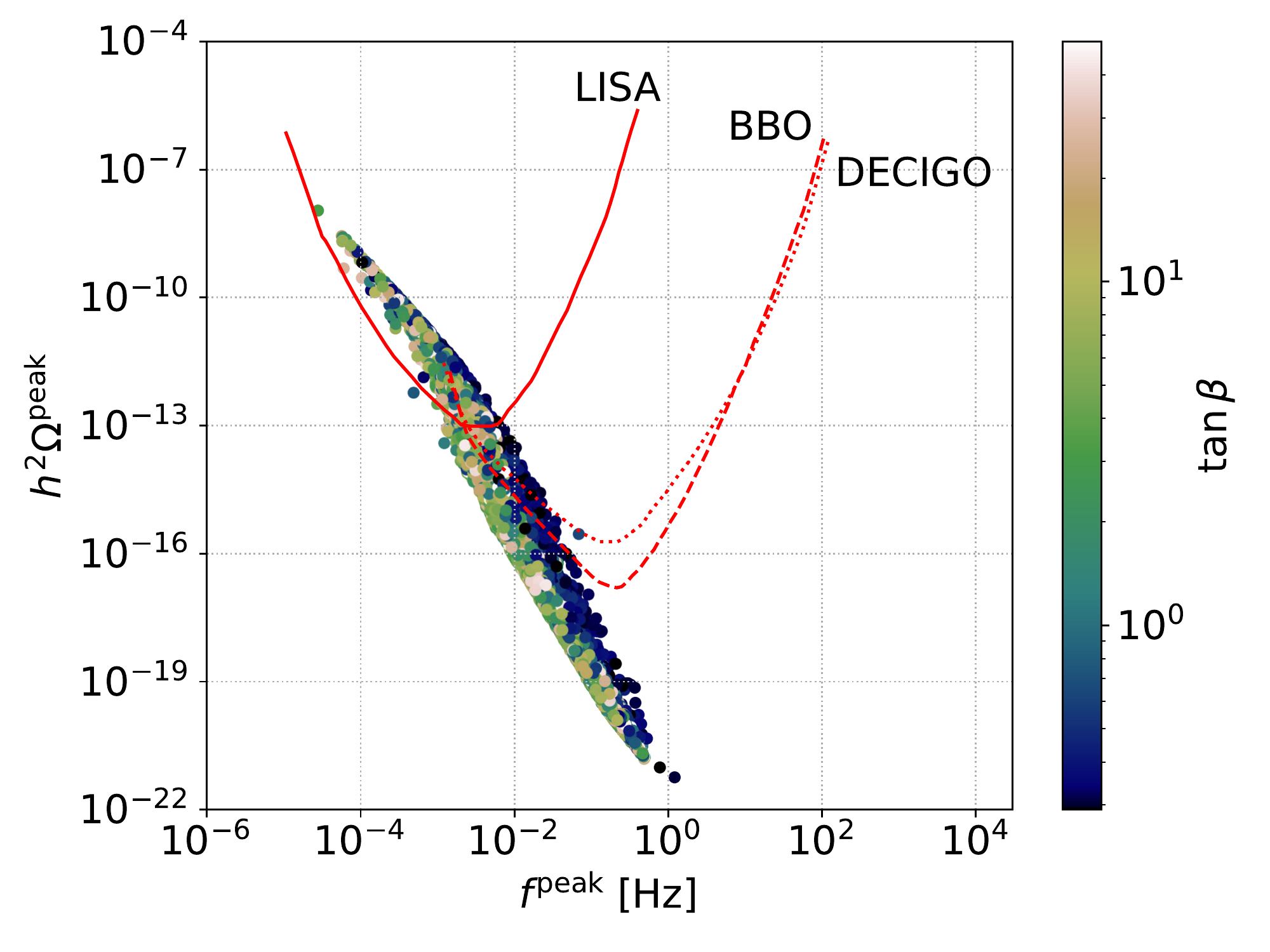}
    \caption{Parameter space with strong FOPTs for pure 2HDM. Left: $\beta/H$ vs $\alpha$. Right: Peak amplitude of the GW signal as a function of peak frequency. Also shown are the sensitivity curves of future detectors LISA, BBO and DECIGO.} \label{fig:beta_vs_alpha_vs_tanBeta_general-2HDM_typeII}
\end{figure}

The left panel of Fig.~\ref{fig:beta_vs_alpha_vs_tanBeta_general-2HDM_typeII} shows that for a fixed  value of $\beta/H_n$, the phase-transition strength $\alpha$ increases when $\tan\beta$ decreases. As for the GW signal, one can see in the right panel of Fig.~\ref{fig:beta_vs_alpha_vs_tanBeta_general-2HDM_typeII} that the population of points is a bit narrower than in the right panel of Fig.~\ref{fig:GW_vs_f_vs_tanBeta_general}. In terms of the range of values for the GW signal and the frequency, the 2HD+a and 2HDM model are quite similar.

All in all, we see that there is a part of the 2HD+a parameter space that can be probed by the future GW detectors such as LISA, BBO or DECIGO, and that considering a singlet pseudoscalar $a$ state in addition to the usual two Higgs doublets of the 2HDM does, indeed, makes a difference.

\clearpage


\section{Conclusions} 

 The extension of the minimal Higgs sector of the SM to include two Higgs doublet fields, leading to four Higgs bosons $A,H$ and $H^\pm$ in addition to the already $h$ state observed, and a relatively light pseudoscalar $a$ boson with significant mixing with the $A$ state of the 2HDM, is interesting in many respects. First, when including a cosmologically stable isosinglet fermion, it allows for a nice resolution of the dark matter problem while not conflicting with present data:  the correct cosmological relic density is essentially obtained through DM annihilation into SM fermions via $s$-channel $a$ boson exchange or into final state involving light $a$ bosons, while elastic DM scattering on nucleon proceeds through loop diagrams which make the cross sections rather small and, hence, DM detection in direct astrophysical experiments more difficult. In addition, it has been recently shown  that the model can resolve two particle physics  ``anomalies": the significant deviations from expectations in the SM of the muon anomalous magnetic moment and the $W$-boson mass reported lately by experiments at Fermilab. Finally, because of its extended particle content, the model has a very rich phenomenology that can be probed in collider experiments, in particular at the LHC.  

Nevertheless, the model is still  subject to severe constraints from both collider and astrophysical data and the first objective of the present paper was to perform a comprehensive study of all the experimental constraints to which it is subject, in addition to the theoretical ones from perturbativity, unitarity and the stability of the electroweak vacuum.  To this end, we have discussed the impact of the high precision measurements of the electroweak observables performed mostly at the LEP and Tevatron colliders, the properties of the already observed SM-like Higgs boson at the LHC, in particular its couplings to fermions and gauge bosons and its invisible decays, and in the flavor sector, with a focus on $B$-meson physics and the muon $g-2$. We have also studied the bounds that one can set on the parameter space of the model from the intensive campaign of searches of the heavy Higgs bosons of the 2HDM and the lighter singlet pseudoscalar $a$ boson that has been performed at the 13 TeV LHC with the full data set, in particular when they are produced as single resonances in gluon (and eventually bottom-quark) fusion and decay into  lepton pairs, $\tau^+\tau^-$ or $\mu^+\mu^-$. We have also studied the impact of the high sensitivity of direct DM detection experiments such as XENON, and very recently LZ, on the mass and couplings of the $a$ boson.  

The combined effect of these constraints on the model turns out to be quite severe. While, indeed, one can explain recent anomalies,  such as the ones affecting the mass $M_W$ and the muon $(g-2)_\mu$, and simultaneously satisfying the DM requirements, with a judicious choice of some key parameters or features (such as the mass splitting between the heavy Higgs bosons in the first case and the mass of the $a$ boson and the value of $\tan\beta$ in the second one), these explanations are made rather difficult in some configurations of the model when other constraints, such as those from Higgs searches at the LHC, are also included. The result strongly depends on the type of configuration which has been chosen for the 2HDM Higgs couplings to fermions in order to avoid flavor changing neutral couplings at tree-level. 

The most studied case, the so-called Type II configuration which also occurs in supersymmetric theories and in which both the $b$-quark and charged-lepton Yukawa couplings are enhanced at high $\tan\beta$ values, is the most constrained one, in particular from searches for single Higgs resonances at the LHC. These searches exclude much of the parameter space that allow for an explanation of the measured value of the $(g-2)_\mu$. The Type X configuration,  in which only the charged-lepton couplings are proportional to $\tan\beta$, is less constrained by these experiments but one needs extremely large values of the latter parameter to comply with the $(g-2)_\mu$ deviation. The other scenarios, namely Type I and Y, have suppressed couplings  to leptons and are thus less severely constrained. All configurations, in turn, allow for an explanation of the recent measurement of $M_W$ performed by the CDF experiment as one simply needs to allow for a sufficient  splitting between the masses of the heavy 2HDM states. Constraints from the signal strengths of the observed light $h$ particle can be easily evaded by enforcing the alignment limit in which the state has SM-like couplings to fermions and gauge bosons, while flavor constraints can be coped with by having a sufficiently heavy charged Higgs and pseudoscalar $a$ bosons.  All these constraints still allow for the additional stable fermionic state to be a good DM candidate, namely to have the correct relic density and to evade the strong bounds from direct detection experiments like XENON and LZ in some areas of the space of the DM and $a$ boson parameters.  
 
In the last part of this work, we have performed a detailed study of the cosmic phase transitions in the 2HD+a model and the  corresponding gravitational wave spectrum which is  generated.  We have calculated the GW signals for phase transitions in the plane of two fields $h_0$ and $H_0$ and have shown that they could be observable by near future experiments such as LISA, BBO and DECIGO. We have also discussed the difference between the 2HD and the 2HD+a models, which  arises from a modified parametrisation of quartic couplings and corrections to thermal masses due to couplings with the additional singlet pseudoscalar $a_0$. The GW signal frequency and amplitude range in the 2HDM and 2HD+a cases are relatively similar, but there is a larger variation in the signal of the latter model. In some cases, it could potentially help distinguish between the two models.

Our calculation accounts for all four 2HDM configurations for the Higgs couplings to fermions. Because the top and bottom Yukawa couplings dominate over others, we do not have to consider Types X and Y separately, since they differ from Types I and II, respectively, only in the lepton Yukawa sector. Moreover, also Types I and II yield a practically identical parameter space of phase transition patterns and GW signals due to the overall small effect of the fermion contribution in the thermal evolution of the effective potential.

In view of its rather rich phenomenology and the fact that it addresses various important issues and anomalies in high-energy physics and in cosmology, the 2HD+a model is an interesting candidate for physics beyond the SM and can serve as a benchmark in the various searches for it at present day and future collider and astroparticle physics experiments. It can be further tested at the present and high luminosity runs of the LHC \cite{Cepeda:2019klc}, the  DM direct detection experiments like XENONnT \cite{XENON:2020kmp} and the new and ultimate one DARWIN \cite{DARWIN:2016hyl}, as well as in future high-precision measurements such as the $W$-mass and muon $(g-2)$. In addition, it is capable of generating  gravitational wave signals which can be tested in planned experiments such as LISA  \cite{Caprini:2015zlo}, 
BBO \cite{Corbin:2005ny} and  DECIGO \cite{Kawamura:2020pcg}. 

In view of all these features, the 2HD+a model deserves further attention and studies. 

\subsubsection*{Acknowledgements}

We would like to thank Joosep Pata for helping us with cluster computation. This work is supported by the Estonian Research Council grants PRG356 and PRG434, by the European Regional Development Fund and programme Mobilitas Pluss grants MOBTT5 and MOBTT86, and the ERDF CoE program project TK133.  AD is in addition supported by the Junta de Andalucia through the Talentia Senior program and the grants PID2021-128396NB-I00, A-FQM-211-UGR18 and P18-FR-4314 with ERDF.

\clearpage 

\setlength{\parskip}{0.1cm}
\bibliographystyle{unsrt}
\bibliography{biblio}

\end{document}